\documentclass[usenatbib]{mn2e} 
\usepackage{lscape}
\usepackage{rotating}
\usepackage{psfig}
\usepackage{amsmath}
\usepackage{amssymb}
\title[YSPs in radio galaxies.]{The properties of the young stellar
  populations in powerful radio galaxies at low and intermediate redshifts}
\author[J. Holt et al.]{J. Holt$^{1}$\thanks{E-mail:
j.holt@sheffield.ac.uk}, C. N. Tadhunter$^{1}$, R. M. Gonz{\'{a}}lez
  Delgado$^{2}$, K. J. Inskip$^{1}$, J. Rodriguez$^{1}$,
\newauthor B. H. C. Emonts$^{3}$,
R. Morganti$^{4}$, K. A. Wills$^{1}$, \\
$^{1}$Department of Physics and Astronomy, University of Sheffield,
Sheffield,  S3 7RH, UK.\\
$^{2}$Instituto de Astrofisica de Andalucia (CISC), Apdto. 3004, 18080
Granada, Spain.\\
$^{3}$Kapteyn Astronomical Institute, University of Groningen,
P. O. Box 800, 9700 AV Groningen, The Netherlands.\\
$^{4}$Netherlands Foundation for Research in Astronomy, Postbus 2,
7990 AA Dwingeloo, The Netherlands\\
}

\newcommand{\kms}{km s$^{-1}$}
\newcommand{\ha}{H$\alpha$}
\newcommand{\hb}{H$\beta$}

\newcommand{\ebv}{E(B-V)}
\newcommand{\lala}{$\lambda\lambda$}
\newcommand{\dipso}{erg s$^{-1}$ cm$^{-2}$ \AA$^{-1}$}
\newcommand{\chisq}{$\chi_{\rm red}^{2}$}

\begin{document}
\maketitle
\begin{abstract}
We present high-quality, wide spectral coverage long-slit optical
spectra for 12 powerful radio sources  
at low and intermediate redshifts ($z < 0.7$) that show evidence for
a substantial 
UV excess. These data were taken using the WHT and VLT
telescopes with the aim
of determining the detailed properties of the young stellar
populations (YSPs) in the host galaxies as part of
a larger project  to investigate evolutionary scenarios for the AGN
host galaxies.  
The results of our spectral synthesis model fits to the
spectra highlight the importance of taking into account AGN-related
components (emission lines, nebular continuum, scattered light) 
and reddening of the stellar
populations in studies of this
type. It is also clear that careful examination of the fits to the
spectra, as well consideration  
of auxilary polarimetric and imaging data, are required to avoid
degeneracies in the model solutions.  
In 3 out of the 12 sources in our sample we find evidence for
broad permitted line components, and a combination of AGN-related continuum 
components and an old (12.5~Gyr) stellar population provides an
adequate fit to the data. However, 
for the remaining 9 sources we find strong evidence for YSPs.
In contrast to some recent studies that suggest relatively old
post-starburst ages for the YSPs 
in radio galaxies (0.3 -- 2.5~Gyr), we deduce 
a wide range of ages for the YSPs in our sample objects (0.02 --
1.5~Gyr), with $\sim$50\% of the sample 
showing evidence for young YSP ages ($\lesssim$0.1~Gyr) in their
nuclear regions. The nuclear 
YSPs are often significantly  reddened ($0.2 < E(B-V) < 1.4$) 
and make up a substantial fraction ($\sim$1 -- 35\%) of the total
stellar mass in the regions  
sampled by the spectroscopic slits. Moreover, in all the cases in
which we have sufficient  
spatial resolution we find that
the UV excess is extended across the full measureable extent of
the galaxy
(typically  5 -- 30~kpc), suggesting galaxy-wide starbursts. 
The implications for photometric and spectroscopic studies of active
galaxies are discussed. 
\end{abstract}

\begin{keywords}
galaxies: active, galaxies: radio galaxies, galaxies: starburst,
galaxies: stellar content
\end{keywords}

\section{Introduction}
\label{sect:intro}
Major merger events are often invoked to explain the origin of nuclear
activity in galaxies.  Indeed, there is much observational evidence to 
suggest that, in a significant fraction of the local radio source
population,  
the activity has been triggered by merger events involving two
or more galaxies where at least one is gas rich. Despite the fact that
the overwhelming majority of radio galaxies are classified as elliptical or S0
types, deep optical imaging
studies  
reveal 
double nuclei, tidal tails, arcs of emission and distorted 
isophotes characteristic of mergers in  $\sim$50\% of nearby radio
galaxies with strong emission lines \citep{heckman86,smith89b}. In addition, 
kinematical studies of the emission line gas 
(e.g. \citealt{tadhunter89,baum92}) further support the idea that the
early-type hosts are interacting and/or merging with companion
galaxies. However, there is a limit to what can be deduced from the imaging 
and kinematical studies alone, especially in terms of gauging the type
of merger (major or minor? dry or gas-rich?), the order of events in the 
triggering of the activity, and the relationship between radio galaxies and
other types of merging systems such as ultraluminous infrared galaxies (ULIRG).

Given that gas-rich mergers can also trigger starbursts, studies of
the stellar  
populations in the early-type host galaxies of radio sources  have
the potential to provide key 
information about the natures and timescales of the triggering 
events. A major challenge  
in such studies is to identify the young stellar populations in the
presence of activity-related 
components such as nebular continuum, direct or scattered AGN light
and emission lines, all of which are likely to be 
significant in galaxies hosting powerful, quasar-like AGN
\citep{tadhunter02}. Therefore, although the broad-band
colours of many radio galaxies show UV excesses compared to normal,
passively evolving elliptical galaxies (\citealt{lilly84,smith89b}),
such excesses may be 
related more to the AGN activity than any 
merger induced star formation. Nonetheless, by taking particular care
to account for all the 
activity related components, it has proved possible to use spectral synthesis
modelling techniques to reveal the presence of young stellar
populations (YSPs) in radio galaxies (e.g. \citealt{tadhunter96}). In
this way \citet{tadhunter02} 
found evidence for YSPs 
in 30 -- 50\% of a complete sample of southern 2Jy radio galaxies at
intermediate redshifts,  \citet{wills02}  deduced the presence of YSPs
in 30\% of their 
complete sample of nearby 3C radio galaxies  
with FRII radio morphologies, and \citet{wills04} detected YSPs in
25\% of a small sample of lower radio 
power FRI radio sources from the 2Jy sample. More recently,
\citet{raimann05}
 found a higher rate of  
incidence of YSPs in a sample of 20 nearby radio galaxies
($\sim$100\%). However, 
the latter authors did not explicitly account for 
all  the AGN-related continuum and emission line components, and
assumed a priori that the optical spectra 
of their radio galaxies could be modelled in terms of 5 stellar
components of varying age. 

Although all the previous studies revealed the presence of young
stellar populations in a significant 
fraction of radio galaxy hosts, the data were not always of
sufficiently high quality in terms of 
spectral range, spectral resolution and S/N to allow the detailed
properties of the YSPs 
to be deduced, and thereby learn about the nature of the triggering
merger events. We have therefore 
 begun a major programme to obtain the requisite data and
investigate the stellar populations 
in all nearby radio galaxies already suspected
of having YSPs on the basis of previous work, 
aiming to analyse the data for all the objects in a
uniform way.  
Some early results from this programme for 5 nearby radio galaxies
were presented in  \citet{tadhunter05,emonts06} and
\citet{holt06}. The latter papers also give a full account of  
the background to the project.
In this paper we present the
results for a further 12 radio galaxies  at low and intermediate redshifts 
($z < 0.7$)\footnote{$H_0 =$ 75 \kms~Mpc$^{-1}$, $q_0 =$ 0.0 assumed
 throughout.}, extending detailed studies of the YSPs in radio
 galaxies up to higher redshifts than the previously published work. The 
implications of the
results of the programme as a whole for our understanding
of the evolution of the host galaxies, triggering of the activity and
the heating 
of the far-IR emitting dust in AGN are discussed in separate papers
(\citealt{tadhunter07}, Tadhunter et al. in prep).

\begin{table*}
\begin{center}
 \begin{minipage}{14.1cm}
   \caption[Properties of the sample.]
{Properties of the sample. $(a)$ radio source; $(b)$ redshift;
$(c)$  Galactic reddening from Schlegel et al. (1998) (\ebv); $(d)$
radio luminosity at 5GHz calculated from the flux densities in Wall \&
Peacock (1985)  
except 3C 285 (from Kellerman et al. 1969); $(e)$
  radio spectral index for  $\alpha_{\rm 2.7 GHz}^{\rm 5
    GHz}$ where 
  F$_{\nu} \propto \nu^{-\alpha}$, 
$(f)$ radio morphology  and $(g)$ optical
  spectral classification (see text for references).
\newline {\bf redshifts}: $\star$ Holt (2005);  $\dagger$ van Breugel
\& Dey
(1993);
$\diamond$ \protect\citet{degrijp92}; $\circ$ \protect\citet{hill96};
$\triangleright$ \protect\citet{smith04};
$\ast$ \protect\citet{tadhunter93};
$\triangleleft$ \protect\citet{devoucouleurs91}.  References to the
other data can
be found in the text.
}
\label{table:sample}
{\footnotesize
\begin{tabular}{lrccc ll} \\ \hline\hline
 \\
 & $z$ & \multicolumn{1}{c}{Galactic} & \multicolumn{1}{c}{Radio} &
\multicolumn{1}{c}{$\alpha$} & \multicolumn{1}{c}{Radio} &
\multicolumn{1}{c}{Spectral}\\
\multicolumn{1}{c}{} & 
\multicolumn{1}{c}{} &
\multicolumn{1}{c}{reddening} & \multicolumn{1}{c}{luminosity (5 GHz)} &
& \multicolumn{1}{c}{morphology} & 
 \multicolumn{1}{c}{class} \\
\multicolumn{1}{c}{} & 
\multicolumn{1}{c}{} &
\multicolumn{1}{c}{(E(B-V))} & \multicolumn{1}{c}{log P (W Hz$^{-1}$)} &
  & \\
\multicolumn{1}{c}{$(a)$}&
$(b)$&$(c)$&$(d)$&$(e)$&\multicolumn{1}{c}{$(f)$}
 &\multicolumn{1}{c}{$(g)$}
 \\
\hline 
\\
3C 218 & 0.05488$\triangleright$& 0.042& 25.91&0.9 & FRI/FRII,
symmetric & NLRG \\
3C 236 & 0.101$\circ$& 0.011&25.44 & 0.7& FRII, double-double &
 NLRG \\
3C 285 & 0.079$\dagger$ & 0.017 & 25.00& 1.27 & FRII &  NLRG \\
3C 321 & 0.096$\diamond$ & 0.044 &25.33 &1.14 & FRII & NLRG \\
3C 381 & 0.161$\ddagger$ & 0.053 & 25.87&0.96  & FRII &  BLRG \\
3C 433 & 0.106$\ddagger$  & 0.044 & 25.94 &1.07& FRI/FRII, x-shaped &
 NLRG \\
\\
PKS 0023-26 & 0.322$\star$ & 0.014 &27.33 & 0.7& CSS & NLRG \\
PKS 0039-44 & 0.346$\ast$ & 0.007 &26.57 &0.93 & FRII & NLRG \\
PKS 0409-75 & 0.693$\ast$ & 0.078 &27.82 & 0.86& FRII &NLRG \\
PKS 1932-46 & 0.231$\ast$ & 0.054 &26.65 &1.03 & FRII & BLRG \\
PKS 2135-209 & 0.636$\star$ & 0.033 & 27.47&0.82 & CSS & BLRG \\
NGC 612 & 0.030$\triangleleft$ & 0.020 &24.85 &0.51 & FRI/FRII,
`hybrid' & NLRG \\
 \\\hline
\end{tabular}
}
\end{minipage}
\end{center}
\end{table*}

\section{The sample}
\label{sect:sample}
The sample considered in this paper
comprises 12 powerful low and intermediate redshift ($z <$ 0.7) radio
galaxies. All have been selected on the basis of having a strong UV excess
that is difficult to explain solely
in terms of AGN-related components (see the discussion in Tadhunter et al.
2002). In essence, all sources in this
sample are suspected of having a large contribution of 
continuum light from a young stellar population (YSP). Therefore they are suitable
for detailed studies of the young stellar populations. 

The sample spans both the northern and southern skies. The
northern objects originate from the 3C catalogue whilst 
the southern objects derive 
from the well-studied 2Jy sample of radio galaxies
(e.g. \citealt{tadhunter93,morganti93,tadhunter02}). Table
{\ref{table:sample}} lists the properties of the sample.

Although many of the objects in our sample have been previously modelled 
\citep{tadhunter96,aretxaga01,robinson01,wills02,wills04,raimann05}, 
for most we have new data with improved
spectral resolution, spectral coverage and S/N. We also fit these data
with the newer,  
higher resolution spectral synthesis models of \citet{bruzual03},
taking full account of reddening 
of the YSPs and potential AGN contributions to the optical/UV continuum.

\section{Observations and data reduction}
The
northern sample was observed using the ISIS dual-beam spectrograph
on the William Herschel Telescope (WHT) on La Palma in
several runs between 1996 and 2005.
The southern
sample was observed during a single run  in 2003 using
the FORS2 spectrograph on the ESO Very Large Telescope (VLT) on
Cerro Paranal, Chile.  Table \ref{tab:obs} summarises the observations
and the  various instrumental setups. 
Note that, in the case of the highest redshift source in our sample (also the faintest) ---
PKS 0409-75 --- we nodded the source along the slit in order to improve
the sky subtraction at the red end of the spectrum.
Although for 3C 321 we use the same data set as that used in
\citet{robinson00}, for this paper, we have improved the flux
calibration and re-modelled the SEDs using the latest models, for the
first time taking into account reddening of the YSP.

This project focusses on accurate modelling of the (often faint)
continuum emission. Hence, all observations were made with a 1.3-1.5 arcsec
slit in order to include as much of the light from the galaxy as possible, whilst 
retaining sufficient wavelength resolution to allow detection and modelling of
stellar absorption features. 
To reduce the effects of differential refraction, all
exposures were taken at low airmass (sec $z$ $<$ 1.1), with the
slit aligned along the parallactic angle, and/or (in the case of the VLT observations)
using a linear atmospheric dispersion compensator (LADC) that was reset
between exposures. The slit PAs  are
listed in Table {\ref{tab:obs}} (for reference, the radio axis PAs are
presented in Table {\ref{table:sample}}).

\begin{center}
\begin{table*}
\begin{minipage}{170mm}
\caption[Observations summary]{Summary of the observations. Note, the
  wavelength ranges given are the {\it useful} wavelength ranges of
  the spectra. Further, where the end/start wavelengths of the blue/red
  frames  are equal, these are where the `trim' was made for the fitting --
  in the majority of sources there was considerable overlap between
  the blue and red frames, of order a few hundred angstroms.}
\label{tab:obs}
\begin{center}
\begin{tabular}{llllcccl}\\  \hline\hline
\multicolumn{1}{c}{Date}&
\multicolumn{1}{c}{Object}&\multicolumn{1}{c}{Setup} &
\multicolumn{1}{c}{Exposure}&  \multicolumn{2}{c}{Slit} &
\multicolumn{1}{c}{Rest $\lambda$}
 & \multicolumn{1}{c}{Seeing} \\
 &  &\multicolumn{1}{c}{(arm/CCD/grating/filter)}  &
 &  \multicolumn{1}{c}{PA} &  \multicolumn{1}{c}{width} &
 \multicolumn{1}{c}{range}\\
 &  &  &
\multicolumn{1}{c}{(s)} & 
\multicolumn{1}{c}{($^{\circ}$)}    &
\multicolumn{1}{c}{(arcsec)} &
\multicolumn{1}{c}{(\AA)}&\multicolumn{1}{c}{(arcsec)} \\ 
   \hline
\multicolumn{8}{l}{\bf WHT/ISIS}\\
13/01/2004 & 3C 218 & B/EEV12/R300B/-- &3*1200 & 10&1.27 &3000-5950&$\sim$1\\
 & & R/MARCONI2/R316R/GG495 & 3*1200&10 &1.27 & 5950-8000&$\sim$1\\
12/05/2001 & 3C 236 & B/EEV12/R300B/-- & 2*1200 & 50& 1.3 &3040-5640&0.7\\
           &        & R/TEK4/R316R/GG495& 2*1200 &50 & 1.3  &5640-6880&0.7\\
06/06/2005 & 3C 285 & B/EEV12/R300B/-- & 3*1200 & 209  &1.23
 &3100-4848& 0.6 \\
 & & R/MARCONI2/R316R/-- & 3*1200 & 209 &  1.23 &4848-6994& 0.6\\
22/07/1996 & 3C 321 & B/LORAL1/R158B/-- & 2*1500 & 130&1.37 &3309-5700&$<$1.0\\
           &        & R/TEK5/R158R/GG495 & 2*1500 & 130 &1.37  &5883-8100& $<$1.0\\
22/07/1996 & 3C 381 & B/LORAL1/R158B/-- & 2*900 & 20& 1.34   &3123-5560&0.8-1.0\\
           &        & R/TEK5/R158R/GG495 & 1*900,1*1300 &20 &1.34
           &5560-8464&0.8-1.0\\  
31/07/2005 & 3C 433 & B/EEV12/R300B/-- &3*1200 & 32 &1.54&2800-4900 &1.1-1.7 \\
 & & R/MARCONI2/R316R/GG495 & 3*1200& 32& 1.54 &4900-7433& 1.1-1.7\\
\\
\\
\multicolumn{8}{l}{\bf VLT/FORS2}\\
24/09/2003 & PKS 0023-26 & B/MIT/G600B/-- & 3*1000 &145 &1.3 & 2850-3800&
0.7-0.8  \\
           &             & R/MIT/G600RI/GG435+81 & 3*1000 &145&1.3 &3800-6100& 0.7-0.8  \\ 
25/09/2003 & PKS 0039-44 & B/MIT/G600B/-- & 3*1000 &72 & 1.3& 2900-4200& 0.7-1.0
\\
           &             & R/MIT/G600RI/GG435+81 & 3*1000 & 72&
           1.3 &4200-6132& 0.7-1.0 \\
24/09/2003 & PKS 0409-75 & R/MIT/G600RI/GG435+81 & 12*600 &-30 & 1.3 &2921-4875& 1.0-1.2\\
           &             & R/MIT/G600Z/OG590+32 & 6*600 &-30 &
           1.3 &4875-5800&1.0-1.2 \\ 
24/09/2003 & PKS 1932-46 & B/MIT/G600B/-- & 3*900 & 9 & 1.3&3000-4900 & 0.7-1.0
\\
           &             & R/MIT/G600RI/GG435+81 & 3*900 & 9 & 1.3 &4900-6700& 0.7-1.0
\\
25/09/2003 & PKS 2135-20 & B/MIT/G600B/-- & 3*900 & 52 &1.3 &2250-3600& 1.3-1.6
\\
           &             & R/MIT/G600RI/GG435+81 & 3*1200 & 52&1.3& 3600-5000& 1.3-1.6
\\
24/09/2003 & NGC 612 & B/MIT/G600B/-- & 3*900 & -13& 1.3 &3500-5855& 0.6-0.9 \\
\hline
\end{tabular}
\end{center}
\end{minipage}
\end{table*}
\end{center}

\subsection{Data reduction}
\label{sect:data-reduction}
The data were reduced in the usual way (bias subtraction, flat
fielding, cosmic ray removal, wavelength calibration, flux
calibration) using the standard packages in {\sc iraf}. The
two-dimensional spectra were also corrected for spatial distortions of 
the CCD. The final wavelength calibration
accuracies, calculated using the standard error on the mean deviation
of the night sky emission line wavelengths from published values
\citep{osterbrock96}, ranged between 0.05-0.5\AA. The spectral resolution, 
calculated using the widths of the night sky emission lines, ranged
between 4-10\AA. Both are dependent on the instrumental setup and the
wavelength range observed.

Comparison of several spectrophotometric standard stars observed with a
wide slit (5 arcsec) throughout each run gave relative flux
calibrations accurate to $\pm$5 per cent. This accuracy was confirmed
by good matching in the flux between the red and blue spectra.  For the  VLT
observations, and for  3C 236, further 
observations of standard stars close in position and time to each
target observation  with a narrow slit, matched to the slit
width used to observe the objects, were used to correct for atmosperic
absorption features (e.g. A and B bands at $\sim$7600 and $\sim$6800
\AA~respectively). The spectra of all
sources were corrected for Galactic extinction using the \ebv~values
from \citet{schlegel98} and the \citet{seaton79} extinction law.

The spectra were extracted and analysed using the {\sc starlink} packages
{\sc figaro} and {\sc dipso}, and {\sc confit}, a customised {\sc idl}
minimum $\chi^2$ fitting programme \citep{robinson01,tadhunter05}.

\section{Results}
The continuum spectra for all sources were analysed using three main
steps: first, the degree of the UV excess and its spatial distribution
were assessed using the 
4000\AA~break; second, the overall spectral energy distribution (SED)
shape was modelled using a combination of nebular, stellar and
power-law continua (which could represent an AGN-like or very young stellar
population continuum); and
finally, making detailed comparisons 
between the better fitting SED models and the data using stellar
absorption lines (e.g. Balmer lines, CaII~K). The individual steps
are described in detail below.

\subsection{The 4000\AA~break}
\begin{figure*}
\begin{minipage}{180mm}
\begin{tabular}{ccc}
\hspace*{-1.3cm}{\bf 3C 218} & \hspace*{-0.8cm}{\bf 3C 236} & \hspace*{-0.8cm}{\bf 3C 285}\\
\hspace*{-1.3cm}\psfig{file=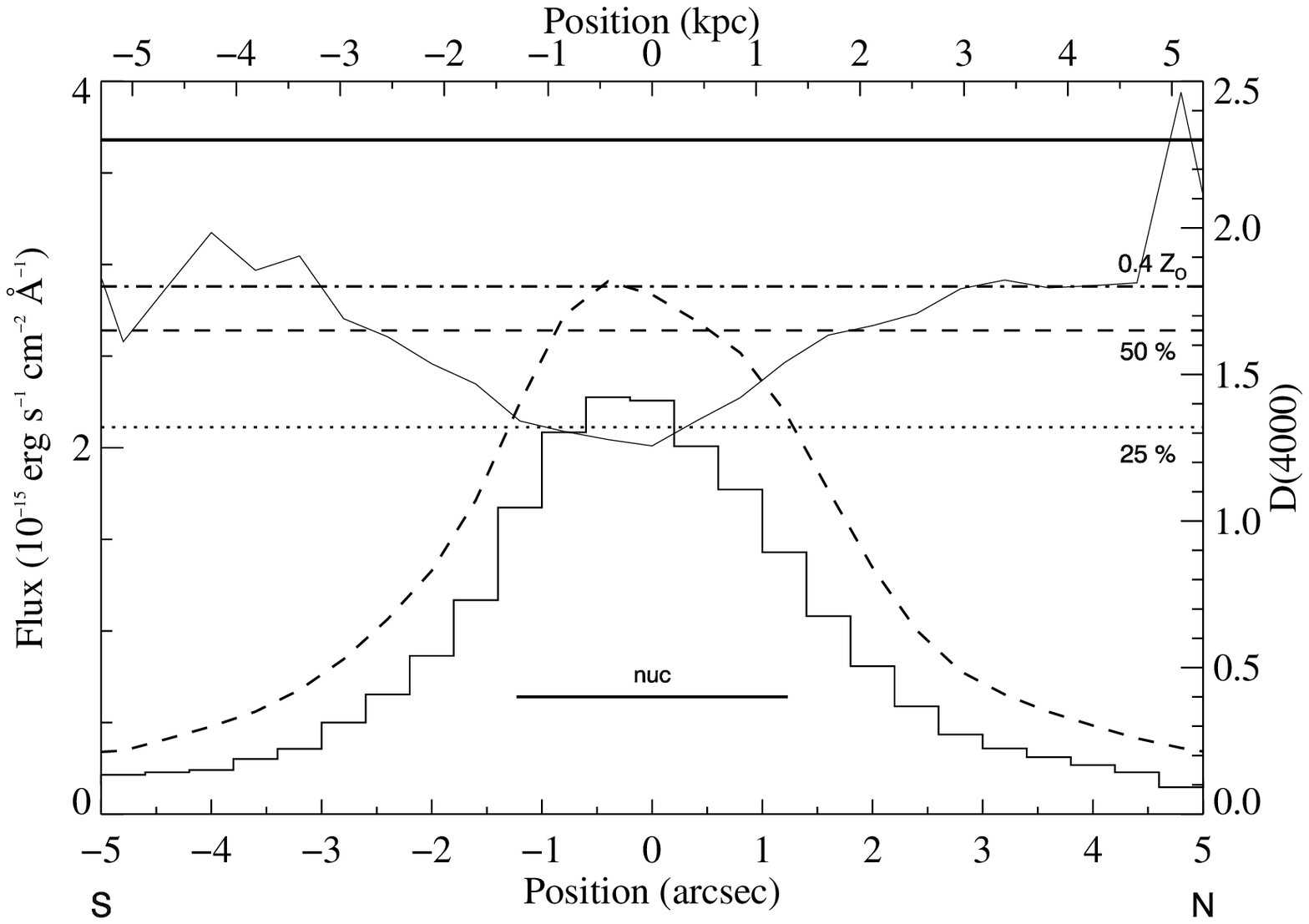,width=7cm,angle=0.} & 
\hspace*{-0.8cm}\psfig{file=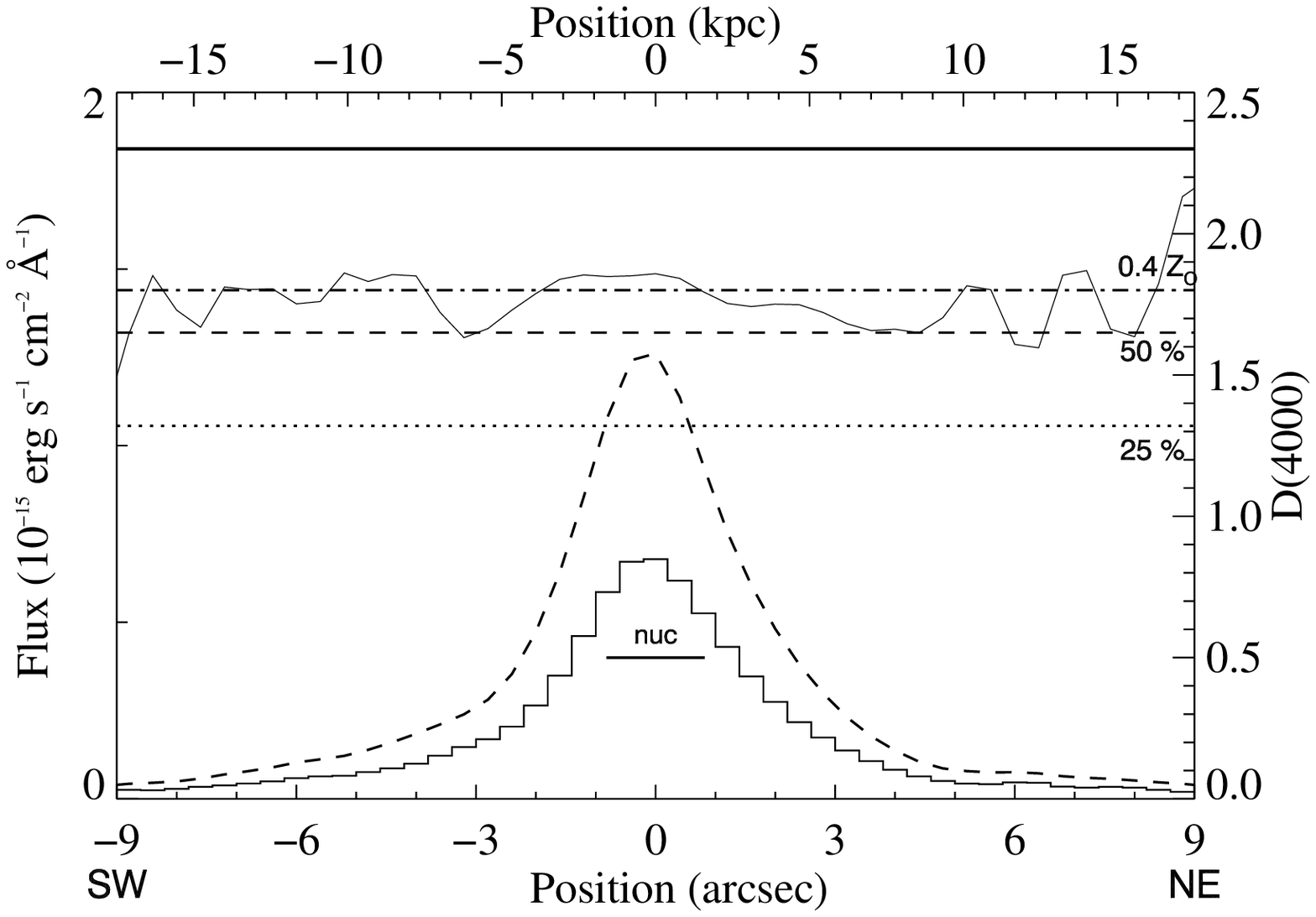,width=7cm,angle=0.} &
\hspace*{-0.8cm}\psfig{file=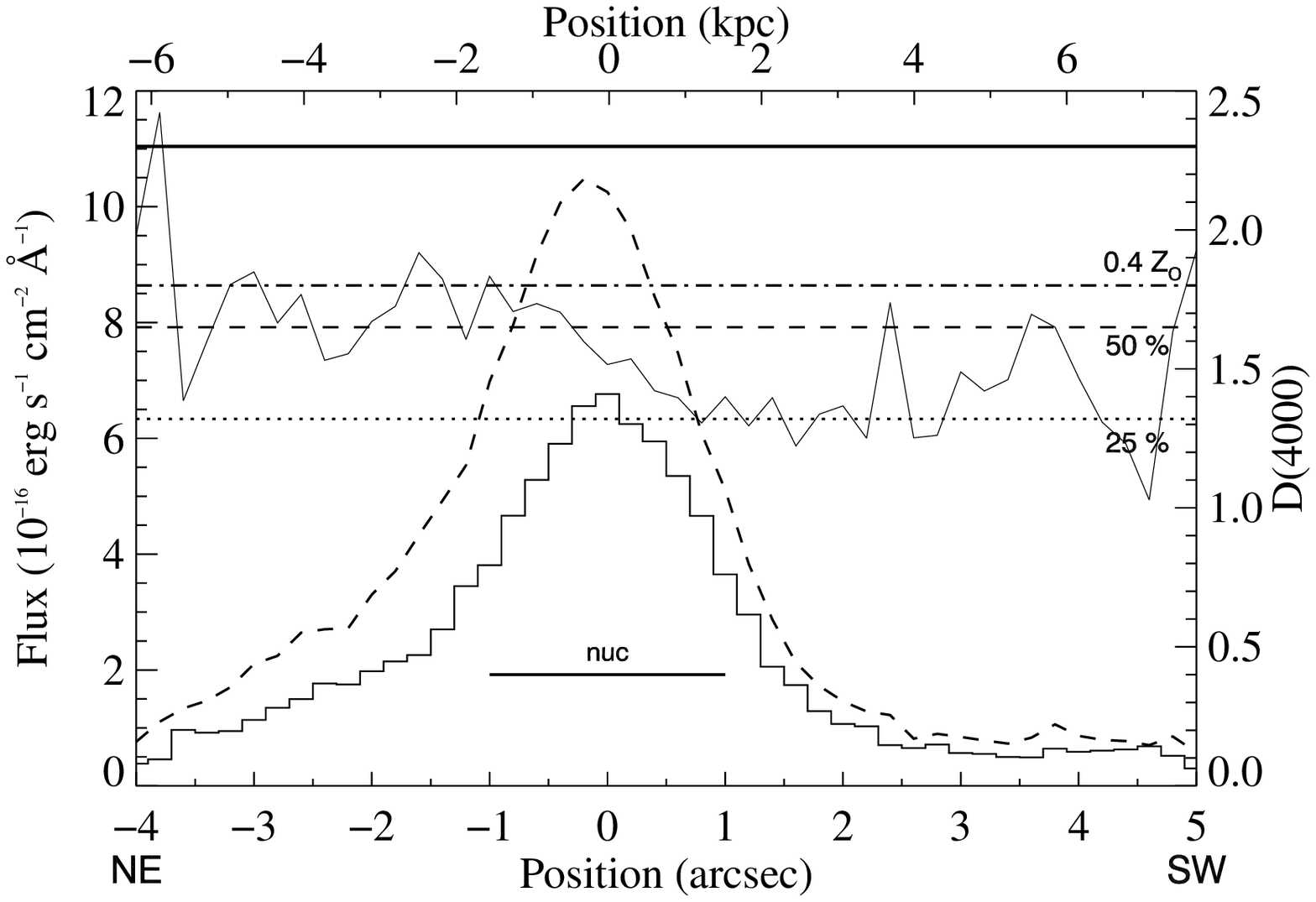,width=7cm,angle=0.} \\
\\
\hspace*{-1.3cm}{\bf 3C 321} & \hspace*{-0.8cm}{\bf 3C 381} &
\hspace*{-0.8cm}{\bf 3C 433}\\
\hspace*{-1.3cm}\psfig{file=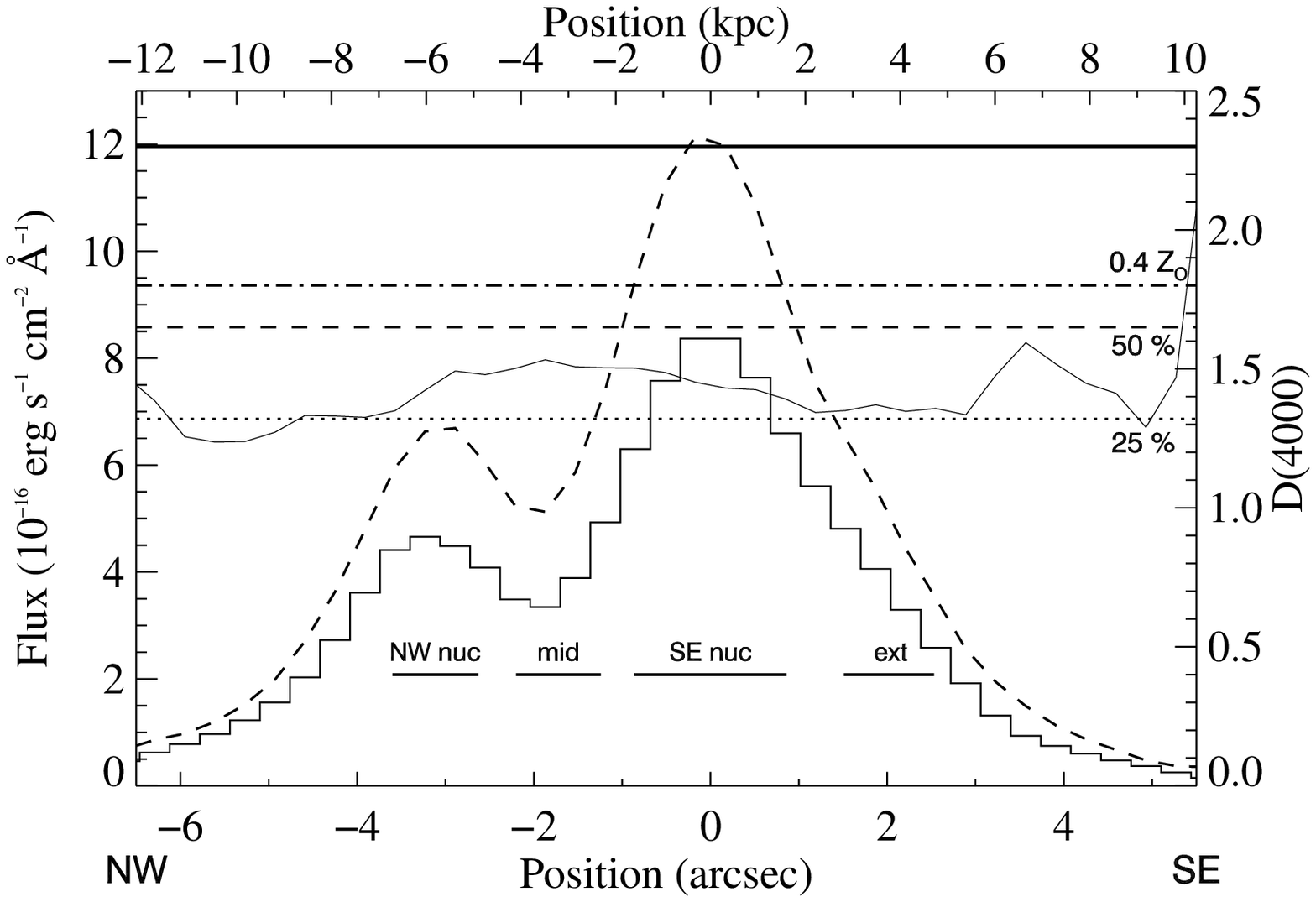,width=7cm,angle=0.} & 
\hspace*{-0.8cm}\psfig{file=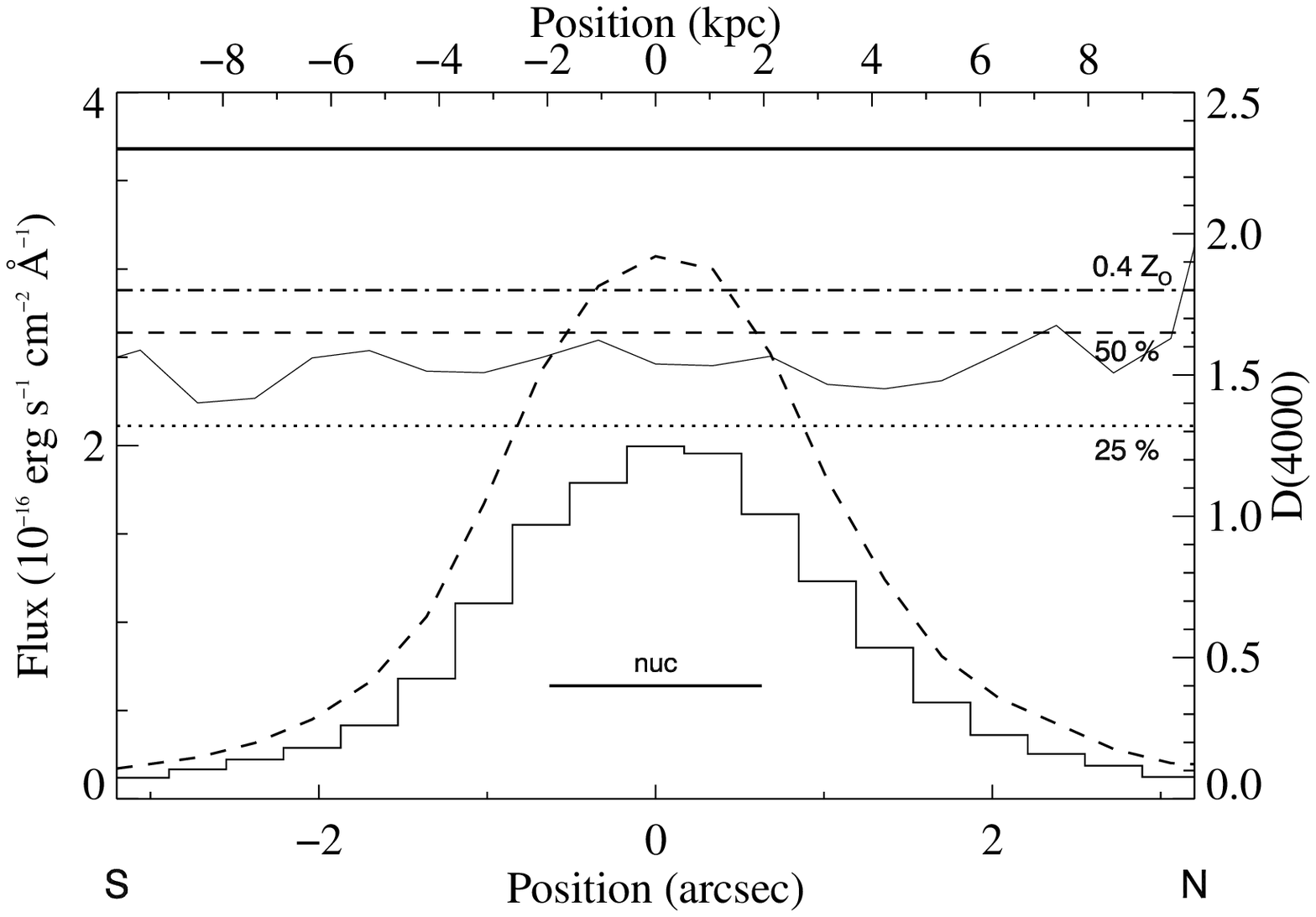,width=7cm,angle=0.} &
\hspace*{-0.8cm}\psfig{file=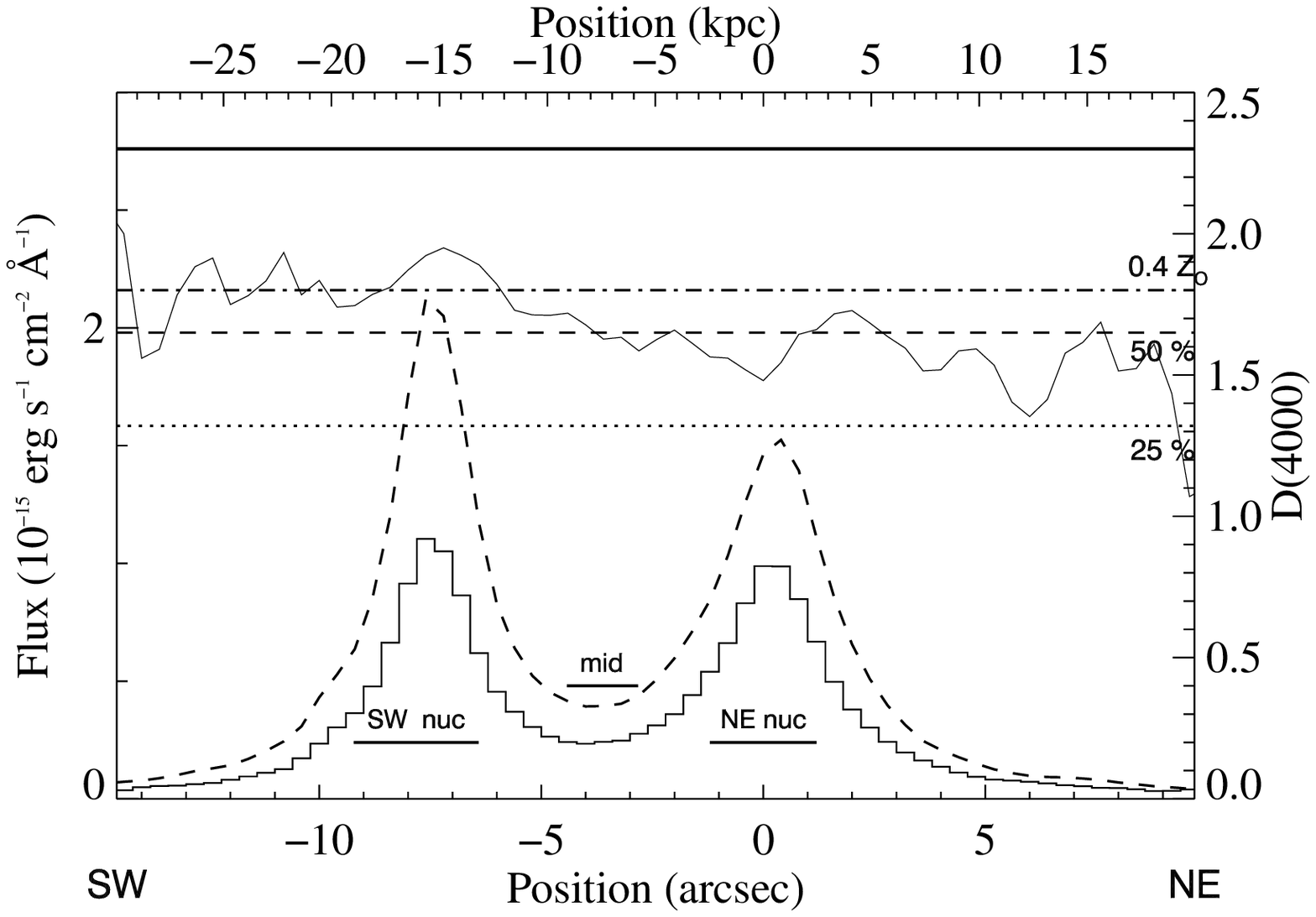,width=7cm,angle=0.}\\
\\
\hspace*{-1.3cm}{\bf PKS 0023-26} &   \hspace*{-0.8cm}{\bf PKS 0039-44}&
\hspace*{-0.8cm}{\bf PKS 0409-75} \\ 
\hspace*{-1.3cm}\psfig{file=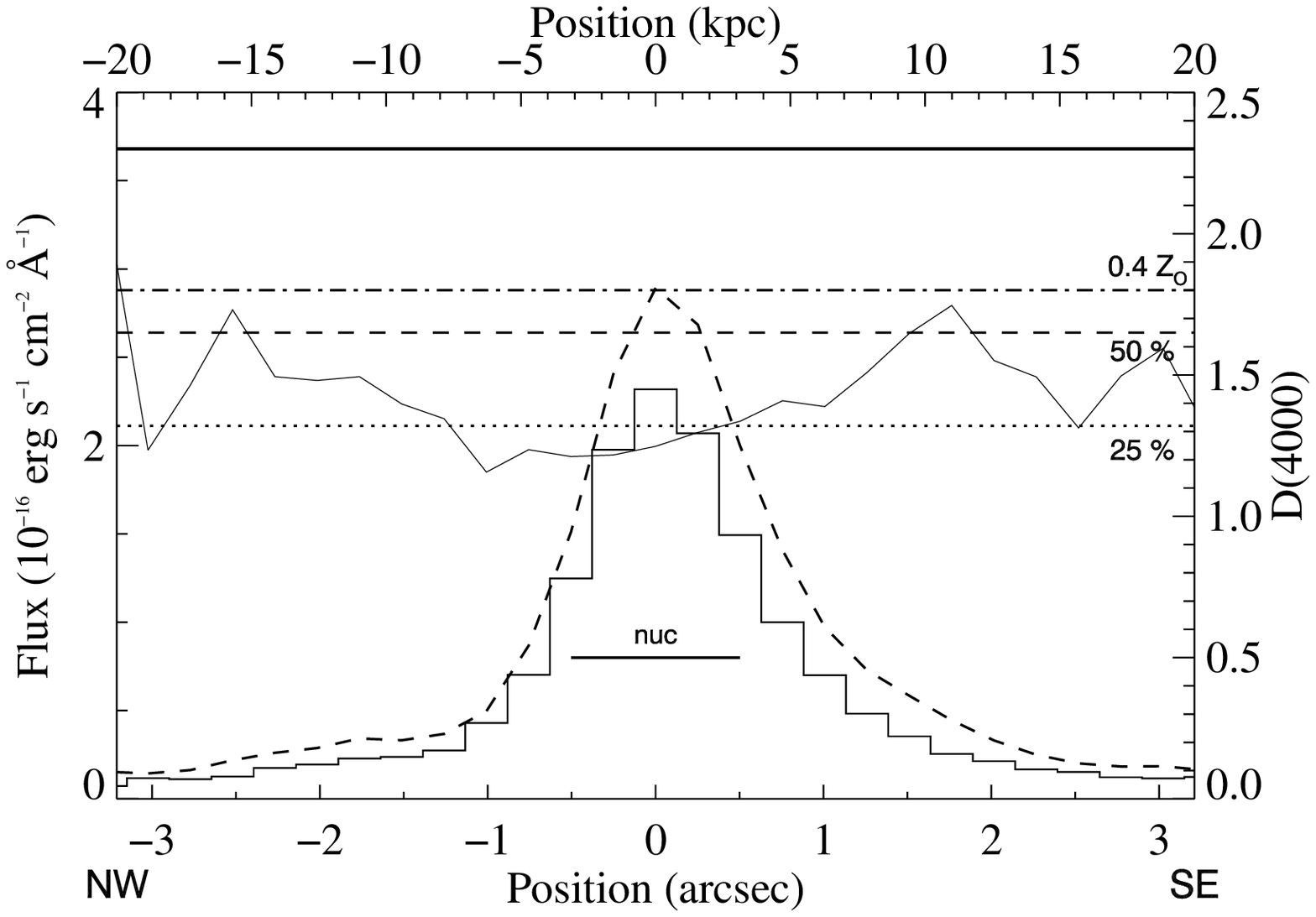,width=7cm,angle=0.} &
\hspace*{-0.8cm}\psfig{file=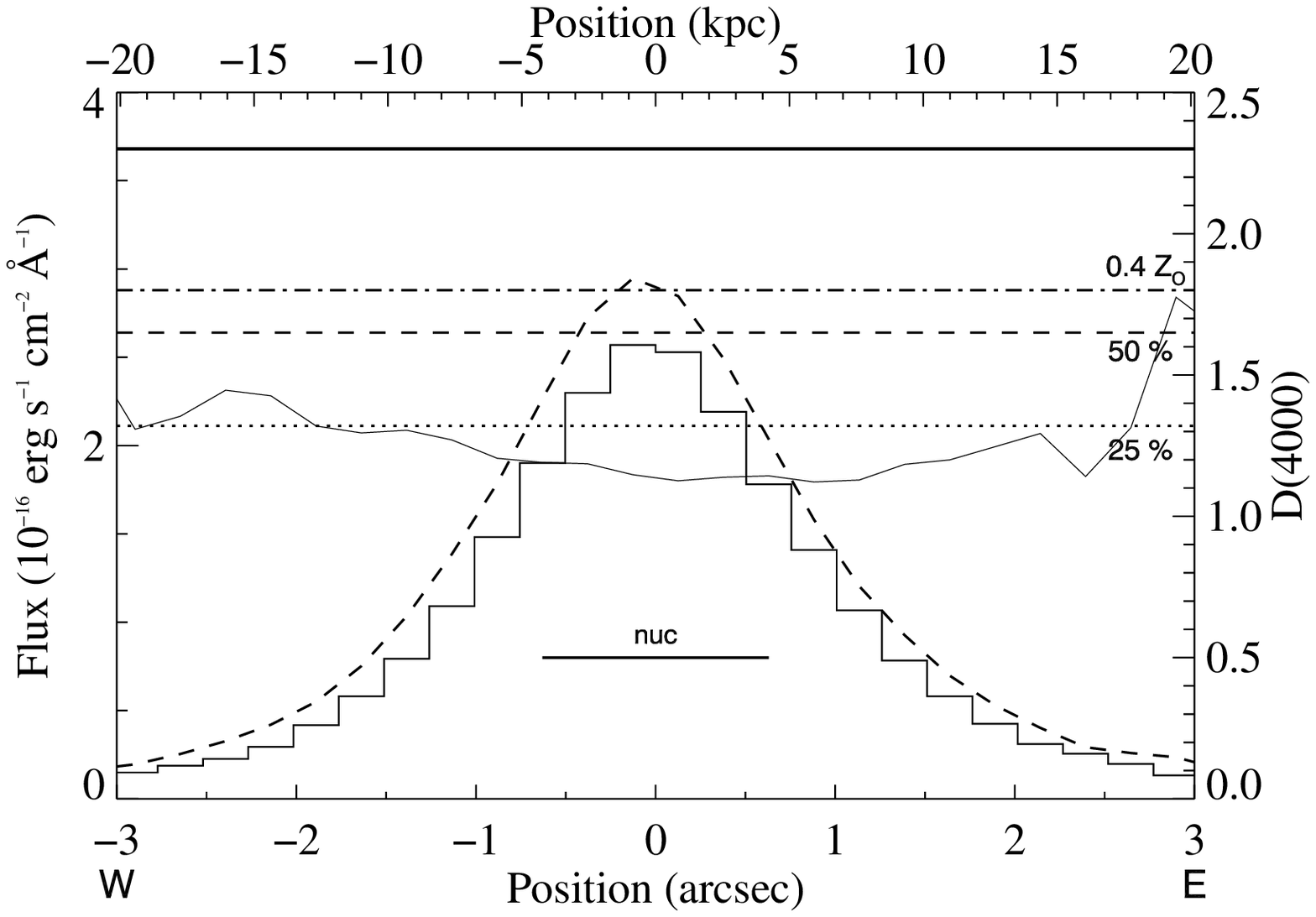,width=7cm,angle=0.} &
\hspace*{-0.8cm}\psfig{file=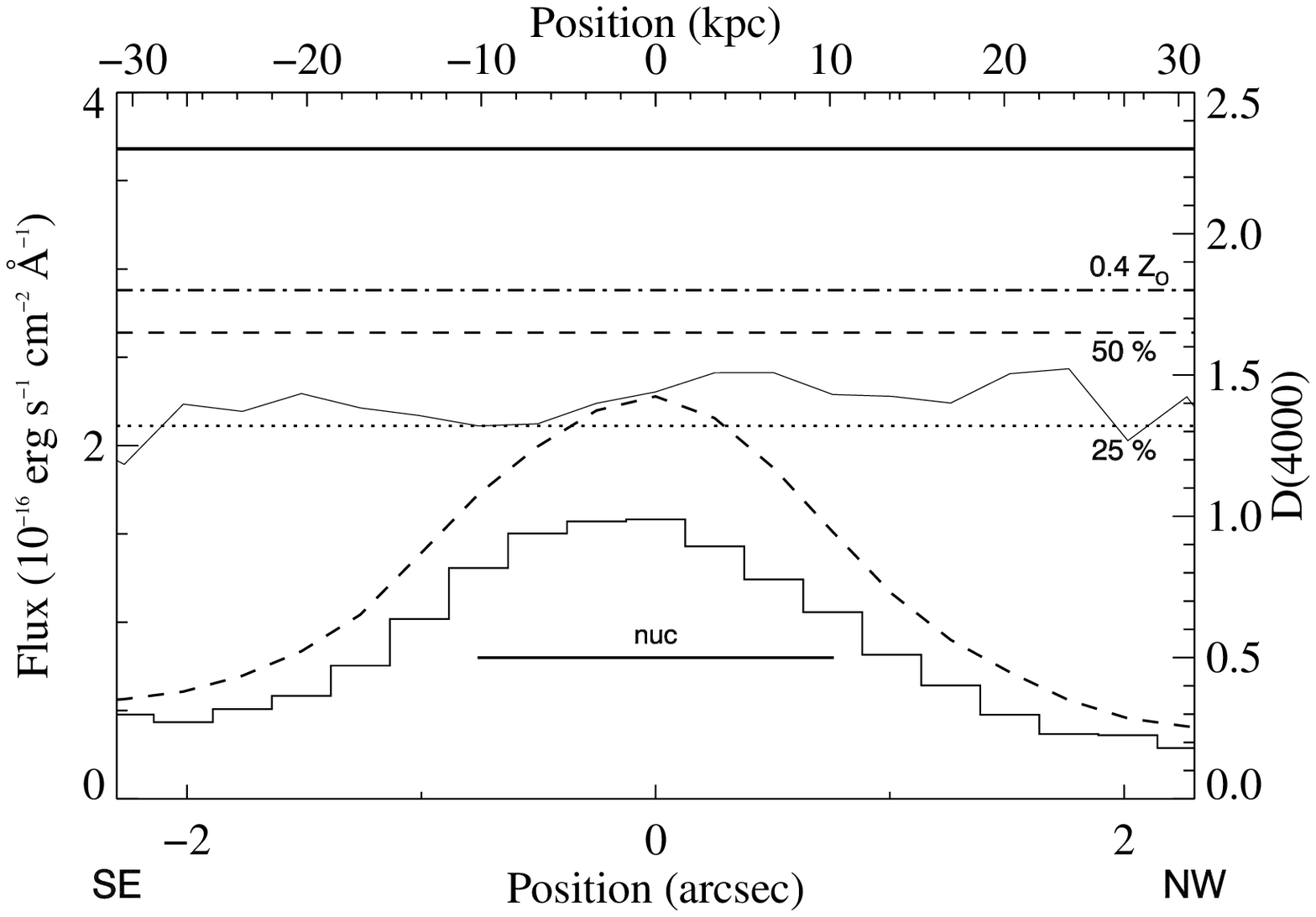,width=7cm,angle=0.}  
\\
\hspace*{-1.3cm}{\bf PKS 1932-46}  & \hspace*{-0.8cm}{\bf PKS 2135-20} &
\hspace*{-0.8cm}{\bf NGC 612}\\
\hspace*{-1.3cm}\psfig{file=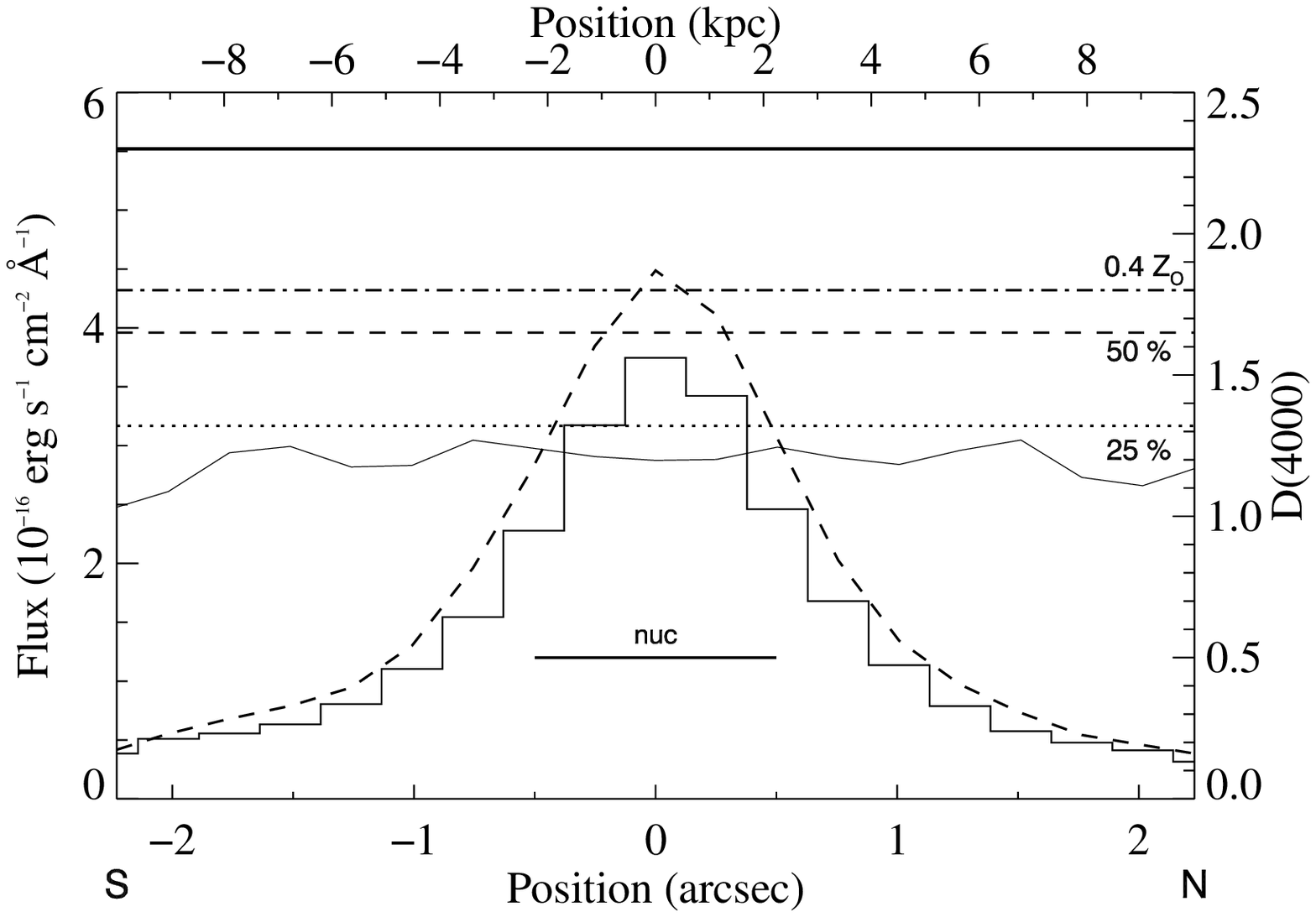,width=7cm,angle=0.} &
\hspace*{-0.8cm}\psfig{file=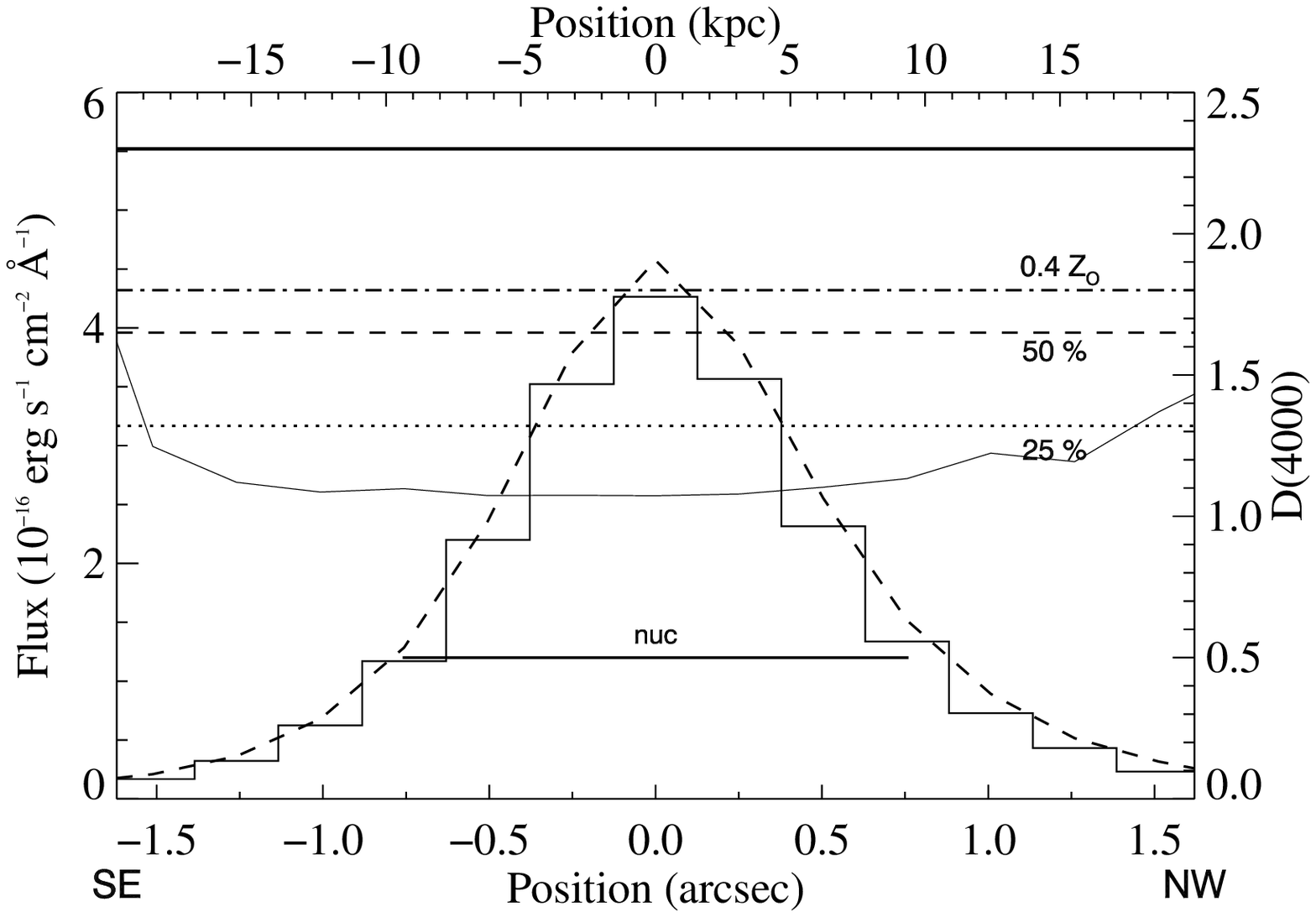,width=7cm,angle=0.} &
\hspace*{-0.8cm}\psfig{file=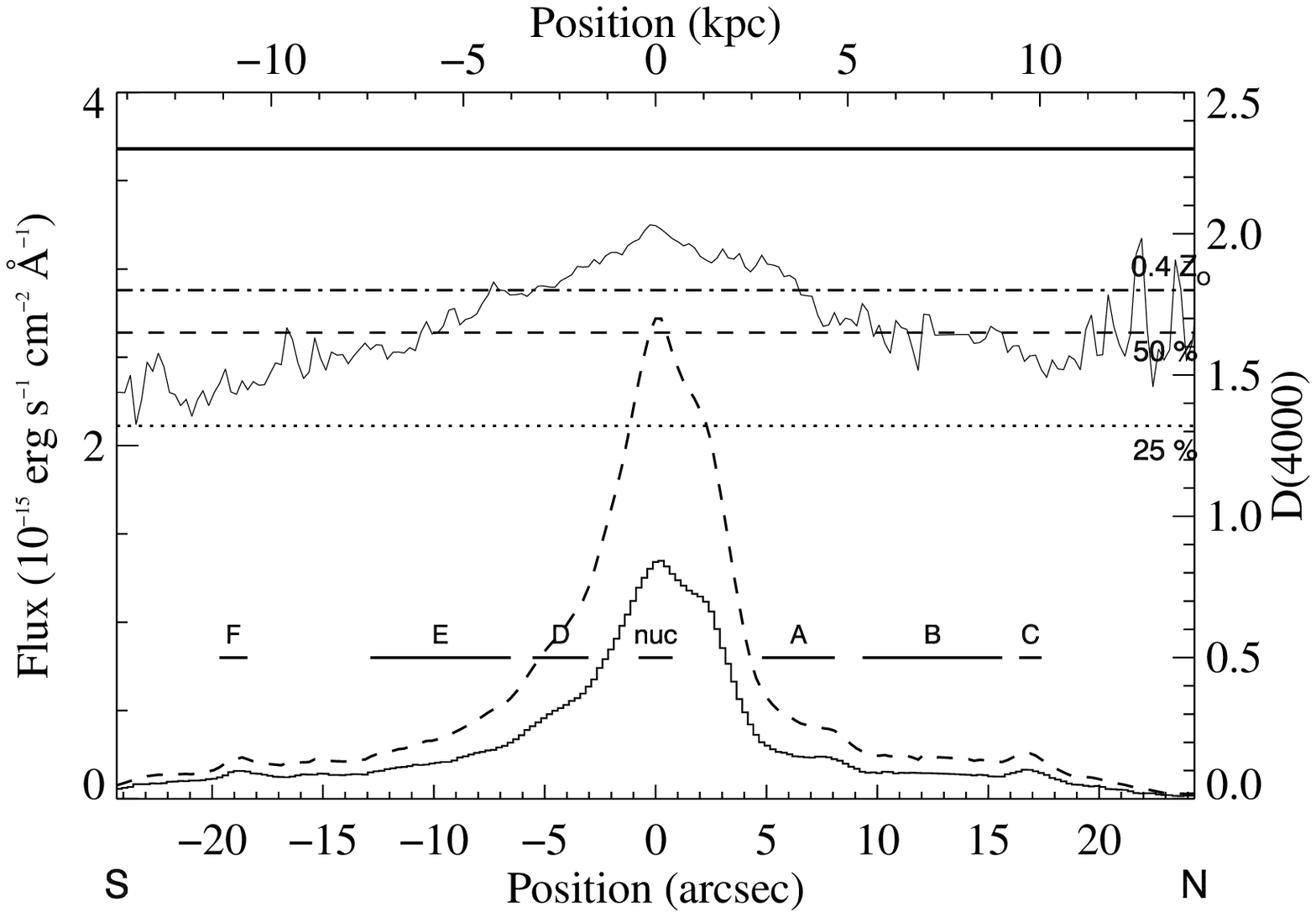,width=7cm,angle=0.} \\
\\
\end{tabular}
\vspace*{-0.5cm}
\caption[D4000A ratio]{$D$4000 ratio for all galaxies in the sample.
  The solid and dashed lines (profiles) represent the integrated
  continuum flux in 
  the ranges 3750-3850\AA~and 4150-4250\AA~respectively. The $D$(4000)
  ratio is overplotted (faint solid line) along with the expected
  ratios for a pure 12.5 Gyr elliptical galaxy (bold
  solid horizontal line) and diluted with a 50 per cent featureless
  continuum in the 3800\AA~bin (dashed horizontal line) and with a 75
  per cent featureless continuum (dotted horizontal line). The ratio
  expected for a subsolar metallicity ($Z$ = 0.4 $Z_{\odot}$)
  12.5 Gyr population is also shown (dot-dashed horizontal
  line). The locations of all apertures used in the analysis are also
  marked.} 
\label{fig:d4000}
\end{minipage}
\end{figure*}

A simple, commonly used diagnostic tool for investigating the
magnitude of the UV 
excess in a galaxy spectrum is to compare the continuum flux on either
side of the 4000\AA~break
(e.g. \citealt{tadhunter02,tadhunter05}). Hence, for each galaxy, 
continuum slices on either side of 4000\AA~were extracted to estimate
the $D$(4000) ratio -- a measure of the 4000\AA~break
amplitude -- following 
the definition in \citet{tadhunter02,tadhunter05}: 
\begin{equation}
D(4000) = \frac{\int_{4150}^{4250} F_{\lambda}
    {\rmn d}\lambda}{\int^{3850}_{3750} F_{\lambda} {\rmn d}\lambda}
\label{eq:4000A}
\end{equation}
This form of the $D$4000 ratio is better suited to active galaxies
than the original form (see e.g.  
 \citealt{bruzual93}) as it is not  contaminated by bright
emission lines (e.g. {[Ne III]}$\lambda$3869). The $D$(4000) ratio is also 
insensitive to both
reddening and redshift ($K$-correction) effects, unlike broad-band
colour comparisons. 
(e.g. \citealt{smith89b}). 

Figure \ref{fig:d4000} shows the spatial distribution of the continuum
flux in the two regions around 4000\AA~for all sources in the
sample along with the derived $D$(4000) ratio. For comparison, 
expected $D$(4000) ratios in the 3750-3850\AA~bin are overplotted.
These ratios were derived using the \citet{bruzual03}
spectral synthesis models (hereafter BC03) for a 12.5 Gyr old stellar
population, formed in an instantaneous burst with solar metallicity
and a Salpeter initial mass function (IMF). The various comparison
$D$(4000) ratios comprise: a) a 12.5 Gyr elliptical with solar
metallicity, b) a 12.5 Gyr elliptical with sub-solar metallicity ($Z$ =
0.4 $Z_{\odot}$), c) a 12.5 Gyr elliptical diluted with a  flat
spectrum such that the 12.5 Gyr population accounts for i) 50\% and
ii) 25\% of the continuum flux.

All sources in the sample show a significant UV excess compared to a
passively evolving old (12.5 Gyr), solar metallicity elliptical
galaxy. For all galaxies 
in the sample, the $D(4000)$ ratio suggests that only 25-50\% of the
flux in the UV originates from the 12.5 Gyr stellar
population. Interestingly, for those sources that are spatially resolved
(3C 236, 3C 218, 3C 321, 3C 285, 3C 433, PKS 0023-26, NGC 612),   {\it
  the UV excess is also 
spatially extended, covering 
the entire spatial extent of the galaxy}, suggesting that, whatever
causes the UV excess (AGN continuum and/or a YSP) is important {\it galaxy
wide}, and is not confined to the nuclear regions. Further, 
some objects show interesting trends in their UV excess. In 3C 218, the
UV excess increases symmetrically towards the centre of the galaxy.
In constrast, the UV excess in NGC 612 decreases symmetrically towards
the nucleus.  It is also notable that, in the case of the double nucleus
systems 3C 321 and 3C 433, the UV excesses extend across both nuclei as 
well the extended haloes of the systems.

The spatial continuum profiles in Figure {\ref{fig:d4000}} have  been
used to identify 
regions of interest in the 2-dimensional spectra for extraction and
continuum modelling (see below). For the majority of objects, only a
central nuclear aperture was extracted. Notable exceptions are NGC
612, 3C 433 and 3C 321.  All apertures are marked on Figure
{\ref{fig:d4000}}. 
Our complete spectra cover a wide wavelength range, vital for accurate
modelling of the 
SEDs, and the observed rest wavelength ranges for each object are
summarised in Table {\ref{tab:obs}}.

\subsection{The nebular continuum}
\label{sect:nebularcontinuum}
Prior to continuum modelling, 
a nebular continuum was generated for each aperture to be
analysed over the entire wavelength range. This comprises the blended
higher Balmer series ($>$ H8) and a theoretical nebular continuum
(generated using the {\sc starlink} package {\sc dipso}, which includes
the effects of free-free emission, free-bound recombination and
two-photon continua) following \citet{dickson95}. Both components of
the nebular continuum were
normalised to the H$\beta$ flux in each emission line component. Note,
H$\beta$ was modelled using the minimum number of Gaussian components
required to give a physically viable good fit and so for many
apertures, in particular the nuclear aperture, the generated nebular
continuum included multiple components, as discussed in \cite{holt03}
for the complex nuclear aperture in PKS 1345+12. 
The spectra of several of  the objects are subject to
significant reddening.  Hence, the nebular continuum was generated using
the {\it de-reddened} H$\beta$ line fluxes and then {\it re-reddened}
before being subtracted from the galaxy spectrum. The degree of
internal reddening was estimated using the Balmer line decrements
(e.g. \ha/\hb, H$\gamma$/\hb) and estimated separately for the
different kinematic components of the lines in each aperture. 

It should be noted that, when generating the nebular continuum, one
must carefully consider the potential pitfalls. In sources
where a significant fraction of the optical/UV light originates from a
young stellar component, the Balmer lines (particularly \hb,
H$\gamma$)  may be affected by underlying 
stellar absorption lines, and the flux therefore under-estimated. To
check on the likely impact of the stellar absorption lines, we have
followed the technique of \citet{holt05} 
and subtracted off the best fitting continuum model and then re-measured the
emission lines. 

Considering these issues, the checks show that the spectra fall  into
three categories: i) apertures in which the original and re-calculated
nebular continua based on \ha/\hb~are
similar (within a few percent in the 3450-3550\AA~bin; e.g. PKS
0039-44) -- we find that, due to this effect, \hb~is typically
under-estimated by  $\lesssim$ 10 per cent. Further, in apertures
where reddening is important,  a
10 per cent correction on the \hb~flux, gives a variation in the
\ebv~value of only $<$0.1 for \ha/\hb~$<$ 10. For a discussion
regarding the extinction law used, see Section
{\ref{sect:reddening}}; ii) apertures 
in which the nebular continuum fraction is so small ($\lesssim$ few
per cent) that even large
differences in nebular reddening would not significantly affect the results (e.g. 3C 218,
3C 236, NGC 612) and iii) apertures in which the reddening estimate
was based solely on weaker lines (e.g. H$\gamma$, \hb) due to \ha~lying
outside of the observed spectral range (e.g. PKS 0023-26, PKS 2135-20,
PKS 1932-46). For case iii), we are less confident of the
reddening estimate due to uncertainties in measuring H$\gamma$ as it
is a weak, blended line which is relatively more affected by
underlying stellar absorption features than \ha~or \hb. In these apertures, 
we have followed the example of PKS 1345+12 as discussed in
\citet{holt03}\footnote{As the technique for correcting for the 
  nebular continuum used here is 
discussed in detail in \citet{holt03}, 
we do not discuss the technique  or show examples of the
generated nebular continua.} which allows us to determine the maximum
nebular continuum fraction, i.e. that which does not produce an unphysical
step-function in the data at the location of the Balmer break when
subtracted.  For these apertures, we have modelled the continuum
twice, with both the maximum and minimum (i.e. 0\%) nebular continuum
fraction (this technique was also used for PKS 1549-79; see
H06). Reassuringly, we find that the modelling results are  similar
for both cases of nebular continuum subtraction (minimum and
maximum). Assuming a zero nebular continuum fraction 
gives a slightly larger UV continuum flux and so would potentially shift the
\chisq~minimum to slightly younger ages. 
For apertures in which the continuum is
dominated by a power-law component (e.g. PKS 1932-46) and/or the YSP
age is poorly constrained (e.g. PKS 2135-20), the modelling results
based on both the minimum and maximum nebular continua are
indistinguishable i.e. the effect on our results due to 
 errors on the subtracted nebular continuum are insignificant. In the
 nuclear aperture of PKS 0023-26, the YSP contribution is large 
(50-60\%) and has a tightly constrained minimum (see Figure
 \ref{fig:contours}). For this case, subtracting off the maximum
nebular continuum (22.5\%) gives a YSP age range of 
0.03-0.05 Gyr whilst a zero nebular continuum model gives a YSP age
range of 0.01-0.03 Gyr. We are therefore
confident that the uncertainties in the nebular continuum subtraction
do not significantly affect our results.

The contribution of the nebular continuum
(reddened where necessary) 
to the flux in the 3540-3640\AA~bin in each aperture ranges from 
$\lesssim$1\% (NGC 612, 3C 381, mid aperture in 3C 321) to $\sim$25\% (extended
aperture of 3C 321, PKS 1932-46) and is summarised in 
 Table {\ref{tab:results}}.  

For the objects with published nebular continuum measurements, the
nebular continuum fraction previously estimated is typically higher than the
estimates here. For the sources in the 2Jy sample (PKS 0023-26, PKS
0039-44, PKS 0409-75, PKS 1932-46 and PKS 2135-20 -- see
\citealt{tadhunter02}), the differences are likely to be
due to the fact that \citet{tadhunter02}
made no correction for
reddening, stating that these nebular continuum fractions were upper
limits. In this paper, we have followed the careful reddening
correction technique described in \citet{holt03} and generally derive
lower nebular continuum fractions.

\subsection{Continuum modelling}
\label{sect:cont}
\begin{figure*}
\begin{minipage}{180mm}
\centerline{\psfig{file=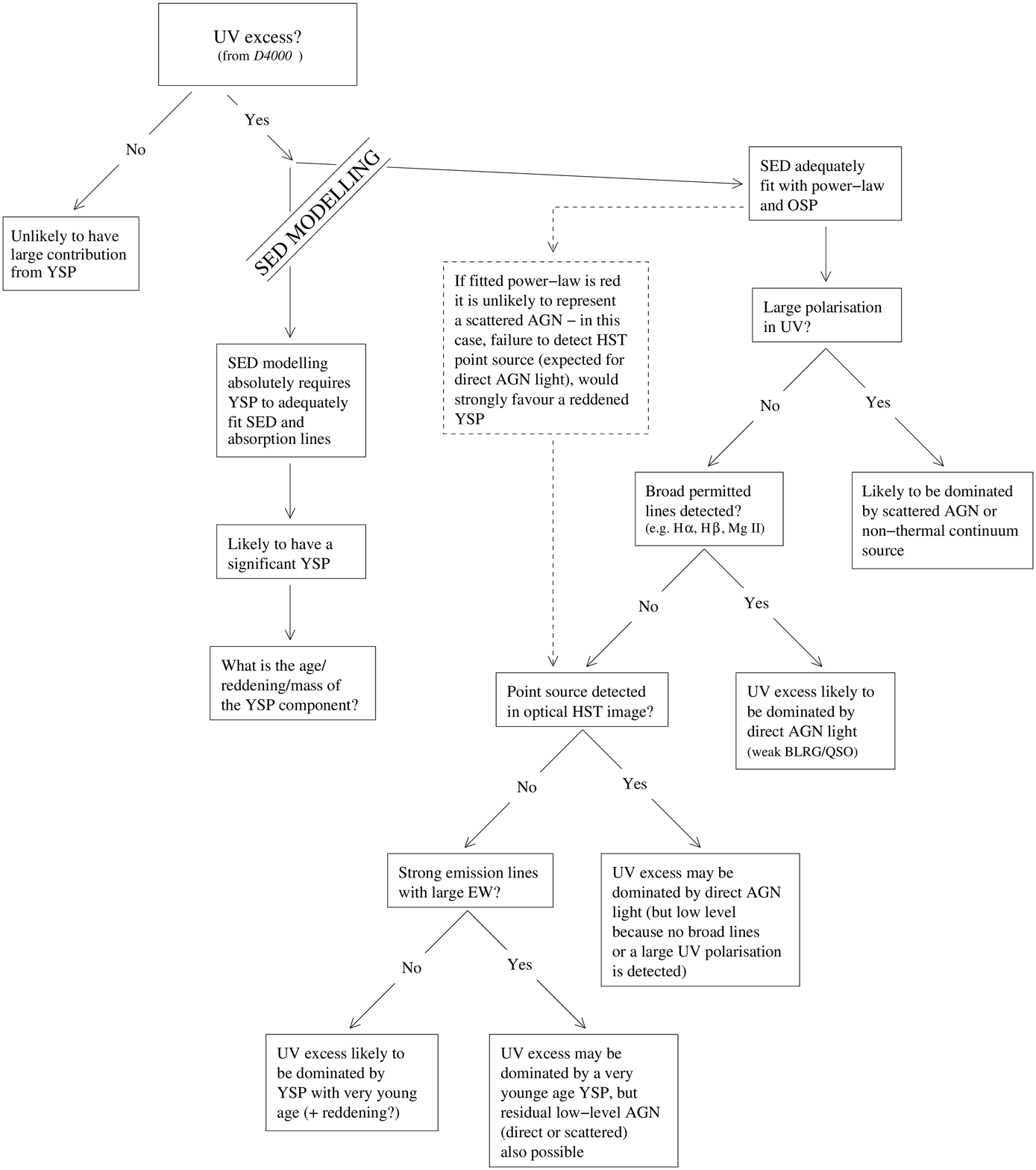,width=17cm,angle=0.}  }
\caption[]{The 'decision tree' used to interpret the modelling
  results. Abbreviations are as follows: SED: spectral energy
  distribution; YSP: young stellar population; OSP: old stellar
  population; EQW: equivalent width; BLRG: broad line radio galaxy;
  QSO: quasar; $D4000$: ratio of continuum fluxes at the
  4000\AA~break. }
\label{fig:decisiontree}
\end{minipage}
\end{figure*}

After subtracting the nebular continuum, the continuum SEDs were modelled
using {\sc confit}, a purpose written {\sc idl} code using a minimum
$\chi^2$ technique \citep{robinson01}. The program allows up to three
continuum components to be combined -- up to two different stellar
populations, with or without a power-law 
component. Note that the power-law could represent an AGN/scattered light
component or a very young stellar population. The stellar
templates used in the modelling were taken from the 2003 version of 
the spectral synthesis
results of Bruzual \& Charlot (2003; hereafter BC03). The stellar population
ages used include
old, evolved stellar populations (OSP: 12.5 Gyr and, in some cases, 7.0 Gyr)
and various young stellar populations (YSP: 0.01-5.0 Gyr in varying
steps). 
All stellar
population models assume a Salpeter initial mass function (IMF), solar
metallicity, instantaneous starburst. The YSP spectra were reddened
over the range 0.0 
$<$ \ebv $<$ 1.6 in steps of 0.1 using the \citet{seaton79} extinction
law (see Section
\ref{sect:reddening}).
Finally, when a power-law component was included (of the form
F$_{\lambda} \propto \lambda^{\alpha}$), the spectral index of the
power-law was allowed to vary between -15.0 and 15.0 (to ensure
reasonable calculation times). 

Various models were tried including: 
\begin{itemize}
\item an OSP  (12.5 Gyr)  with zero reddening.
\item an OSP (12.5 Gyr, zero reddening) with a
  power-law  component.
\item an OSP (12.5 Gyr, zero reddening) with a
 YSP (0.01 $<$ t$_{\rm ysp}$ $<$ 5.0 Gyr, 0.0 $<$ \ebv $<$ 1.6).
\item an OSP (12.5 Gyr, zero reddening) with a
  YSP (0.01 $<$ t$_{\rm ysp}$ $<$ 5.0 Gyr, 0.0 $<$ \ebv $<$ 1.6) and a power-law
   component. 
\item for the higher redshift sources (e.g. PKS 2135-20, PKS 0039-44,
  PKS 0409-75 and PKS 0023-26), `younger' old stellar
  populations were also tried for all four combinations above
  including 7.0 Gyr, 9.0 Gyr and 10.0 Gyr because, at the redshifts of
  these objects, current cosmological models show the universe is younger
  than 12.5 Gyr.  
\end{itemize}

Our general philosophy is to model the data with the minimum
number of stellar and/or power-law components that provide an adequate 
fit to the spectra. Given that the galaxies we are modelling are all 
early-type in terms of their optical morphologies, we start by assuming 
that all galaxies contain an old (7 -- 12.5~Gyr) stellar population (OSP) at 
some level, and that any young stellar populations were formed in a 
single instantaneous burst as a result of a merger, and hence have a 
particular age. In our models very young, reddened YSPs
($<$10~Myr) may be simulated by adding a  power-law component, provided
that they can be distinguished from AGN-related components (see below).
Clearly more components would be required to model more 
complex star formation histories, but the greater the numbers of components,
the greater the likelihood of degeneracies in the solutions.

One of the major ambiguities in our model solutions is that we cannot
readily distinguish  
between the case of  a highly reddened very young YSP ($<$10~Myr) plus an OSP,
and an AGN-related power-law plus an OSP, based on the SED modelling
alone. This 
is because the spectra of the very young YSPs are relatively
featureless in terms of 
the strengths of the absorption lines and the amplitude of the Balmer
break (e.g. \citealt{gonzalezdelgado99,gonzalezdelgado05}). Therefore 
it is necessary to use auxilary information, such as UV polarisation,
detection of point sources 
in HST images, detection of broad permitted lines, and emission line
luminosity to distinguish 
between these cases. For example, if the UV excess is predominantly
caused by a scattered AGN  
component, revealed as a power-law in the SED fits, we expect to
measure significant UV  
polarisation and a relatively flat or blue power-law slope ($\alpha <
0.5$)\footnote{Such 
blue power-law slopes have been measured for all cases with
significant scattered AGN 
components \citep{tadhunter96,tran98,cohen99,vernet01} and strong
emission lines -- the objects with the strongest scattered 
AGN components tend to be those with the most luminous emission lines
(see \citealt{tadhunter02})}. On the other hand we expect a direct AGN
component (from a low-luminosity 
or partially obscured quasar) to be associated with the detection of
broad permitted 
lines and a point source in HST images. Figure
{\ref{fig:decisiontree} presents a flow diagram of the 
decision making process involved in interpreting the modelling 
results for such cases.

For the modelling, continuum bins were selected to avoid emission
lines and, in the case of 3C 285, 3C 433, 3C 218/Hydra A, 3C 321 and 3C 381,
the uncorrected atmospheric 
absorption bands. Typically, 40-50 bins were chosen for each aperture
to ensure good sampling over a large spectral range as evenly
distributed in wavelength 
as  possible. In addition, a normalising bin in the blue part of the
spectrum, common to all objects, was
chosen with rest wavelength 4720-4820 \AA. The {\sc confit} program then
generated the models by scaling the different components to the flux
in the normalising bin so the total model flux was $<$125 per cent of
the observed flux\footnote{The maximum allowed model flux is greater
  than 100 per cent of the measured flux to allow for uncertainties in
  the measured flux and models in the bin.}, finding the model
combination with the minimum reduced $\chi^2$ (\chisq). For the
\chisq~determination, we assume a relative flux calibration error of $\pm$5 per
cent.  

Combining the best fitting  model for each combination of
OSP with reddened or unreddened YSPs of various ages
(both with and without a power-law component) forms a
surface in \chisq~for YSP age versus \ebv~from which contour plots can
be generated.  Examples
of the contour plots of the \chisq~space for each aperture requiring a
YSP component are shown in Figure {\ref{fig:contours}}. Using these plots,
we identified minima\footnote{Note, whilst the main discriminating
  factor is the {\it minimum} in the \chisq~space, all fits with
  \chisq~$\lesssim$1 are deemed to be a `good fit'. In addition, fits 
  with \chisq$\lesssim$1.2 are also considered for the detailed
  comparisons. See detailed discussion in \protect\citet{tadhunter05}.}
 in the \chisq~space to be further investigated
using detailed comparisons of the stellar absorption features known to
be sensitive to age. The
CaII~K$\lambda$3968, the G-band
(4306\AA) and the Mg~Ib band (5173\AA) absorption features 
are particularly important for
such comparison because they are not significantly infilled by
emission lines. In some cases with weak emission lines it
was also possible to use the higher order Balmer line features for
detailed comparisons.  Along with examples of the
overall SED fits, detailed comparisons of the region containing the
Balmer series and CaII~H+K are shown in Figure
{\ref{fig:SED}}. 

Note that, when discounting models on the basis of detailed fits to CaII~K
feature, consideration must be made of
the potential contribution of intrinsic ISM absorption to the
to this feature, especially in highly reddened cases. A contribution
from the ISM would 
strengthen the CaII~K line in the observed spectrum, but
not in the model spectra. Therefore models that underpredict the
observed strength 
of CaII~K may be viable, whilst models which significantly overpredict
the strength of CaII~K can be discounted. This technique proved
particularly useful in distinguishing between two clear minima in the
\chisq~space (e.g. 3C 285).  In some cases it was necessary to
apply smoothing to the models to obtain good matching in the
absorption lines because the models and the data have different effective
resolutions. 

Hence, in order to be considered a viable fit to the data, the various models
had to provide {\it both a good fit to the overall SED and  the stellar
absorption lines.} When relying on one technique alone, the solutions
are often degenerate in age/reddening space for the YSP (see also
\citealt{tadhunter05}).

\begin{figure*}
\begin{minipage}{180mm}
\begin{tabular}{ccc}
\hspace*{-1.2cm}\psfig{file=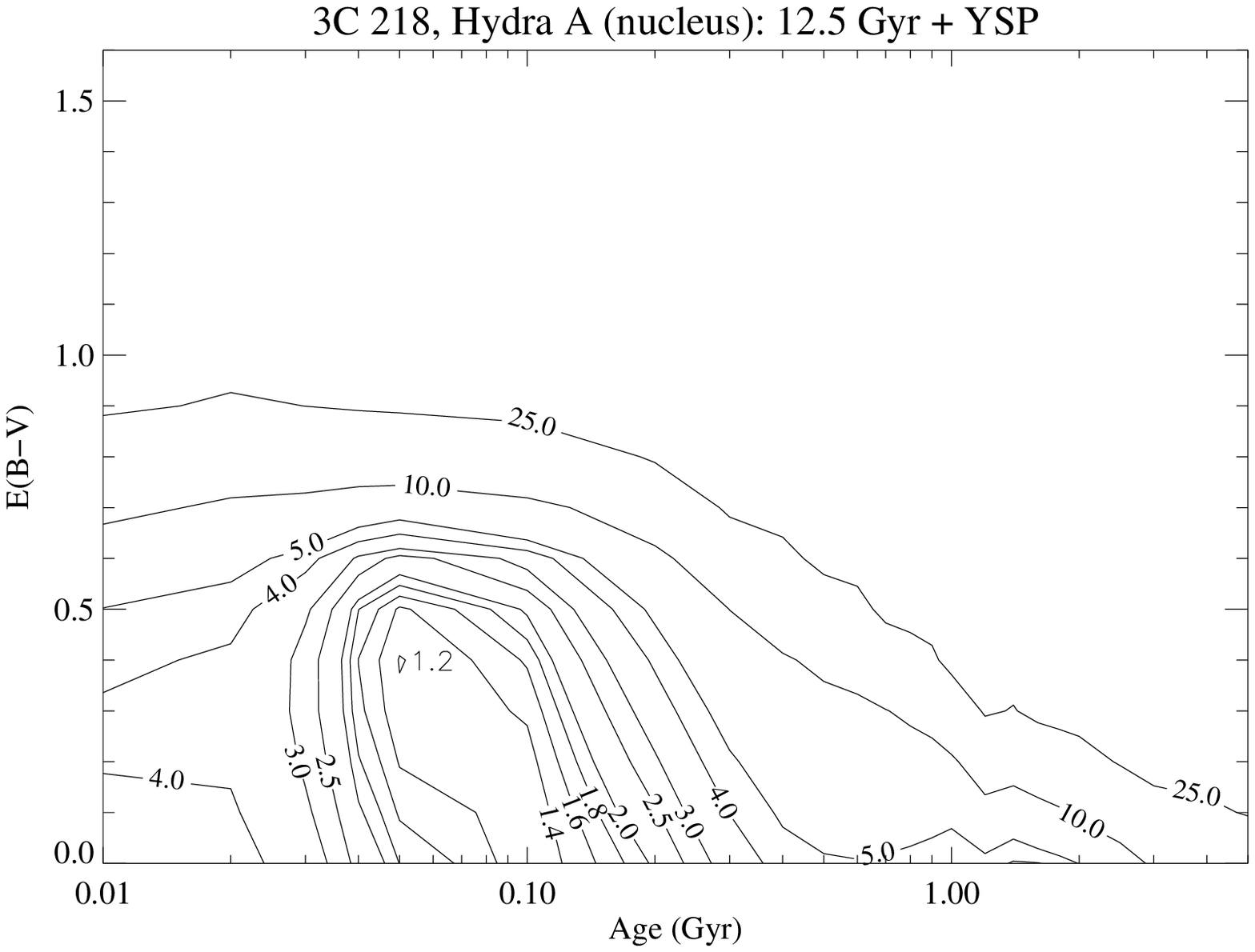,width=7cm,angle=0.}
&
\hspace*{-0.8cm}\psfig{file=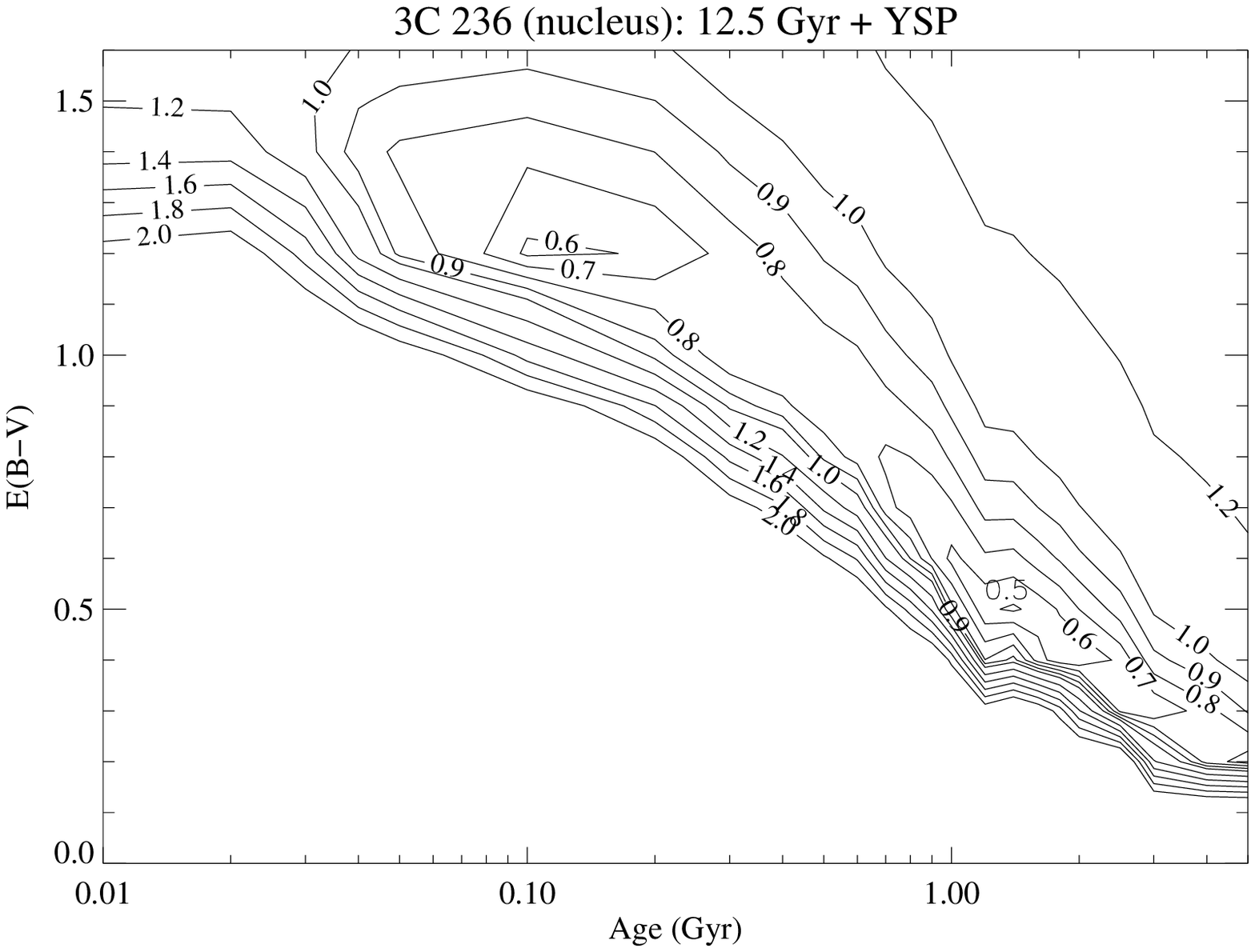,width=7cm,angle=0.} &
\hspace*{-0.8cm}\psfig{file=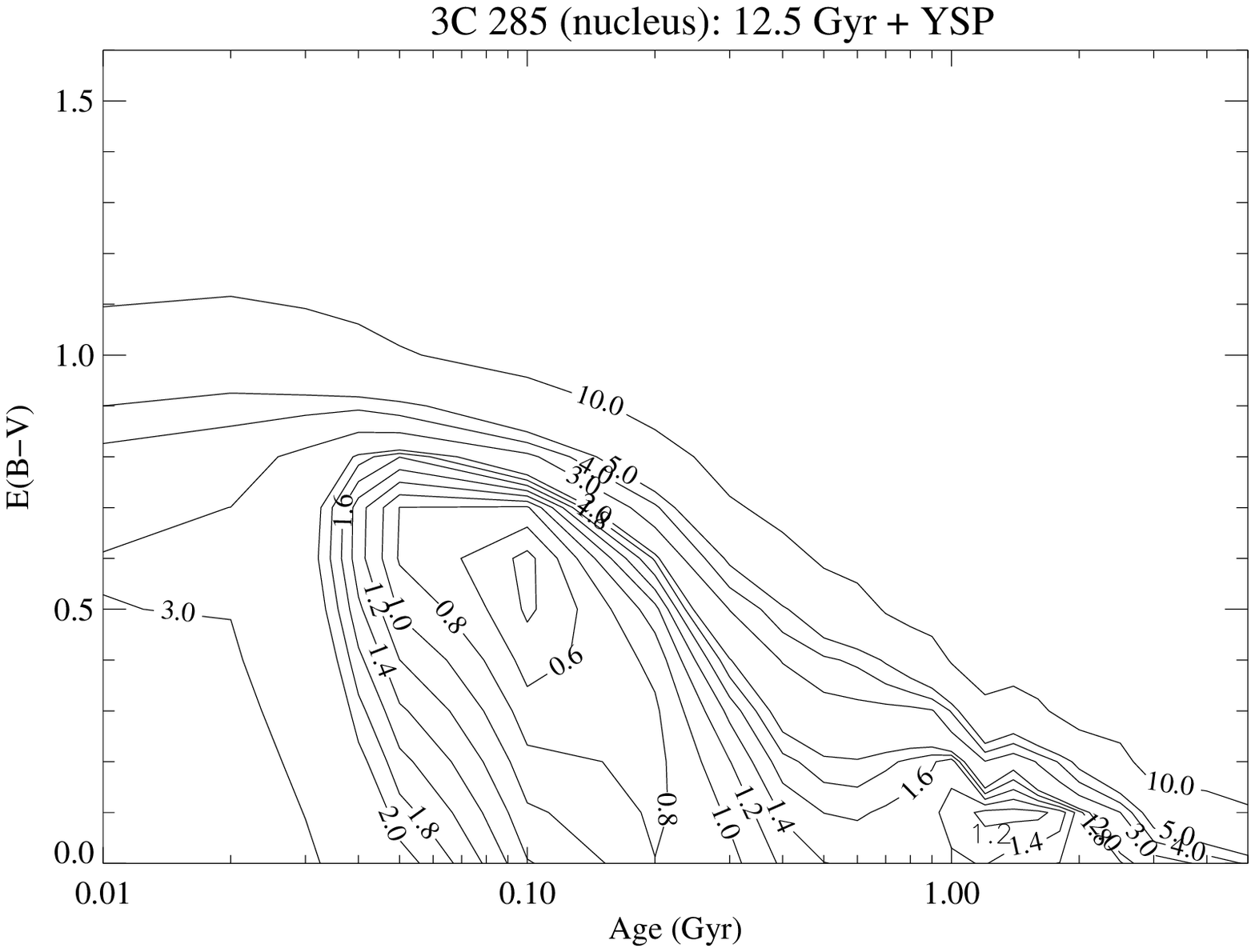,width=7cm,angle=0.}\\
\hspace*{-1.2cm}\psfig{file=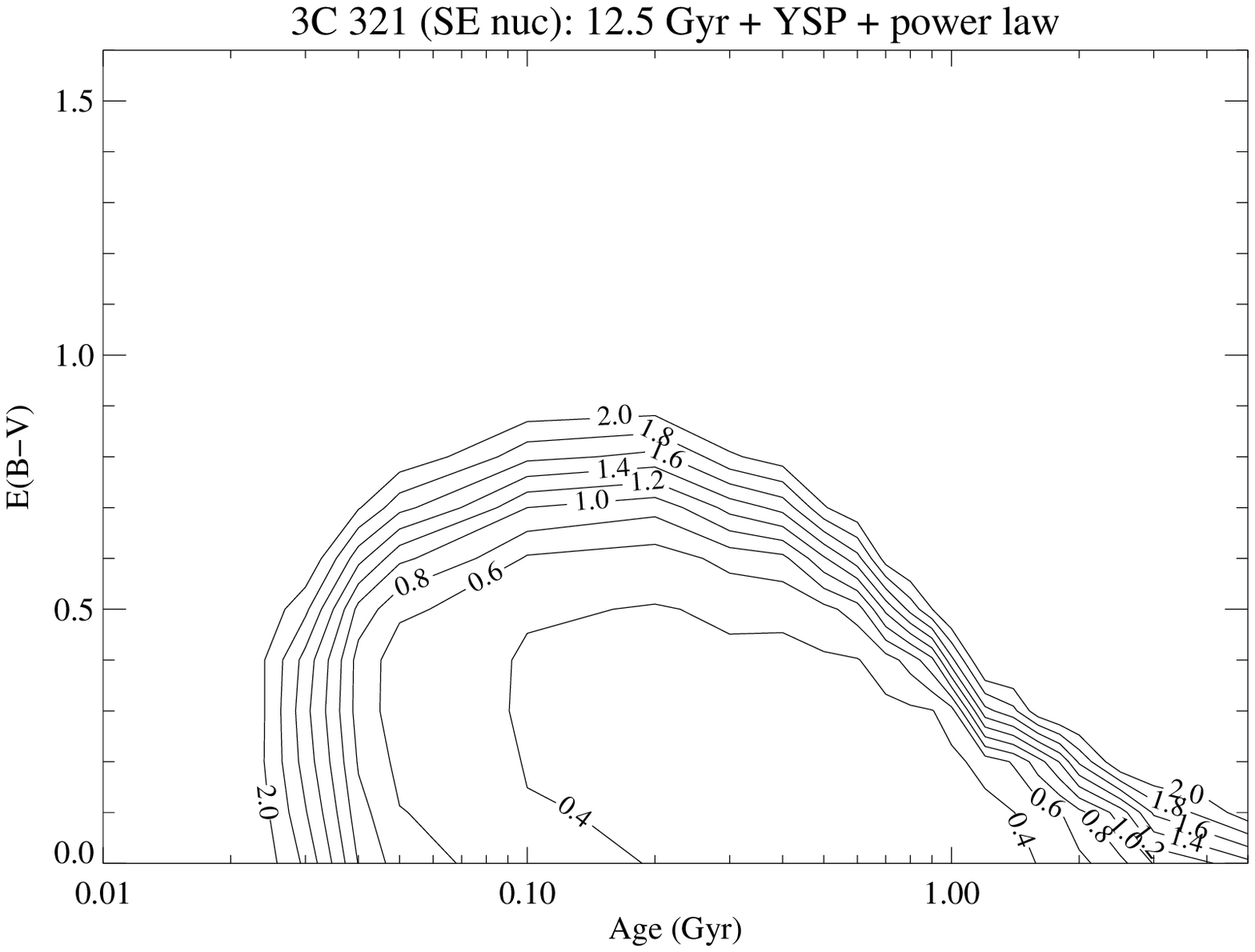,width=7cm,angle=0.} &
\hspace*{-0.8cm}\psfig{file=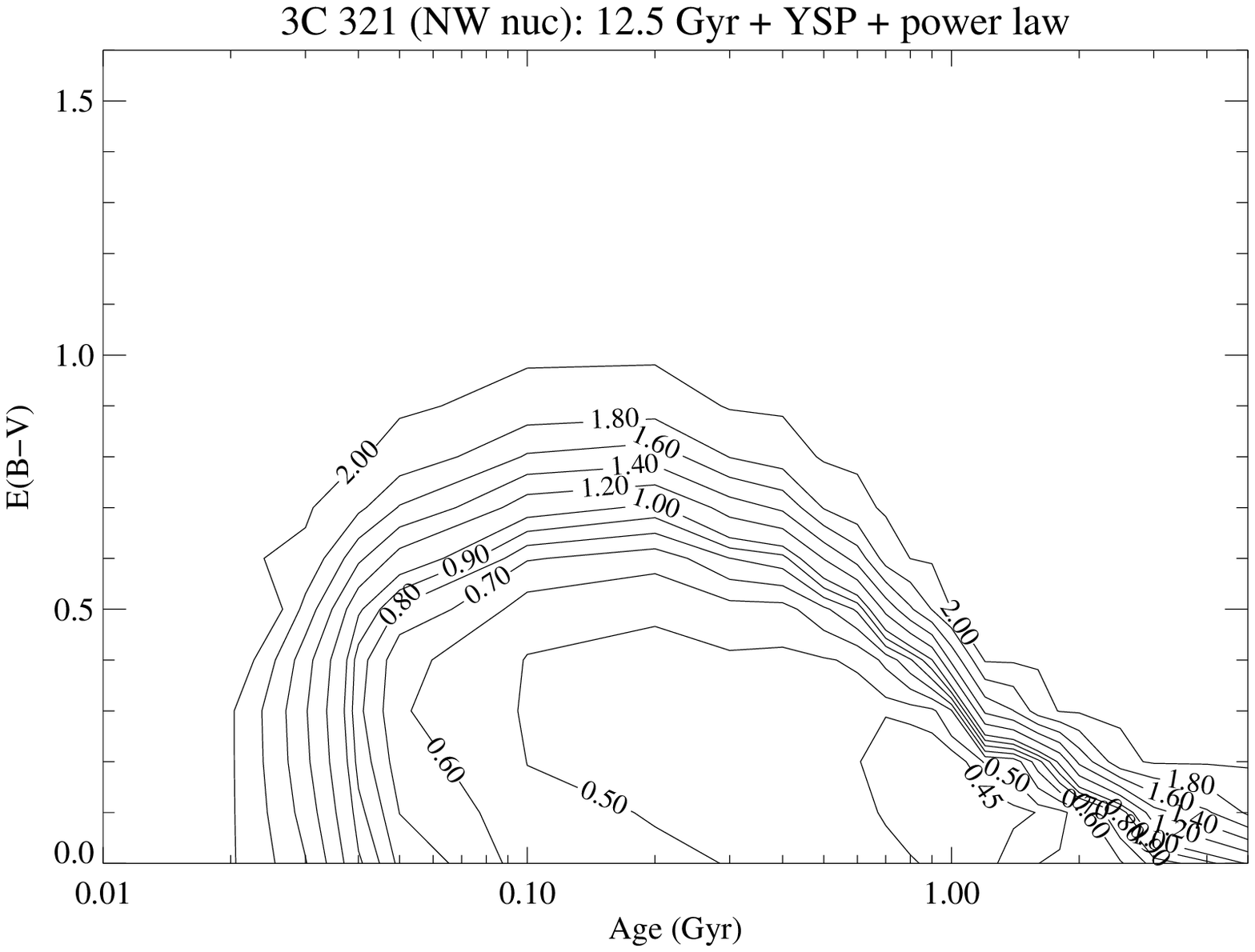,width=7cm,angle=0.} &
\hspace*{-0.8cm}\psfig{file=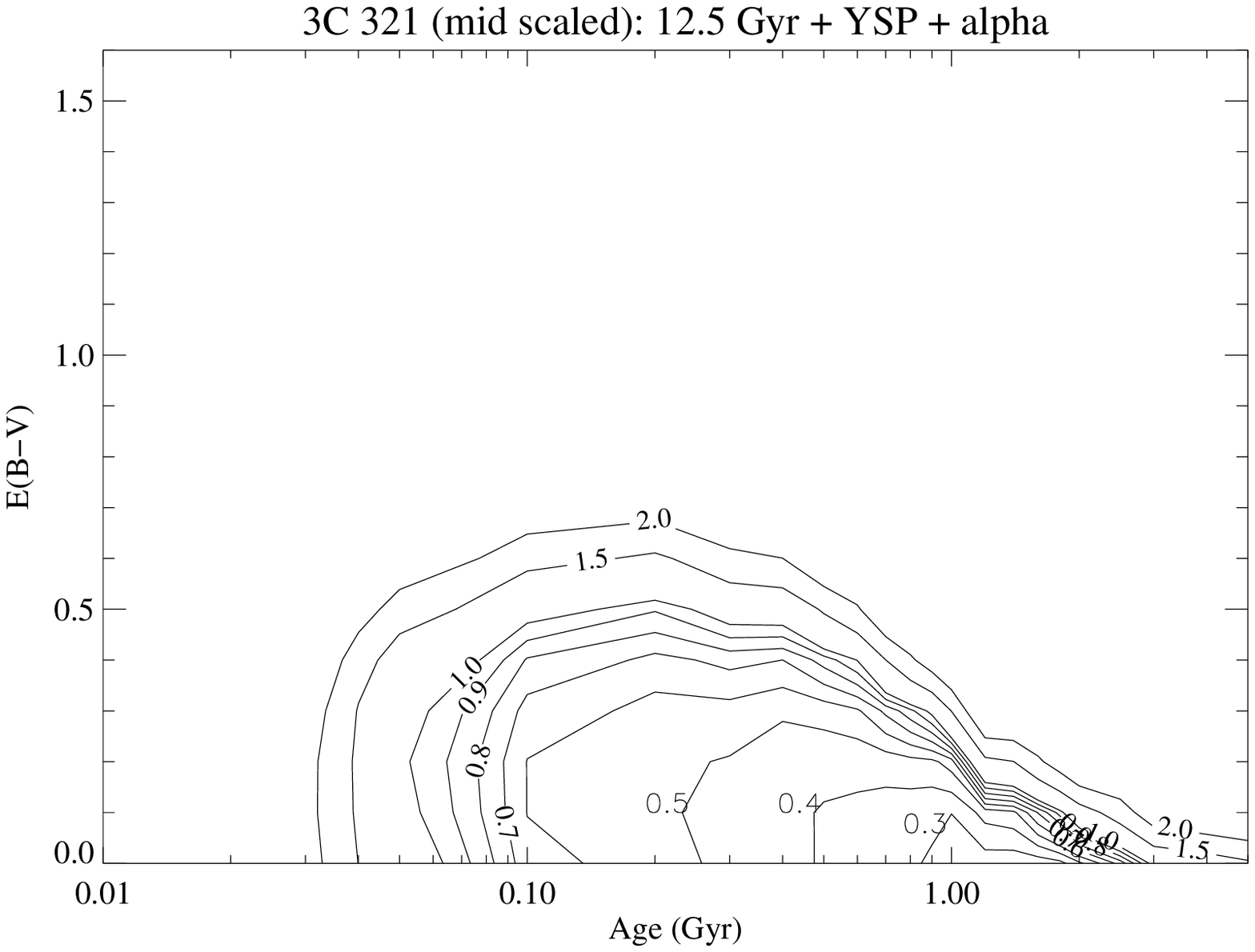,width=7cm,angle=0.} \\
\hspace*{-1.2cm}\psfig{file=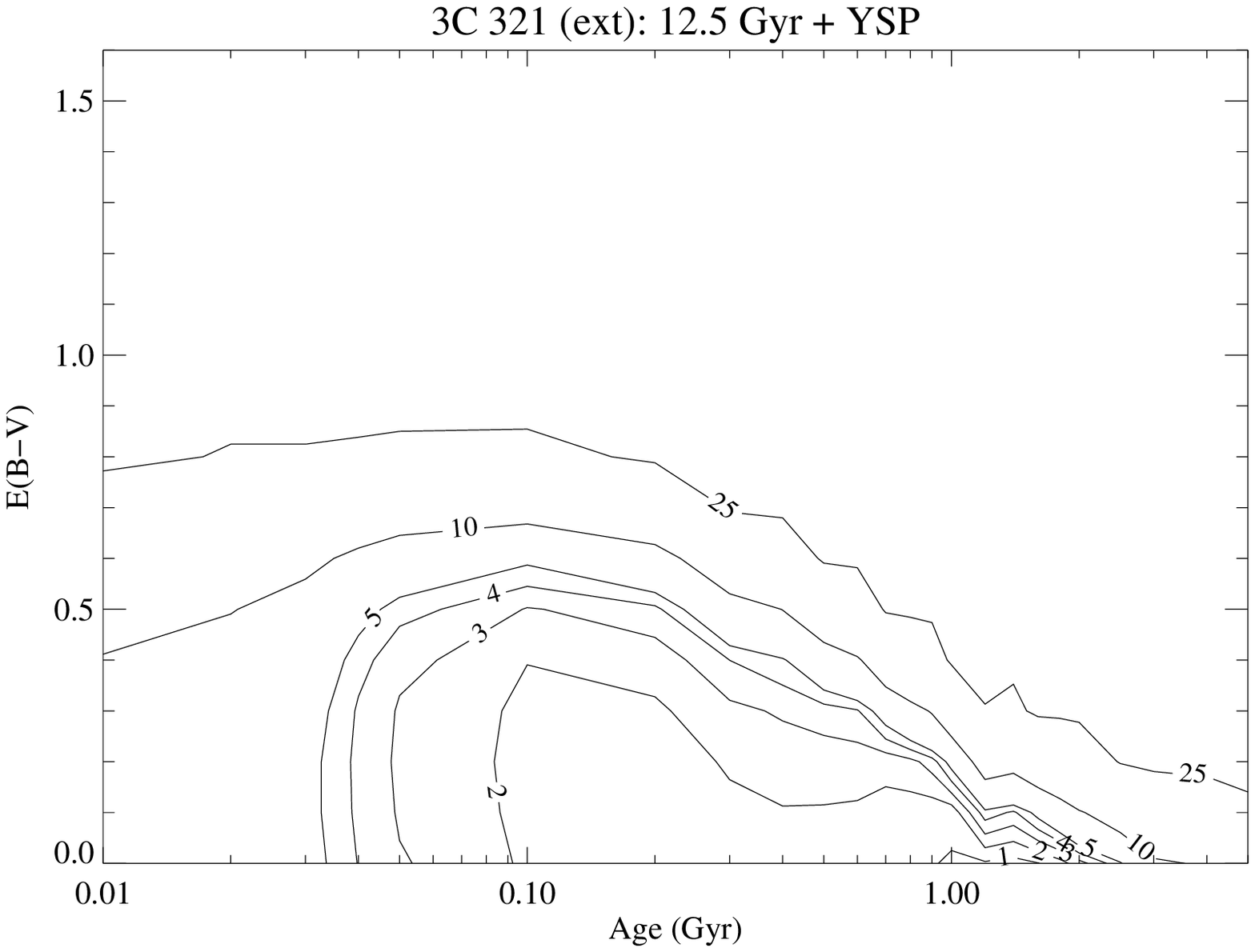,width=7cm,angle=0.} &
\hspace*{-0.8cm}\psfig{file=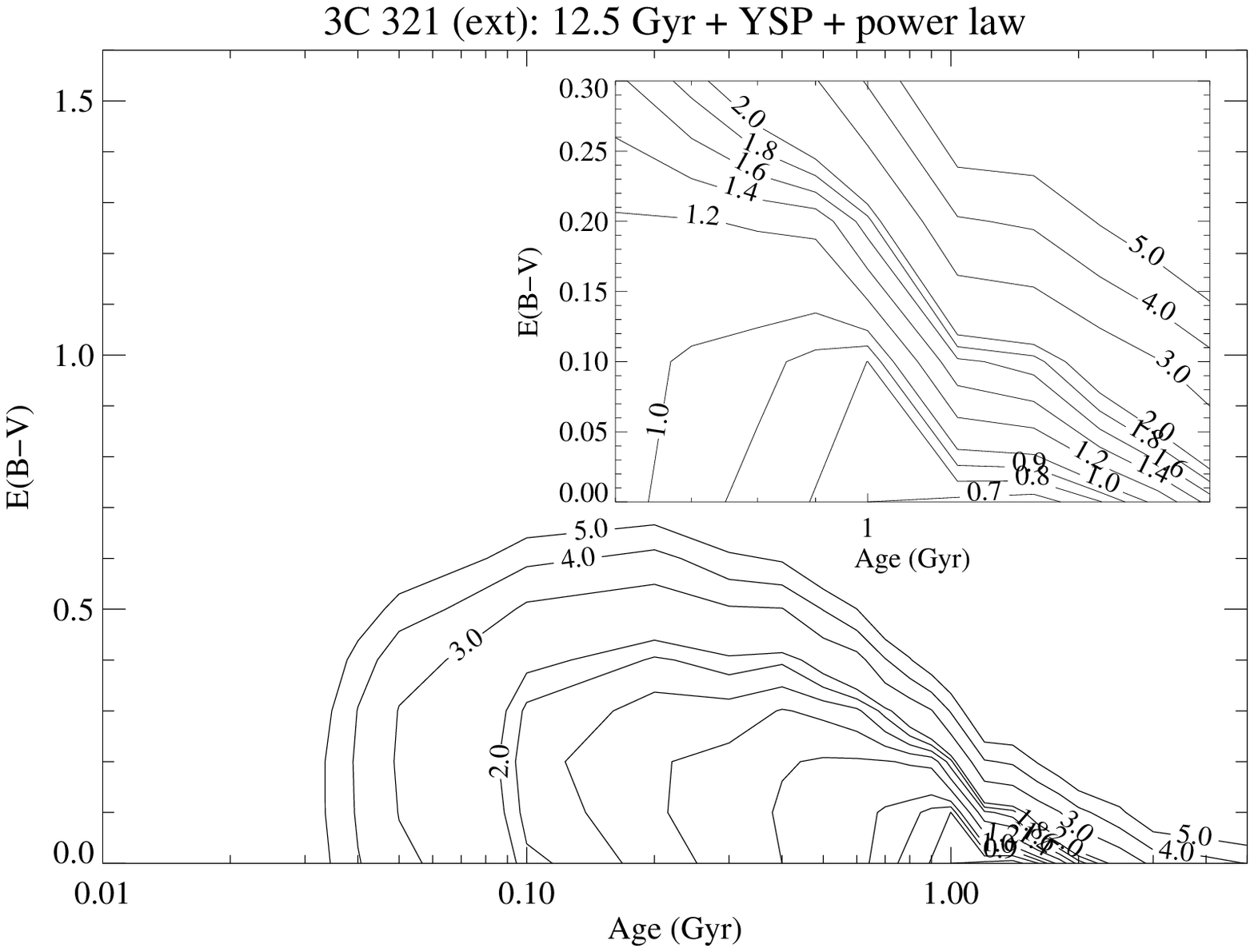,width=7cm,angle=0.}&
\hspace*{-0.8cm}\psfig{file=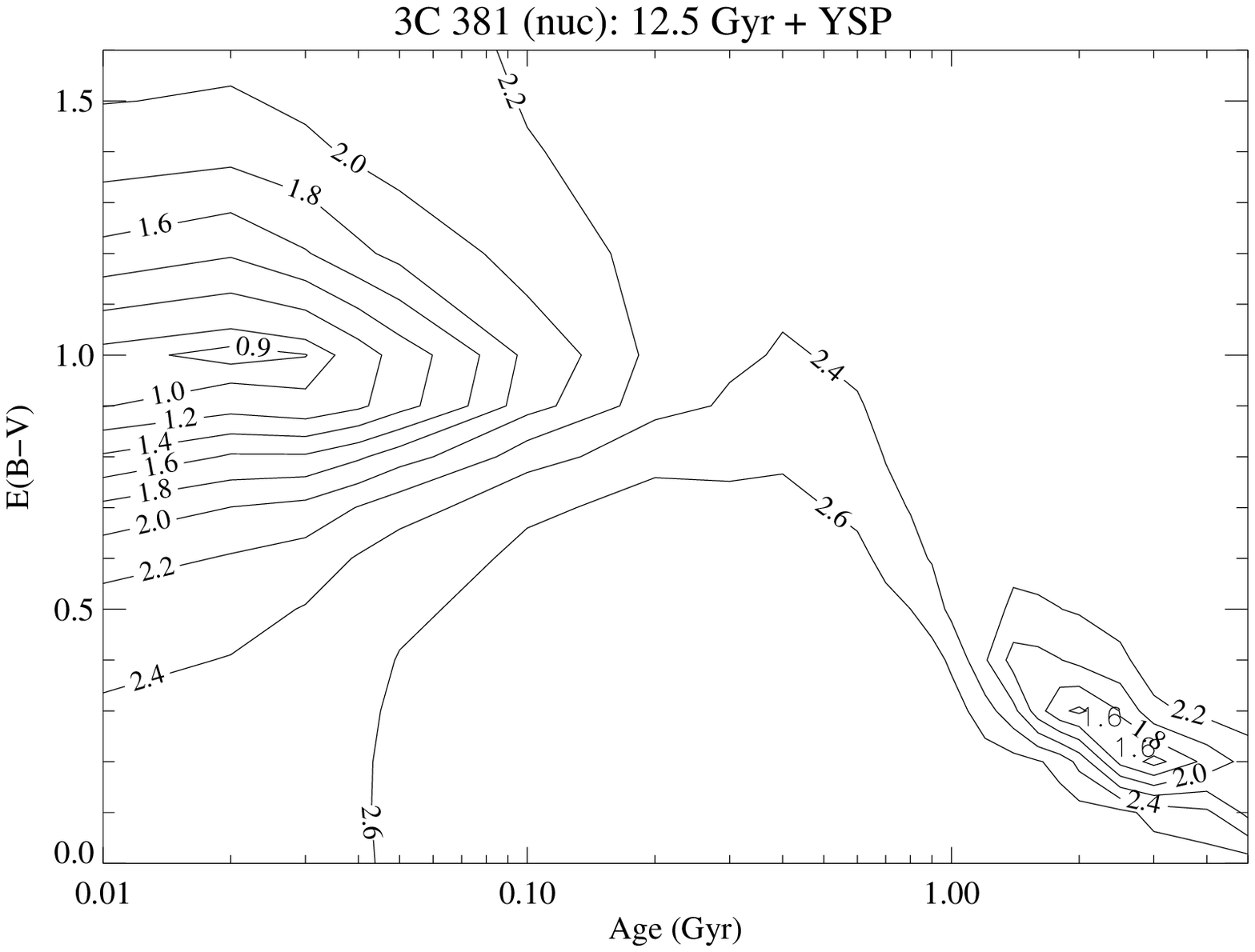,width=7cm,angle=0.} \\
\hspace*{-1.2cm}\psfig{file=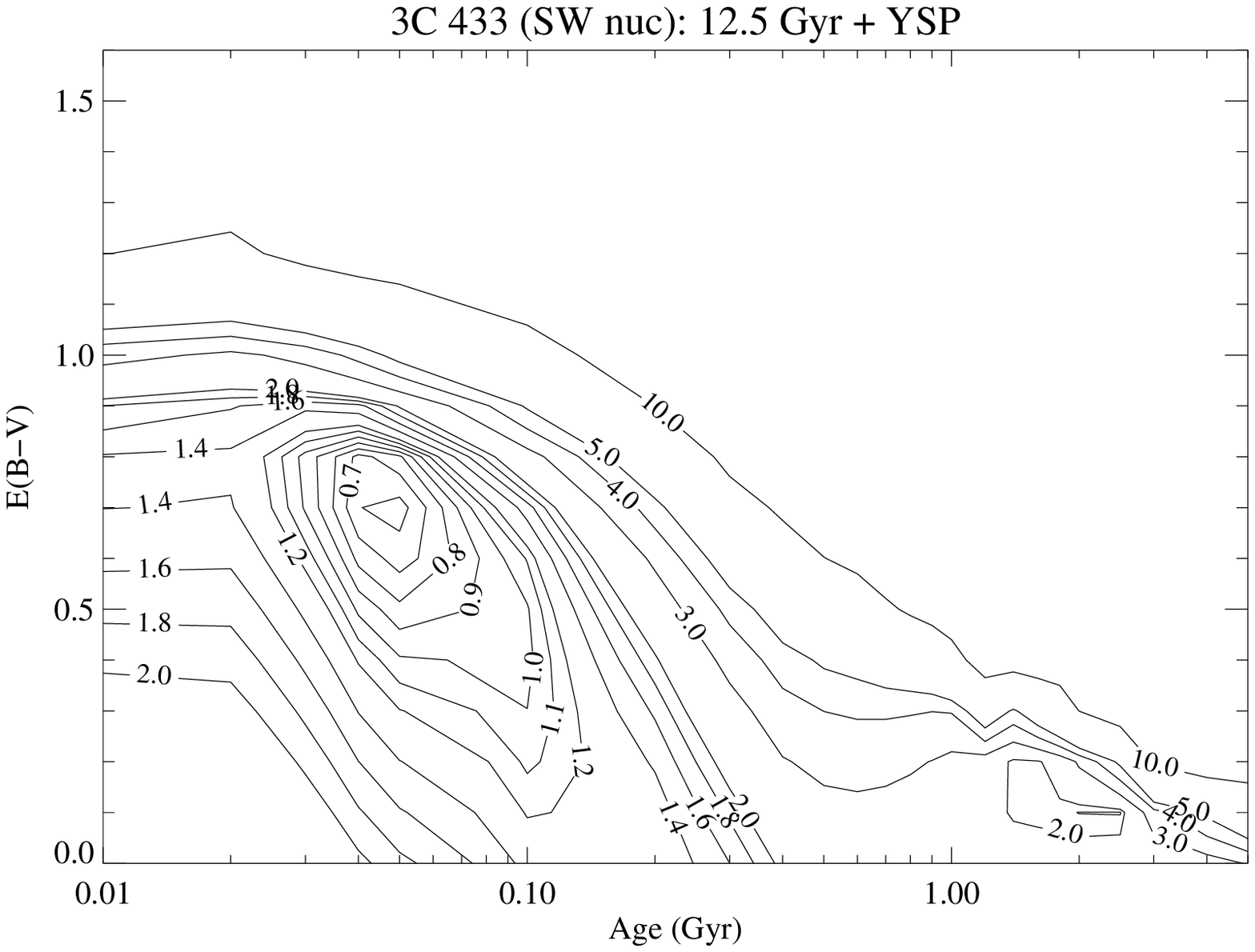,width=7cm,angle=0.}&
\hspace*{-0.8cm}\psfig{file=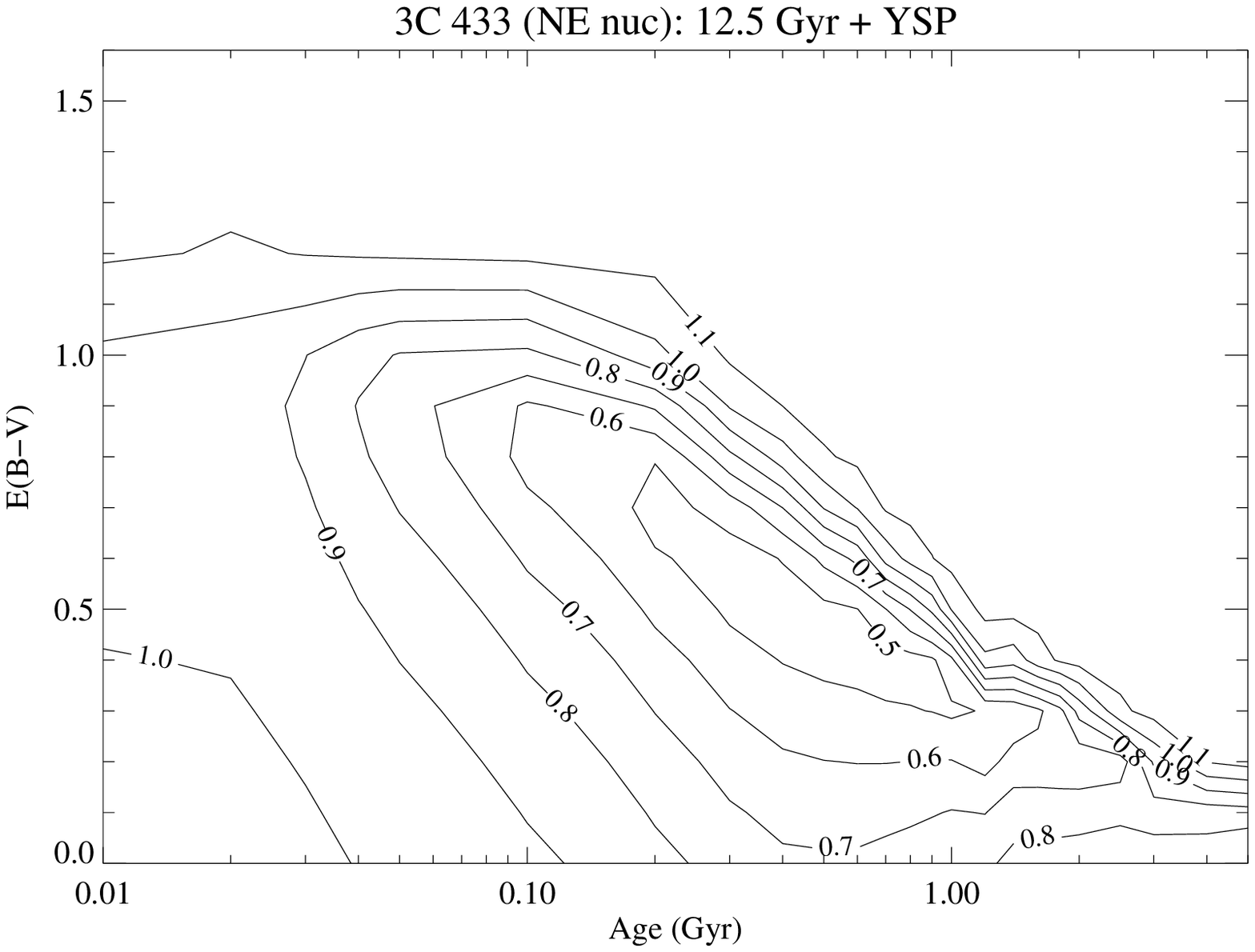,width=7cm,angle=0.}&
\hspace*{-0.8cm}\psfig{file=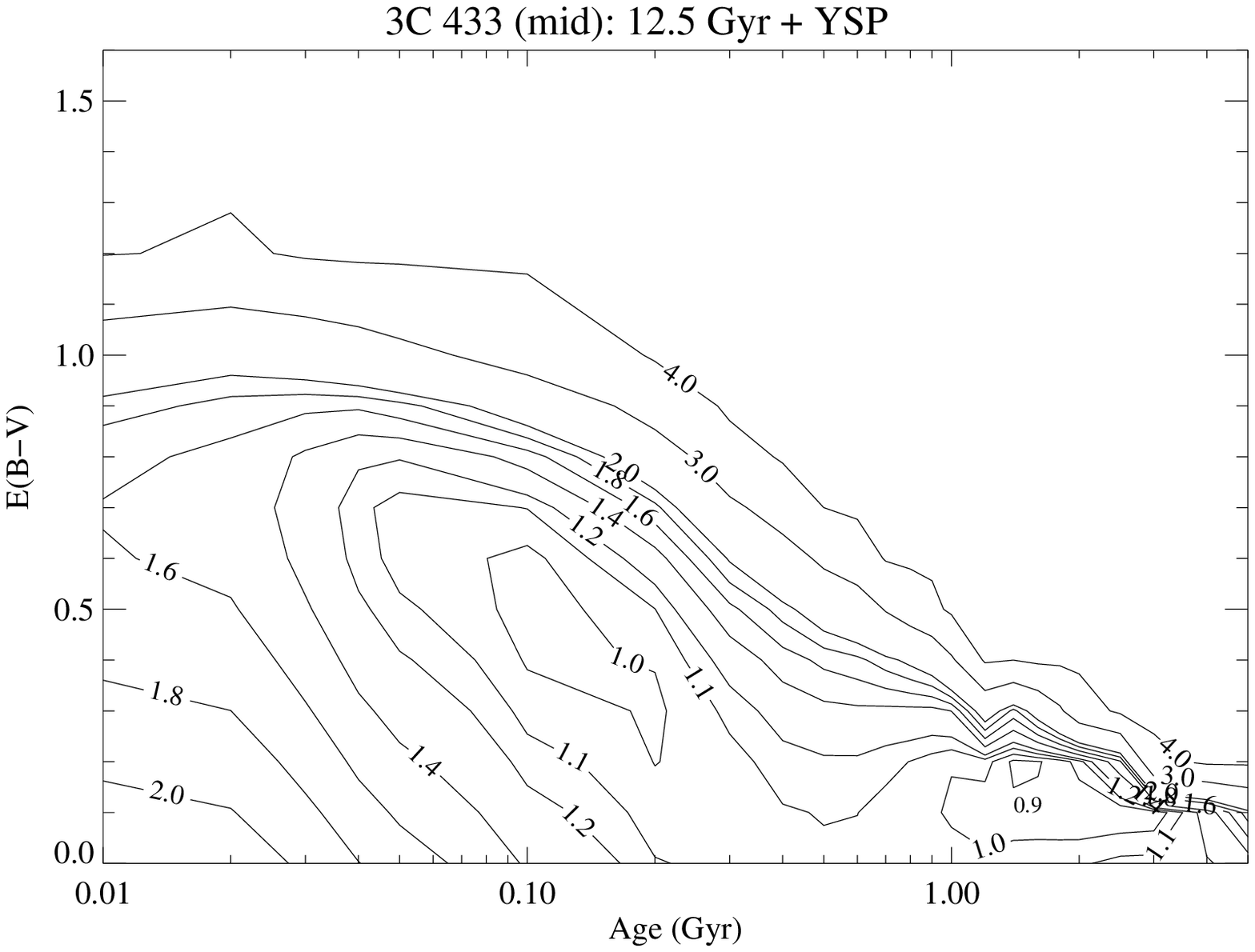,width=7cm,angle=0.}\\
\end{tabular}
\caption[]{Contour plots of the reduced chi squared space (\chisq) for
all combinations of age (x axis) and reddening (y axis) of the YSP for 
  all apertures requiring a YSP componet. All plots
assume the {\protect{\citealt{seaton79}}} reddening law {\it except}
for the second (labelled) plot for the nuclear aperture of NGC 612 which uses the 
\protect\citealt{calzetti00} reddening law. For descriptions of the
plots see Section {\ref{sect:cont}}. In general, 
we show one contour plot for
  each aperture. For apertures where a power-law component is
  expected, the plot is for models with an OSP, a
  YSP and a power-law component. For apertures in which a power-law
  component is not expected (e.g. due to location, the weakness of
  emission lines), the contour plot is for models with an OSP and a
  YSP only. Where the contours in the region of the 
  minimum are crowded, a zoom of this region is shown inside the
  plot. Note, a contour plot is not shown for
  PKS 0039-44. Two plots for the nuclear aperture of NGC 612 are
  presented, the first using the \protect\citep{seaton79} extinction
  law and the second using the \protect\citep{calzetti00} law. These
plots illustrate how the  different reddening laws may influence the result.
 The plots clearly show that,
  whilst the details of the models may vary, the differences between
  the reddening laws do not change the overall result within a certain
  tolerance.  We discuss the different
  reddening laws further in Section {\ref{sect:reddening}} and Figure
  {\ref{fig:reddening}}  }
\label{fig:contours}
\end{minipage}
\end{figure*} 
\setcounter{figure}{2}
\begin{figure*}
\begin{minipage}{180mm}
\begin{tabular}{ccc}
\hspace*{-1.2cm}\psfig{file=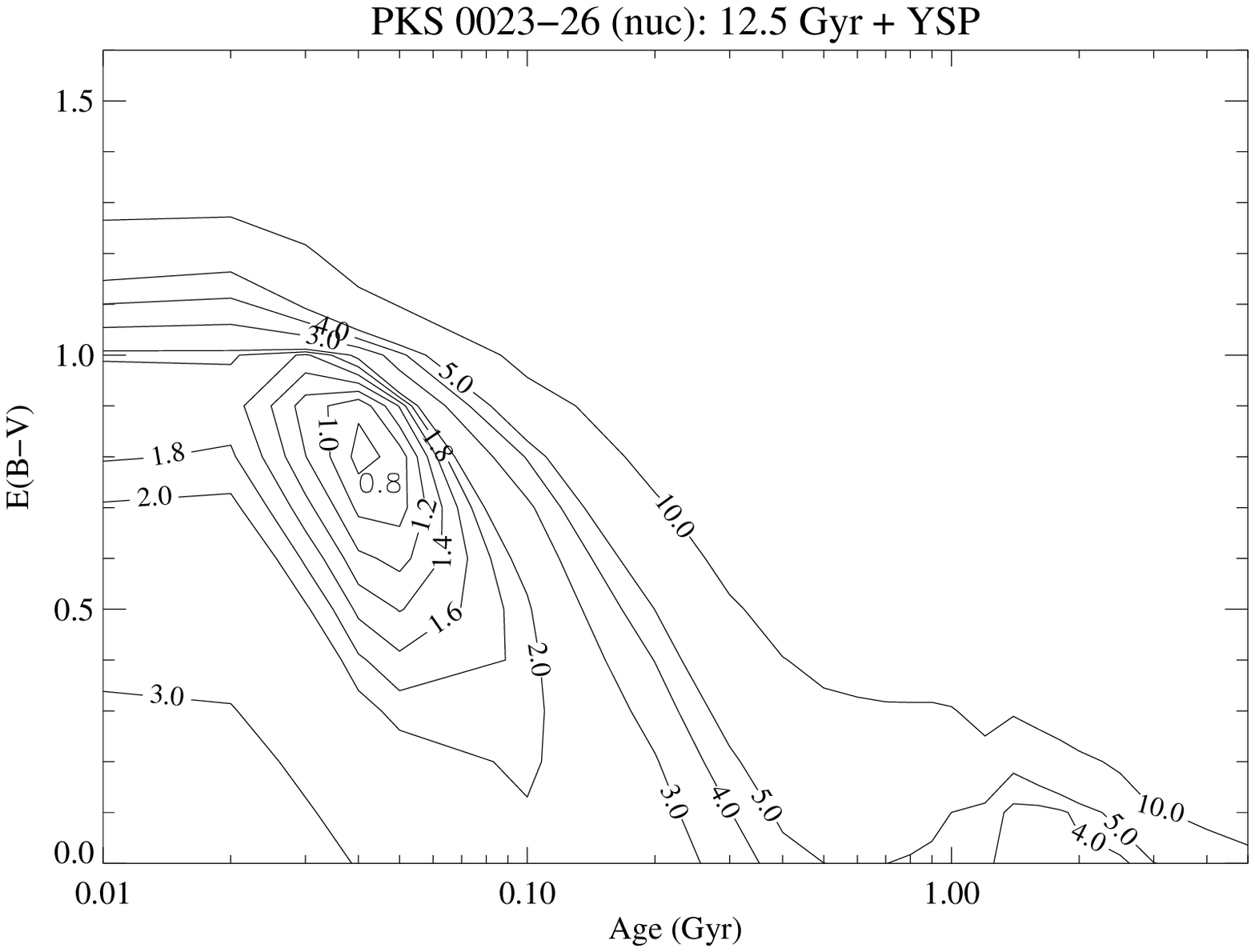,width=7cm,angle=0.} &
\hspace*{-0.8cm}\psfig{file=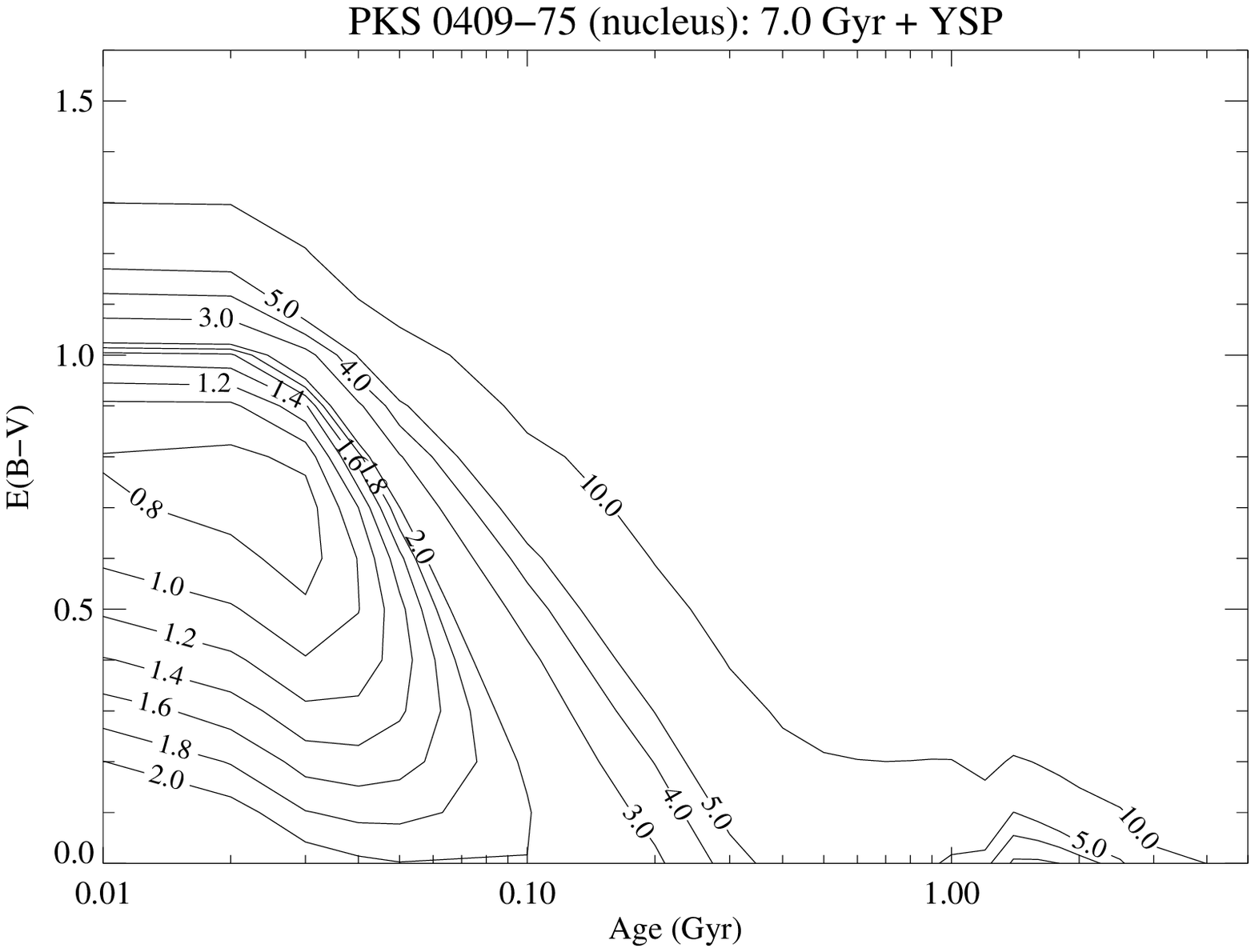,width=7cm,angle=0.}&
\hspace*{-0.8cm}\psfig{file=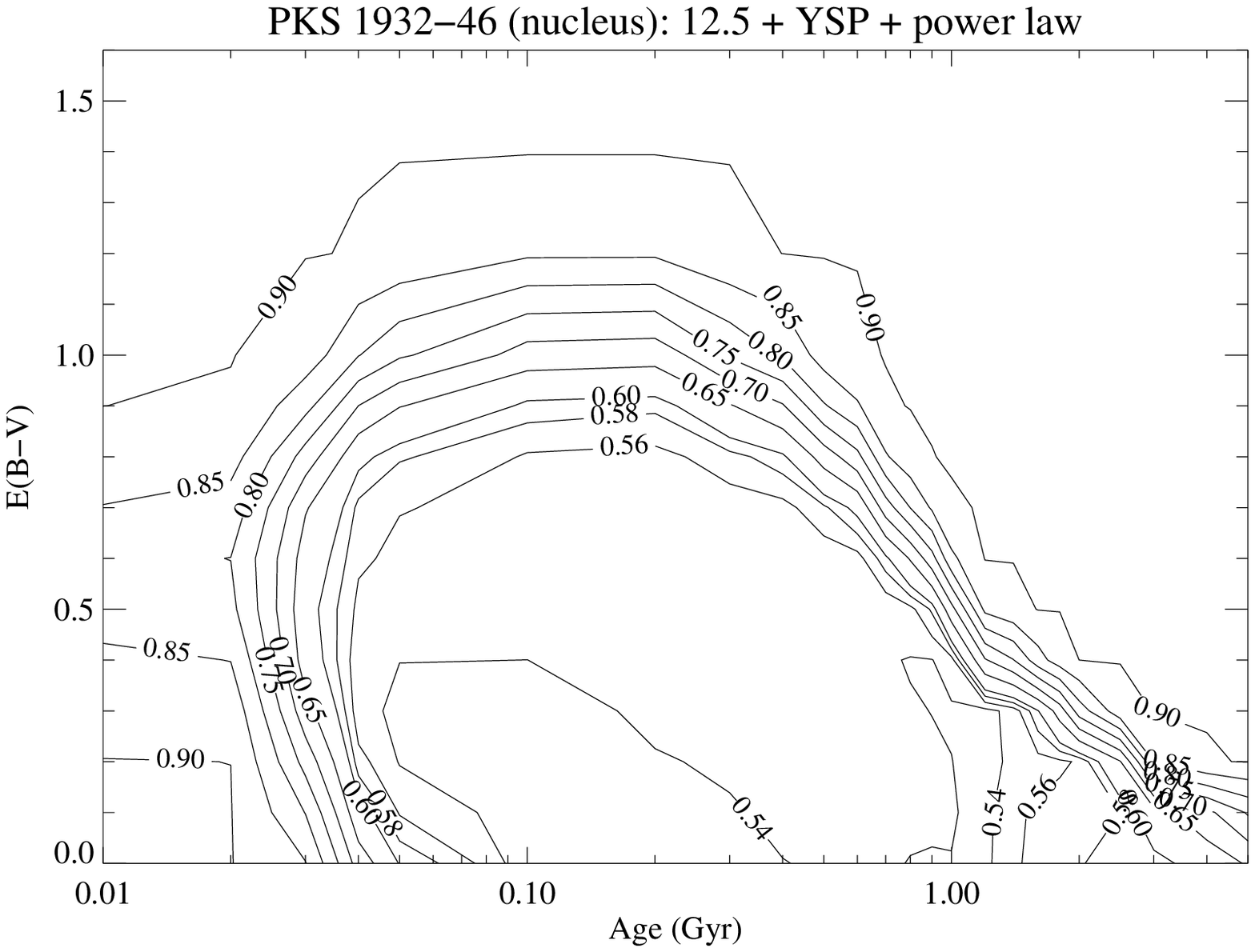,width=7cm,angle=0.}\\
\hspace*{-1.2cm}\psfig{file=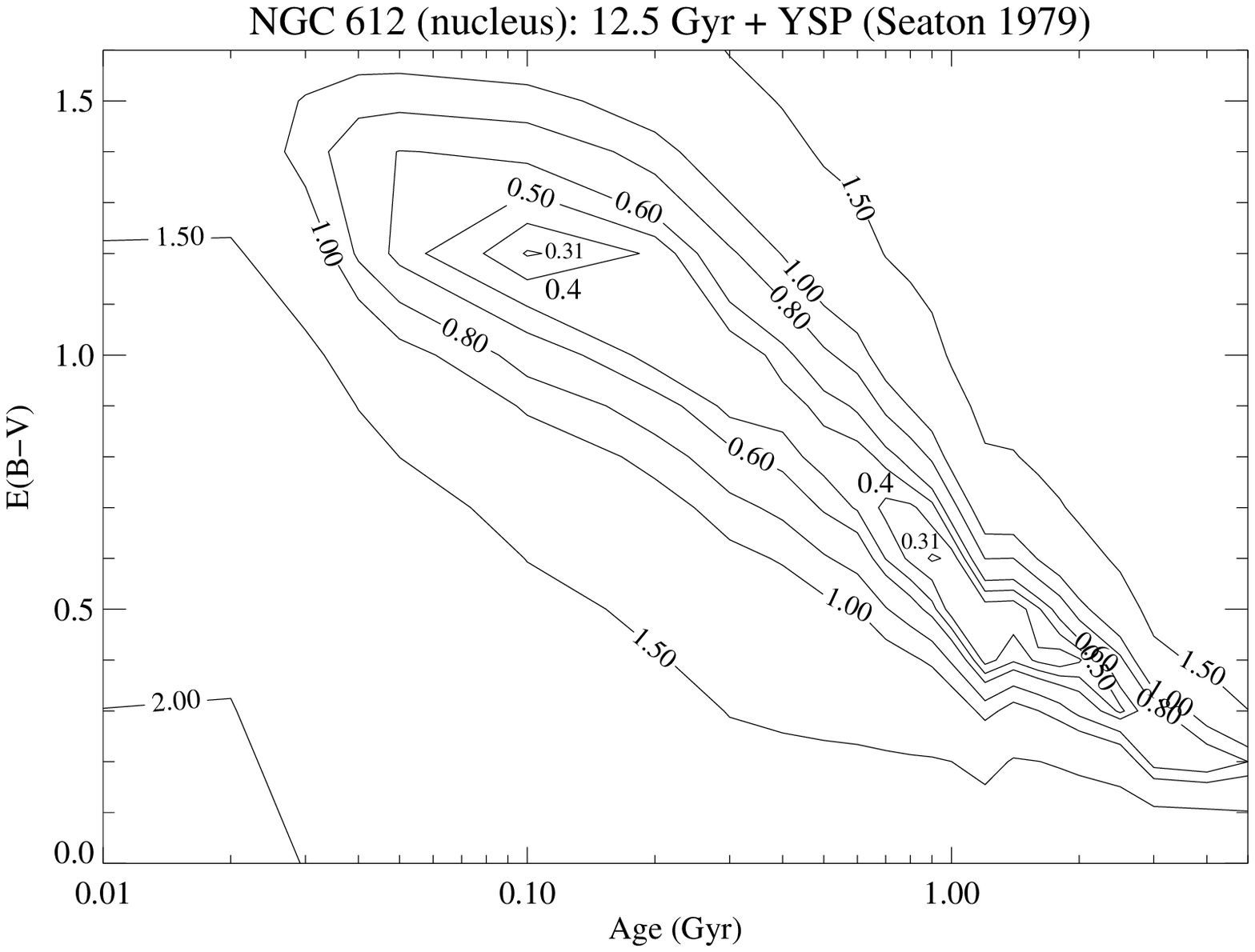,width=7cm,angle=0.} &
\hspace*{-0.8cm}\psfig{file=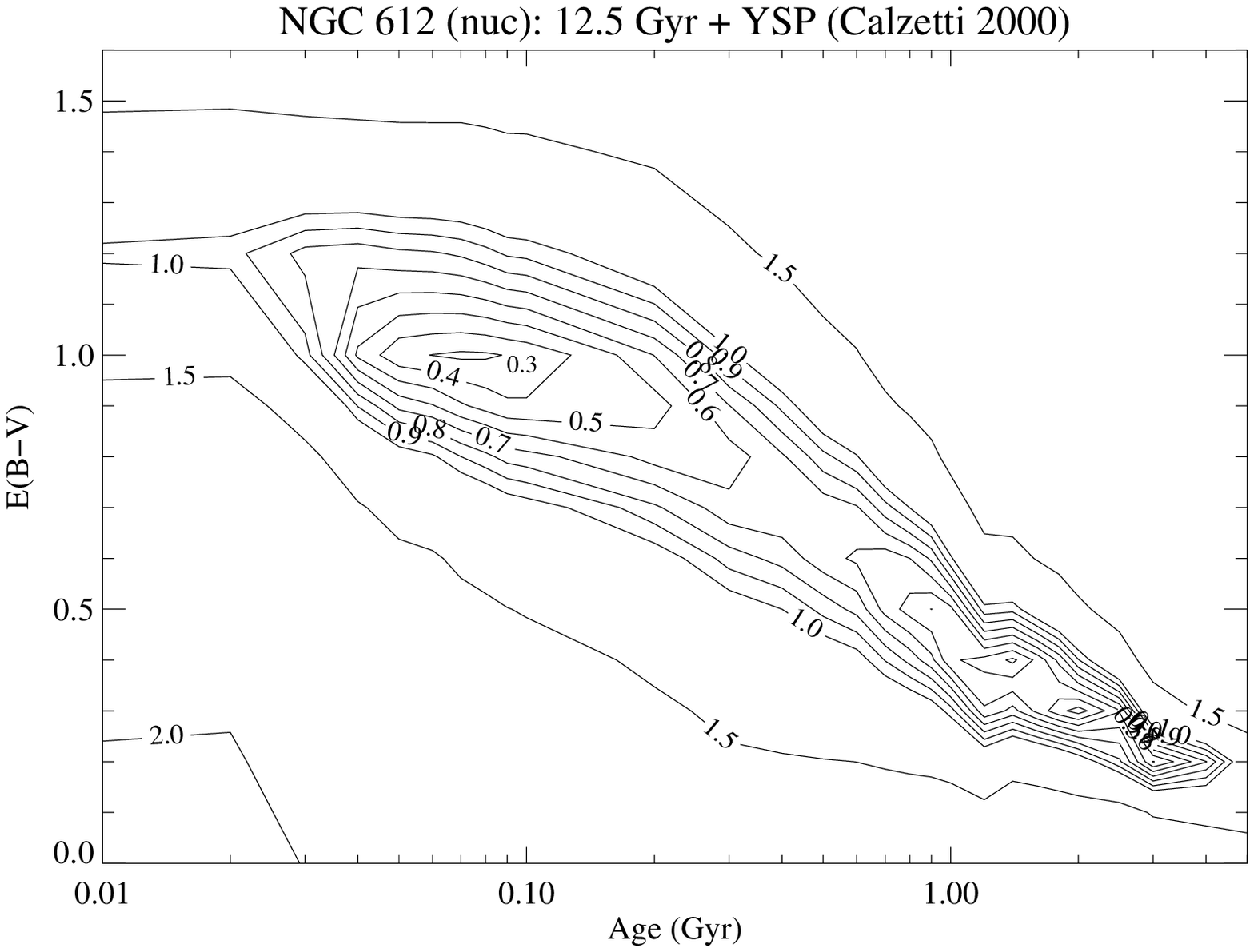,width=7cm,angle=0.} &

\hspace*{-0.8cm}\psfig{file=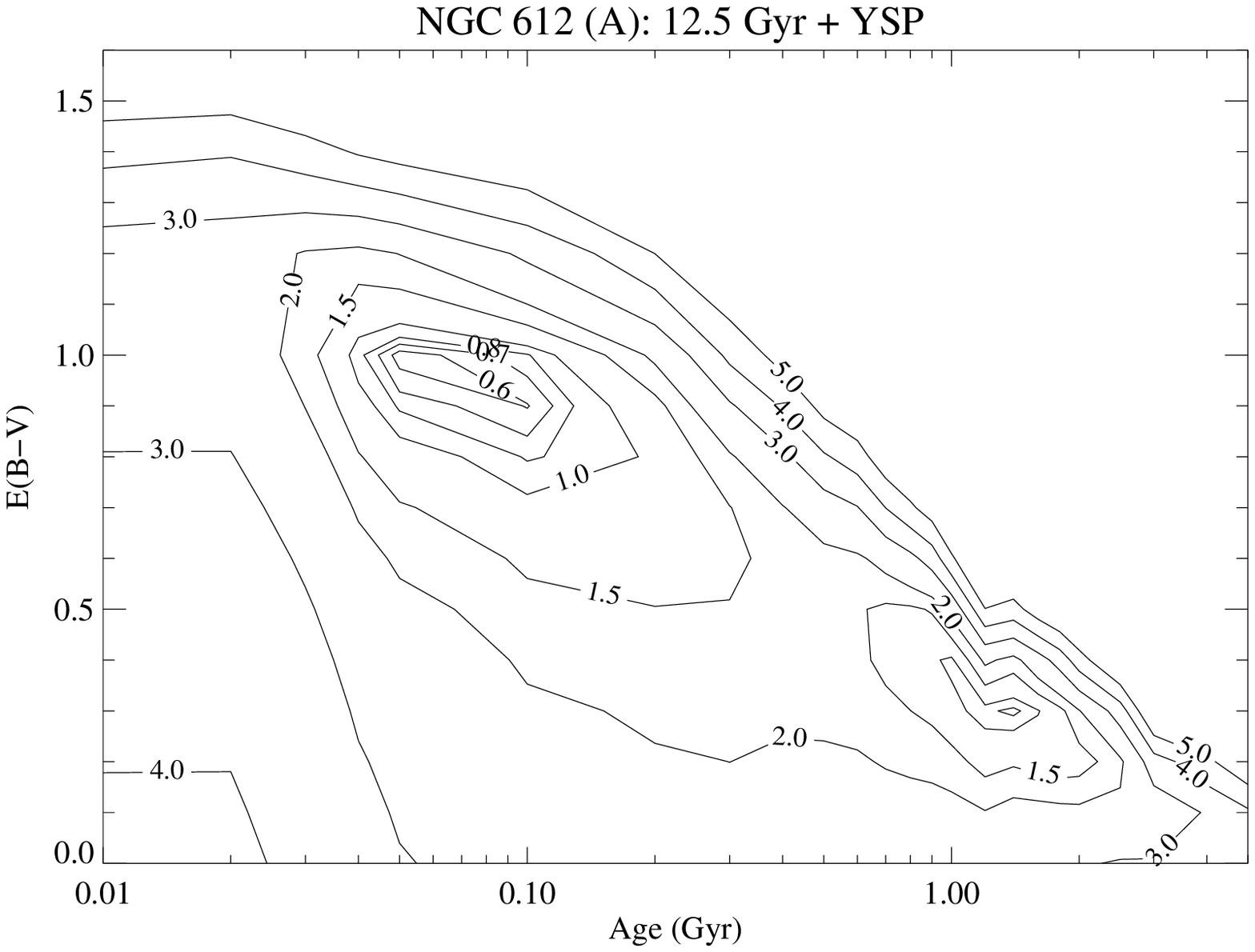,width=7cm,angle=0.} \\
\hspace*{-1.2cm}\psfig{file=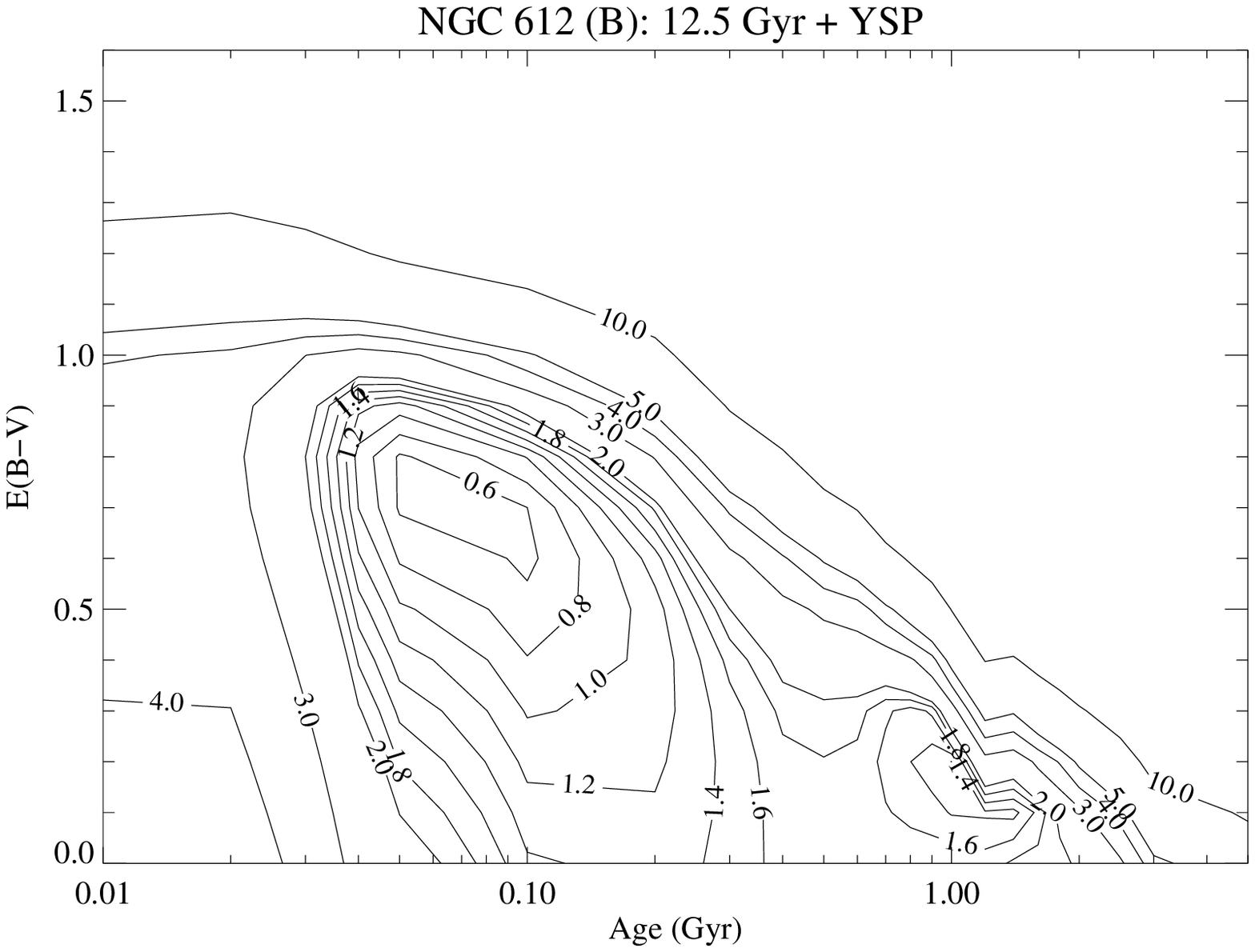,width=7cm,angle=0.}&
\hspace*{-0.8cm}\psfig{file=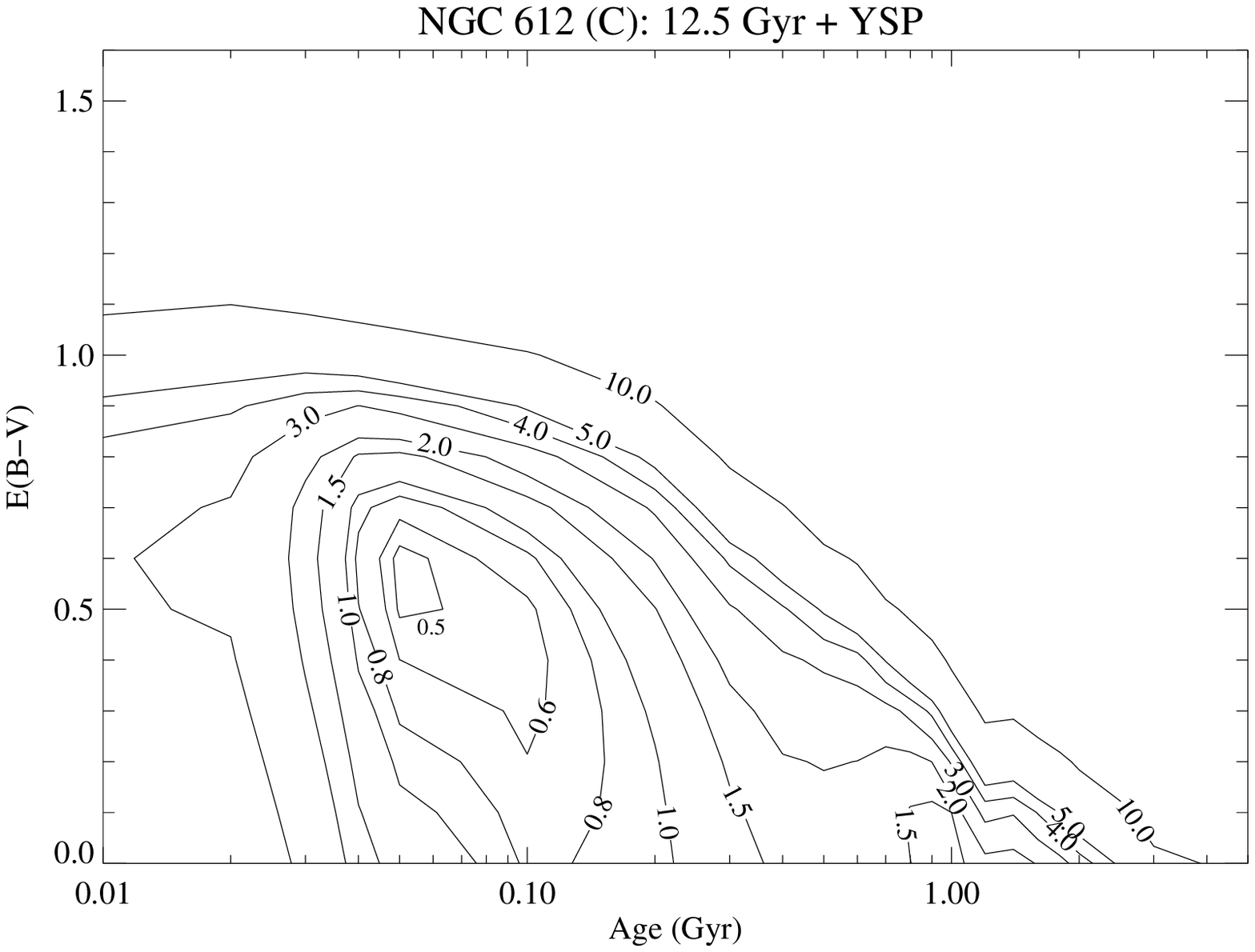,width=7cm,angle=0.} &
\hspace*{-0.8cm}\psfig{file=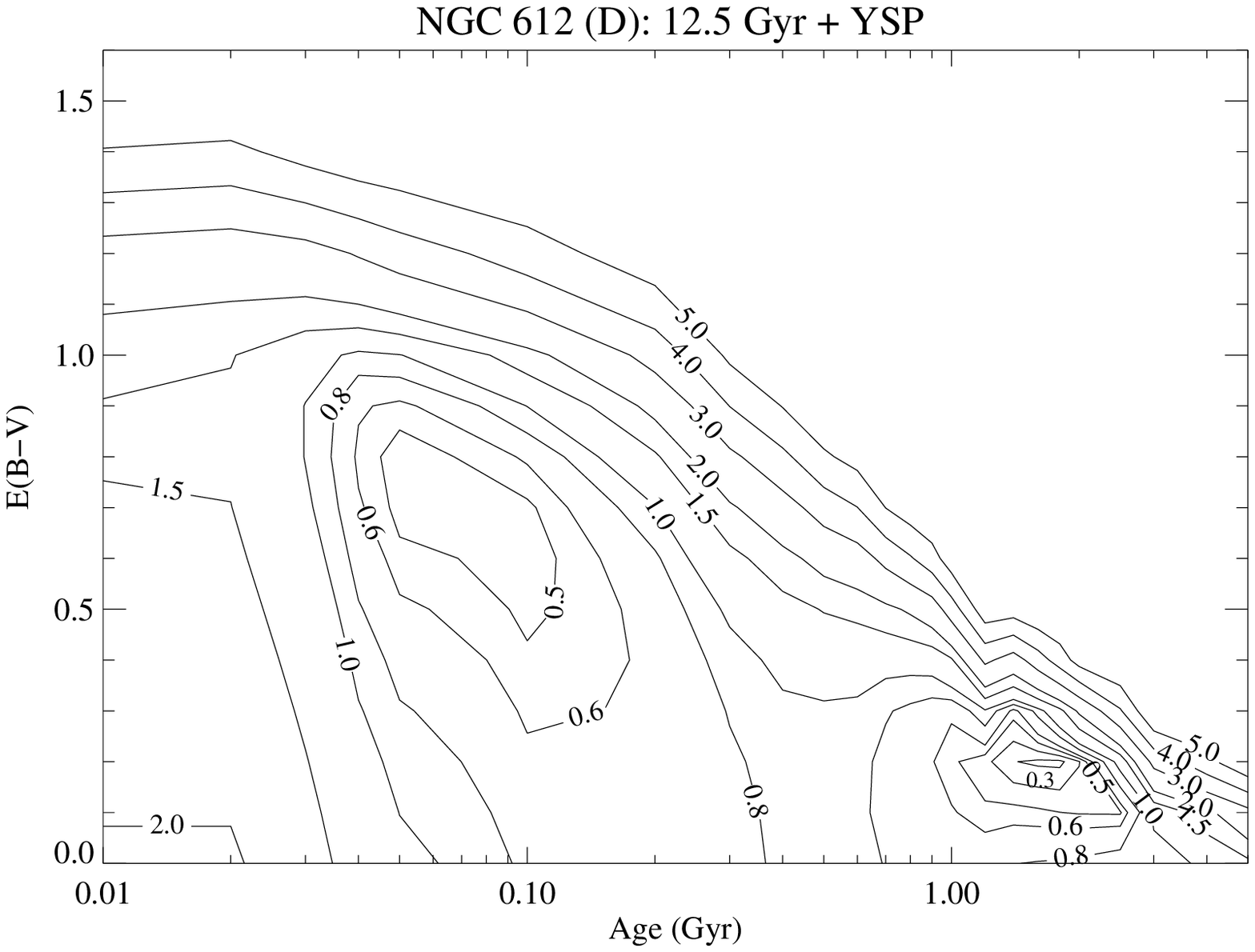,width=7cm,angle=0.} \\
\hspace*{-1.2cm}\psfig{file=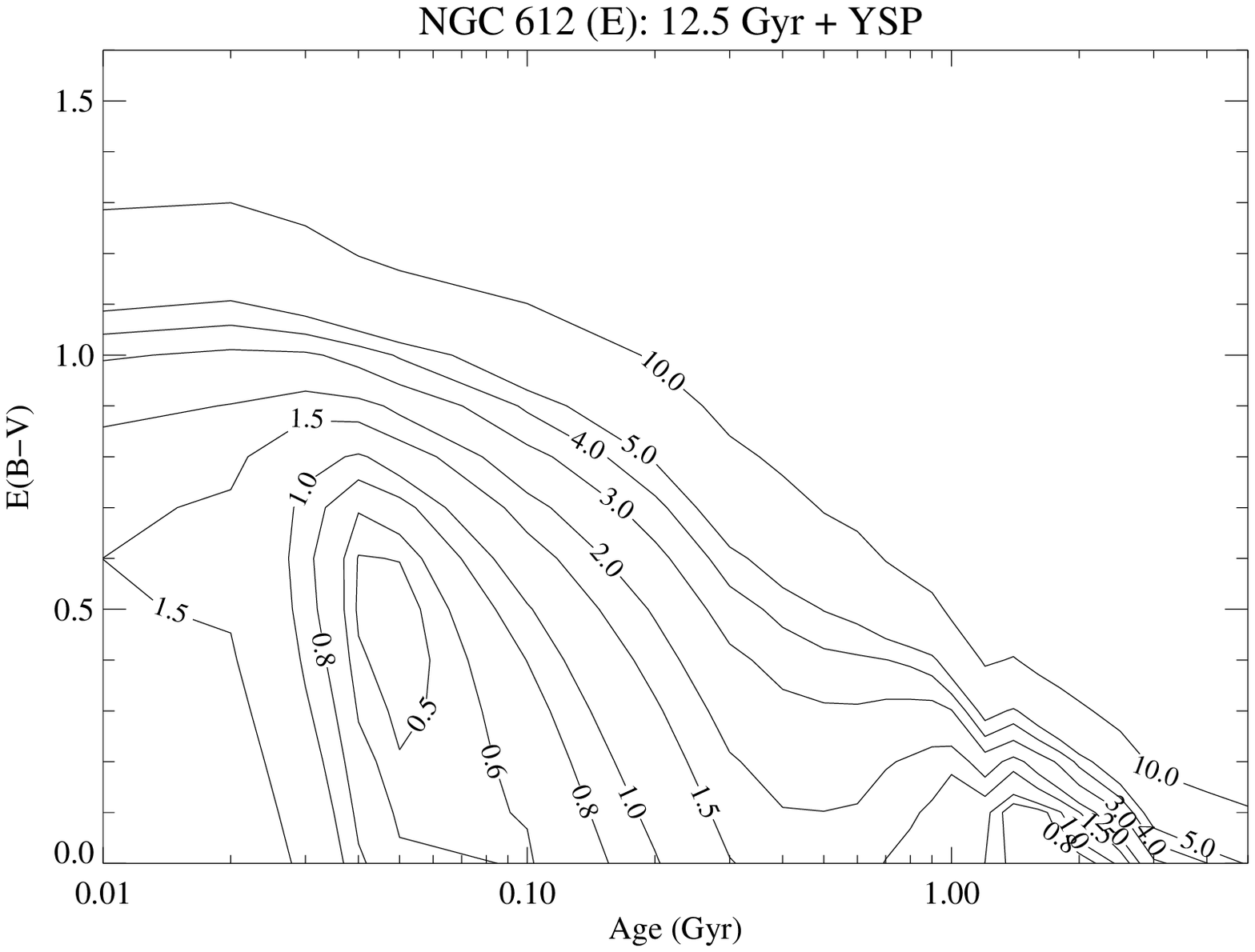,width=7cm,angle=0.} &
\hspace*{-0.8cm}\psfig{file=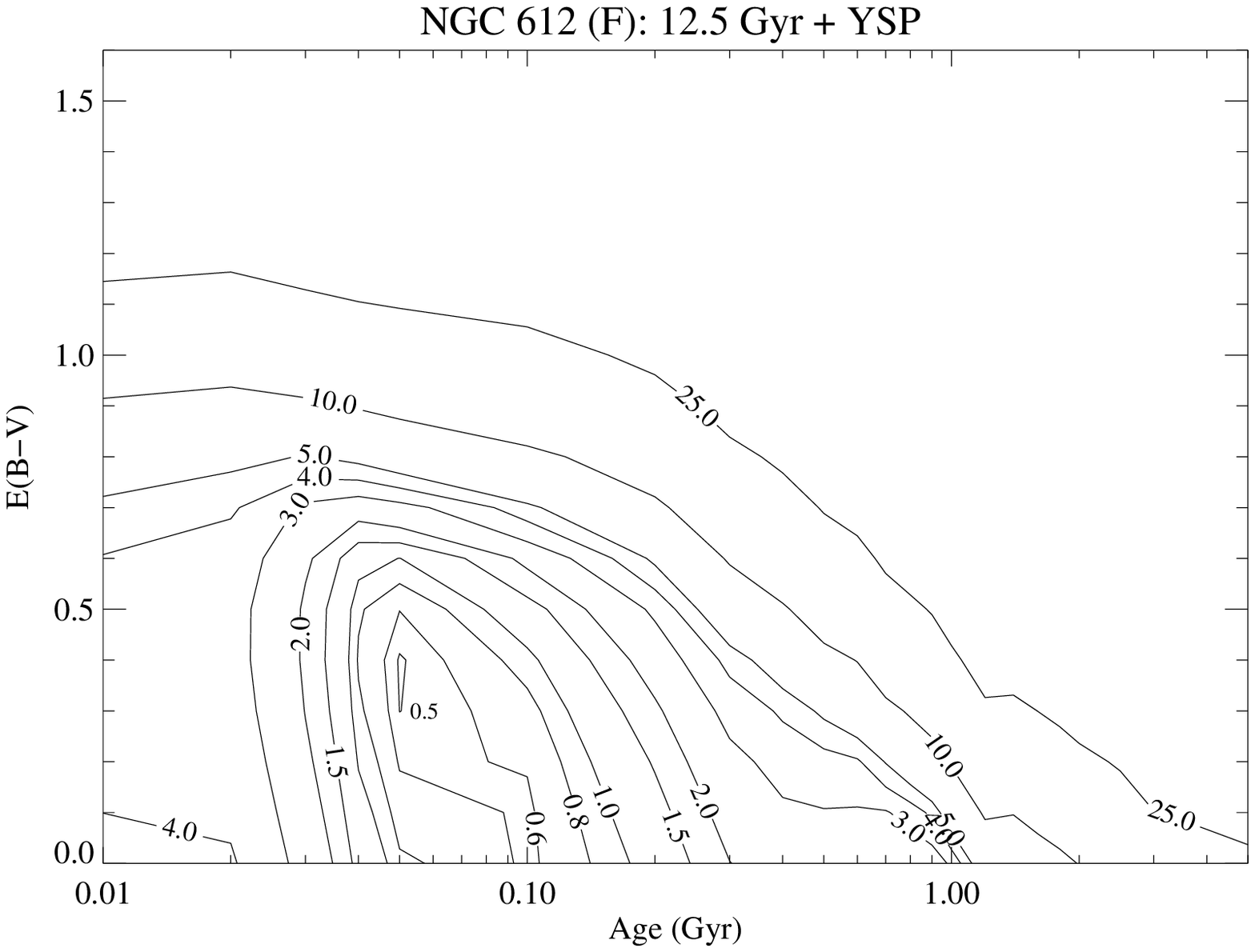,width=7cm,angle=0.} &
\hspace*{-0.8cm}\psfig{file=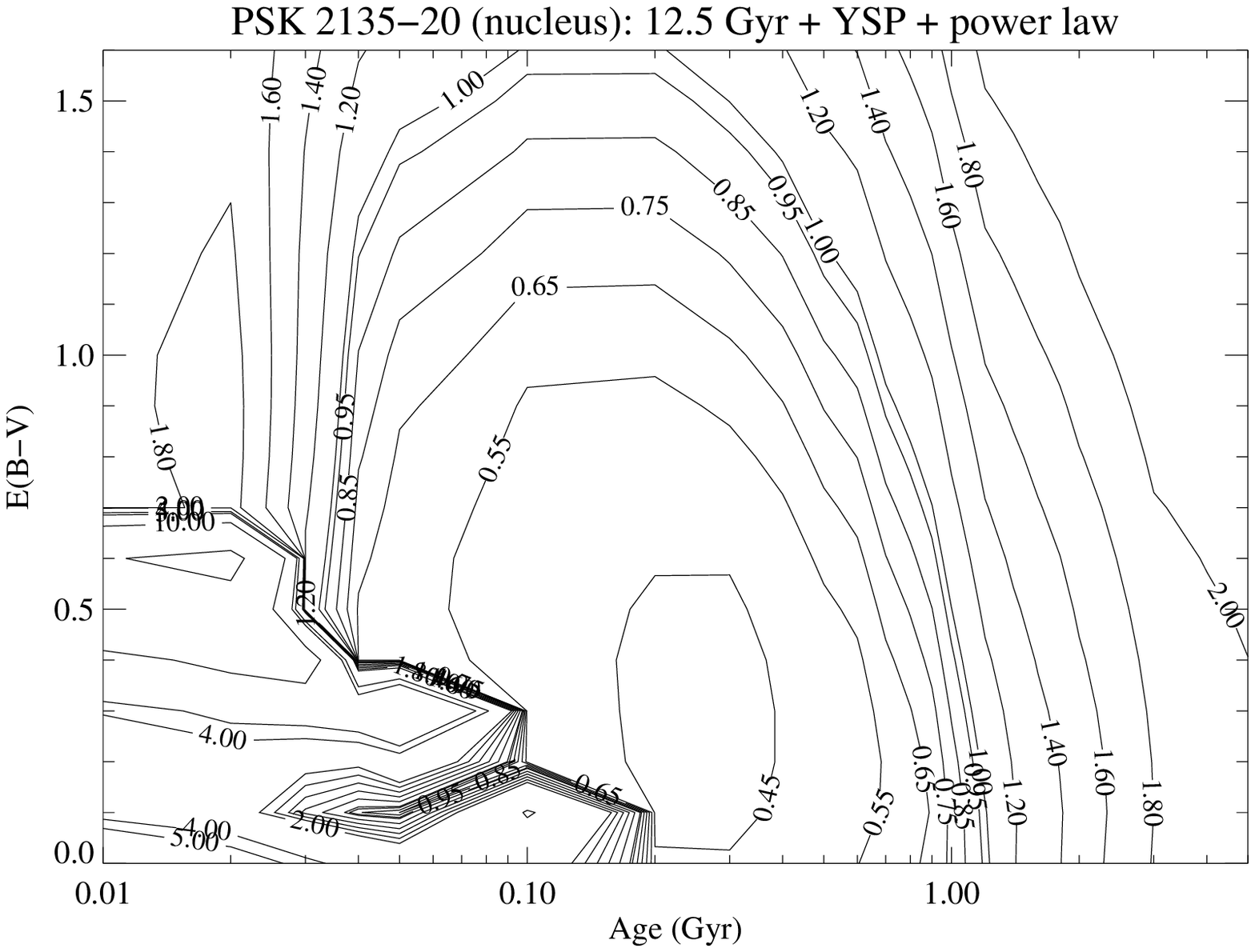,width=7cm,angle=0.} \\
\\
\end{tabular}
\caption[]{Contour plots {\it continued}. }
\end{minipage}
\end{figure*}

\subsubsection{Hydra A (3C 218)}
3C 218 is one of the most powerful radio galaxies in the local 
universe, however, despite having a radio luminosity more than an order
of magnitude
larger than the characteristic break luminosity between the FR I and
FR II classes, its morphology is edge darkened --- more characteristic of
FR I sources (e.g. \citealt{baum88}).

3C 218 is  optically identified with Hydra A \citep{simkin79},
the dominant cD2 galaxy in the centre of a poor but X-ray luminous
(e.g. \citealt{david90}) cluster.  Hydra A has a double nucleus with
the second nucleus lying $\sim$ 7 arcsec SE of the active nucleus,
within the common envelope \citep{hansen95}.  
Like a significant number of radio galaxy hosts, the optical
colours of 3C~218 are blue in comparison to normal elliptical galaxies (see also
Figure \ref{fig:d4000}). Several
authors have attributed these colours to a massive (10$^8$-10$^9$
M$_{\odot}$) burst of star formation occuring $\sim$ 50Myr ago
(e.g. \citealt{mcnamara95,hansen95}). Optical spectroscopy supports
this view with the detection of a prominent Balmer series in the
nuclear spectrum
\citep{hansen95,melnick97,aretxaga01,wills04}. Using a variety of
line and break indices, \citet{aretxaga01} attribute the blue
bulge-subtracted continuum to a 0.007-0.04 Gyr old starburst dominated
by B-type giant/supergiant stars.  Spectral synthesis modelling by
\citet{wills04} finds results consistent with \citet{aretxaga01}
although their best-fit model suggests the starburst occurred at the
earlier end of this range, 0.05 Gyr ago.

In this paper we have repeated the optical spectral modelling of
\citet{wills04} using new, deeper spectra with wider spectral coverage,
and using  
the latest generation of high resolution spectral models of
\citet{bruzual03}.

As discussed above, and shown in Figure {\ref{fig:d4000}}, 3C 218 has a
clear UV excess and the prominent Balmer series makes it a strong
starburst candidate. Indeed, as reported by \citet{wills04}, we find
that the SED cannot be adequately modelled without a YSP
component. Hence, we have modelled the SED with an OSP and a YSP both
with and without a power-law component. In Figures {\ref{fig:contours}}
and {\ref{fig:SED}}
we only show the case of an OSP plus a YSP -- in common with many other
cases, adding a
power-law gives a similar shape to the \chisq~space with a similar
position for the \chisq~minimum although the number of solutions
giving a viable fit to the overall shape of the SED is dramatically
increased.  Whilst including a power-law  component gives many viable
fits, there  is no strong observational evidence for a power-law
component in 3C  218. Despite harbouring a powerful radio source, the optical
emission lines are  relatively weak (we have not corrected for the
nebular continuum) and there is no strong point
source in any of the published images (e.g. \citealt{mcnamara95} --
ground based). 

For fits without a power-law component, one clear minimum is observed
at 0.05~Gyr with significant reddening (\ebv~= 0.4; \chisq~=
1.2) accounting for 50\% of the continuum flux. In fact, only YSPs
with ages 0.05-0.1 Gyr with similar reddening 
provide acceptable fits to the SED modelling and, even when the
detailed fits are used, it is difficult to distinguish between
them. By combining the fits to the lines and the continuum around
them, overall the best fit is for the OSP plus a 0.05~Gyr YSP. 
This result is entirely consistent with the findings of
\citet{wills04}, although our results require a significantly larger
contribution of the YSP (c.f. 10-30\% in \citealt{wills04}), and
towards the older end of the range reported by \citet{aretxaga01}. 

Including a power-law component dramatically increases the number of
models providing a good fit, marginally improves the \chisq~and shifts
the minimum to slightly older ages; OSP + YSP (0.1 Gyr, \ebv~= 0.2,
47\% contribution). 
Hence, for the nuclear region of Hydra A we find the best fitting
model to be an old 
(12.5 Gyr) stellar population (45\%) with a YSP (0.05 Gyr, \ebv~= 0.4,
50\% contribution). This is entirely consistent with the findings of 
both \citet{wills04} (the difference in flux percentage is likely to
be due to differences in the seeing and extraction aperture) and
towards the older end of the range reported  by \citet{aretxaga01}. 
\citet{wills04} also attempted to model the spectrum with  {\it
  younger} age  `old' stellar populations. They found the best fit
  comprised a 2 Gyr population (89\%) with a 0.05 Gyr YSP
  (11\%). We have also attempted to model the nuclear spectrum of
  Hydra A with `younger' old populations and found this made no
  difference to our overall result.

\subsubsection{3C 236}
The radio source 3C 236 and its host galaxy are peculiar at both 
optical and radio wavelengths. The radio source exists both on the
large scale (a classical double source extends to $\sim$ 4 Mpc, making
it the largest known radio source, e.g. \citealt{schilizzi01}) and yet
the nuclear regions harbour a 
second, compact (2 kpc) radio source suggesting it belongs to the class of
`double-double' radio sources
(e.g. \citealt{kaiser00}).   On the basis
of radio spectral ageing techniques and the detection of different
star forming regions, two of which are relatively young (of order
10$^7$ yr) and two which are significantly older (of order
10$^8$-10$^9$ yr, comparable to the age of the giant radio source),
\citet{odea01} argue that this strange radio morphology is
the result of a period of fuel starvation of the AGN which lasted
around 10$^7$ years before new fuel was provided and the radio source
restarted. Note, the relative locations of the radio jets and the star
forming regions preclude jet-induced star formation, but the timescales
are consistent with a common origin of fuel such as the infall of gas
to the host galaxy \citep{odea01}. 

At optical wavelengths the deep ground-based images of Smith \& Heckman
(1989) reveal the presence of shells and twisted isophotes
that are suggestive of a past merger or interaction 
with a companion galaxy. On smaller scales, HST imaging shows the isophotes of
the host galaxy to be 
flattened and inclined, with a broad asymmetric dust lane on scales of
a few kpc which is itself misaligned with the inner disk
(e.g. \citealt{martel99,dekoff00}). They also show a second,
$\sim$1~kpc scale dust 
feature \citep{dekoff00}. In the UV, three clear knots of emission are
detected which trace an arc coinciding with the SE edge of the larger dust
lane \citep{allen02,odea01}. 

The integrated nuclear spectrum of 3C 236 has a clear  UV excess
(Figure \ref{fig:d4000}).  
We find that the overall SED and the
stellar absorption features can be satisfactorily reproduced (\chisq~=
1.1) with an old (12.5 Gyr) stellar population (87\%) and a red power-law
component ($\alpha$ = 3.6, 12\% contribution) and this model is shown in
Figure {\ref{fig:SED}}. 

The power-law component required in this model is red. 
Following our `decision tree' (Figure {\ref{fig:decisiontree}}), a red
power-law component is unlikely to be a scattered AGN component and is
more likely to be either direct AGN light or a young reddened YSP
component. The HST images of \citet{martel99,dekoff00} and
\citet{odea01} do not show a point source in the nucleus. Hence a
large direct 
AGN light contribution is unlikely. Further, after subtracting the
best-fit model we see no evidence for broad emission line components
also expected if we are viewing the AGN directly. Hence, the UV
excess is likely to be dominated by a very young YSP component which
is heavily reddened. 

Figure {\ref{fig:contours}} shows the \chisq~space for an OSP plus a
YSP component. Two clear minima are observed for for YSPs with ages 0.1 Gyr
(\ebv~= 1.2, \chisq~= 0.58) and 1.4 Gyr (\ebv~ = 0.5, \chisq~= 0.48),
although viable models (\chisq~$<$ 1.0) to the SED are  found across
all ages.  Detailed comparisons with the stellar absorption features
rule out models with ages $\gtrsim$1.0~Gyr because such models
overpredict the depth  
of CaII~K feature, 
although it is difficult to discriminate further. An example of a two
component model is shown in 
Figure {\ref{fig:SED}}: 12.5 Gyr (84\%) plus 0.05 Gyr YSP (15\%,
\ebv~= 1.4) with \chisq~= 0.78. Including a power-law component 
expands the number of viable fits. In summary, for the OSP
plus YSP 
models, both with and without a power-law component, the YSP
contributes typically 10-30\% of the flux in the normalising bin for
YSP ages $<$ 1.0 Gyr.  Hence, our spectral fitting results are
consistent with both age estimates derived from HST photometry of 
\citet{odea01} -- $\sim$ 10$^7$ yr and $\sim$ 10$^{8-9}$ yr -- for 
different bright knots in the halo. 

\subsubsection{3C 285}
The double-lobed FRII radio galaxy 3C 285 is identified with an
elliptical galaxy, 
the brightest in a group of galaxies \citep{sandage72}. A chaotic
morphology consisting of a distorted
S-shaped envelope aligned with another galaxy  $\sim$40 arcsec to the
NW \citep{heckman86,roche00}, tidal tails and fans, irregular dust
lanes and knots 
of emission \citep{allen02} and a $B-V$ colour significantly bluer
than a normal elliptical galaxy \citep{sandage72} all
provide evidence that the galaxy has undergone a recent merger
or interaction with a companion.  The UV/optical continuum of 3C 285
has been previously modelled by \citet{aretxaga01}. By fitting the
4000\AA~break, they conclude the blue light in the inner 2 arcsec is
dominated by a starburst component
with an age of 10-12
Myr. In addition, \citet{vanbreugel93} have investigated the knot of
UV emission located halfway between the nucleus and the eastern
radio lobe and found the $UBV$ colours and the 4000\AA~break to be
consistent with a 70 Myr starburst, that may have been triggered by the
passing radio jet.

The nuclear  spectrum of 3C 285 
 is
highly indicative of a major contribution to the continuum
flux by light from young stars -- the Balmer sequence is clearly
observed in absorption and the galaxy displays a clear UV excess (Figure
\ref{fig:d4000}). 
Indeed, SED modelling
excluding a YSP contribution failed to provide an adequate fit to
the overall shape of the continuum spectrum. 

Whilst it is necessary to have a YSP component to model the SED, good
fits are obtained both with and without an additional power-law
component. However, the emission lines are relatively weak and HST imaging does
not reveal a strong point source; rather, the
nucleus is obscured by one of the dust lanes \citep{allen02,madrid06}. 
Models including a power-law can therefore
 be excluded for two reasons: i) models requiring a very
young YSP do not require a power-law component and ii) models with an
older age YSP require a large power-law contribution. We have 
therefore focussed on models  which exclude a
power-law component. The contour plot for the \chisq~space for a 12.5
Gyr OSP  and a YSP is shown in Figure {\ref{fig:contours}}. 

3C 285 is an ideal example to highlight the
degeneracy in the solutions when relying on SED modelling
alone --  Figure {\ref{fig:contours}} shows two clear minima, one with
a young (0.04-0.2 Gyr), reddened (\ebv~$<$ 0.5) YSP contributing
30-50\% of the flux in the normalising bin and another for a
significantly older (1.0-3.0 Gyr), unreddened YSP (\ebv~$<$
0.2) contributing $\sim$100\% of the flux in the normalising
bin. 
Detailed comparisons reveal that, whilst the fits to the 
majority of the diagnostic lines are indistinguishable, the viable
models with an older YSP ($\gtrsim$ 1.0 Gyr) significantly overpredict
the strength of  the
CaII~K line.  
Further, models with YSPs $<$
0.1 Gyr do not fit either the absorption lines or the continuum in the
3700-4300\AA~region  or the G band as well as  other models.

Hence, by combining both techniques, it is likely that the
nuclear region of 3C 285 contains a significant YSP with age 0.1--0.5
Gyr and reddening 0.2 $<$ \ebv~$<$0.0,  contributing 33\%--39\% of
the continuum flux in the normalising bin.  An example of
a good fit is shown in Figure {\ref{fig:SED}}: 12.5 Gyr (72\%) + 0.2
Gyr with \ebv~=0.2 (33\%); \chisq~= 0.74. The age we
estimate is at least an order of 
magnitude larger than that found by \citet{aretxaga01} using the Balmer
and 4000\AA~breaks and line indices (10-12 Myr). \citet{aretxaga01}
give no indication of the scale of the YSP contribution to the flux.

\subsubsection{3C 321}
\begin{figure*}
\begin{minipage}{170mm}
\begin{tabular}{cc}
\multicolumn{2}{c}{\bf 3C 218 (Hydra A): 12.5 Gyr OSP + 0.05 Gyr YSP
  (\ebv~= 0.4)} \\
\hspace*{-1cm}\psfig{file=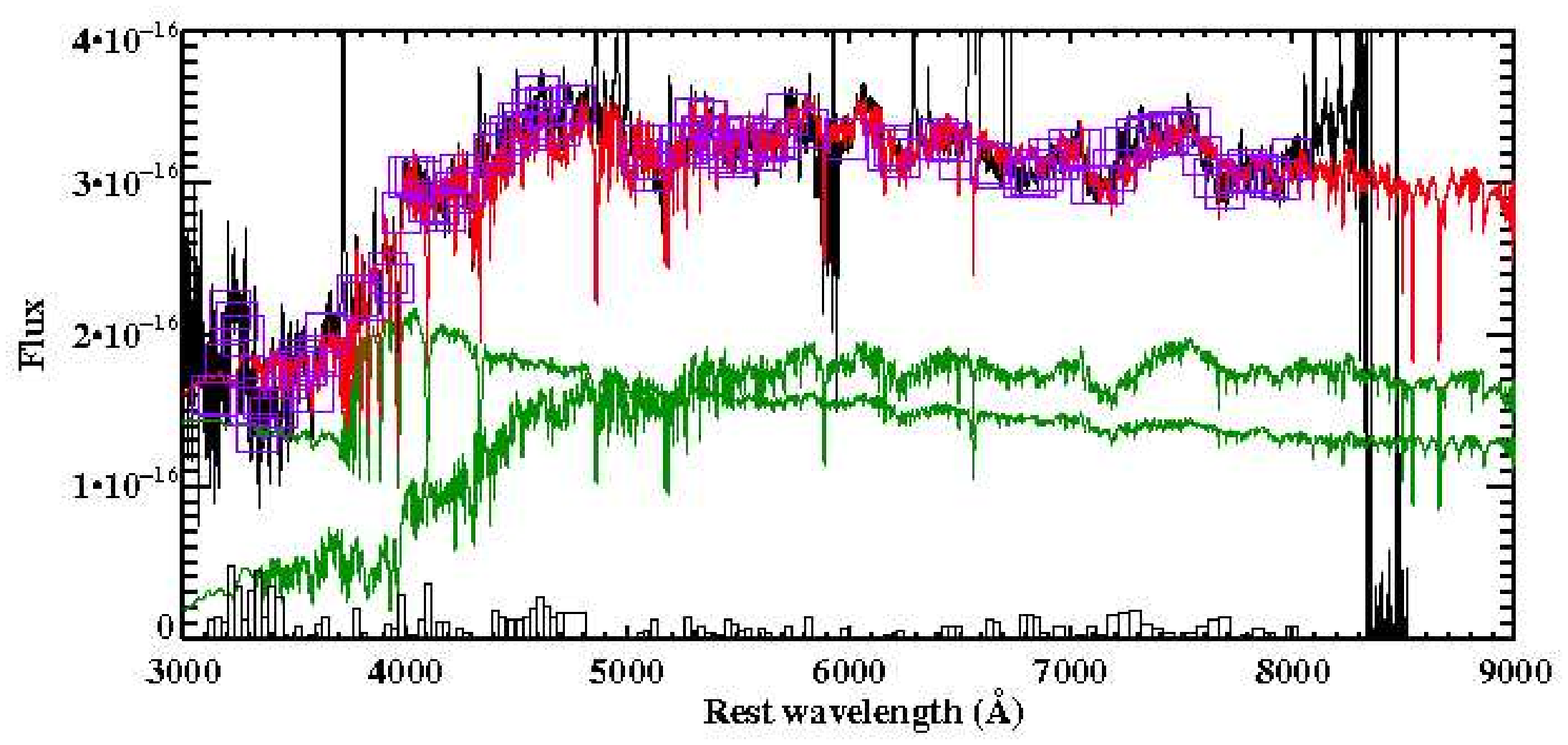,width=9cm,angle=0.} &
\psfig{file=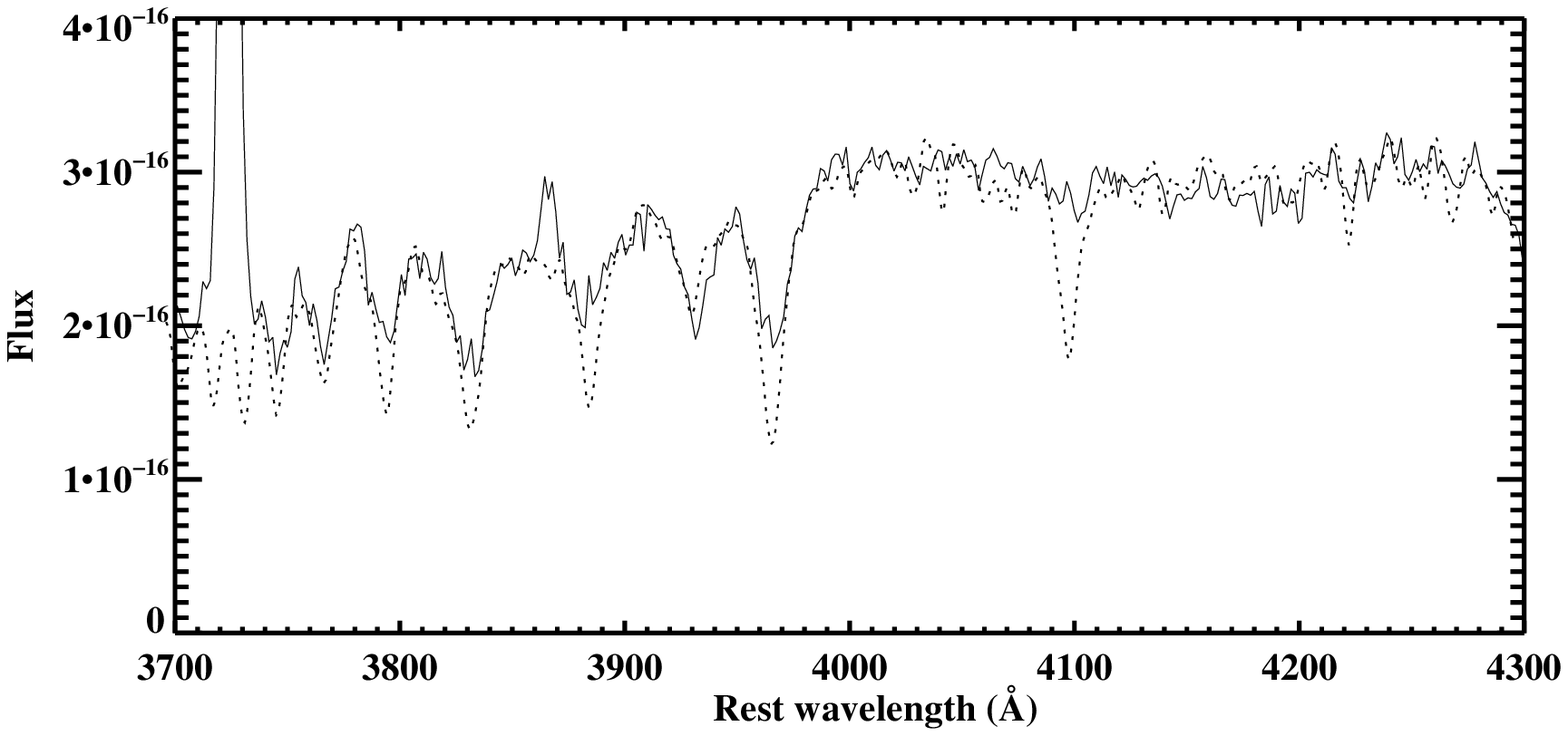,width=9cm,angle=0.}\\
\multicolumn{2}{c}{\bf 3C 236:  12.5 Gyr OSP  + power law} \\
\hspace*{-1cm}\psfig{file=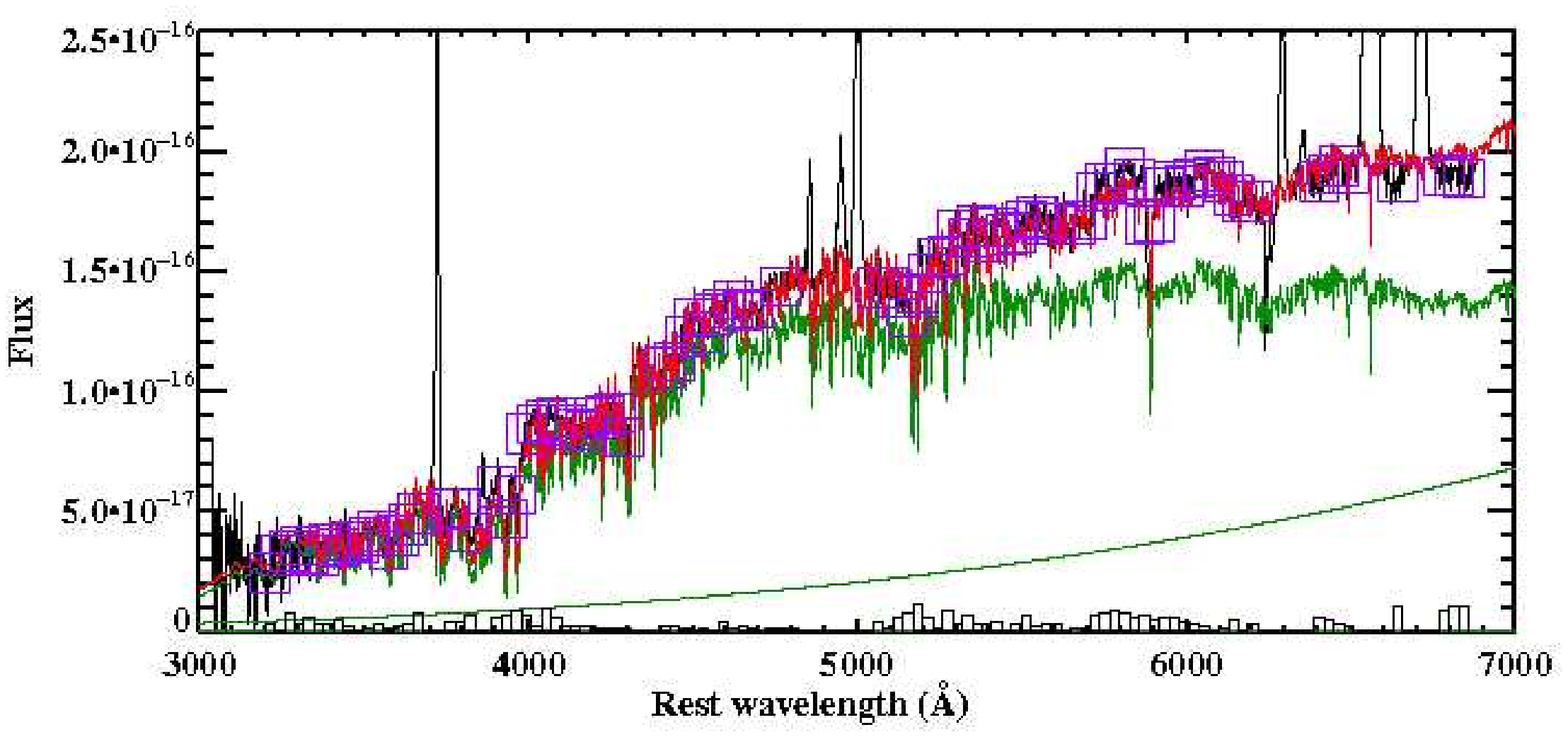,width=9cm,angle=0.} &
\psfig{file=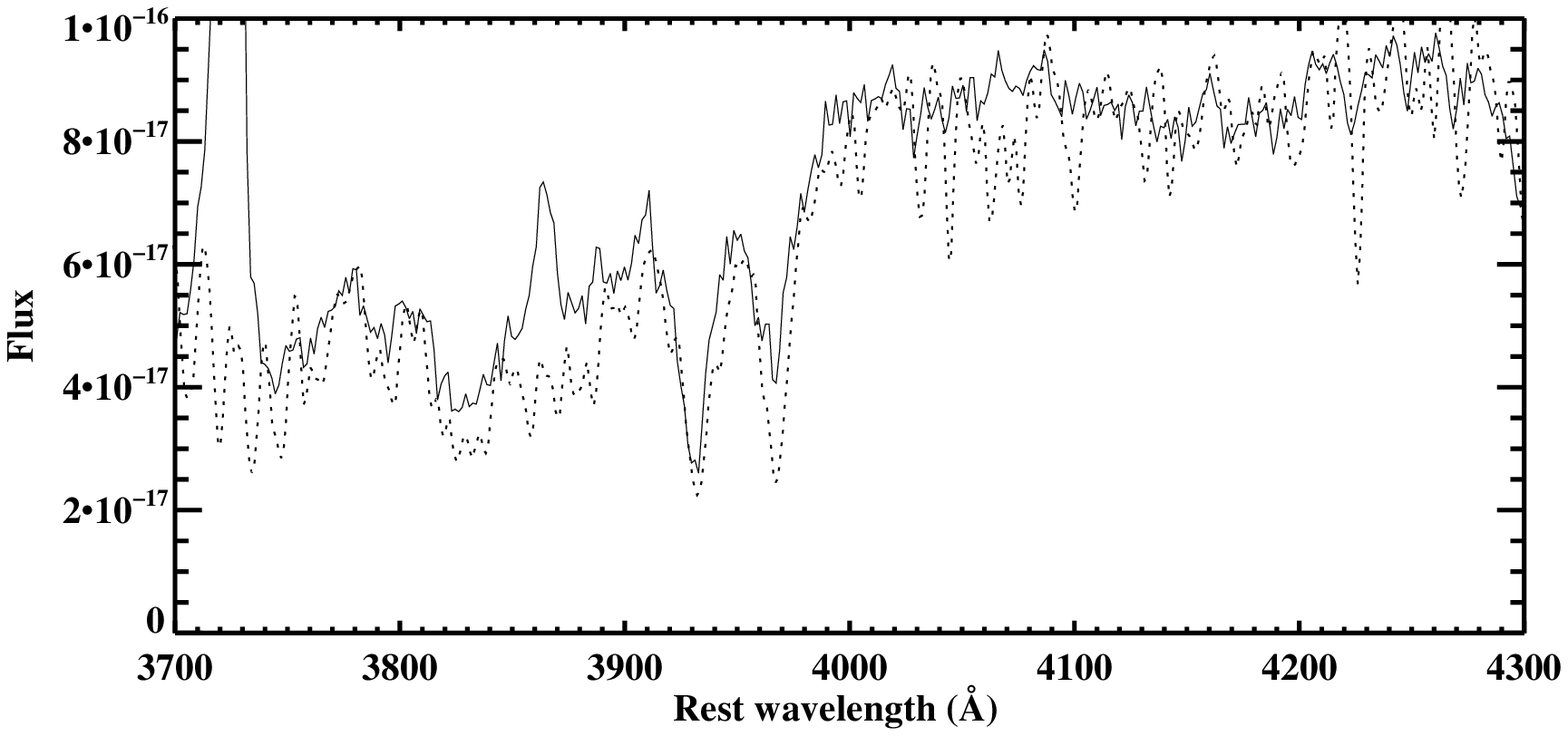,width=9cm,angle=0.}\\
\multicolumn{2}{c}{\bf 3C 236: 12.5 Gyr OSP + 0.05 Gyr YSP (\ebv~=
  1.4)} \\
\hspace*{-1cm}\psfig{file=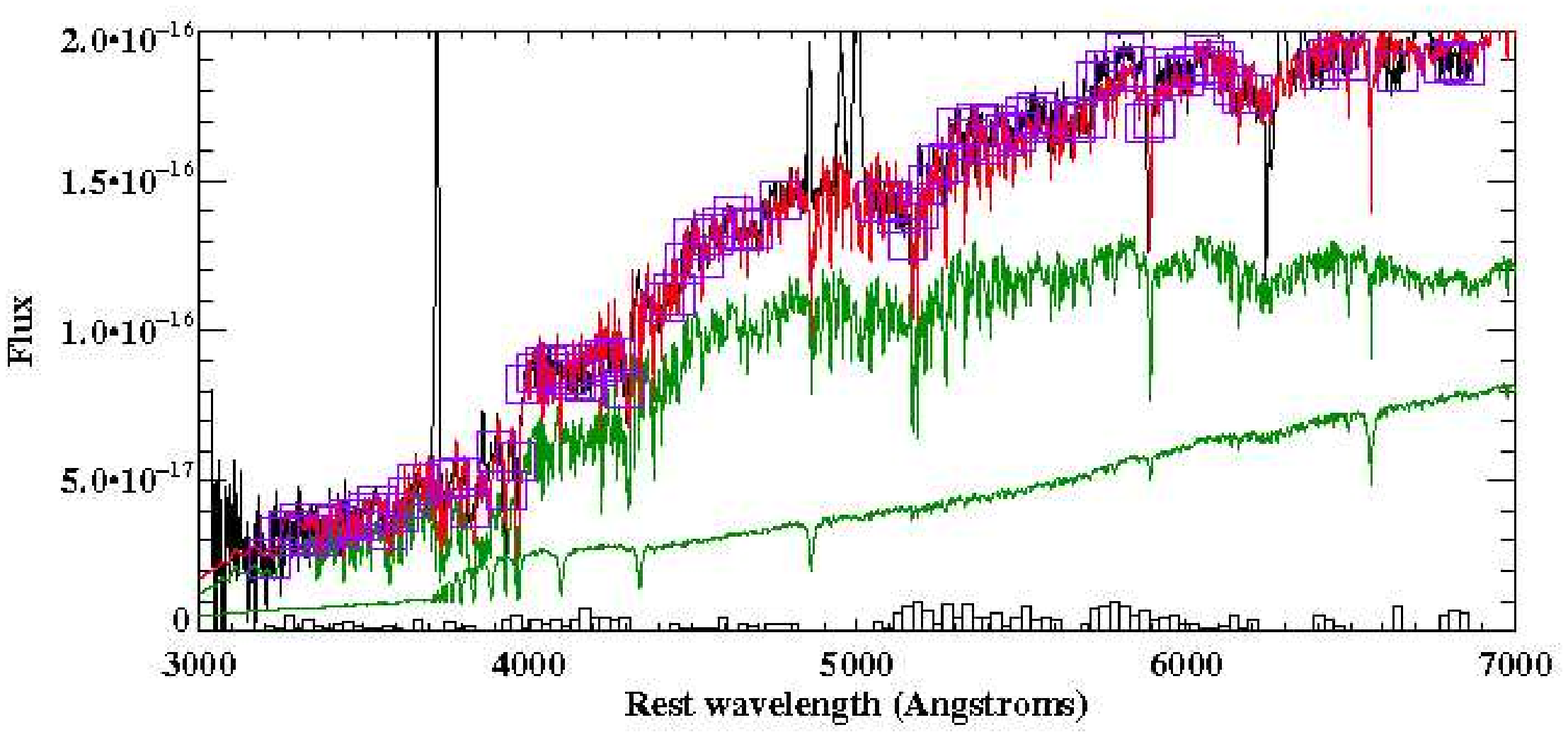,width=9cm,angle=0.} &
\psfig{file=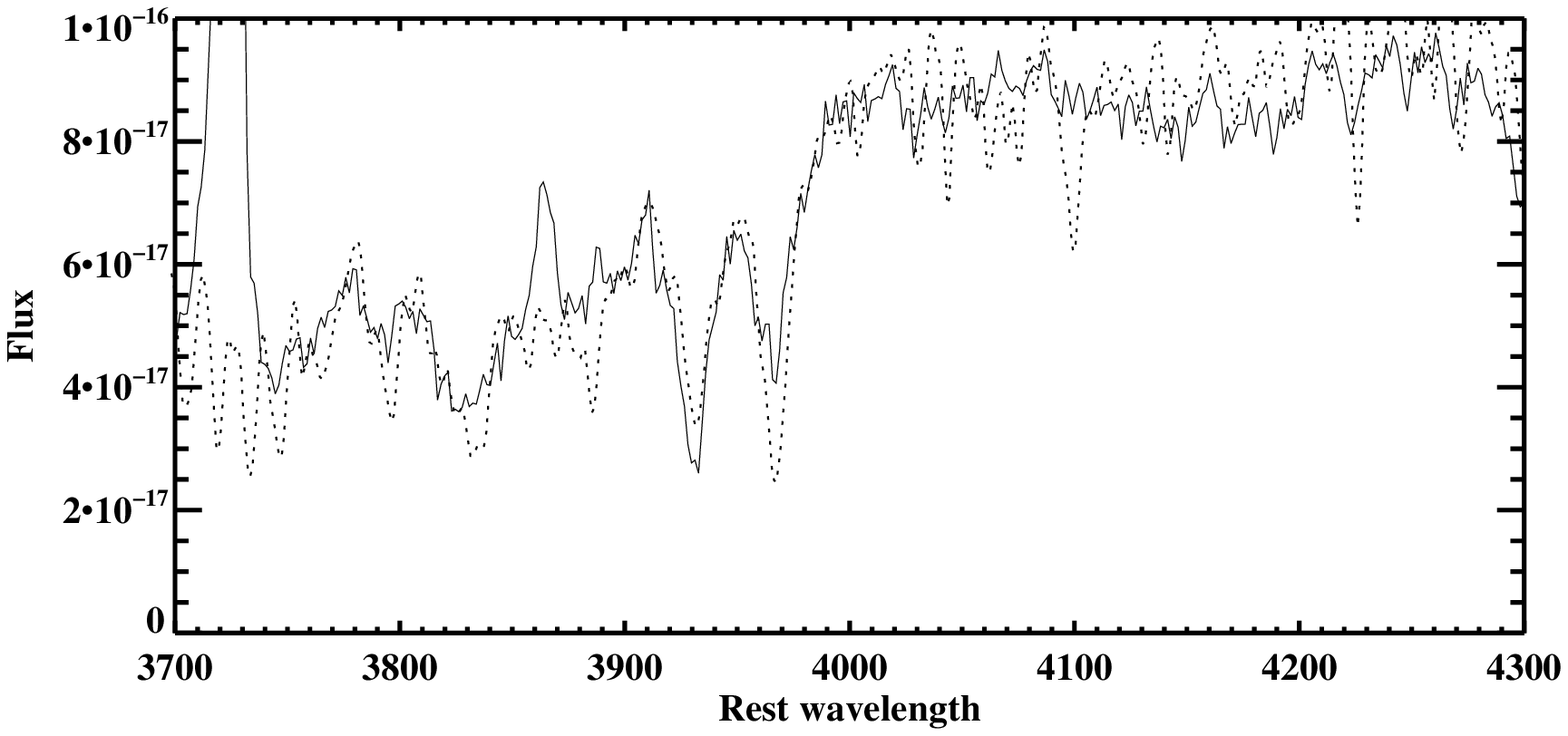,width=9cm,angle=0.}\\
\multicolumn{2}{c}{\bf 3C 285: 12.5 Gyr OSP + 0.2 Gyr YSP (\ebv~=
  0.2)} \\
\hspace*{-1cm}\psfig{file=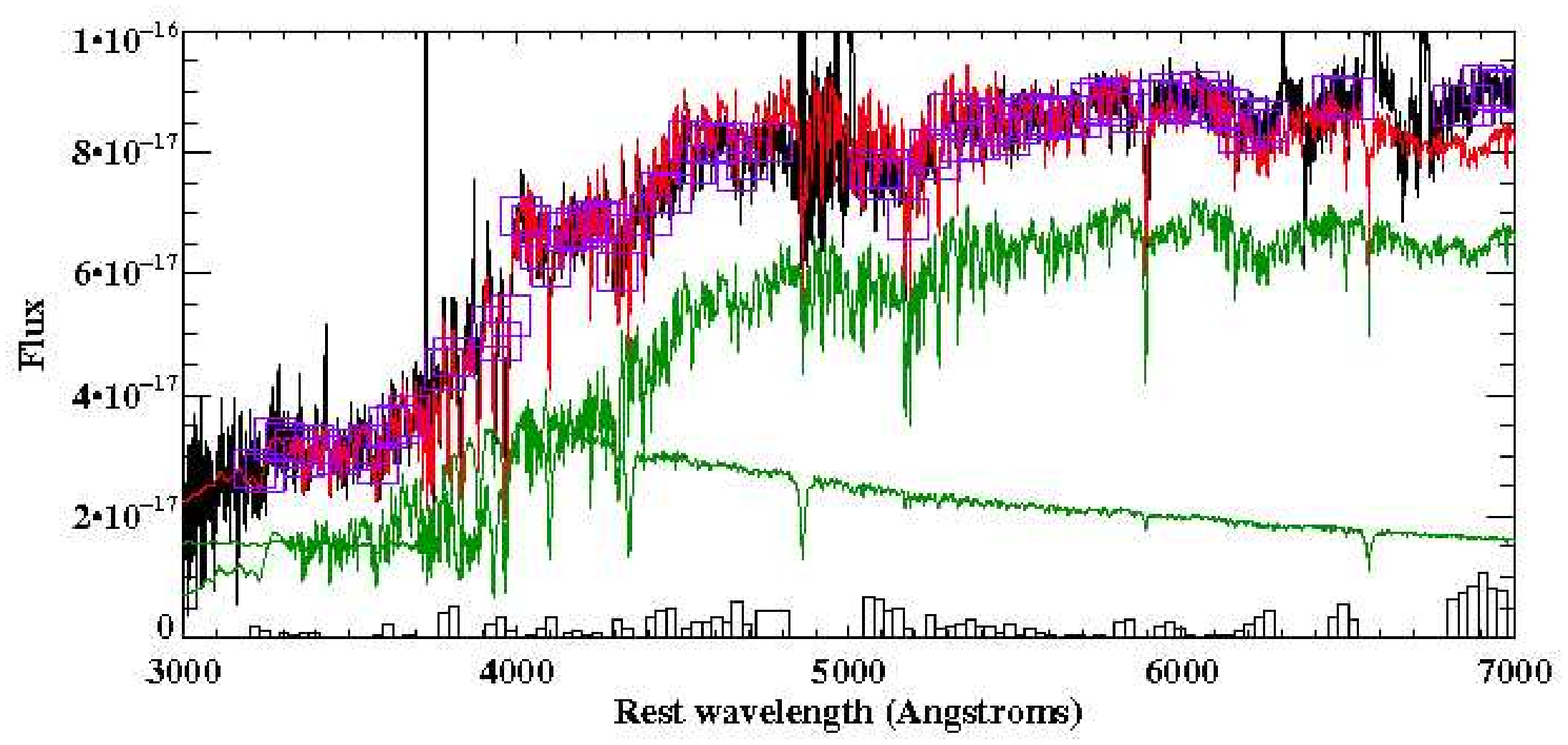,width=9cm,angle=0.} &
\psfig{file=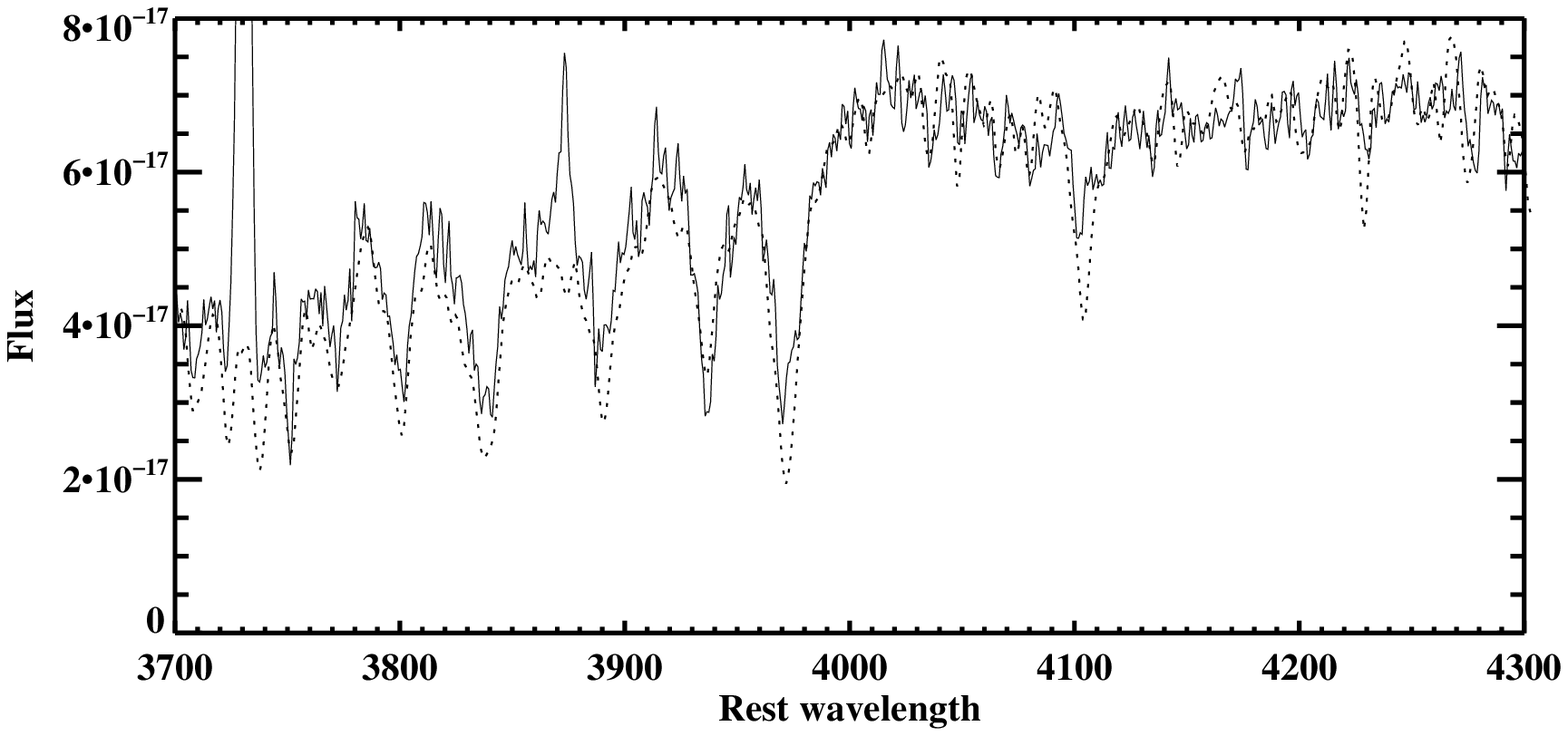,width=9cm,angle=0.}\\
\end{tabular}
\caption[]{SEDs and detailed fits. Left: overall fit to SED where the
  black line traces the data, the red line is the overall model with
  one or more components (OSP, YSP, power-law) represented by the
  green lines. The wavelength bins used to evaluate the fit are
  identified by the purple boxes (not to scale) and the histogram
  gives the residuals in each bin. Right: detailed fits in the regions
  of the Balmer and Ca II lines. The solid line traces the data and
  the dotted line represents the model (red line in the left hand
  plots). 
}
\label{fig:SED}
\end{minipage}
\end{figure*} 

The powerful FRII radio source 3C 321 is identified with a peculiar
galaxy with two nuclei separated by $\sim$3 arcsec
\citep{spinrad85,heckman86,baum88} surrounded by an  extended halo of
line emission \citep{mccarthy95}, as well as continuum arcs and tails suggestive of a
recent merger \citep{heckman86,smith89b,roche00}. Similar to a
significant number of powerful radio sources, 3C 321 is blue in colour
compared to normal elliptical galaxies \citep{smith89b}. Figure
{\ref{fig:d4000}} also highlights the spatially resolved UV excess.

The SE  nucleus is brighter and more closely coincident with the radio core
whilst the NW nucleus is clumpy \citep{zirbel98} but also associated
with a peak in the radio emission which then extends as a long (35 arcsec)
jet beyond this nucleus \citep{roche00}. However, it is
interesting to note that the narrow emission line spectra of the
two nuclei have similar line strengths and surface brightnesses
\citep{filippenko87,mccarthy95}.  

3C 321 is a case in which the scattered quasar component makes a 
major contribution to the optical continuum, at least in the 
SE nucleus \citep{draper93}. Spectropolarimetry observations
of the SE nucleus reveal a high degree of polarisation in the near UV
\citep{tadhunter96},  
as well as broad quasar-like permitted lines in the polarised intensity
spectrum \citep{cohen99}. However \citet{tadhunter96} and \citet{robinson00}
have demonstrated that it is not possible to model the continuum SED
solely in terms of 
an OSP plus a power-law or quasar template, but that a significant contribution
from a YSP is required to fit the spectrum, particularly in the region of the
Balmer break. Although they detected a YSP component, \citet{tadhunter96} and 
\citet{robinson00} considered only unreddened YSP models, and were unable
to pin down the age of the YSP.

3C 321 is spatially well-resolved although, in contrast to
\citet{robinson00}, we have extracted only four apertures. These are
shown in Figure {\ref{fig:d4000}}, which also highlights the
significant UV excess observed across the {\it entire} galaxy. Despite
the fact that \citet{robinson00} have already presented work on these
spectra, we have re-calibrated the data to improve the flux
calibration and our analysis uses the latest high resolution
models from \citet{bruzual03} as well as consideration
of the reddening of the YSP component.

{\bf  The SE nucleus}.  
The spectrum of the SE nucleus of 3C 321 
 shows evidence for the Balmer sequence in absorption and so a
YSP contribution is likely to be important. Indeed, previous studies
of the continuum emision of 3C 321 (see above) have shown that at
least three distinct components are required to model the SED: an OSP,
a YSP and a power-law component (the latter representing the
scattered quasar indicated by the polarimetry measurements). Note
that we cannot obtain viable fits with an OSP and power-law alone
(\chisq $>$ 2) in this aperture, consistent with the results of 
Tadhunter et al. (1996). 
Hence, we have focussed on the three component fits to the
SED.

When including a power-law component, as for many of the sources
studied, acceptable fits to the SED were found for most ages of
YSP -- see Figure {\ref{fig:contours}}. Hence, the detailed comparisons
proved vital for distinguishing between the models. Using the CaII~K
line (Figure {\ref{fig:SED}}), we can confidently rule out models
including a YSP component older than $\sim$ 1.0 Gyr -- all models with
a YSP age $\gtrsim$ 1.0 Gyr significantly over estimate the depth of the Ca
K line -- this can not be overcome by smoothing of the models to
match the resolution of the data.  
The youngest ages (0.04, 0.05 Gyr) are also considered unlikely  
as they do not require a
contribution from  the power-law component (i.e. 0\% contribution in
the fits) -- previous polarimetric observations combined with the
detection of a broad component to H$\alpha$ suggest a contribution of
at least 10-20\% at the position of our normalising bin (see above). 
Hence, the SE nucleus in
3C 321 can be well modelled  by an OSP
(15-55\%) plus  YSP (age: 0.1-1.0 Gyr, \ebv: 0.1-0.3, 30-60\%) and power
law (10-20\%) components with $\sim$0 $< \alpha <$ $\sim$ 0.6. Figure
{\ref{fig:SED}} shows an example of a model in the best fitting range
comprising a 12.5 Gyr OSP (46\%), a 0.4 Gyr YSP (\ebv~= 0.2, 30\%) and
a power-law
component ($\alpha$ = 0.1, 22\%). These results are
consistent with those of \citet{tadhunter96} and
\citet{robinson00} in terms of the contribution of the power-law component
and the YSP contribution. Note, \citet{robinson00} do not find
evidence for a YSP in the SE nucleus, only in the more extended
regions. However, our SE nuclear aperture is larger than that used by
\citet{robinson00} and therefore includes light from more extended
regions which do have a YSP component.

{\bf The NW nucleus}. 
As discussed above, whilst the SE nucleus is that identified with the
radio source, the second nucleus $\sim$3 arcsec to the NW also shows signs
of activity -- it is close to an enhancement in the radio emission
along the jet
and has an emission line spectrum similar in strength and ionisation state
to the SE nucleus. Further, as demonstrated by Figure
{\ref{fig:d4000}},  its observed UV excess is  similar to that
observed in  the SE nucleus. 

As for the SE
nucleus, it was not possible to obtain a good fit with 
an OSP, with or without a power-law   (\chisq~$\sim$
2.0), and so a 
YSP component is likely to be important. 

It is possible to model the SED of the NW nucleus
with an OSP plus a YSP without a power-law. Viable fits
(\chisq~$\lesssim$ 1.2) are found for 
all ages of YSP with the familiar 2 minima degeneracy. The `younger'
minimum is for ages $\sim$0.1~Gyr (\ebv~$<$ 0.3) whilst the second
minimum is
for much larger ages ($>$~1.0~Gyr). Around the 
`younger' minimum, the best fits to the lines {\it and} the continuum
are for ages 0.04-0.2 Gyr (12.5 Gyr: 70-75\%, YSP: 25-30\%)
-- ages $>$ 0.3 Gyr do not fit the continuum
in the region 4000-4300\AA~well. Solutions for all ages of YSP can provide
good fits to the G-band and Mg~Ib features.

Unlike in other apertures we have not
been able to distinguish between the two minima. At first glance, the 
models including the older YSPs  (1.0-1.4 Gyr) appear to significantly
overpredict the strength of CaII~K. However, smoothing of the models
to better match the resolution of the data
show that for this aperture, the models including older age YSPs can
not be confidently ruled out

The strong emission
lines detected in the NW nucleus suggest that it also harbours a powerful
AGN. Moreover, the results of \citet{draper93} show that it is 
significantly polarised. We have therefore
attempted to model the SED of the NW nucleus with a three component
model. In common with the two component OSP plus YSP model, we find
viable fits for most ages of the YSP (0.05-2 Gyr)
with a minimum consistent with the
older ages in the two component models at $\sim$ 1 Gyr. The contour
plot of the \chisq~space is shown in Figure {\ref{fig:contours}}. 
Detailed fits show that the models including the oldest age YSPs ($>$
1.6 Gyr) can be ruled out due to the large over-estimate of the Ca K
line and the models including the youngest age YSPs ($<$0.1 Gyr)
are ruled out as they do not include a zero power law component. The
remaining models, however, are indistinguishable. Hence, the viable
fits have YSP ages 0.1-1.4 Gyr with 0.1 $<$ \ebv~$<$ 0.3
contributing  25-75\% of the SED flux. An example of a good fitting
model is shown in Figure {\ref{fig:SED}}: 12.5 Gyr OSP (52\%) plus 0.6 Gyr YSP
(\ebv~= 0.2; 30\%) plus a power law (15\%; $\alpha$ = -0.1) with
\chisq~= 0.45.

{\bf Extended apertures}
Because the UV excess is spatially resolved, we have followed the work of
\citet{robinson00} and extracted a further two apertures, one between
the two nuclei and one to the south east of the SE nucleus, marked on Figure
{\ref{fig:d4000}}. Indeed, \citet{robinson00} showed that, in the case
of 3C 321, the YSP component makes a relatively larger
contribution off-nucleus than
in the regions of the two nuclei.

{\bf Between the nuclei}.
It is not possible to model the region between the nuclei with an OSP
or OSP plus power-law component (\chisq~$>$ 2). This is not surprising as
the spectrum 
shows clear evidence for a Balmer break and Balmer absorption features,
suggesting a strong YSP component. For  OSP plus YSP models, 
viable fits (\chisq~$\lesssim$ 1.2)
can be obtained for ages 0.05-2.0 Gyr with \ebv~$\lesssim$
0.4, dependent on age -- younger YSPs require more reddening. There is
also a clear degeneracy with 
two minima at ages $\sim$
0.1-0.2 Gyr and $\sim$ 1.0-2.0 Gyr.
However, examination of the overall fit
to the SED shows that whilst the value of \chisq~is low, two regions
are not modelled well -- $\sim$5000-5500\AA~and 
$\gtrsim$7000\AA. 

As the two component models fail to successfully fit these two
regions, we have attempted to model the SED including a power
law. Whilst for most off-nuclear apertures we have assumed that a
power law component is unlikely, for 3C 321, based on the previous
observational results, this is not necessarily the case. 

Figure {\ref{fig:contours}} shows the \chisq~space for the three
component models. When including a power law, the \chisq~reduces
significantly and a broad minimum is found for ages 0.04-2.0 Gyr with
reddening \ebv~$\lesssim$ 0.5. Combining the large scale and detailed
fits, the youngest ages can be ruled out on the basis of the large
scale (5000-5500\AA~and $\gtrsim$7000\AA~regions) fits (0.04-0.1 Gyr)
and the fits to the CaII~K line ($\lesssim$ 0.3 Gyr). Older age YSPs
($>$ 1.0 Gyr) can be ruled out as the CaII~K line is significantly
over-predicted. 

Hence, the best fitting models for the region between the nuclei
comprise a 12.5 Gyr OSP (30-50\%) plus a 0.4-1.0 Gyr YSP (\ebv~=
0.0-0.1, 40-60\%) and a power law component (5-8\% with $\alpha$
$\sim$ -1.6 -- -1.7). An example of a model in this range is shown in
Figure {\ref{fig:SED}}: 12.5 Gyr plus 0.4 Gyr YSP (\ebv~= 0.1, 39\%)
plus a power law (8\%, $\alpha$ = -1.6) with \chisq~= 0.42.

{\bf Extended aperture}. As for the aperture between the nuclei, it is
not possible to model the SED 
with
an OSP or OSP plus power-law alone (\chisq~$>$ 6). 
Modelling the SED using two and three components, each incorporating a YSP,
shows that, whether or not a power-law component is
included, the best fits to the overall SED consistently required a
$\sim$0.7 -- 2.0 
Gyr YSP with zero reddening. Figure {\ref{fig:contours}} shows the
\chisq~space for both two and three component models.  This is
reassuring and shows our results are 
 robust against the fine details of the modelling. Detailed comparisons
of the absorption lines show that, when a power-law component is
excluded, the CaII~K line is {\it always} significantly over-predicted. 
Including a power-law component has the effect of diluting
the absorption features. Indeed, this produces a significantly better
fit to the CaII~K line for a YSP age of 0.7-1.0~Gyr, whilst older and younger
ages give worse fits to either the CaII~K line and/or the continuum
surrounding the absorption features. 

The modelling suggests that three components are required to model the
spectrum. A significant YSP component (54-72\% of the flux in
the normalising bin) is essential to provide a good fit to the
SED and detailed comparisons reveal that a small (5-6\% of the flux in
the normalising bin) power-law component improves the fit to the
absorption features. An 
example of a good fitting model, 12.5 Gyr (21\%) plus 1.0 Gyr YSP (\ebv =
0.0; 72\%) plus power-law (5\%; $\alpha$ = -1.2) with \chisq~= 0.7 is shown
in Figure \ref{fig:SED}. This aperture contains 
the largest 
contribution from a YSP ($>$50\% in the normalising bin). 

{\bf Summary for 3C 321}. 
The good spatial resolution of the continuum
structure in 3C 321 allows us to investigate the spatially extended UV
excess in four distinct regions including both nuclei, the region
between the nuclei and an extended aperture to the SE of the active
nucleus. Despite the large spatial scale and the different
environments analysed, the results are generally consistent. All
apertures require a significant contribution ($\sim$30-70\%) from a
$\sim$ 0.5-1.0 Gyr YSP with little reddening (\ebv$\lesssim$ 0.3)
although we can not rule out significantly younger ages in the two nuclei.
In addition, all apertures require a relatively small contribution
(few-20\%) from a power law in all apertures, the largest contribution
in the active  nucleus.

\setcounter{figure}{3}
\begin{figure*}
\begin{minipage}{170mm}
\begin{tabular}{cc}
\multicolumn{2}{c}{\bf 3C 321 (SE nuc): 12.5 Gyr OSP + 0.4 Gyr YSP
  (\ebv~= 0.2) + power law} \\
\hspace*{-1cm}\psfig{file=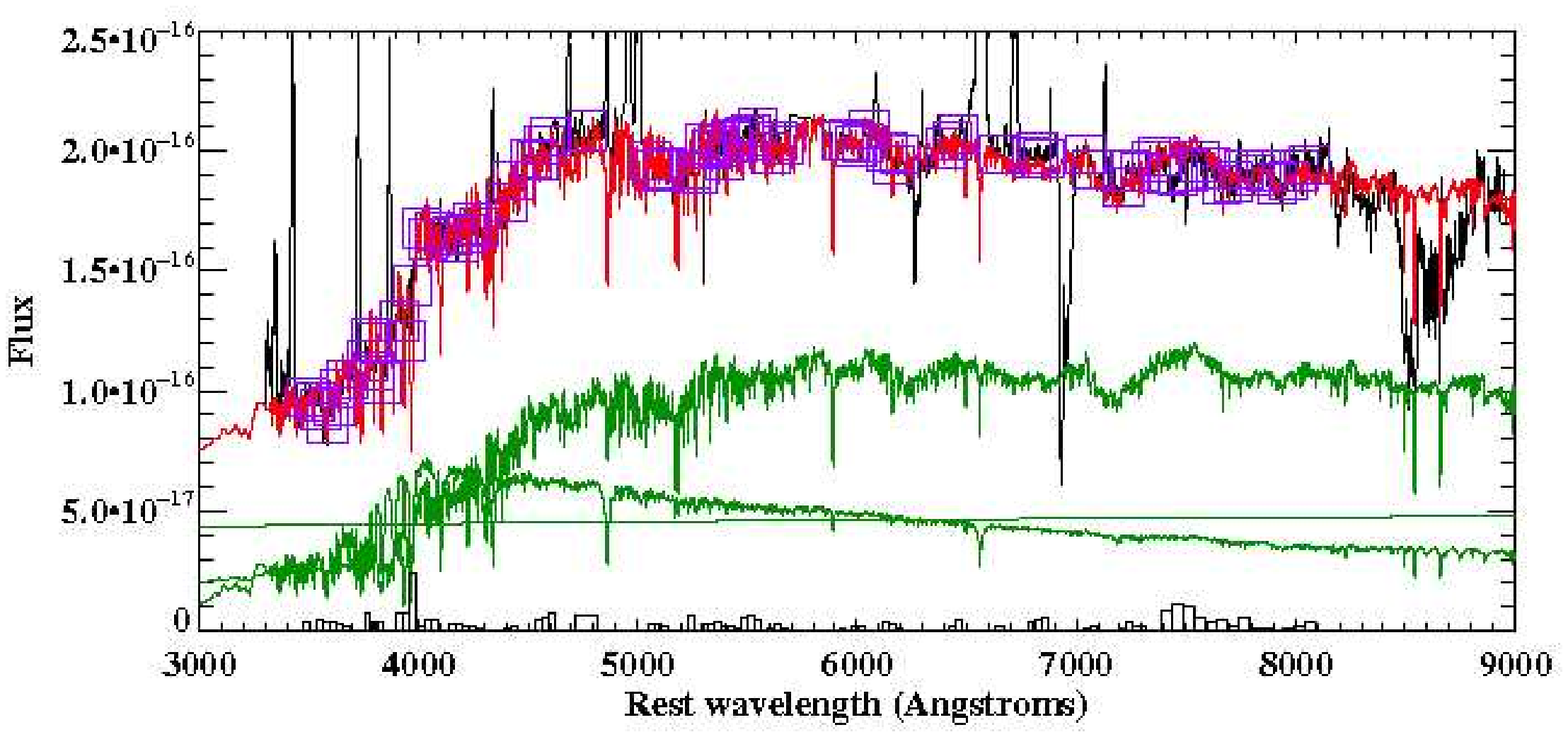,width=9cm,angle=0.} &
\psfig{file=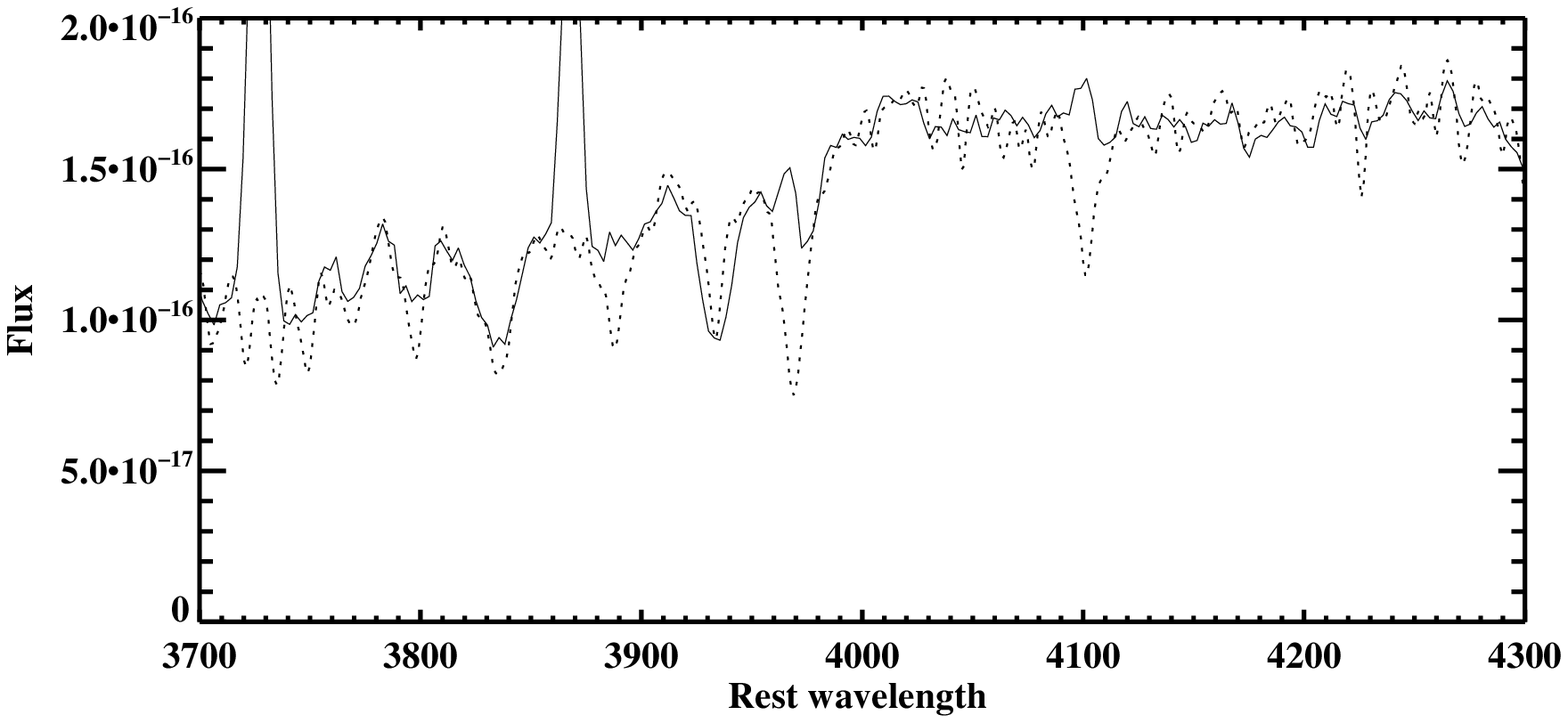,width=9cm,angle=0.}\\
\multicolumn{2}{c}{\bf 3C 321 (NW nuc): 12.5 Gyr OSP + 0.6 Gyr YSP
  (\ebv~= 0.2) + power law} \\
\hspace*{-1cm}\psfig{file=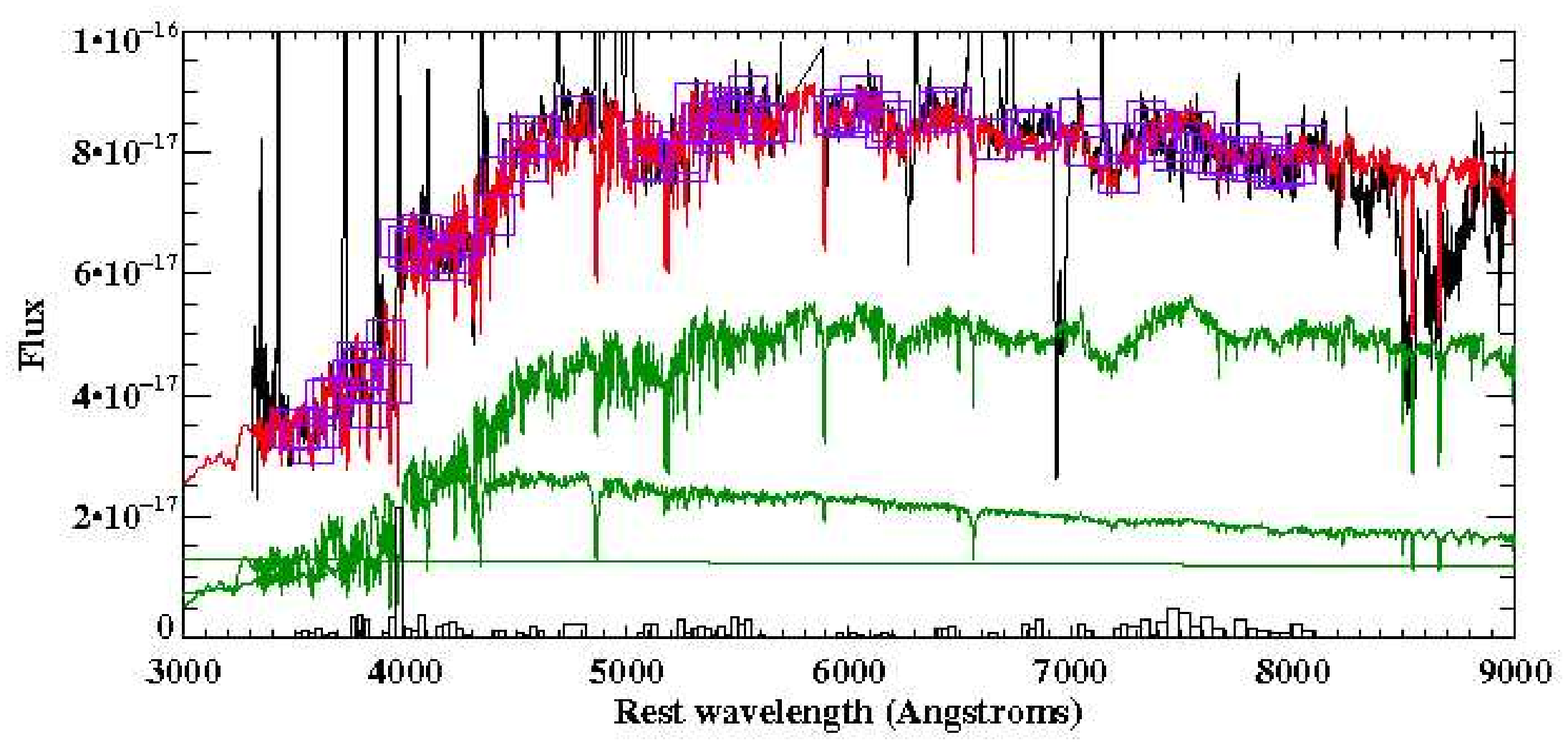,width=9cm,angle=0.} &
\psfig{file=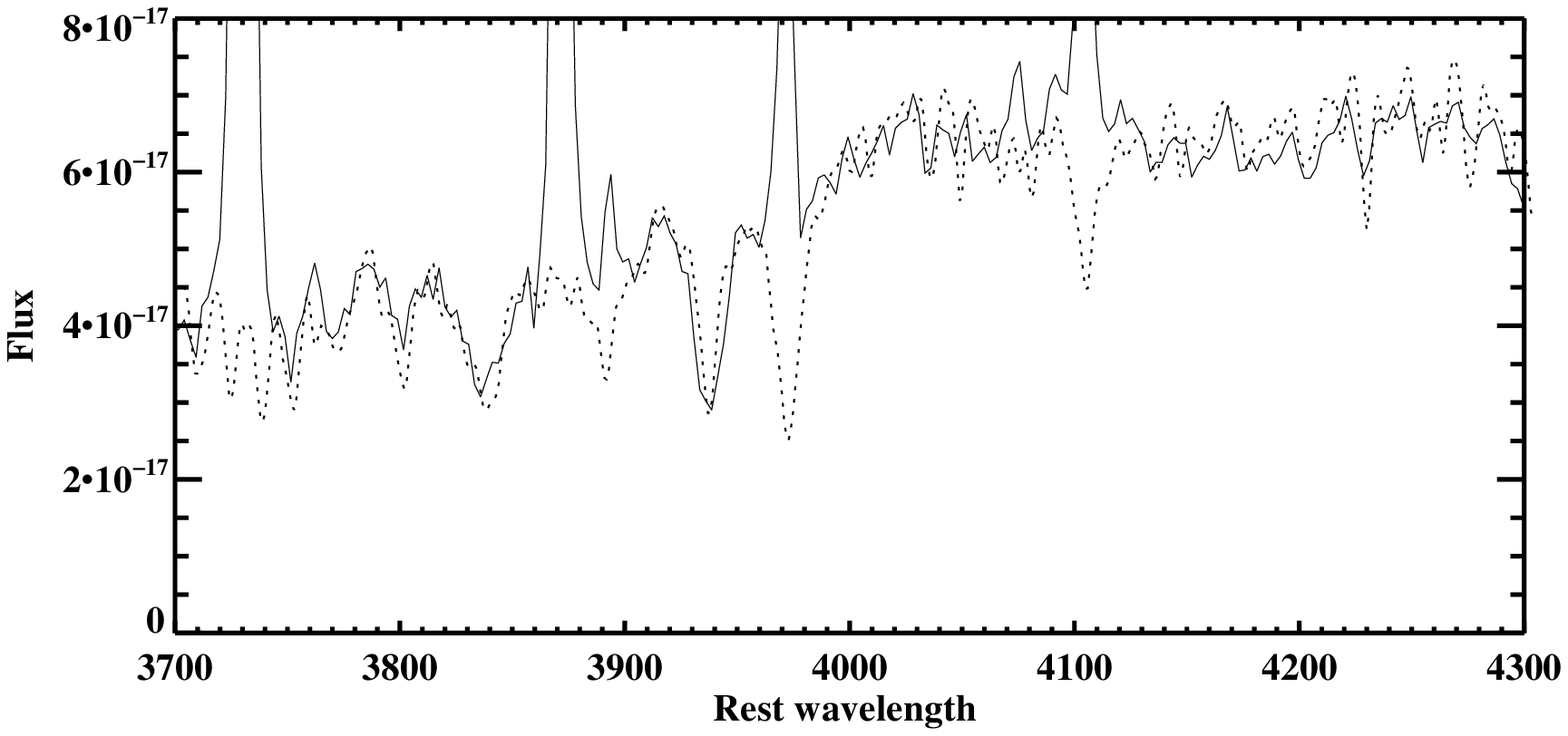,width=9cm,angle=0.}\\
\multicolumn{2}{c}{\bf 3C 321 (mid, scaled): 12.5 Gyr OSP + 0.4 Gyr
  YSP (\ebv~= 0.1) + power law} \\
\hspace*{-1cm}\psfig{file=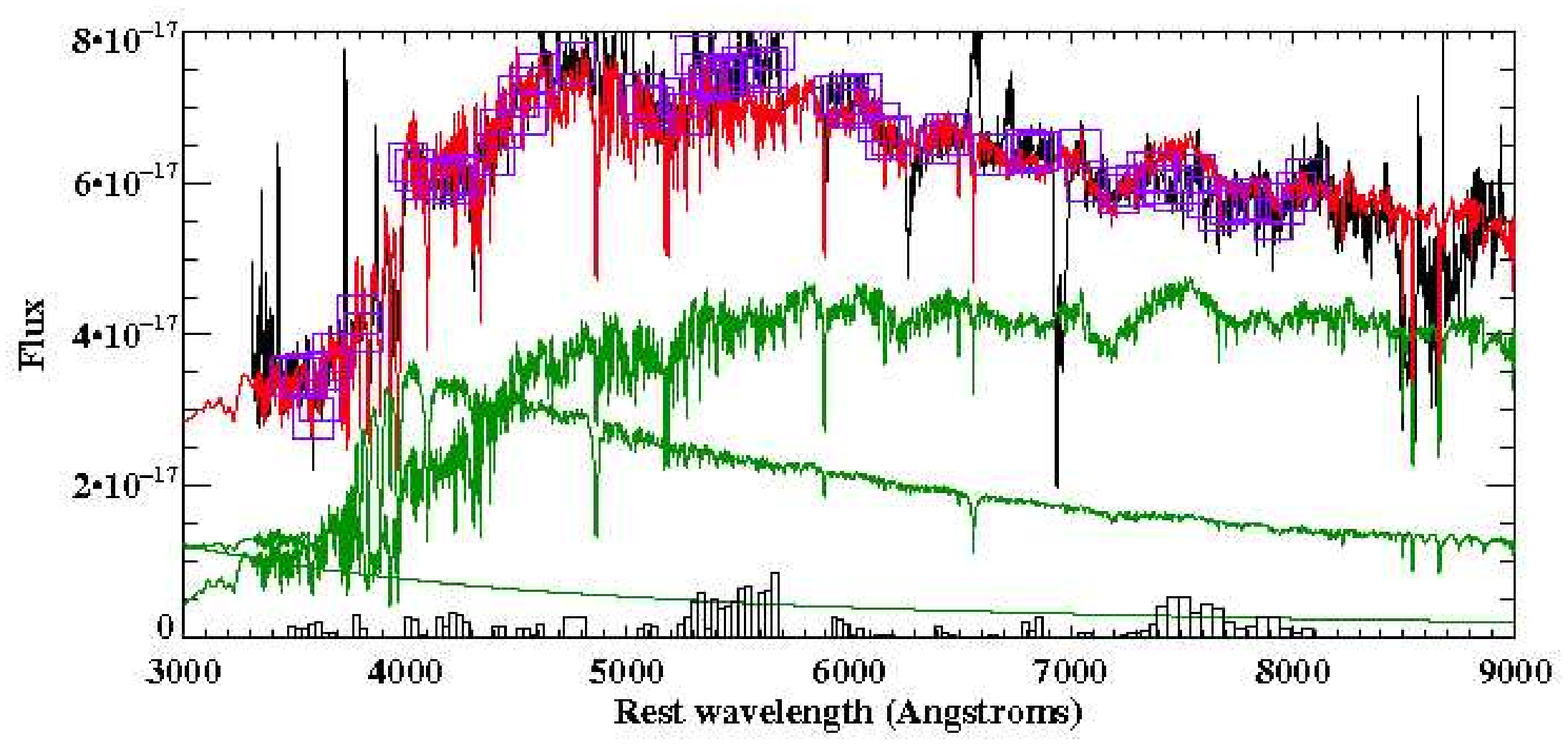,width=9cm,angle=0.} &
\psfig{file=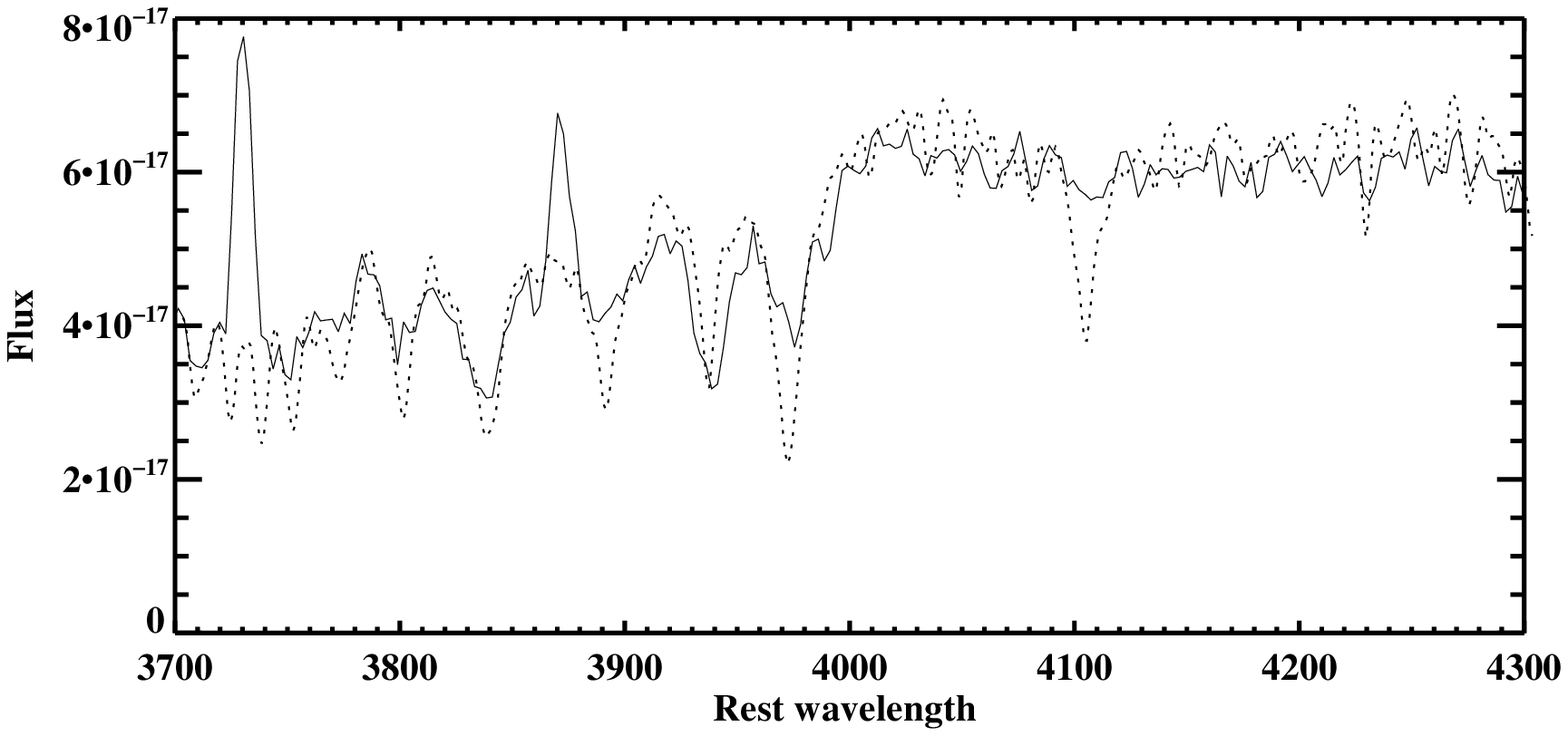,width=9cm,angle=0.}\\
\multicolumn{2}{c}{\bf 3C 321 (ext): 12.5 Gyr OSP + 1.0 Gyr YSP
  (\ebv~= 0.0) + power law} \\
\hspace*{-1cm}\psfig{file=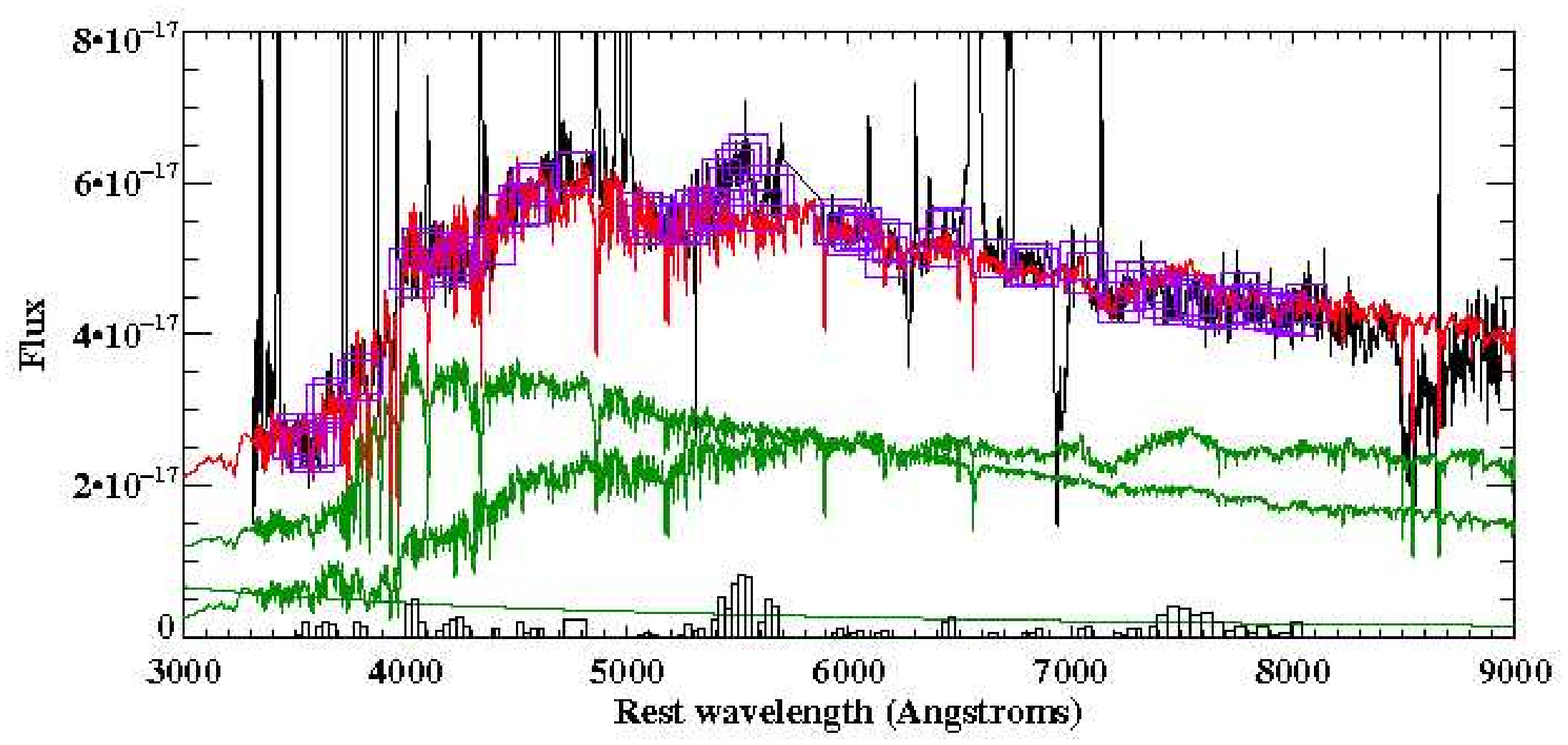,width=9cm,angle=0.} &
\psfig{file=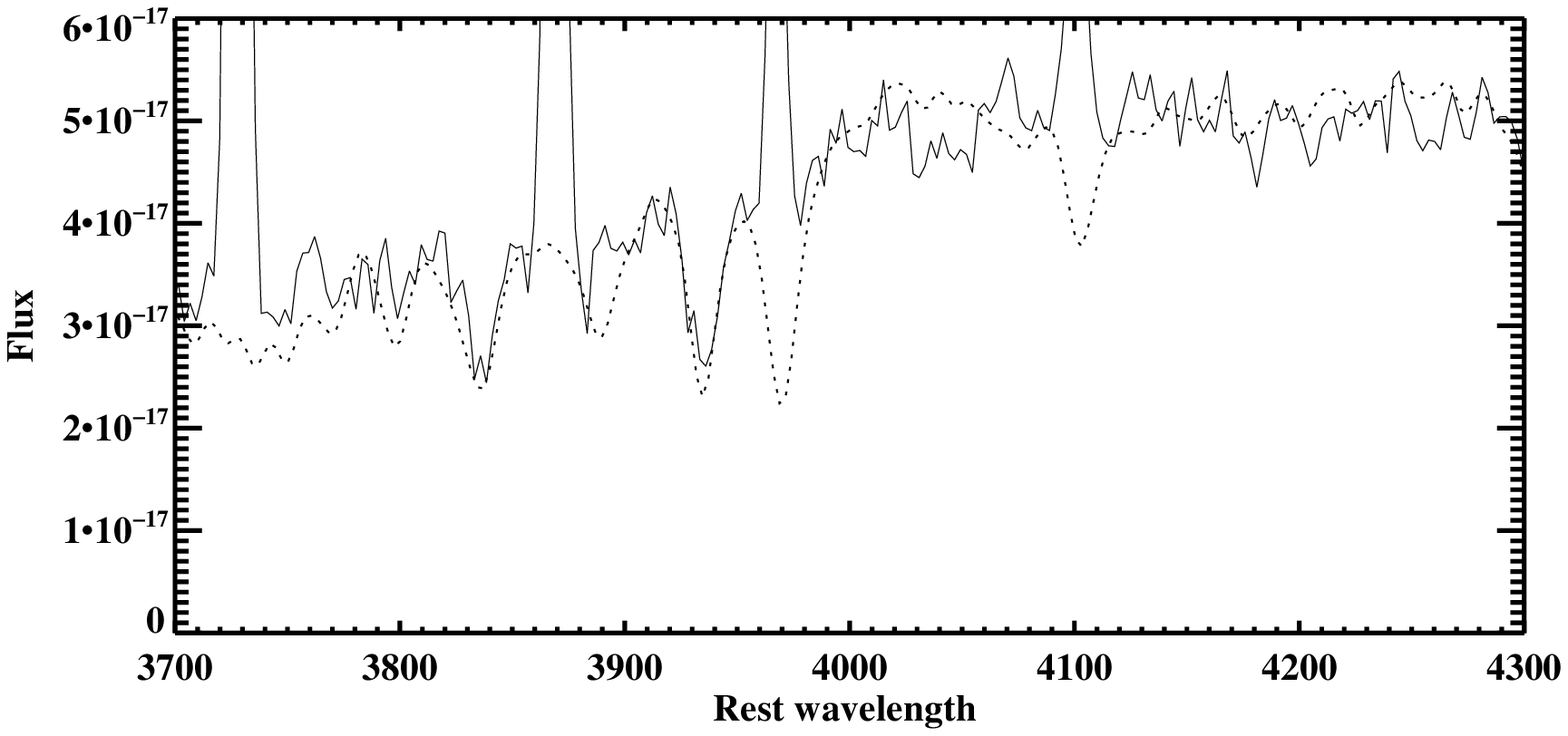,width=9cm,angle=0.}\\
\end{tabular}
\caption[]{ SEDs and detailed fits \it continued
}
\label{fig:SED}
\end{minipage}
\end{figure*}

\subsubsection{3C 381}
\setcounter{figure}{4}
\begin{figure}
\begin{minipage}{85mm}
\centerline{\psfig{file=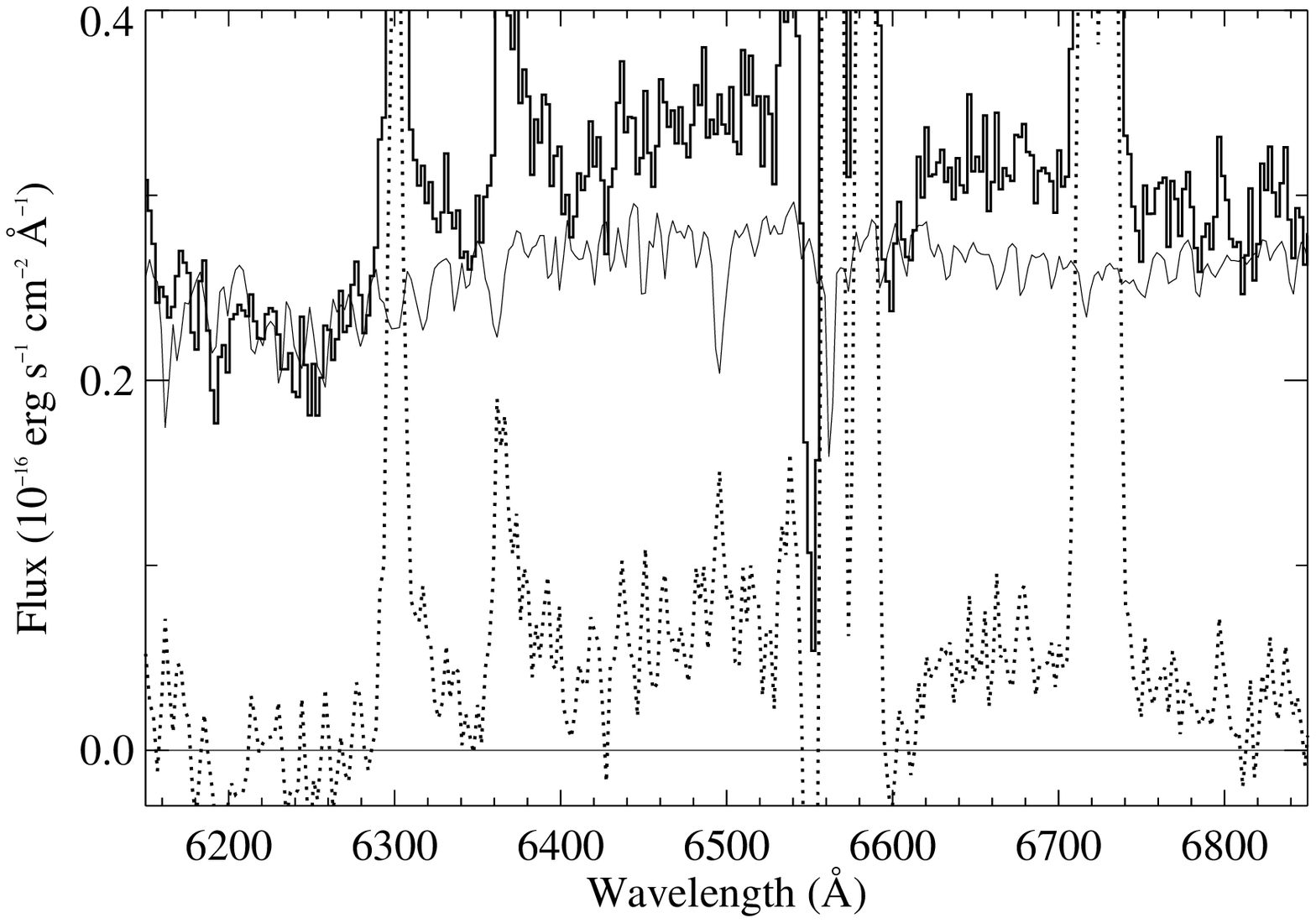,width=9cm,angle=0.}}
\caption[]{Detection of a broad component to H$\alpha$ in the nuclear
  aperture of 3C 381 with FWZI $\sim$ 300\AA~or $\sim$ 14,000 \kms.
The lines plotted are: nebular continuum subtracted nuclear
  spectrum (bold solid line; offset by -0.4 $\times$ 10$^{-16}$
  \dipso), best fitting (OSP+power law; offset by -0.4 $\times$ 10$^{-16}$
  \dipso) SED model 
  (solid line) and the continuum (OSP+power law) subtracted nuclear
  spectrum (dotted line). Note, the spectrum has not been corrected
  for atmospheric absorption features and the A band (at $\sim$ 7600
  \AA) is coincident with the \ha/{[N II]} blend in the observed frame.}
\label{fig:3c381halpha}
\end{minipage}
\end{figure} 
The ultra-steep spectrum radio source 3C 381  is a classical double
radio source extending more than 70 arcsec along PA 90
\citep{riley75}.  Optically it 
is identified with a round high surface brightness galaxy with a
compact nucleus.  A   lower  
surface brightness companion lies 5.5 arcsec away along PA 65 and the
detection of tidal distortions and tails extending N and S confirms 3C
381 is  currently interacting with this companion
\citep{dekoff96,roche00}.  Although
\cite{grandi78} reported the detection
of weak broad wings to H$\alpha$ along with high ionization narrow lines,  
\citet{dennettthorpe00} \& \citet{eracleous04} failed to find evidence
for broad H$\alpha$ in their higher S/N spectra. Moreover,
\citet{dennettthorpe00}  
argue against the classification of this object as a BLRG because the
radio core 
and jet are relatively weak, and the near-IR colours are consistent
with NLRGs rather 
than BLRGs (c.f. \citealt{lilly82}). \citet{rudy83} measure
a relatively low polarisation for 3C 381 at optical wavelengths:
 1.8$\pm$0.6\%, consistent
with the observations of other BLRGs in 
their sample. 
On the basis of Figure {\ref{fig:d4000}},
3C 381 has a clear UV excess. Given that it 
is only marginally resolved (Figure {\ref{fig:d4000}}), we
have extracted only a nuclear aperture for this source.

We find that an adequate fit to the SED and the discrete features can
be obtained using a two component 
(OSP plus power-law) model (\chisq~= 0.90) where 80\% and 20\% of the flux
originates from the OSP and power-law respectively. This model is shown in
Figure {\ref{fig:SED}}.
In this model, the power-law
component is red ($\alpha$ = 1.65) which suggests that, if the light
originates from the AGN, it is direct rather than scattered, consistent
with the low optical polarisation. 

In  Figures
{\ref{fig:SED}} and \ref{fig:3c381halpha} 
there is clear evidence for a broad component to
\ha~in the form of an excess in the continuum region around H$\alpha$
relative 
to our best fitting model. This excess has a FWZI of $\sim$300\AA~or
$\sim$14,000 \kms. However, 
due to the strong atmospheric A band ($\sim$ 7600\AA)
coincident with the \ha/{[N II]} blend, it is not possible to model
this component in detail. We further note that, although
\citet{chiaberge02} present only an upper limit on the flux of a
compact nucleus 
in this source based on HST images, their upper limiting flux
is consistent with the flux in the power-law component
required by our continuum modelling. Hence, our observations are
consistent with a BLRG classification for 3C 381.

Despite the possible evidence for a broad line nucleus in this object,
the evidence is not conclusive.
Therefore, we have modelled the SED including a YSP,
both with and without a power-law component. Figure {\ref{fig:contours}}
shows the \chisq~space for the OSP plus YSP models. For the two
component  models, 
we find one mininmum (\chisq~= 0.85) for an OSP plus a 0.02 Gyr YSP
reddned by \ebv~= 1.0 contributing 82\% and 17\% of the flux in the
normalising bin respectively. 
This model also produces a viable fit to the stellar absorption lines.

\subsubsection{3C 433}
The  FR I/FR II radio galaxy 3C 433 is peculiar at both radio and
optical wavelengths. The radio source extends over $\sim$ 100 kpc
and is highly asymmetric in both the strength and shape of the radio
emission, overall resembling an `x' shape
(e.g. \citealt{vanbreugel83,black92}).  Such  complex radio
morphology, often associated with 
lower power radio sources, is unexpected 
due to the relatively high power of the radio source \citep{vanbreugel83}.

Optically, 3C 433 is identified with a NLRG \citep{koski78}, 
with a bright companion  galaxy $\sim$10
arcsec to the NE \citep{vanbreugel83,baum88}, and the nuclei of both galaxies
are surrounded by a common envelope \citep{matthews64}. A fainter galaxy is
observed $\sim$25 arcsec to the SW. The
large-scale optical structure is complex, with emission line 
filaments clearly detected in ground-based (H$\alpha$) images
(e.g. \citealt{baum88,mccarthy95}), and patchy dust features, some
trailing in the direction of the radio jet, also
detected in higher resolution HST images \citep{dekoff00}.

Modelling of the
spectrum of the SW nucleus by \citet{wills02} suggests that the large
UV excess  observed  is 
attributable to  a significant ($\sim$ 20-40\% contribution to the
  flux at 4780\AA), highly reddened   (\ebv~$\sim$ 0.5-0.7) YSP
  (0.05-0.1 Gyr).  Note that, on the  basis of the lack of a 
broad component to H$\alpha$ and the low UV polarisation  ($<$ 8\%;
\citealt{wills02}),  
reddened YSP models are favoured
over scattered AGN models for this source.
Moreover, while this object shows unusually red colours at near-IR
wavelengths (\citealt{lilly85}), and a relatively high surface
brightness in the 
complex nuclear structure revealed by the optical HST images,
the nucleus is  resolved and there is no evidence for
a point source at optical wavelengths (Chiaberge, Capetti \& Celotti 1999). 
 
In our new WHT observations the slit position was chosen to cover both nuclei.
We find that the UV
excess covers the entire
measureable extent of the double nucleus system (see Figure
{\ref{fig:d4000}}).  
Hence, we have extracted three apertures: one for the SW nucleus coincident
with the radio source,  one for the NE nucleus, and one sampling the region 
between the nuclei. All apertures are shown on Figure {\ref{fig:d4000}}.

{\bf The SW nucleus}. 
The SW nucleus of the 3C 433 system, coincident with the main radio source, is
likely to have a YSP component (see Wills et al. 2002).  However,
it is possible to obtain a marginally acceptable fit (\chisq~= 1.40) to the SED
and absorption features  using an OSP (74\%) plus a power-law component
(24\%; $\alpha$ = 1.0) only, consistent with the two component fitting
by \citet{wills02}. Since the UV polarisation is low
($<$8\%; \citealt{wills02}), the large power-law is unlikely to be due
to a scattered AGN 
component. Although the fitted power-law is red,  which might be consistent 
with a direct AGN component, given the failure to detect a point source in
broad-band HST images \citep{chiaberge99} and broad wings to the
H$\alpha$ in our spectra, an interpretation in terms of a highly reddened
YSP component is more plausible. 

Figure
{\ref{fig:contours}} shows the plot of the \chisq~space for models
that only include YSP and OSP components. 
\citet{wills02} found a two component OSP plus YSP model gave as good
or better fits than an OSP plus a power-law and the results
presented here are consistent with this -- there is a clear mininmum
in the \chisq~space at YSP age $\sim$ 0.05 Gyr with reddening
\ebv~$\sim$ 0.7 and \chisq~$\sim$ 0.6 contributing $\sim$ 20-30\% of the
flux in the normalising bin. Models with YSP ages greater than 0.2~Gyr
are ruled out by our SED modelling. An example of a two component model is
shown in Figure {\ref{fig:SED}} 
 with: 12.5 Gyr OSP (67\%) plus 0.05 Gyr YSP (\ebv~ = 0.7, 32\%) with
\chisq~= 0.6. It is clear that this model also provides an acceptable fit to
the stellar absorption features.

{\bf The NE nucleus}.
The SED of the NE nucleus is markedly different to that of the active
SW nucleus -- the 4000\AA~break is more prominent, as are the stellar
absorption features (e.g. CaII~H+K), whilst there is little evidence for
line emission. However, it is possible to model the SED with an OSP,
both with and without a power law (\chisq~= 0.99 and 1.1
respectively). Figure {\ref{fig:SED}} shows the OSP (100\%) model.

Despite being able to model the SED adequately without a YSP
component, because there is evidence for a significant UV excess (see
Figure \ref{fig:d4000}}),
we have attempted to model the SED including a YSP to see
how the fit will change. In the two component OSP plus YSP models, the
SED fit improves significantly when including a YSP
and good fits (\chisq~$<$ 1) are obtained for all ages of YSP (see
Figure {\ref{fig:contours}}). The inclusion of the YSP component also
improves the fit to the absorption lines, although the younger
($\lesssim$ 0.03) and older ($\gtrsim$ 1.6) ages tend to over predict
some of the absorption features. An example of the 12.5
Gyr OSP (89\%) plus a 0.1 Gyr YSP (\ebv~= 0.8, 10\%) with \chisq~=
0.57 is shown in Figure {\ref{fig:SED}}.  Despite the fact that the
models are virtually indistinguishable, all consistently have OSP and YSP
contributions of 95-65\% and 5-30\% respectively with \chisq~$\sim$
0.45-0.9.

{\bf Between the nuclei}.
As highlighted in Figure {\ref{fig:d4000}}, the UV excess is spatially
resolved and extends across the entire system, with a similar $D4000$
ratio.  We have therefore also investigated an aperture located
between the two nuclei. The location of this aperture is shown on
Figure {\ref{fig:d4000}}. 

It is not possible to model this aperture with an OSP alone, with or without
a power-law component. Moreover, an AGN component is not expected in this
aperture as it is located far from the active nucleus and does not
contain any line emission -- the UV excess is therefore likely to be
due to a population of young stars. 

Figure \ref{fig:contours} shows the \chisq~space for  two component
OSP plus YSP models. Two minima are clearly seen at $\sim$ 0.1-0.2 Gyr
(\ebv~$\sim$ 0.5) and $\sim$ 
1-2 Gyr (\ebv~$\sim$ 0.2) -- note that viable fits are also found between
the minima. As for other apertures displaying such a
degeneracy in the fits, we can rule out older age YSPs on the basis of
the strength of CaII~K -- models including YSPs older than $\sim$ 1.0
Gyr significantly over-predict the strength of CaII~K. 
For all models in this range, the OSP
and YSP contributions are typically 80-95\% and 10-25\% respectively with
\ebv~$\sim$ 0-0.7. Figure {\ref{fig:SED}} shows an example of a good
fitting model: 12.5 Gyr OSP (91\%) plus 0.1 Gyr YSP (\ebv~= 0.5, 15\%)
with \chisq~= 0.95.

\subsubsection{PKS 0023-26}
\setcounter{figure}{3}
\begin{figure*}
  \begin{minipage}{170mm}
\begin{tabular}{cc}
\multicolumn{2}{c}{\bf 3C 381: 12.5 Gyr   OSP + power law} \\
\hspace*{-1cm}\psfig{file=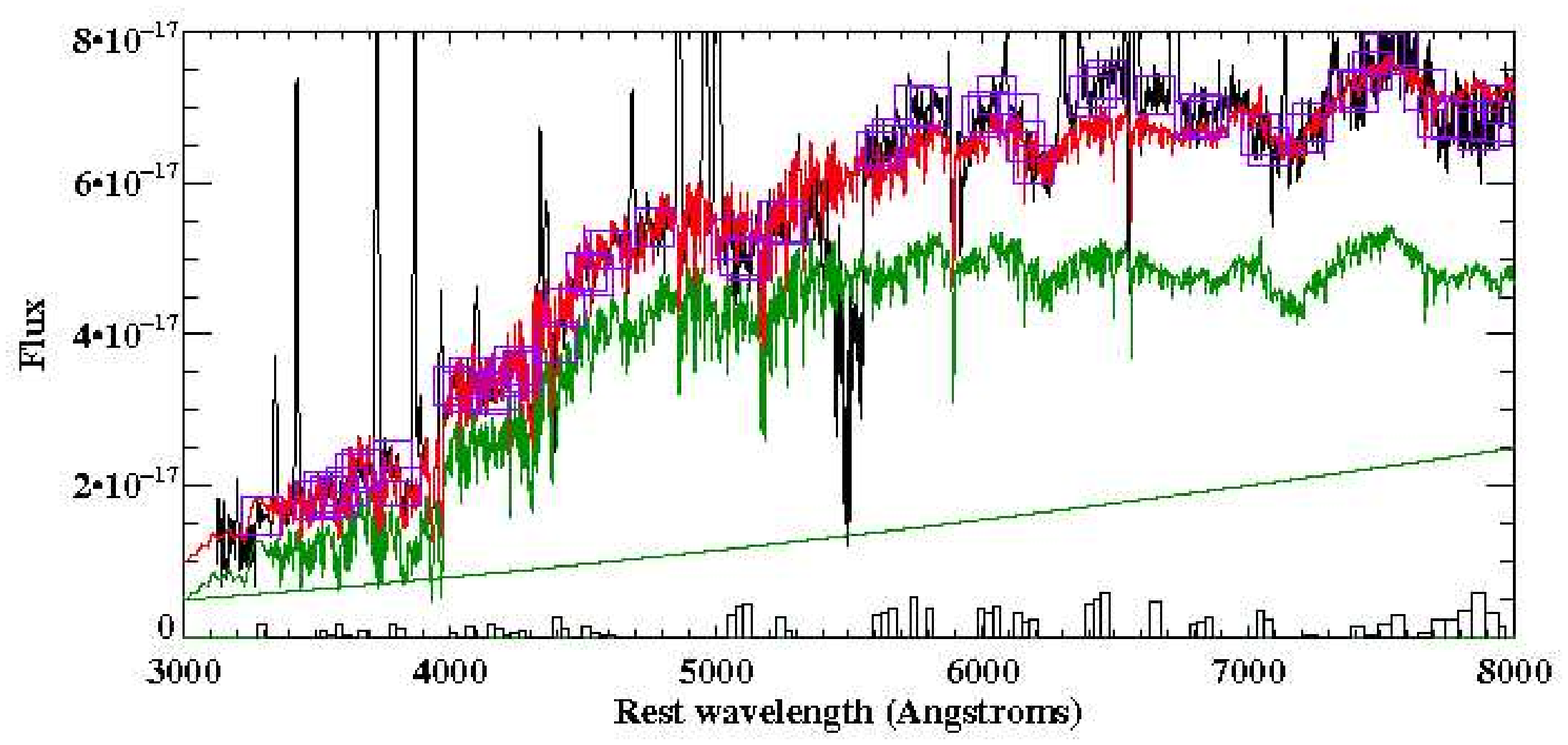,width=9cm,angle=0.} &
\psfig{file=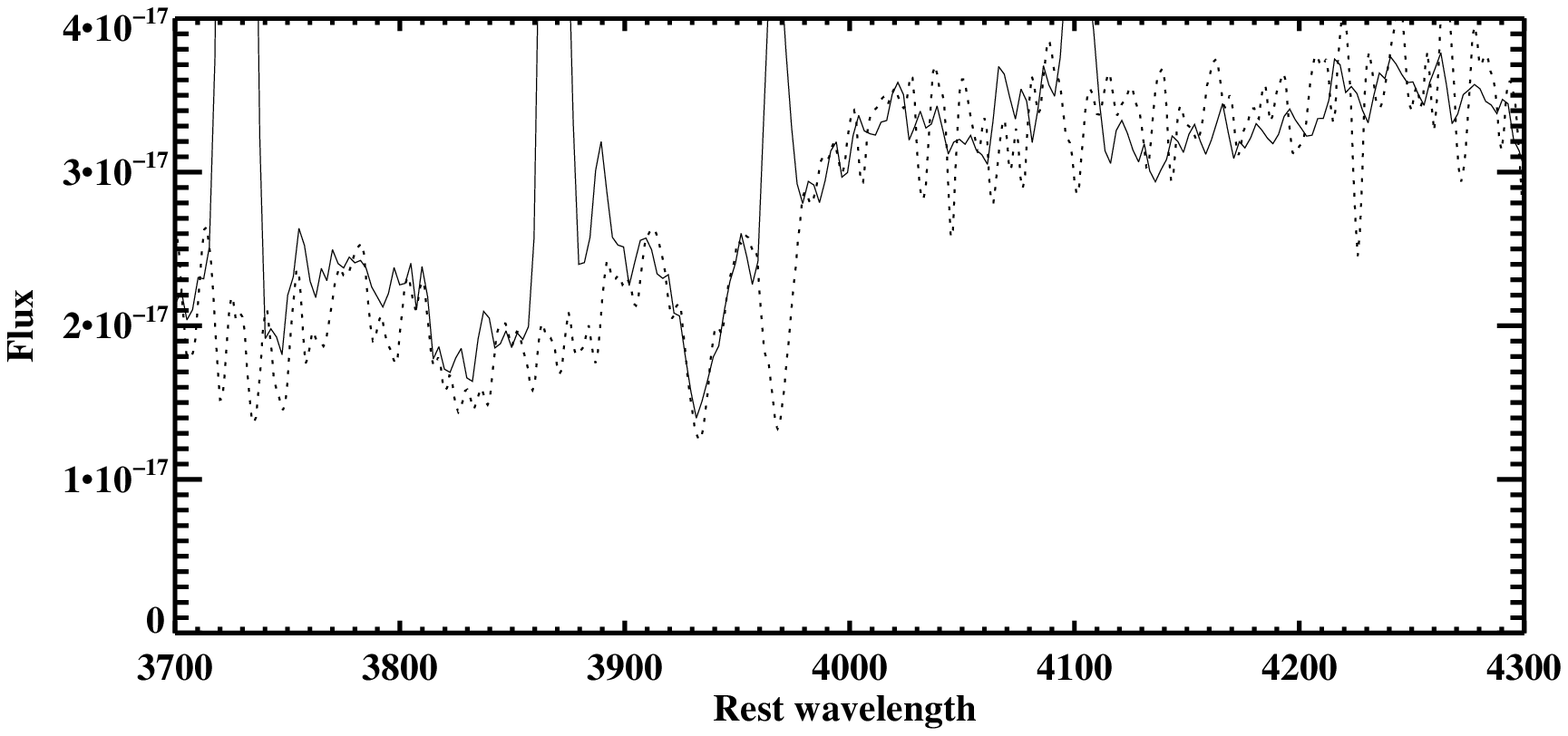,width=9cm,angle=0.}\\
\multicolumn{2}{c}{\bf 3C 381: 12.5 Gyr OSP + 0.02 Gyr YSP (\ebv~= 1.0)} \\
\hspace*{-1cm}\psfig{file=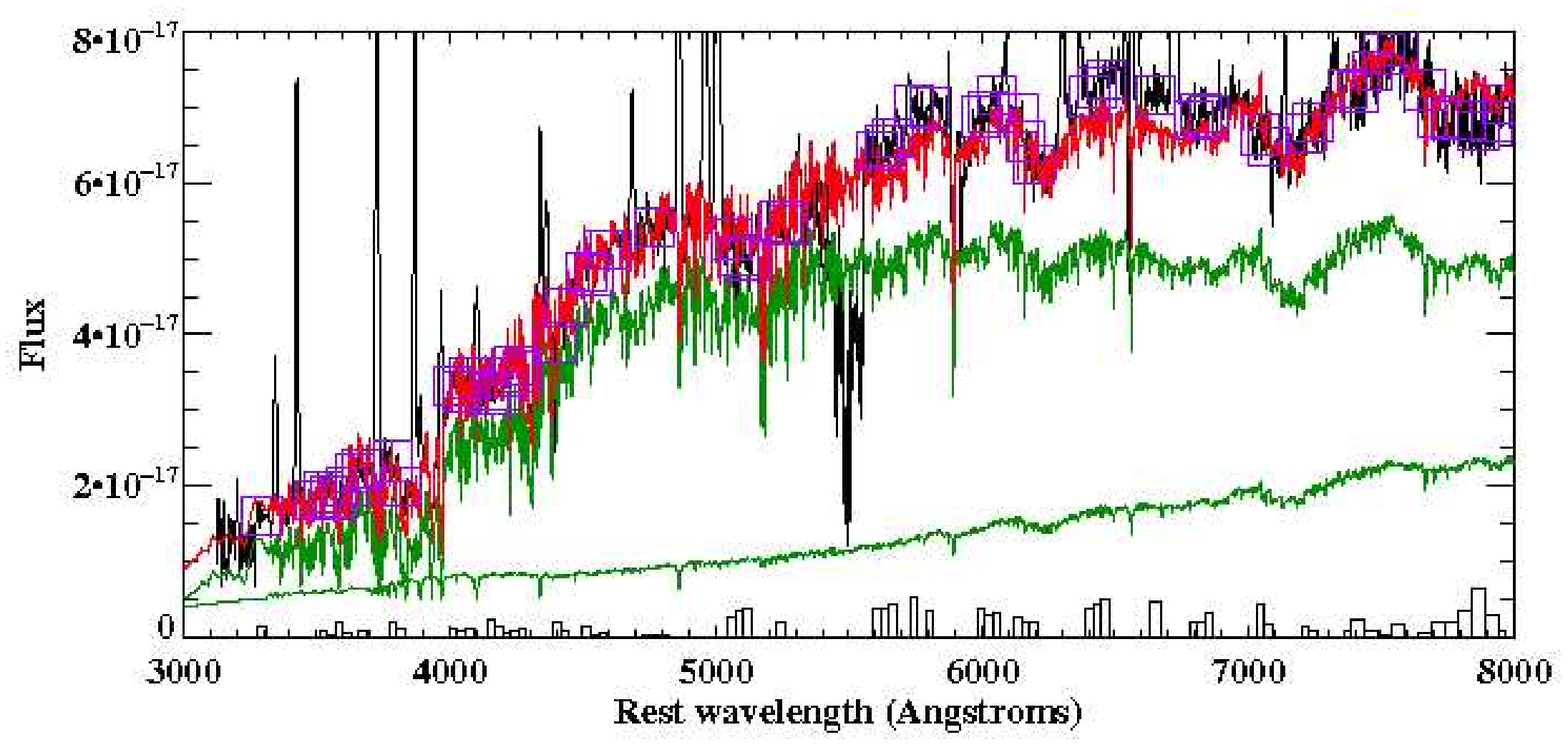,width=9cm,angle=0.} &
\psfig{file=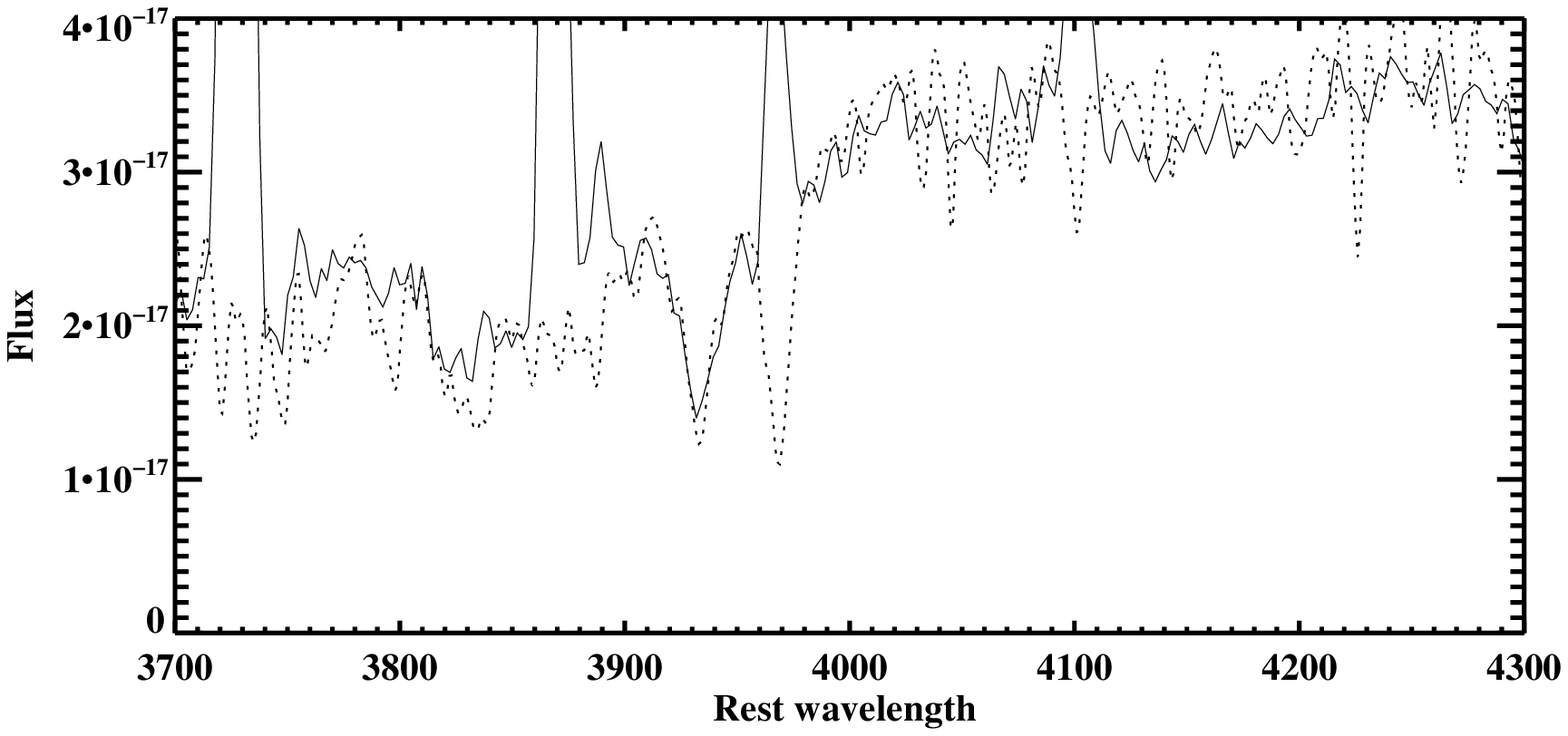,width=9cm,angle=0.}\\
\multicolumn{2}{c}{\bf 3C 433 (SW nuc): 12.5 Gyr OSP + 0.05 Gyr YSP
  (\ebv~= 0.7)} \\
\hspace*{-1cm}\psfig{file=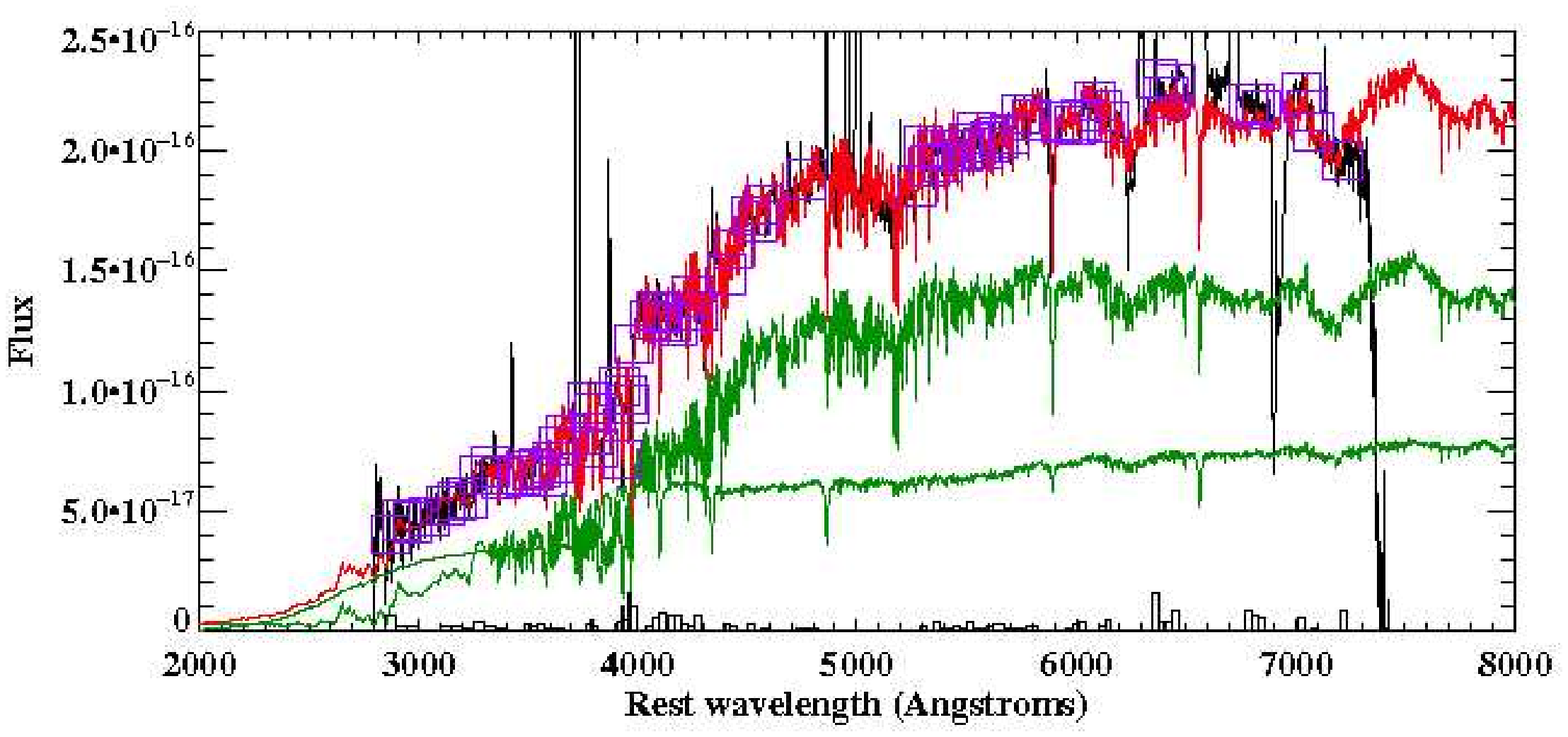,width=9cm,angle=0.} &
\psfig{file=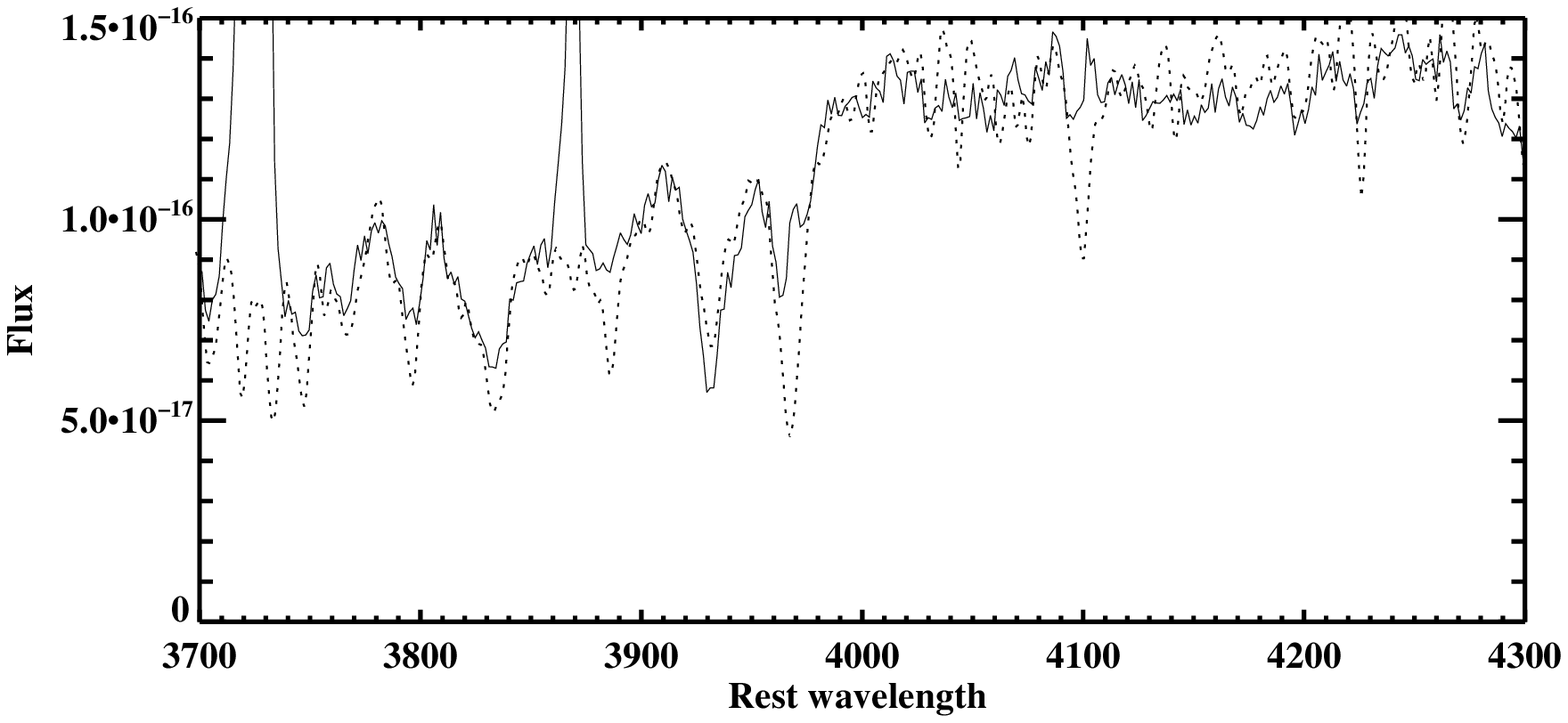,width=9cm,angle=0.}\\
\multicolumn{2}{c}{\bf 3C 433 (mid): 12.5 Gyr OSP + 0.1 Gyr YSP
  (\ebv~= 0.5)}
  \\
\hspace*{-1cm}\psfig{file=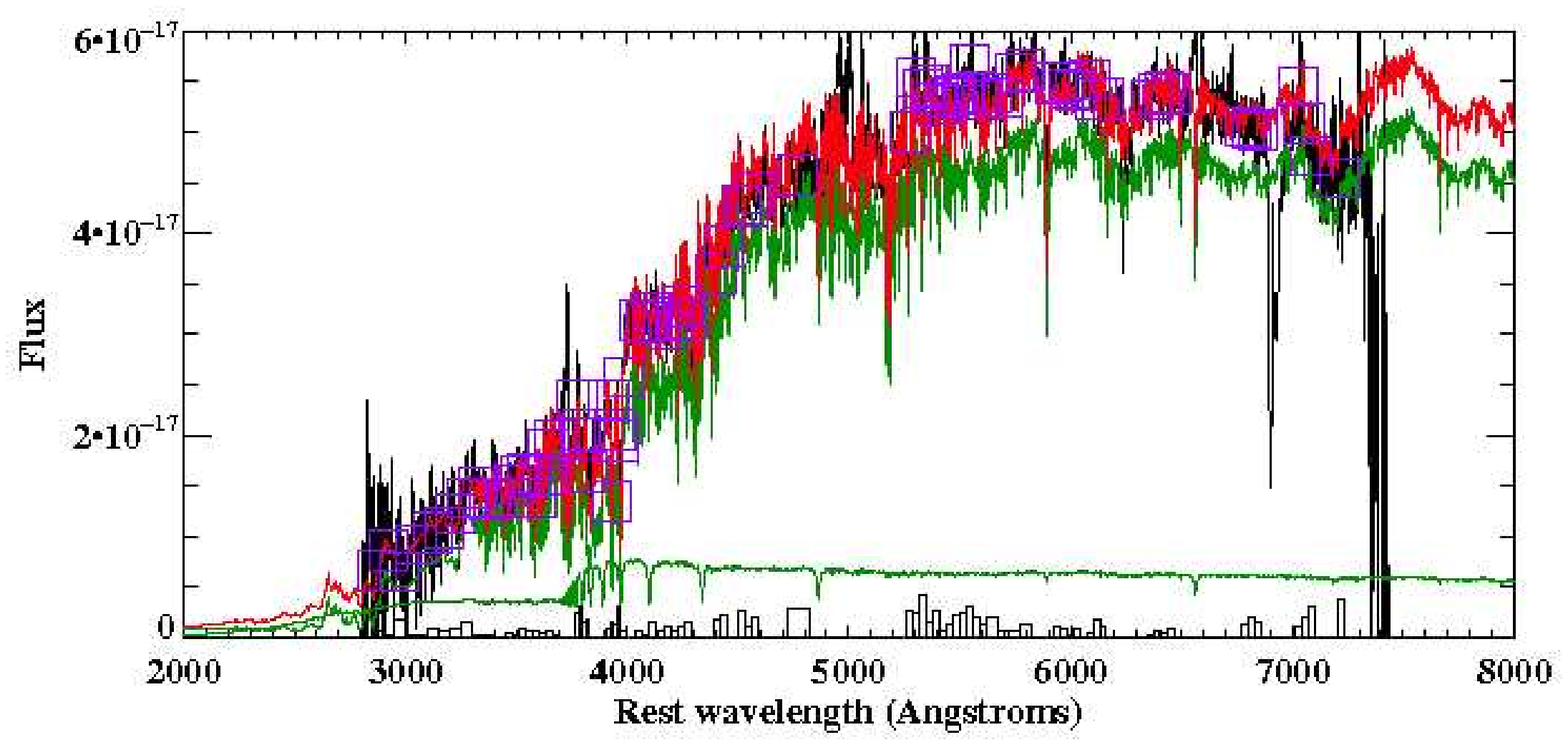,width=9cm,angle=0.} &
\psfig{file=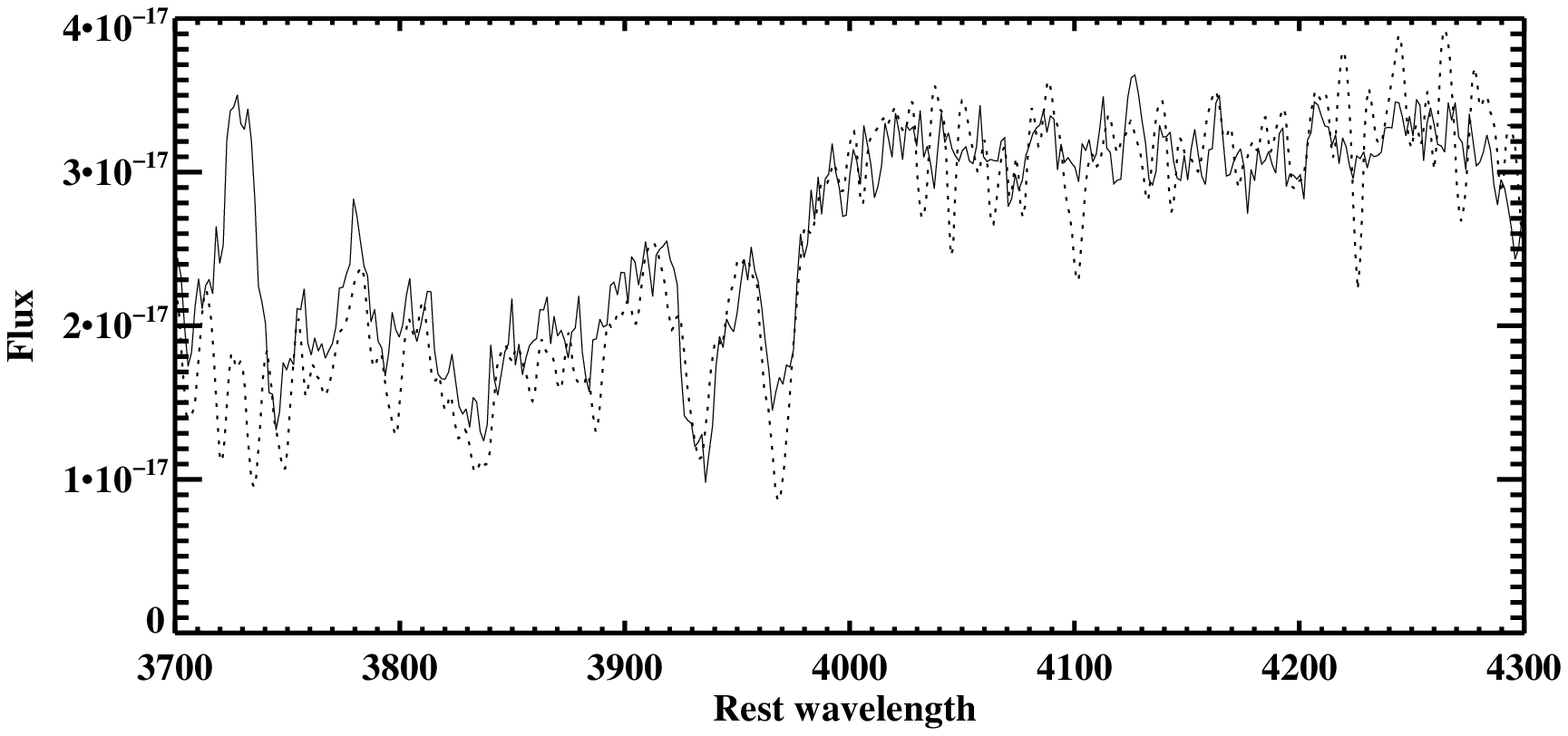,width=9cm,angle=0.}\\
\end{tabular}
\caption[]{SEDs and detailed fits {\it continued}
}
\label{fig:SED}
\end{minipage}
\end{figure*} 
The powerful (log P$_{\rm 5 GHz}$ = 27.33 W Hz$^{-1}$)
Compact Steep Spectrum (CSS) radio source PKS 
0023-26 is identified with an  m$_v$ = 19.5
galaxy \citep{prestage83,wall85}. The double lobed radio source has a
maximum size of $\sim$ 680 mas ($\sim$ 4.25 kpc), aligned along PA
-34 \citep{tzioumis02}. 
Spectroscopically it is classified
as an NLRG and its emission line spectrum has intermediate ionisation
with relatively strong  {[O II]}\lala3727 and  {[N I]}$\lambda$5199
emission 
\citep{tadhunter93,morganti97}.   
The optical continuum displays 
   a large UV excess with low UV polarisation ($P_{UV} < 2.0$\%) and
   no sign of 
  broad permitted lines. This source is therefore a  possible
  starburst candidate 
  \citep{tadhunter93,morganti97,tadhunter02}.  

Figure {\ref{fig:d4000}} shows that the UV excess is extended
beyond the immediate nuclear regions (indicated in figure).
We were unable to obtain adequate fits to the spectrum
of the nuclear aperture of PKS 0023-26 with an
OSP alone or an OSP plus power-law component. Indeed, the spectrum shows
clear evidence for the high order Balmer lines in absorption (see
Figure {\ref{fig:SED}}. 

For models including an OSP and a YSP we find a clear minimum in the
\chisq~space for a young YSP age ($\sim$ 0.03-0.05 Gyr) with
significant reddening 
(\ebv~$\sim$ 0.8) and a $\sim$ 50-60\% 
contribution from the YSP component (see Figure
{\ref{fig:contours}}). Detailed comparisons for this 
aperture are difficult as many of the absorption features are infilled
by strong emission lines. However,  the CaII~K line is not filled and on
the basis of the fit to CaII~K and the surrounding continuum, the best
fitting model is for an OSP plus a young (0.03 Gyr) YSP with
significant reddening (\ebv~= 0.9);  this model is shown in Figure 
{\ref{fig:SED}}: 12.5 Gyr (44\%) plus YSP (0.03 Gyr, \ebv~= 0.9, 54\%)
with \chisq~= 1.1. Models including older age YSPs fail to reproduce the
continuum in this region and, as the age increases, the CaII~K line is
over-estimated. 

As for  many of the other objects/apertures, including a power-law 
component significantly
increases the range of viable YSP fits. A  single broad minimum is observed 
covering a large range of 
 YSP ages (0.05-2.0 Gyr) and reddenings (\ebv~$\lesssim$ 1.2). YSP models
 within this minimum   are
indistinguishable due to the power law contribution (10-60\%). 
However, previous studies of PKS 0023-26
have found that, whilst the 
UV excess is large, there is little evidence for a significant AGN
component -- no broad permitted lines are observed
(making a significant direct component unlikely, \citealt{tadhunter02}
and this paper) and the UV  
polarisation is low, ruling out a large scattered component
\citep{tadhunter02}. 

It is encouraging to note that all of the better fits
including a power-law component require at least 20\% of the flux to
originate from the YSP, often as high as 50\%, although the dominance
of the power law component for some models make distinguishing between
the ages in the three component models difficult.

\subsubsection{PKS 0039-44}
PKS 0039-44 is one of the higher redshift sources in our sample ($z$ =
0.346) and has been previously studied as part of the 2Jy survey by 
Tadhunter et al. (e.g. \citealt{tadhunter93,tadhunter02}). This
FR-II double-lobed radio source  is identified with a galaxy displaying a
rich, high ionisation emission line spectrum \citep{tadhunter93} and
is classified as an NLRG.

PKS 0039-44 shows evidence for a significant UV excess
(see  Figure {\ref{fig:d4000}}) and
\citet{tadhunter02} have previously  investigated the optical/UV
continuum using lower resolution data and spectral synthesis
models. Whilst their observations show PKS 0039-44 to be significantly
polarised ($P_{UV} = 4.8\pm1.2$\%), suggesting a scattered AGN
component is likely to be  
significant, \citet{tadhunter02} believe this component is not necessarily the
{\it dominant} component in the UV. This is because the {\it intrinsic} polarisation
of the continuum component causing the UV excess is
relatively low ($P_{UV} < 10$\%), and they detected no broad emission
lines.  Tadhunter et al.
therefore believe that PKS 0039-44 is a good candiate for a
significant contribution from a YSP. \citet{tadhunter02} also report a
19\% contribution of the 
nebular continuum in the UV. 

As found by \citet{tadhunter02}, we can adequately model the SED using
an OSP and a power-law component (\chisq~= 0.4) contributing 57 and 40\% to
the flux in the normalising bin respectively. As well as reproducing
the SED, this fit also provides an adequate fit to the stellar
absorption features, and the slope of the required 
power-law ($\alpha \sim -0.5$) is consistent with a
scattered AGN component. Further evidence for a
scattered AGN component is provided by the clear detection
of a  broad  MgII feature (FWHM $\sim$ 9000 \kms: see Figure
\ref{fig:pks0039mg2}).

Although the detection of significant UV polarisation and 
a broad MgII feature suggest a major contribution from a scattered AGN
component, 
this does not by itself rule out a significant contribution from
a YSP component. Hence,  we have modelled the SED including both  YSP
and power-law components. Unsurprisingly, given the contribution of the scattered AGN,
the models do not provide strong constraints on the age of the YSP:
{\it all} ages of YSP provide a
viable fit to the SED (\chisq~$\lesssim$ 0.4) and the detailed fits to
the absorption lines are indistinguishable. However, by comparing the
percentage contributions of the components, some models can be ruled
out if it is assumed that both a YSP and a power-law are required
 -- many
of the viable fits do not  require a YSP component (0\%
model contribution). Ruling these out, all of the results are broadly
consistent with component contributions of $\sim$ 20-50\%, $\sim$
few-20\% and  25-40\% for the OSP, YSP and power-law (-0.7 $<$
$\alpha$ $<$ -0.2)
respectively. However, the this only rules out certain regions of the
{\it reddening} space (\ebv~$<$ 1.0 with larger reddenings required
for younger YSPs) and viable models are still obtained for all
ages of YSP. Overall, while the optical
continuum of PKS 0039-44 clearly has a large contribution from a scattered 
AGN component, we cannot rule out a significant contribution from a YSP
conponent. Indeed, such a component would help to explain the fact that the
UV polarisation is lower than expected if the UV excess is solely
due to scattered AGN light \citep{tadhunter02}.

PKS 0039-44 is one of the higher redshift objects in our sample
($z$ = 0.346). Therefore we  have further modelled the SED using different OSPs
in addition to the `standard' 12.5 Gyr population used for the
majority of the sources in the sample. Based on the redshift, a Hubble
constant of 75 \kms~Mpc$^{-1}$ gives ages of the universe at the
source redshift between 10.5
and 13 Gyr for an open and flat universe respectively. We have
therefore also modelled the data with an OSP of age 10 Gyr. This
gives similar results
(50-55\% old, 40-45\% power-law with 0.4 $< \alpha <$ 0.5)
to models with a 12.5 Gyr OSP. Therefore we conclude that, given
the large contribution of the power-law and/or YSP components, our
modelling does not provide strong constraints on the age of the 
OSP component. 

Because the SED of PKS 0039-44 can be adequately modelled by an OSP and a
power-law component and there is the hint of a possible broad component to
Mg II, we do not show a contour plot for the YSP models. As an example of the
fits obtained, we show the model comprising an OSP and a power-law component
with the best chi-squared, comprising a 10Gyr OSP
(57\%) and a power-law component (41\%; $\alpha$ = -0.5) in Figure
{\ref{fig:SED}}.

\subsubsection{PKS 0409-75}
\setcounter{figure}{3}
\begin{figure*}
\begin{minipage}{170mm}
\begin{tabular}{cc}
\multicolumn{2}{c}{\bf 3C 433 (NE nuc): 12.5 Gyr OSP } \\
\hspace*{-1cm}\psfig{file=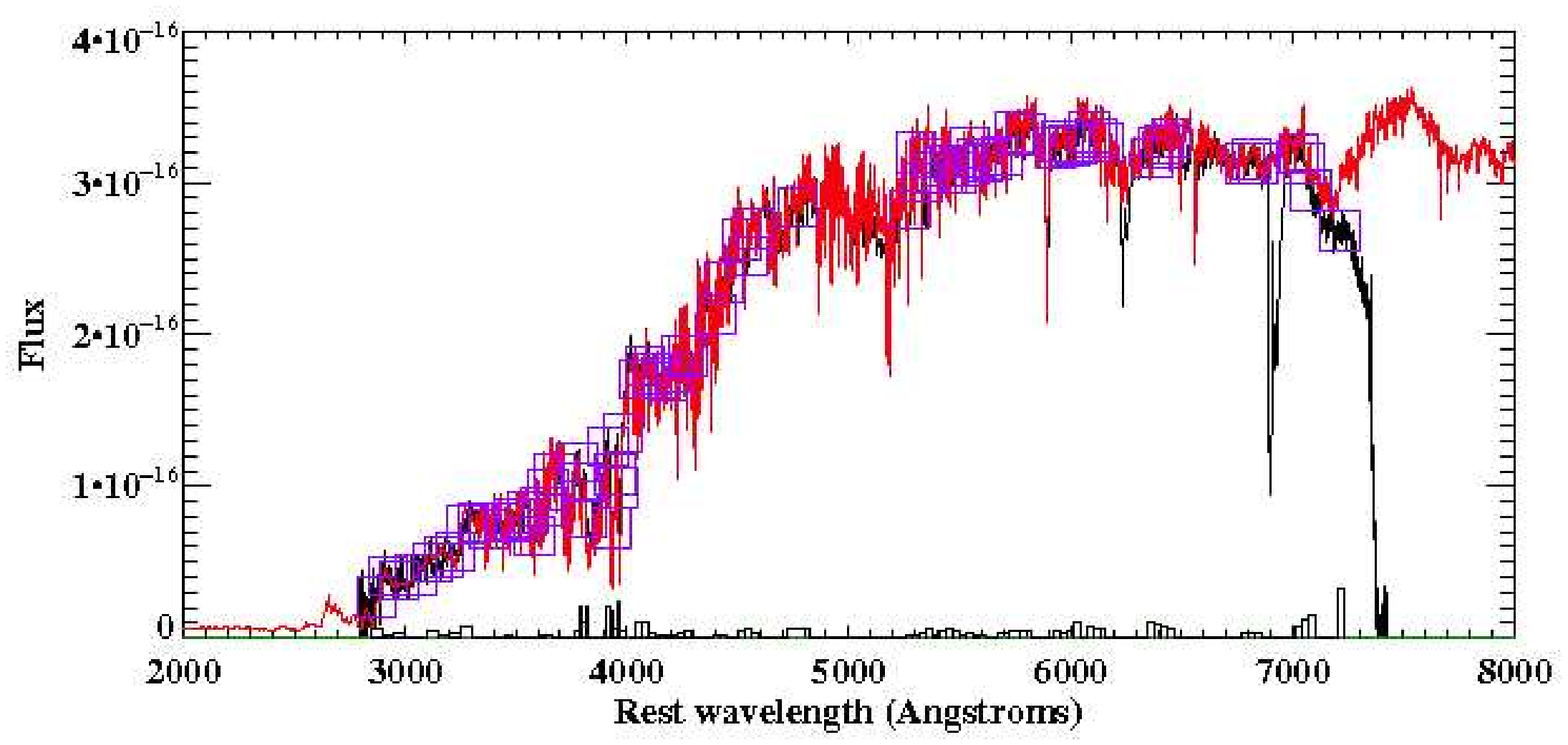,width=9cm,angle=0.} &
\psfig{file=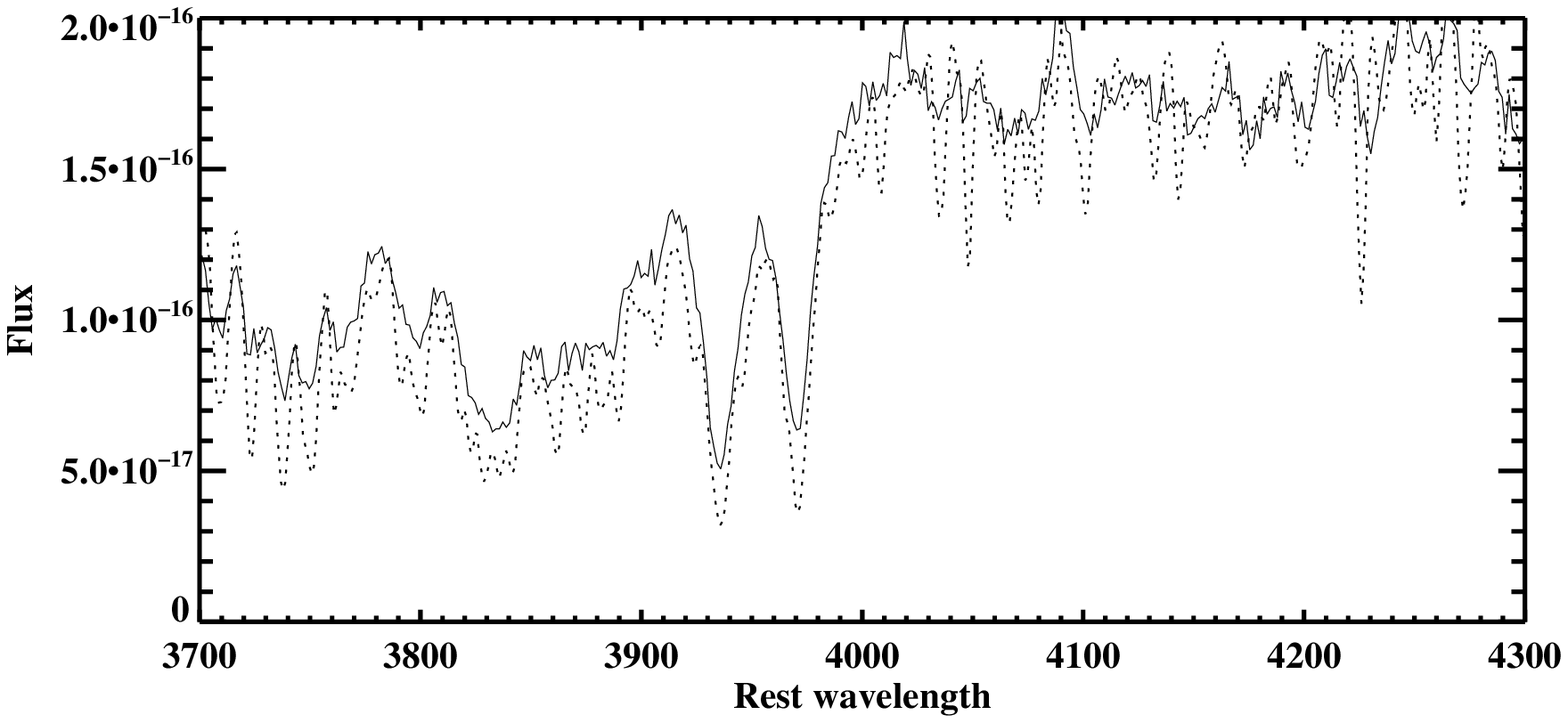,width=9cm,angle=0.}\\
\multicolumn{2}{c}{\bf 3C 433 (NE nuc): 12.5 Gyr OSP + 0.1 Gyr YSP
  (\ebv~= 0.8) } \\
\hspace*{-1cm}\psfig{file=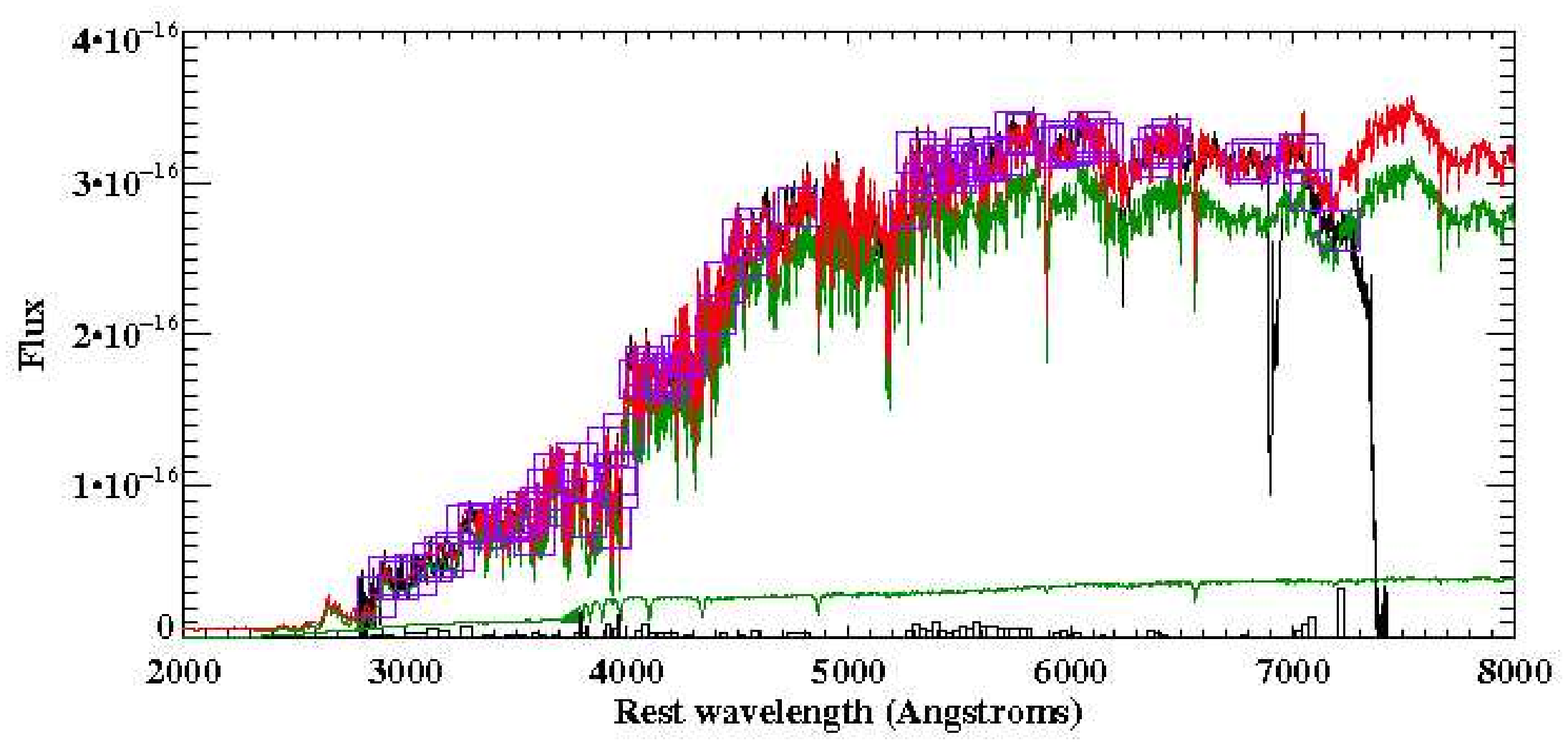,width=9cm,angle=0.} &
\psfig{file=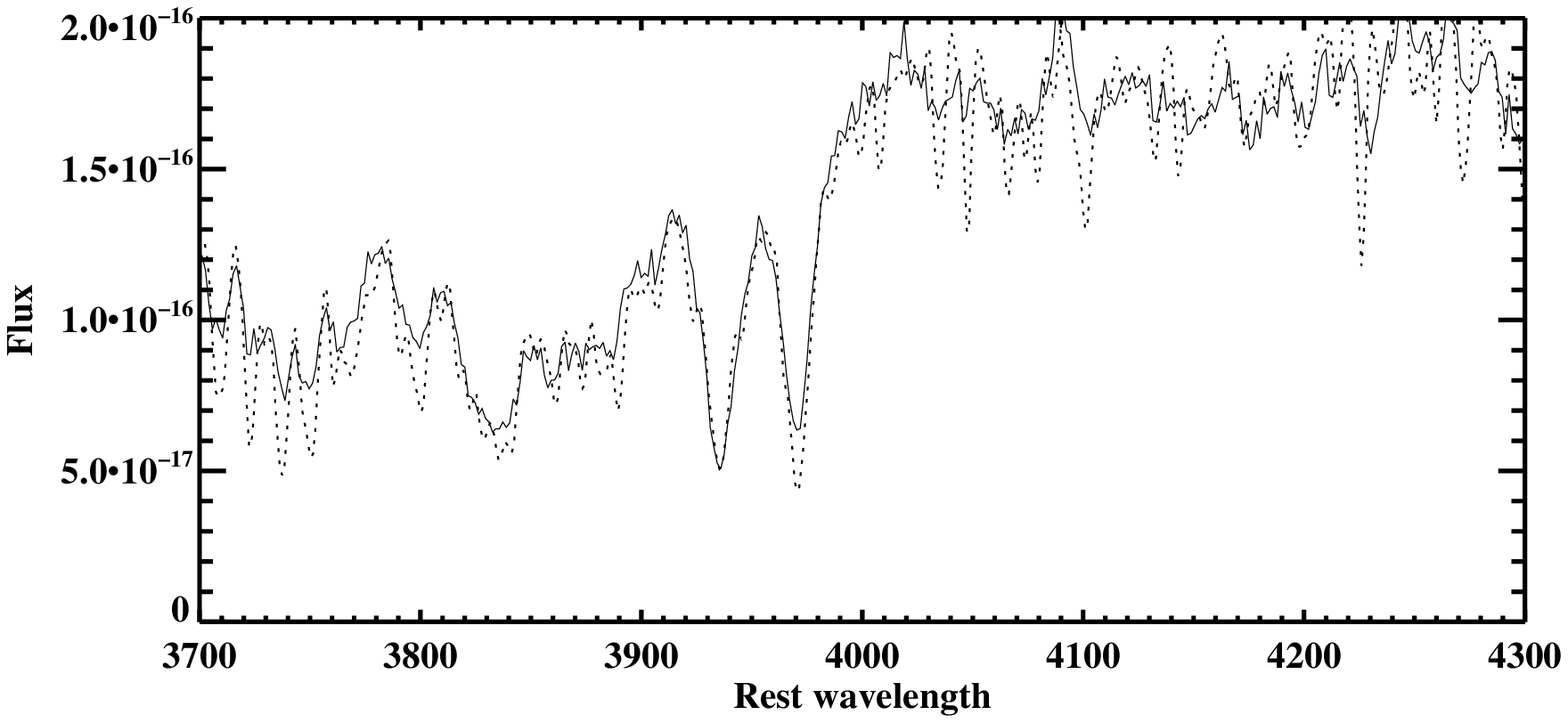,width=9cm,angle=0.}\\
\multicolumn{2}{c}{\bf PKS 0023-26: 12.5 Gyr OSP +
  0.03 Gyr YSP (\ebv~= 0.9)}\\
\hspace*{-1cm}\psfig{file=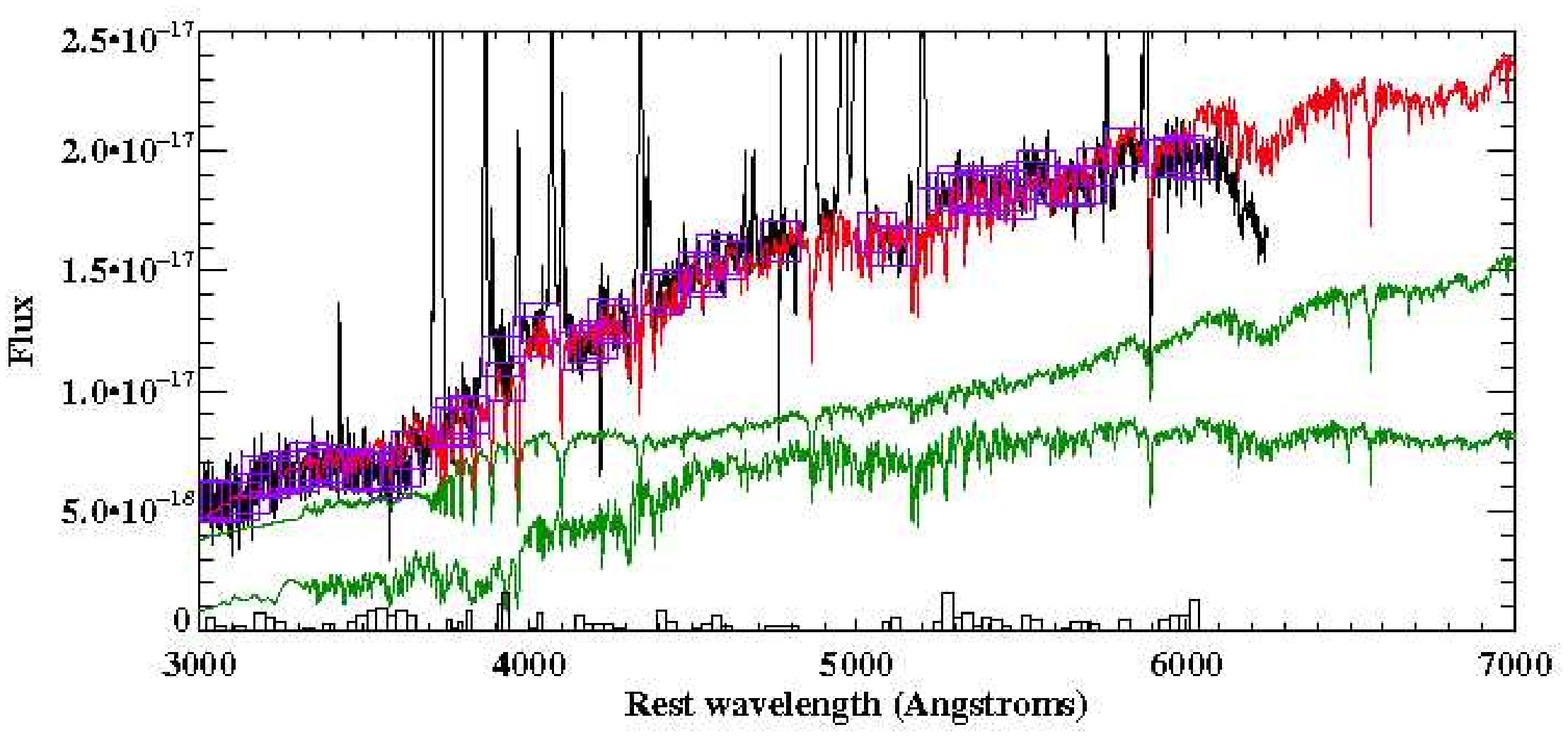,width=9cm,angle=0.} &
\psfig{file=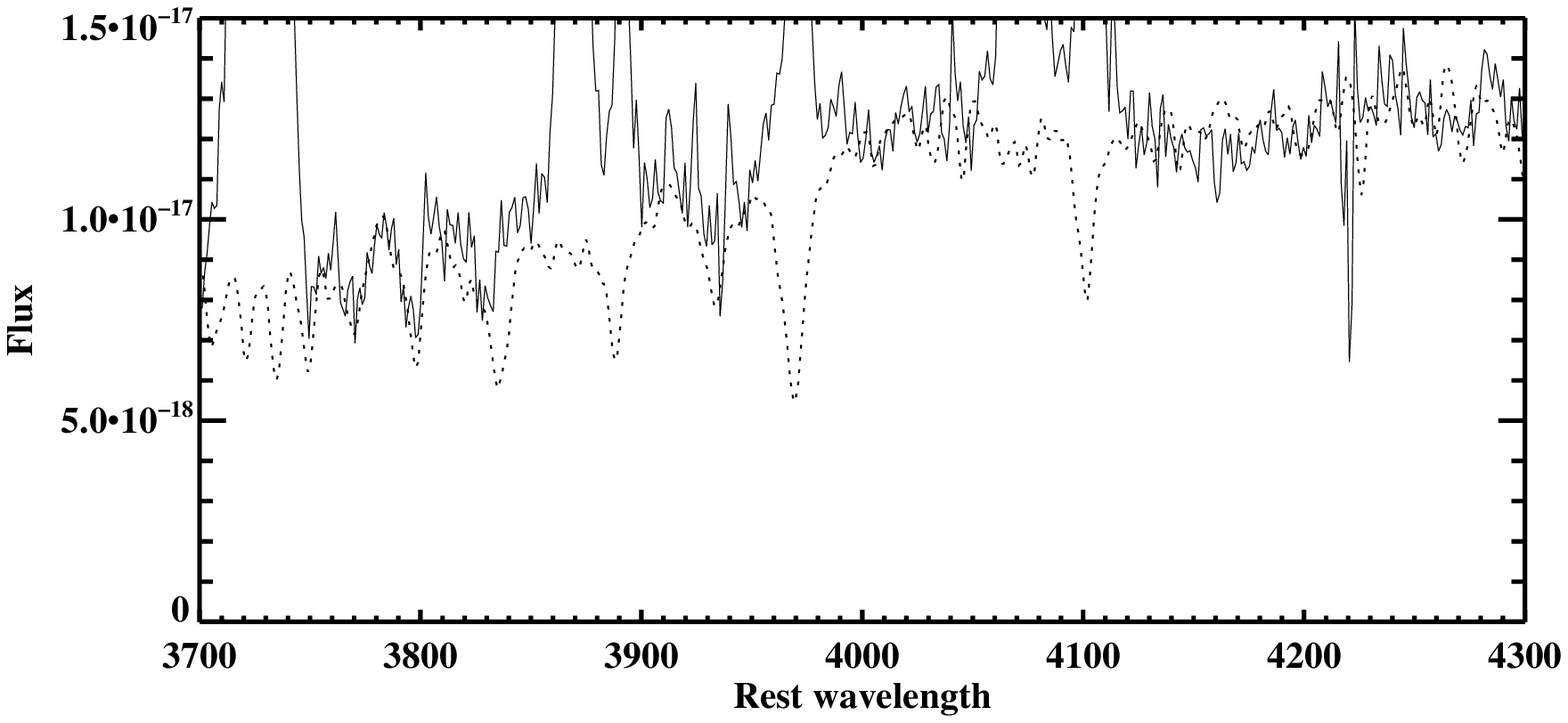,width=9cm,angle=0.}\\
\multicolumn{2}{c}{\bf PKS 0039-44: 10 Gyr OSP + power law}\\
\hspace*{-1cm}\psfig{file=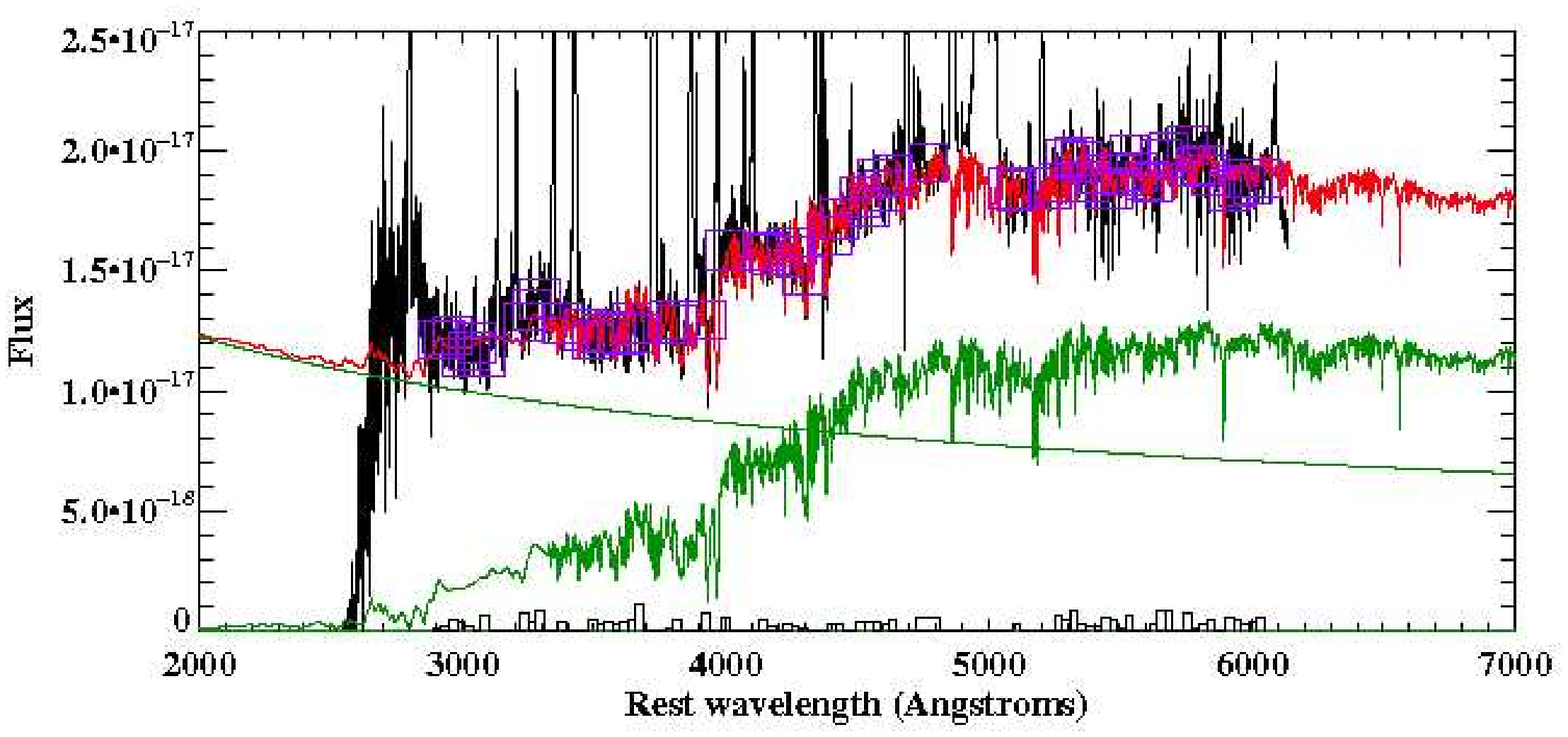,width=9cm,angle=0.} &
\psfig{file=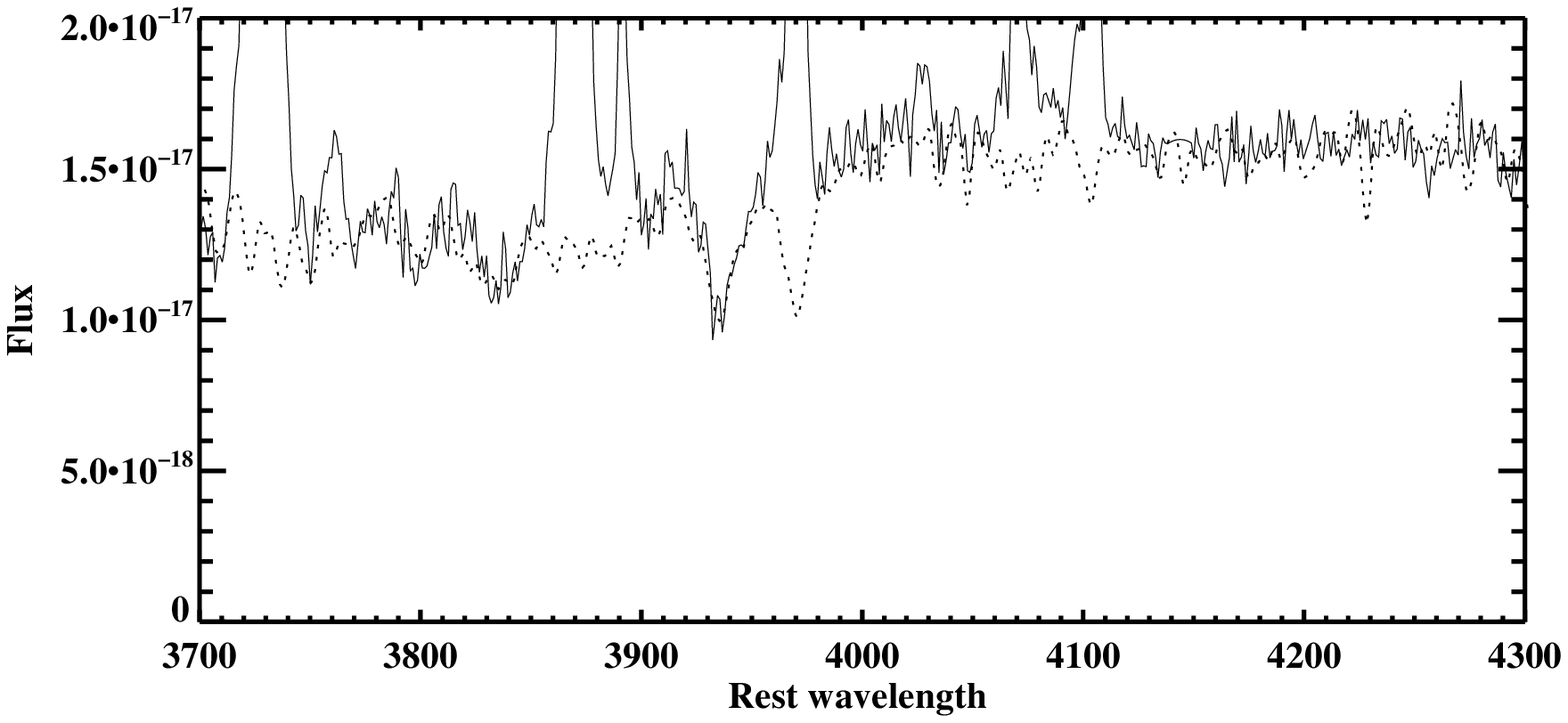,width=9cm,angle=0.}\\
\end{tabular}
\caption[]{SEDs and detailed fits {\it continued}
}
\label{fig:SED}
\end{minipage}
\end{figure*} 
At $z = 0.693$, PKS 0409-75 is both the most distant radio source in
our sample and the most powerful at radio wavelengths. Indeed, it is
one of the most powerful 
intermediate redshift radio sources in the southern hemisphere
(L$_{\rm 45 MHz - 4.8 GHz}$ = 
10$^{45.19}$ erg s$^{-1}$; \citealt{alvarez93}). The 
FR II radio source is relatively small (LAS = 9 arcsec or 85 kpc) and is
dominated by two lobes of emission extending along PA130
\citep{morganti99}. PKS 0409-75 is 
optically identified with an m$_{\rm V}$ = 21.6 galaxy which is
possibly resolved into two components
 and displays extended emission
in {[O II]}\lala3727 \citep{diserego94}.  Low resolution optical
spectra of PKS 0409-75 reveal a significant UV excess compared
with passively evolving elliptical galaxies at similar redshifts
\citep{tadhunter02}. This is also highlighted in our new
measurements shown in Figure
      {\ref{fig:d4000}}. The narrow emission line luminosity
and ionisation state are relatively low for the radio power -- {[O
	  III]} is barely detected whilst {[O II]} is relatively
      strong in the spectrum presented by \citet{tadhunter93}.
\citet{tadhunter02} attempted to model the
optical/UV continuum but were unable to obtain a result due to poor
sky subtraction. They do report a measurement of the
4000\AA~break which suggests 95\% of the light in in the
3750-3950\AA~region is not emitted by the OSP. Further, due to the low
UV polarisation ($P_{UV} < 6.3$\%) and non-detection of broad
permitted lines in the UV spectrum (e.g. Mg II),
Tadhunter et al. propose PKS 0409-75 to be a strong starburst
candidate. 

By using the VLT and paying careful attention to the calibrations
and sky subtraction, we
have obtained new deep, higher resolution optical spectra of PKS 0409-75
which are of sufficient quality to model the SED. 
Using the simplest
models, we find the SED shape can be adequately modelled using an OSP
and a power-law component.  As for PKS 0039-44, we have also considered OSP
ages younger than 12.5 Gyr and find the best fitting model is for a 7
Gyr OSP plus a power-law component  (\chisq~= 1.2), each contributing 60
and 45\% of the flux in the normalising bin respectively. This fit is
shown in Figure {\ref{fig:SED}}. 

However, whilst an adequate fit can be gained without a YSP component,
no significant polarisation has
been detected for this source in the UV \citep{tadhunter02}, the required
power-law spectral index is relatively red ($\alpha \sim$0.86), no 
broad permitted lines have been detected, and the emission line luminosity
is relatively low. Together, these features make a major contribution from
a scattered AGN component unlikely, although without high resolution 
HST imaging we cannot absolutely rule out a contribution from
a direct AGN component. 

Given the lack of clear evidence for an AGN contribution
to the UV excess, we have  also attempted to model the SED of PKS 0409-75
including a YSP component.

For models with an OSP (7 Gyr) and a YSP only, viable fits can be obtained
with a young ($<$ 0.04 Gyr), reddened (\ebv~$\sim$ 0.8) YSP
contributing $\sim$ 25-45\% of the flux in the normalising
bin. Figure \ref{fig:contours} shows the \chisq~space for models
including an OSP and a YSP and Figure {\ref{fig:SED}} shows an example
of a good fitting two component model: 7 Gyr OSP (73\%) plus 0.02 Gyr
YSP (\ebv~= 0.7; 32\%) with \chisq~= 0.75).
Inspection of the detailed fits shows that, due to the noise in
the data, all of the models close to the \chisq~minimum provide good
fits to the absorption 
features and are indistinguishable. 

\subsubsection{PKS 1932-46}
\setcounter{figure}{5}
\begin{figure*}
\begin{minipage}{170mm}
\begin{tabular}{cc}
\hspace*{-1cm}\psfig{file=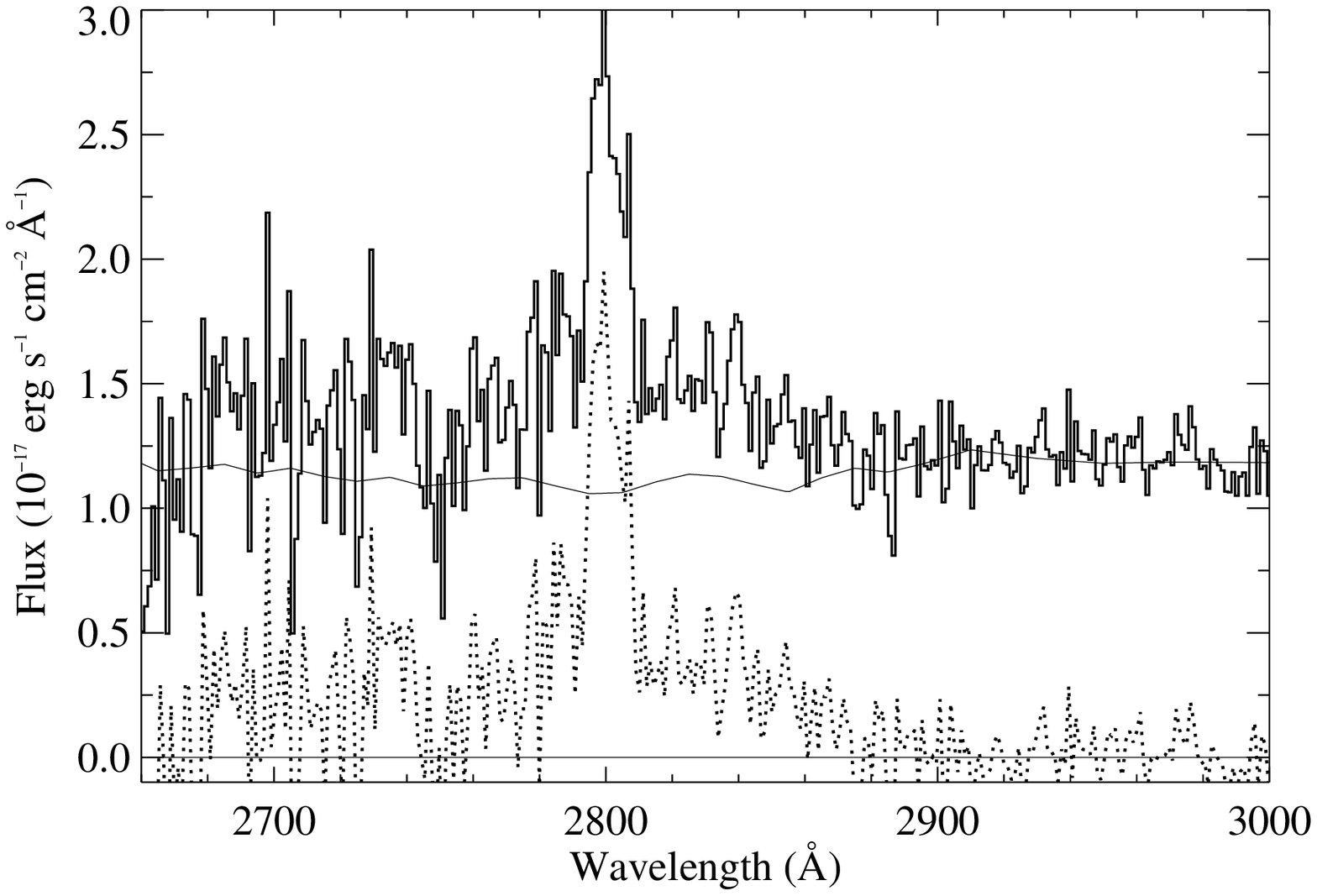,width=9cm,angle=0.}&
\psfig{file=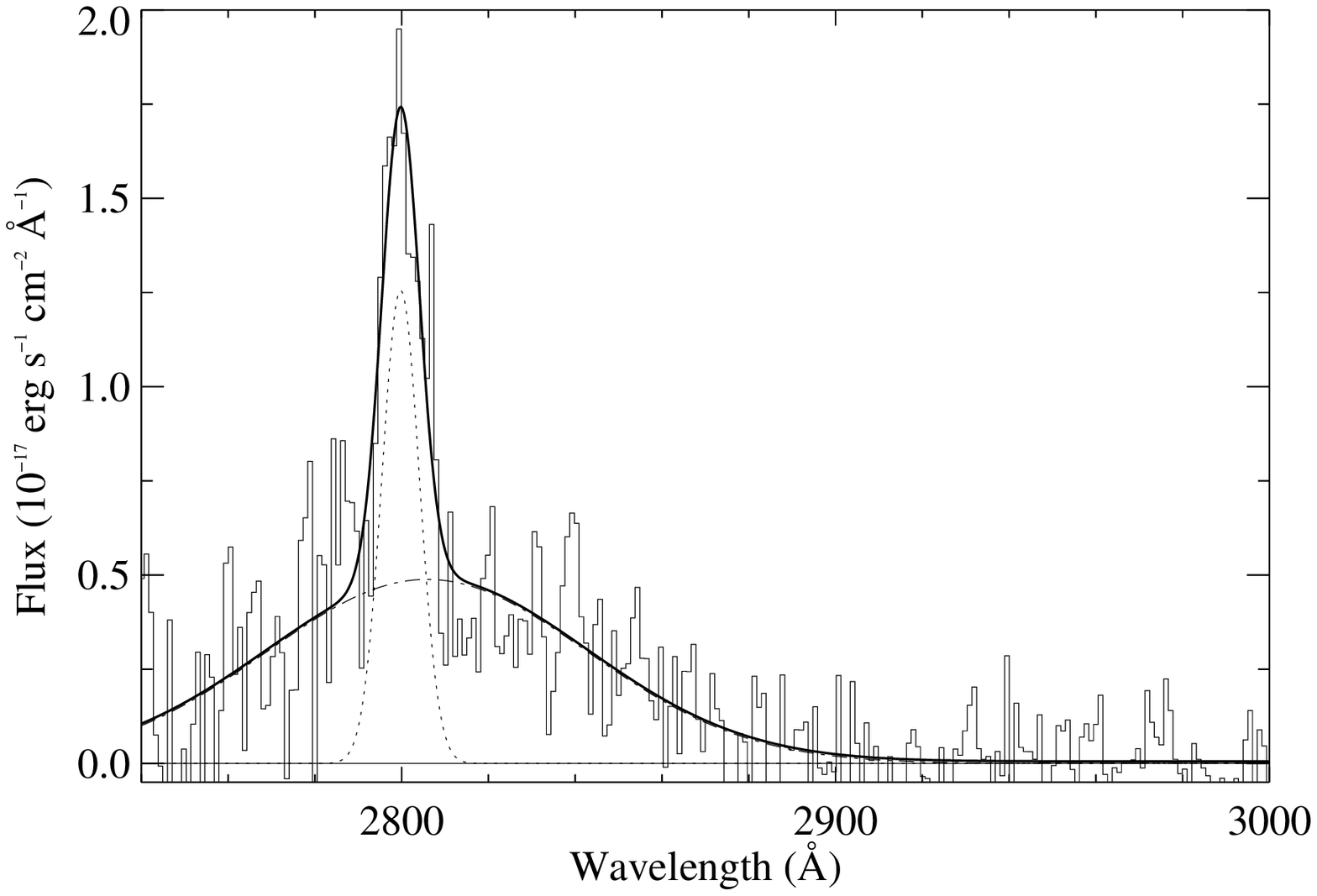,width=9cm,angle=0.}\\
{\it a)} & {\it b)} \\
\end{tabular}
\caption[]{Detection of a broad component to Mg II in the nuclear
  aperture of PKS 
  0039-44. {\it a)} Plot showing a zoomed in region of the nuclear
  spectrum of PKS 0039-44 highlighting the underlying broad component
  to Mg II. The lines plotted are: nebular continuum subtracted nuclear
  spectrum (bold solid line), best fitting (OSP+YSP+power law) SED model
  (solid line) and the continuum (OSP+YSP+power law) subtracted nuclear
  spectrum (dotted line). In plot {\it b)} we model the emission line
  components of the Mg II blend. The 2 component model
  comprises an intermediate ($\sim$ 1100 \kms) component (dotted lines)
  and a broad ($\sim$ 9300 \kms)
  component (dashed lines). }
\label{fig:pks0039mg2}
\end{minipage}
\end{figure*} 
\setcounter{figure}{6}
\begin{figure*}
\begin{minipage}{170mm}
\begin{tabular}{cc}
\hspace*{-1cm}\psfig{file=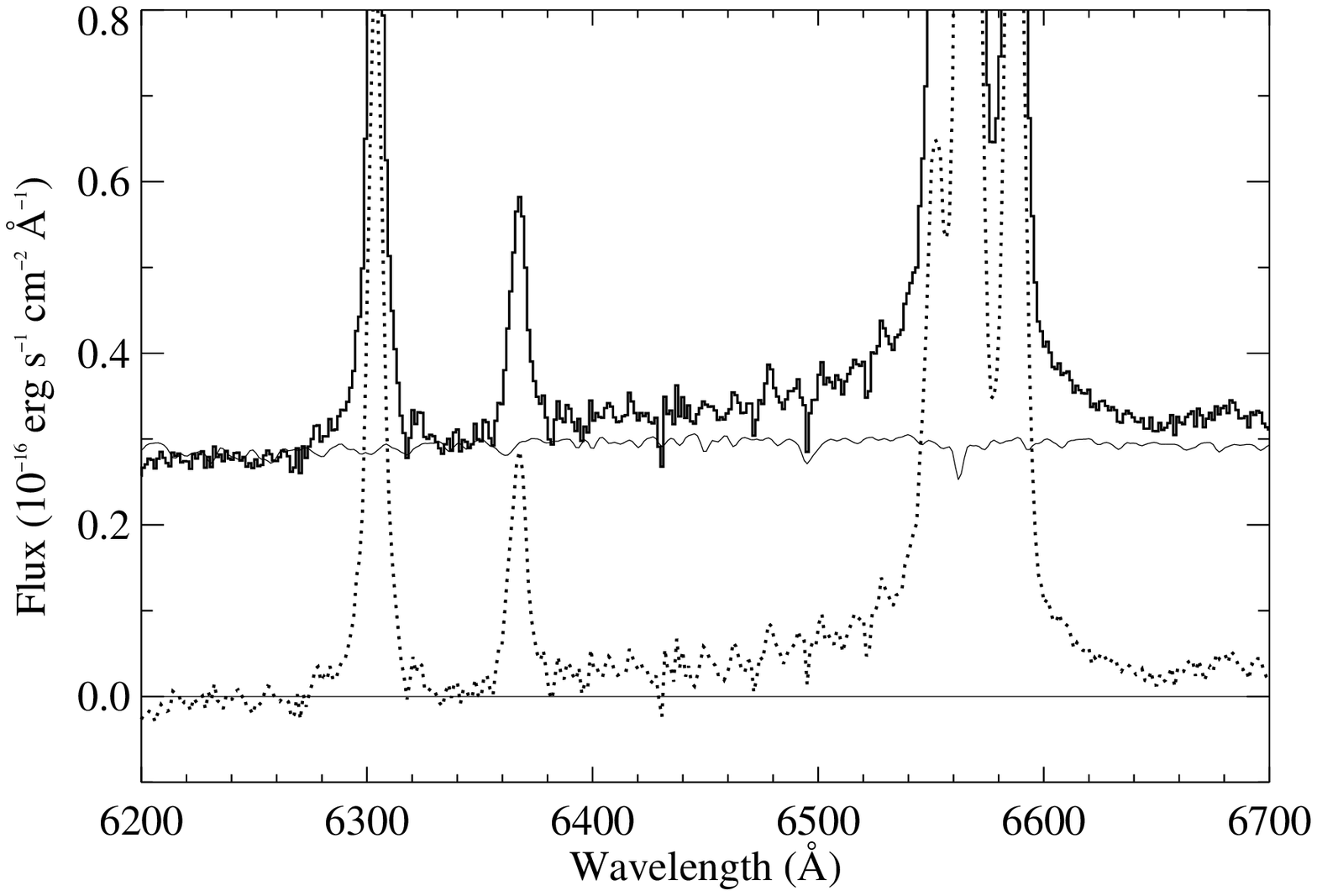,width=9cm,angle=0.}&
\psfig{file=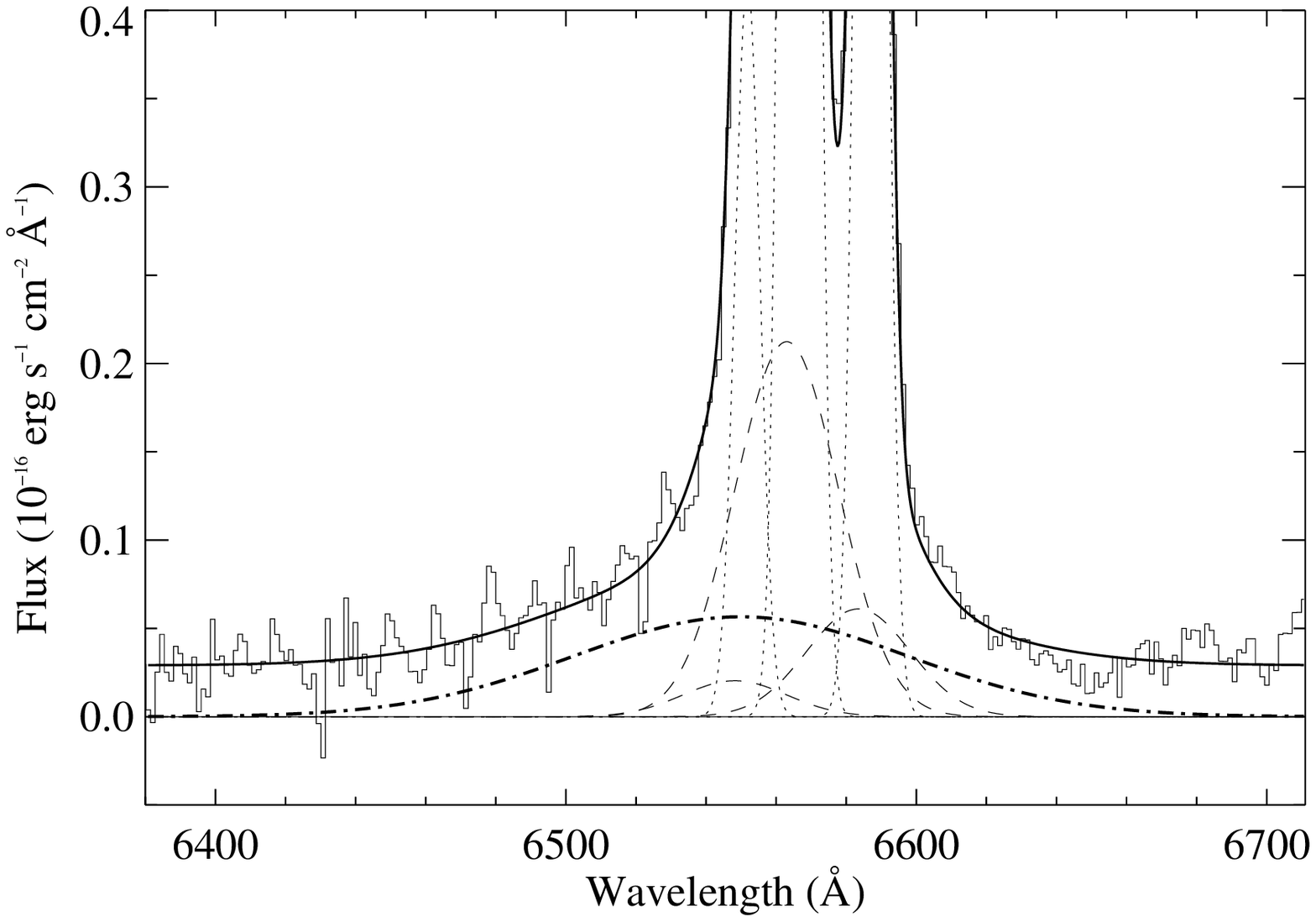,width=9cm,angle=0.}\\
{\it a)} & {\it b)} \\
\end{tabular}
\caption[]{Detection of a broad component to the emission lines in
  the nuclear aperture of PKS 1932-46.  {\it a)} Plot
  showing a zoomed in region of the nuclear 
  spectrum of PKS 1932-46 highlighting the underlying broad component
  to \ha. The lines plotted are: nebular continuum subtracted nuclear
  spectrum (bold solid line), best fitting (OSP+power-law) SED model
  (solid line) and the continuum (OSP+power-law) subtracted nuclear
  spectrum (dotted line). In plot {\it b)} we model the emission line
  components of the \ha/{[N II]} blend. The 7 component model
  comprises a narrow (180 \kms) component (dotted lines)
  and a broad (1550 \kms)
  component (dashed lines) to both \ha~and the {[N II]}\lala6548,6583
  doublet,  and a further very broad (5130 \kms) component (dot-dashed line)
  to \ha. 
}
\label{fig:pks1932halpha}
\end{minipage}
\end{figure*} 
The double-lobed FR II radio source PKS 1932-46 is optically
identified with an early-type galaxy at $z = 0.231$ located between
the two radio lobes but closest to the eastern lobe
\citep{diserego94}.  The galaxy has a double structure in {[O III]}
emission with the two components separated by 5 arcsec (25 kpc). The 
 SW component is coincident with the optical continuum
emission. The main galaxy structure is surrounded by a giant
(up to $\sim$ 160 kpc) emission line 
nebula, making this one of the largest
emission line haloes ever detected around an active galaxy at any redshift
 \citep{villar98,villar05}. Note,
\citet{villar05} used the same data set presented in this paper. 
The halo is complex in structure with clumps and filaments of emission
and evidence for stellar photoionization of
the emission line regions on scales of more than 100~kpc in a direction
close to perpendicular to the radio axis
\citep{villar05}. The nuclear spectrum 
shows a clear UV excess (\citealt{tadhunter02}; see also Figure
{\ref{fig:d4000}}), and a rich emission line spectrum with moderate
ionisation \citep{tadhunter93, villar05}. The detection of weak broad permitted
lines (FWHM $\sim$ 2400 \kms)
was reported by \citet{villar98} although \citet{villar05} retracted
this detection after analyses of higher quality data showed H$\alpha$
could be modelled by the same kinematic components as {[O
    III]}\lala4959,5007. \citet{tadhunter02} have modelled a lower
resolution spectrum of the nuclear regions and suggest that, in addition
to an OSP and power-law component, a contribution from a YSP is also likely.
PKS 1932-46 is not significantly 
polarised in the UV ($P_{UV} < 1.6$\%; \citealt{dickson97,tadhunter02}). 

We find that it is possible to model the nuclear SED of PKS 1932-46
adequately with an OSP and a flat($\alpha$ = 0.23) power-law component
(\chisq~= 0.82) with 59\% and 41\% of the flux in the normalising bin
originating in the two components respectively. The best fitting model
is shown in 
Figure {\ref{fig:SED}}.

Given that the UV polarisation
is low \citep{tadhunter02} and the modelled power-law component is red, this
suggests that any light originating from the AGN will be direct. 
The best
(ground-based) images of PKS 1932-46 known to us are presented in
\citet{villar98} and are of insufficient resolution to determine
whether a strong point source is detected. However, further evidence
for or against a significant AGN contribution can be gained from the
detection/non-detection of broad permitted lines.  As
discussed above, previous searches for broad lines in the nuclear
spectrum give contradictory results. After subtracting off 
the best fitting model (12.5 Gyr OSP plus power-law), 
we find evidence for a  broad component to
H$\alpha$ (see Figure {\ref{fig:pks1932halpha}}) although
this component is significantly broader (FWHM $\sim$ 5100 \kms)
than the original detection (FWHM $\sim$ 2400 \kms). Whilst
\citet{villar05} used the same spectra as presented here and searched
for such a broad component, due to its broadness their non-detection
is not surprising for a number of reasons. First,  although the flux in the
component is relatively large, the 
large FWHM reduces the peak flux  to the point that it is only clear
when a large spectral range is studied. Second, analysis in the
\citet{villar05} paper focussed on a region around H$\alpha$ only
250\AA~wide in the observed frame whilst the component detected here
has FWHM $\sim$ 140\AA~and a total width of $\sim$ 370\AA~in
the observed frame. Finally, \citet{villar05} did not model and
subtract the continuum -- a step which has been shown to be vital for
studying the broader emission line components in radio galaxies
(e.g. in PKS 1345+12; \citealt{holt03}). 
We therefore support the classification of PKS 1932-46 as a BLRG. 

Given the strong evidence for broad lines in the nuclear spectrum of
PKS 1932-46 it is likely that a direct AGN component makes
a major contribution to the optical/UV continuum. However, there
is strong evidence from the HII-region like emission line ratios
for ongoing star formation in the extended halo of this object
\citep{villar05}   
Moreover, when fitting a two component model (OSP plus
power-law), \citet{tadhunter02} remained confident that a YSP component was
required after the subtraction of their model left significant
residuals. Indeed, detailed inspection of the SED in the region
$\sim$3700-4100\AA~reveals a narrow absorption line at $\sim$
3830\AA -- most likely H9 $\lambda$3835. 
Whilst the two component (OSP plus power-law) model reproduces the
majority of the SED well, it fails to provide a good fit to this
feature.  Hence we have also attempted
to model the SED using a three component fit (as the evidence suggests
a significant AGN component is required) to include a
YSP; the plot of the \chisq~space is shown in Figure
{\ref{fig:contours}}. Unsurprisingly, the three component models
produce viable fits 
for all ages of YSP with a large minimum in the range 0.05-3.0 Gyr
with \ebv~$\lesssim$0.8 and it is virtually impossible to distinguish between
these models. However, it is clear that a YSP component is required to
model the H9 line. 
The models have varying contributions from the three
components: $\sim$ 0-60\% OSP, $\sim$ 10-80\% YSP and $\sim$ 5-40\%
power-law with $\alpha$ $\gtrsim$ -0.6. Including the YSP  improves the
fit to H9, with the marginally better fits for
the younger age YSPs suggesting that a YSP component may be important
on the 10-20\% level in the normalising bin. In addition to the OSP
plus power-law fit, we also show an example of a three
component model in Figure {\ref{fig:SED}}: 12.5 Gyr (54\%) plus a 0.05
Gyr YSP (19\%; \ebv~= 0.4) and a power-law component (28\%; $\alpha$ =
0.9). This model has a \chisq~of 0.44.

\setcounter{figure}{3}

\setcounter{figure}{3}
\begin{figure*}
\begin{minipage}{170mm}
\begin{tabular}{cc}
\multicolumn{2}{c}{\bf PKS 0409-75: 12.5 Gyr OSP + power law}\\
\hspace*{-1cm}\psfig{file=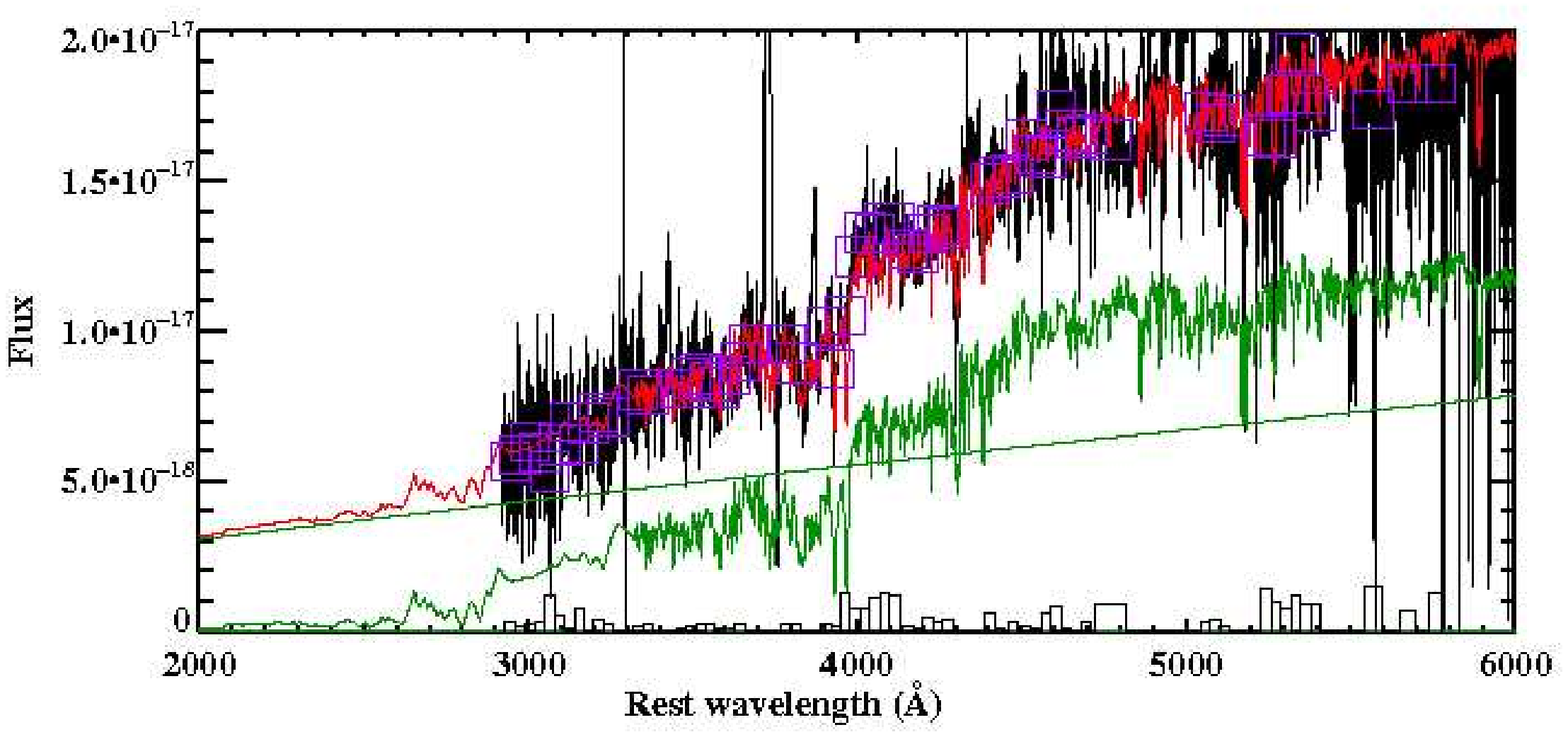,width=9cm,angle=0.} &
\psfig{file=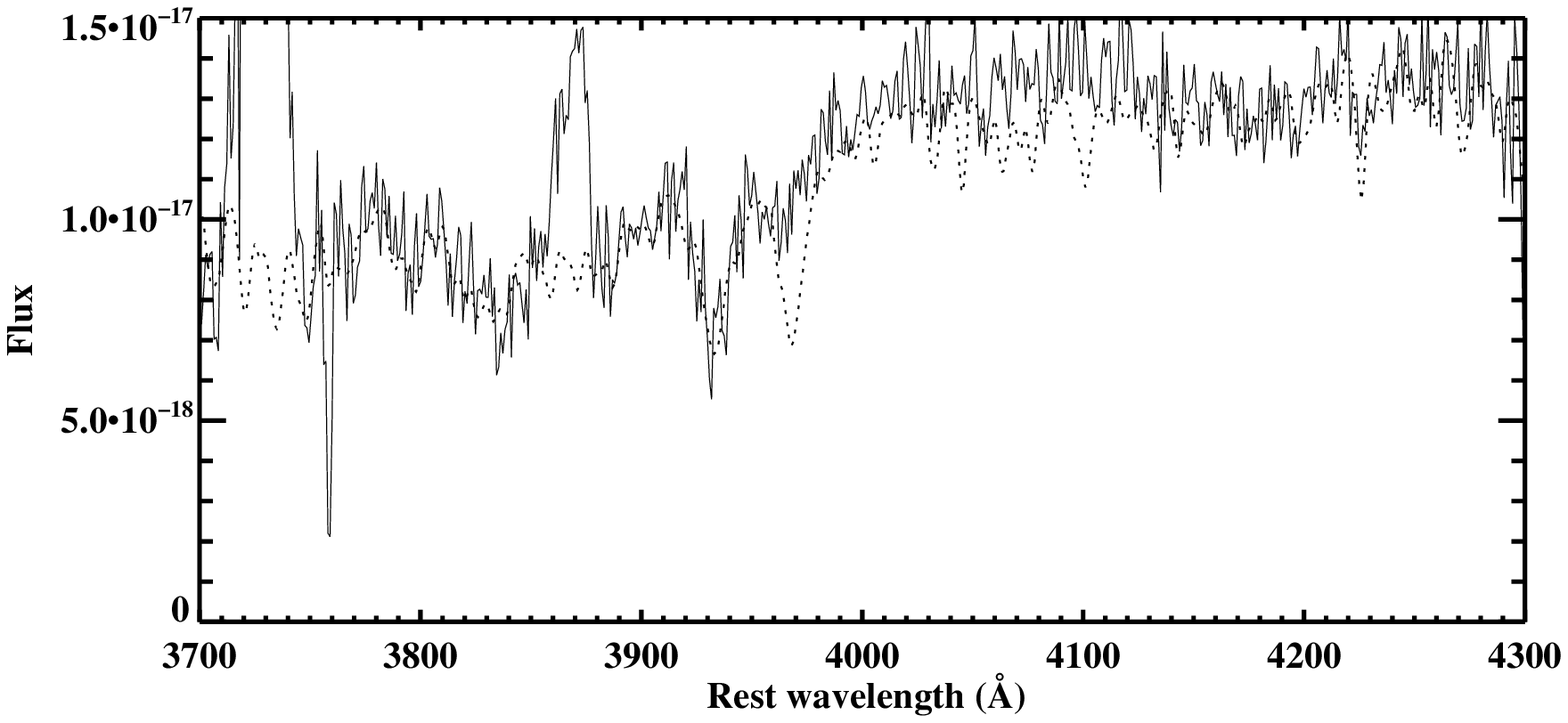,width=9cm,angle=0.}\\
\multicolumn{2}{c}{\bf PKS 0409-75: 12.5 Gyr OSP + 0.02 Gyr YSP
  (\ebv~= 0.7)}\\
\hspace*{-1cm}\psfig{file=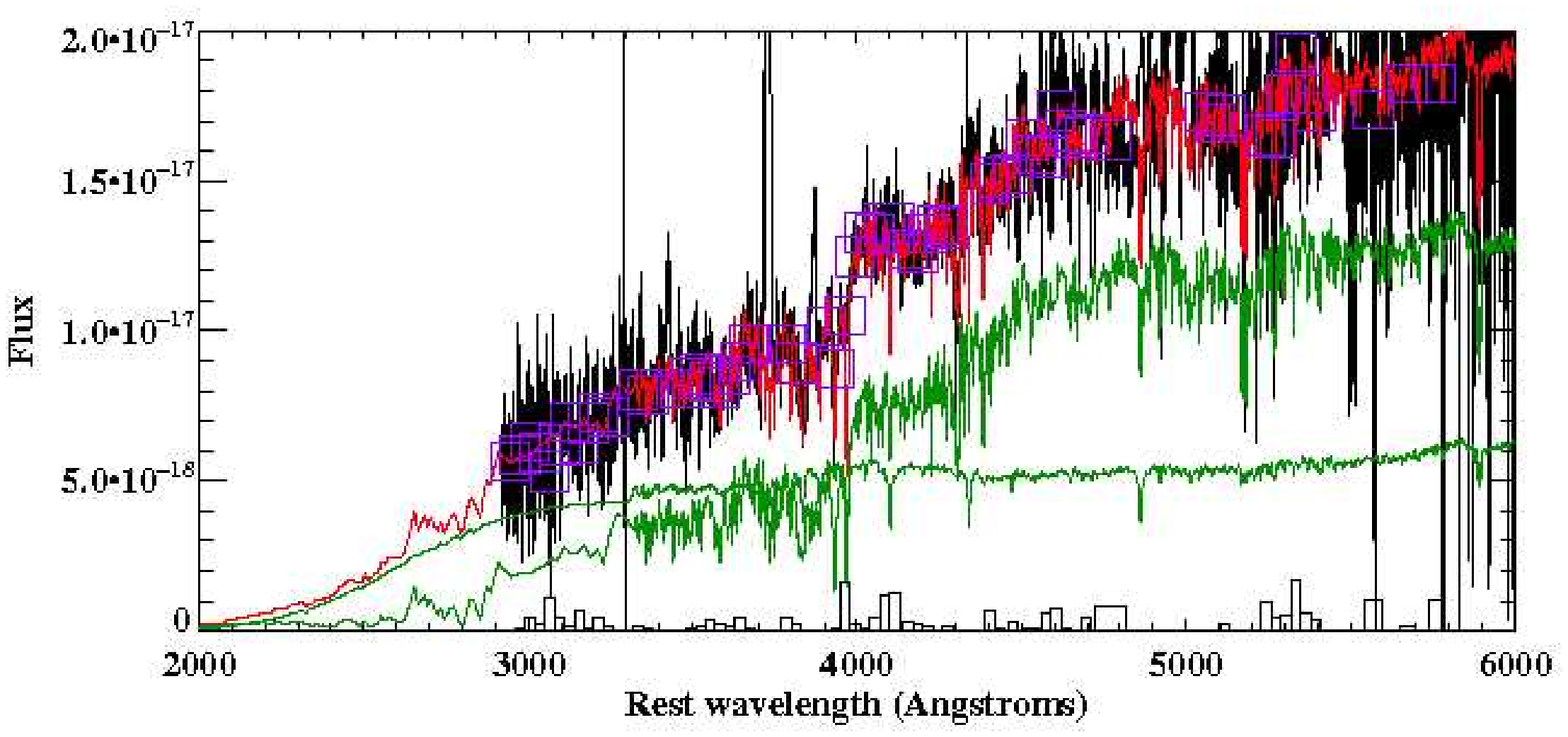,width=9cm,angle=0.} &
\psfig{file=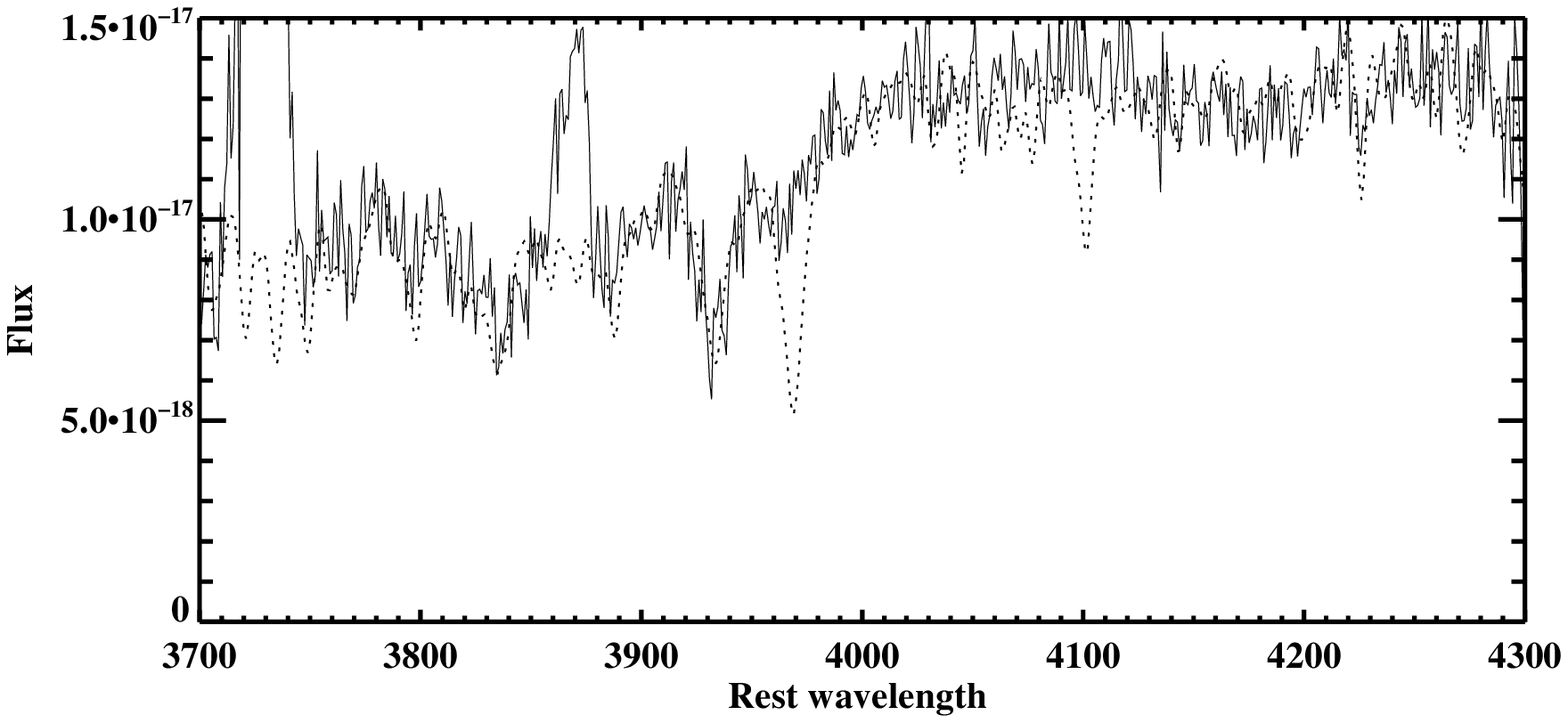,width=9cm,angle=0.}\\
\multicolumn{2}{c}{\bf PKS 1932-46: 12.5 Gyr OSP + power law}\\
\hspace*{-1cm}\psfig{file=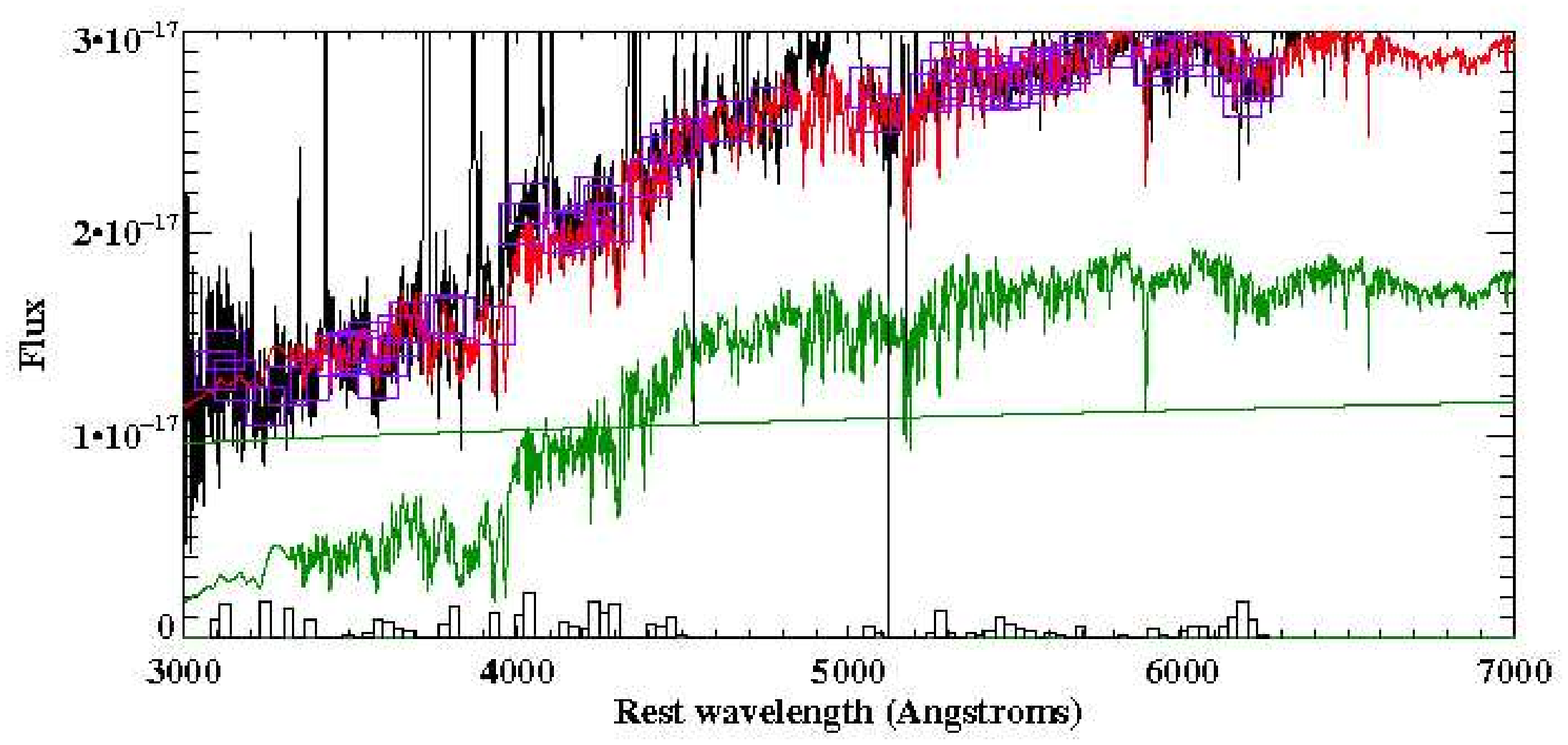,width=9cm,angle=0.} &
\psfig{file=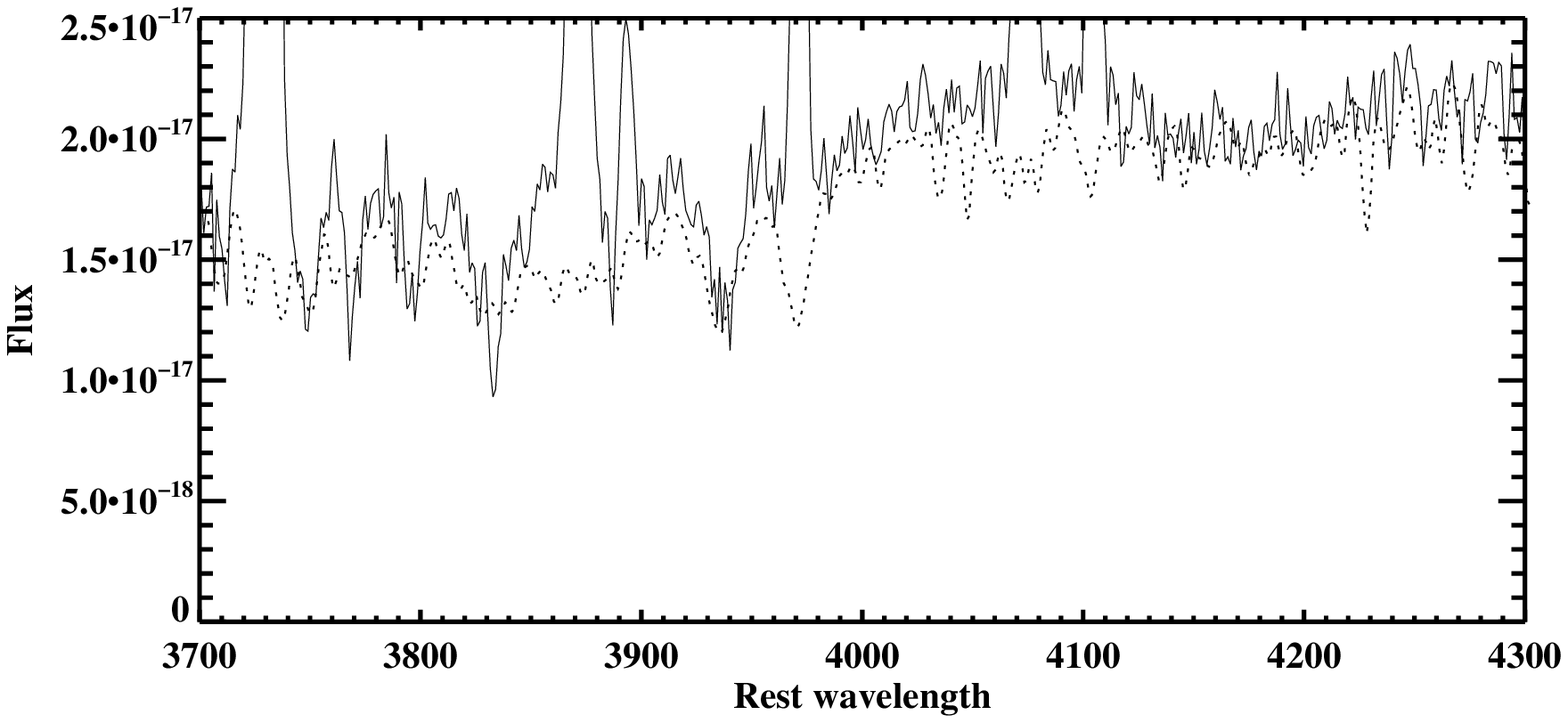,width=9cm,angle=0.}\\
\multicolumn{2}{c}{\bf PKS 1932-46: 12.5 Gyr OSP + 0.05 Gyr YSP (\ebv~= 0.4) + power law} \\
\hspace*{-1cm}\psfig{file=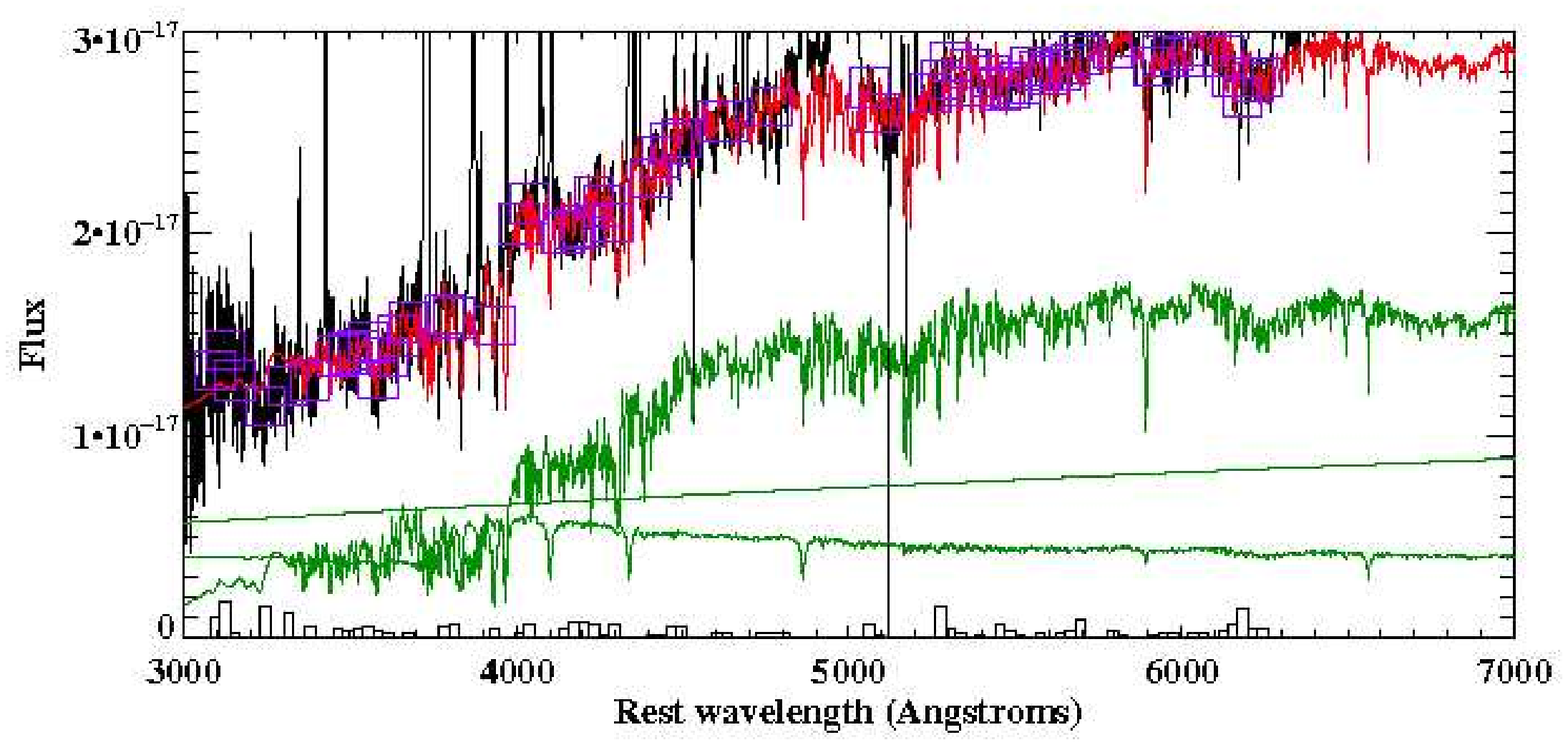,width=9cm,angle=0.} &
\psfig{file=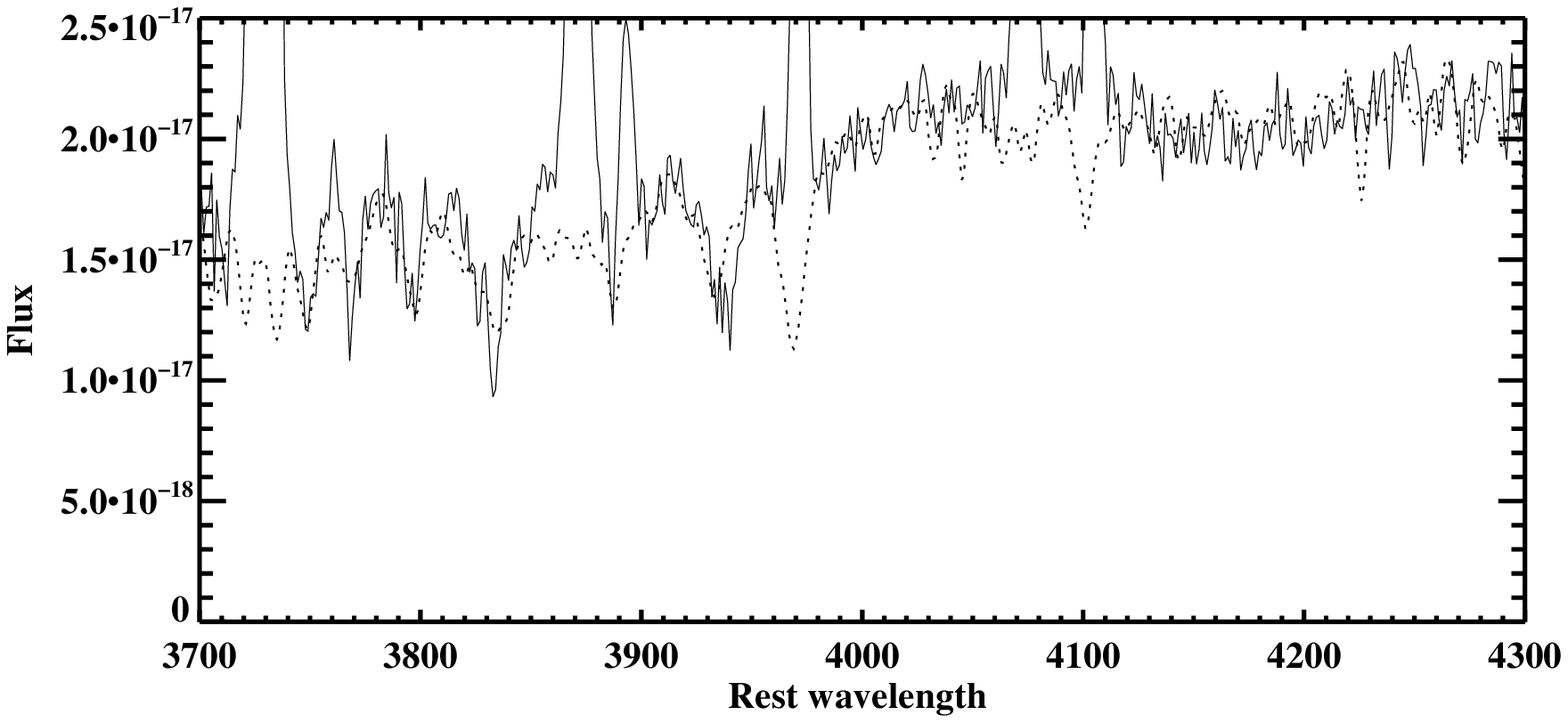,width=9cm,angle=0.}\\
\end{tabular}
\caption[]{SEDs and detailed fits {\it continued}
}
\label{fig:SED}
\end{minipage}
\end{figure*}

\subsubsection{PKS 2135-20}
\setcounter{figure}{7}
\begin{figure}
\centerline{\psfig{file=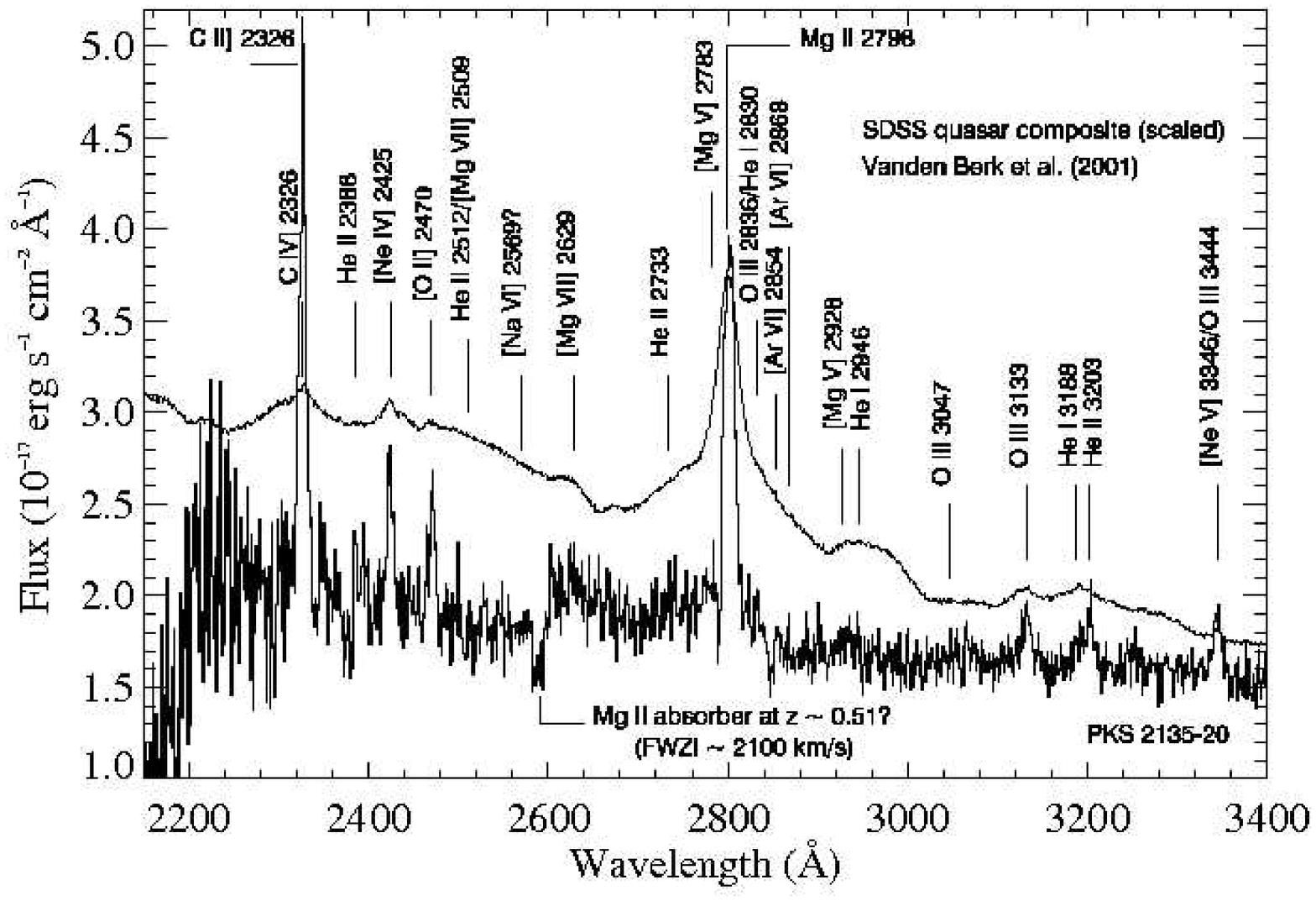,width=9cm,angle=0.}}
\caption[]{Identification of the features in the nuclear aperture of
  PKS 2135-20  
  in the UV in the range 2200-3400\AA. Line IDs are taken from
  \protect\citet{tran98}. 
The nebular continuum subtracted spectrum
  of the nuclear aperture of PKS 2135-20 is plotted along with a
  scaled version of the SDSS quasar template taken from
  \protect\citet{vandenberk01} and line identifications for the known
  emission lines in this region. From this, it seems likely that the
  broad 
  feature at $\sim$ 2600-2700\AA~is the result of several different
  emission line components, 
  namely  a broad component to Mg II$\lambda$2798 plus emission from
  He II$\lambda$2734 and [Mg VII]$\lambda$2632. The dip observed in
  the spectrum of PKS 2135-20 at $\sim$ 2600\AA,which is not observed
  in the quasar template spectrum, is possibly at least partially caused by the
  combination of the quasar with the light from the stellar
  populations -- there is a small `break' in all of the stellar templates at
  $\sim$ 2600\AA~present which appears stronger for older age populations.
}
\label{fig:pks2135}
\end{figure} 
The powerful (log P$_{5 \rm GHz}$ = 27.47 W Hz$^{-1}$)
Compact Steep Spectrum (CSS) radio source PKS 2135-20 is identified with
an m$_V$ = 19.4 galaxy \citep{wall85} at redshift $z$ = 0.635
\citep{diserego94} and has the second highest redshift in our sample.
At radio wavelengths, the  MERLIN 5 GHz image
shows a double-lobed, highly asymmetric 
structure covering $\sim$250 mas  ($\sim$3 kpc) along PA 52
\citep{tzioumis02}. All of the radio flux is detected in the MERLIN image suggesting
there are no extended components.
 
Broad-band imaging 
of PKS 2135-20 by \citet{shaw95} reveals a
compact structure, while spectra and polarisation measurements
demonstrate that
this object has a large UV excess combined with a low UV polarisation
(P$_{UV} < 2.7$\%: \citealt{tadhunter02}).  
Moreover,  \citet{shaw95} and \citet{tadhunter02} report
the detection of high order Balmer lines in absorption, consistent with
the presence of a significant YSP component.

Spectroscopically, the classification of the AGN component in PKS 2135-20 is
uncertain. Originally classified as an NLRG on the basis of its rich
spectrum of narrow emission lines with moderate ionisation
(e.g. \citealt{tadhunter93}), deep EFOSC spectra reveal evidence for broad
permitted lines (e.g. MgII, \hb) suggesting PKS 2135-20 is instead a
BLRG \citep{shaw95}.  
Further, the relatively low UV polarisation of
the source suggests that we are detecting
direct rather than scattered AGN light. 
Similarly, the classification of PKS 2135-20 as a
radio galaxy has been brought into question by \citet{tadhunter98}
who suggest that its absolute magnitude of M$_V <$ -23.0 is consistent with
the classification of PKS 2135-20 as a QSO using the criteria of
\citet{veroncetty93}. The presence of both quasar and stellar spectral
features in PKS 2135-20 suggests that PKS 2135-20 is a radio-loud
example of a post-starburst quasar (e.g. \citealt{brotherton99}).

Our new VLT spectrum of PKS 2135-20
is
consistent with the detection of broad permitted lines -- see Figure
{\ref{fig:pks2135}} for a detailed look at the features in the UV. We
therefore confirm the classification of this object as a
BLRG or quasar. Given the detection of the broad lines 
it is likely that a significant power-law component is
required. However, it is not possible to  model the SED solely
with an OSP and a
power-law (\chisq~$\sim$2). Hence, we have also attempted
to model the SED including a YSP component, consistent with the
presence of the higher order Balmer lines in absorption
and a significant Balmer break (see Figure {\ref{fig:SED}}). 
In addition, because PKS 2135-20 has one of the
highest redshifts in the sample, and we observe down to $\sim$
2300\AA~in the rest frame, we have also investigated the effects of
different reddening laws on the YSP. 

Figure {\ref{fig:contours}} shows the \chisq~space for three component
(OSP, YSP and power-law) models. Consistent with the results for the other sources
studied in this paper, a three component model provides a
large range of viable fits to the SED. For PKS 2135-20, a clear
minimum (\chisq~$\sim$ 0.45) is observed for YSPs with age $\sim$
0.2-0.3~Gyr and reddening  \ebv~$\sim$ 0.1-0.5.

Due to the strong power-law-like continuum shape and broad emission
lines, it is difficult to distinguish between the models using the
stellar absorption lines -- the CaII~K and the G-band are
relatively weak. A few of the higher order Balmer lines are
visible around 3800\AA~but these appear to be fitted well by all of
the models. A possible discriminator could be the continuum shape
between $\sim$ 3800\AA~and $\sim$ 4100\AA~-- the older age YSPs
($\gtrsim$ 0.7 Gyr) increasingly underpredict this region and the CaII~K
line,  whereas the younger age YSPs (0.04-0.6 Gyr) reproduce this
region and CaII~K well.  For this range of ages of YSP, the models
typically comprise: OSP (12.5 Gyr; 0-24\%), YSP (0.04-0.6 Gyr,
\ebv~$\sim$ 0.2-0.5; 45-77\%) and power-law ($\alpha$ $\sim$ -1.3 -- -1.4;
3-42\%) with \chisq~$\sim$ 0.4-0.6. Note, the OSP and YSP contributions
decrease with increasing age of the YSP, but the power-law component increases. An
example of a model in the best fitting range is shown in Figure
{\ref{fig:SED}}: 12.5 Gyr OSP (22\%) plus 0.2 Gyr YSP (\ebv~= 0.2;
45\%) plus power-law ($\alpha$ = -1.47; 34\%) with \chisq~= 0.42.

As for some of the other sources at high redshift in our sample, we
have also attempted to model the SED using a younger OPS with age
7~Gyr --- consistent 
with the age of the Universe at the redshift of the source. This
gave similar results to the 12.5 Gyr OSP models -- the level, pattern
and location of the minimum of the \chisq~space contours are similar,
as are the detailed fits.

\subsubsection{NGC 612}
\setcounter{figure}{3}
\begin{figure*}
\begin{minipage}{170mm}
\begin{tabular}{cc}
\multicolumn{2}{c}{\bf PKS 2135-20: 12.5 Gyr OSP + 0.2 Gyr YSP (\ebv~=
 0.2) + power law} \\
\hspace*{-1cm}\psfig{file=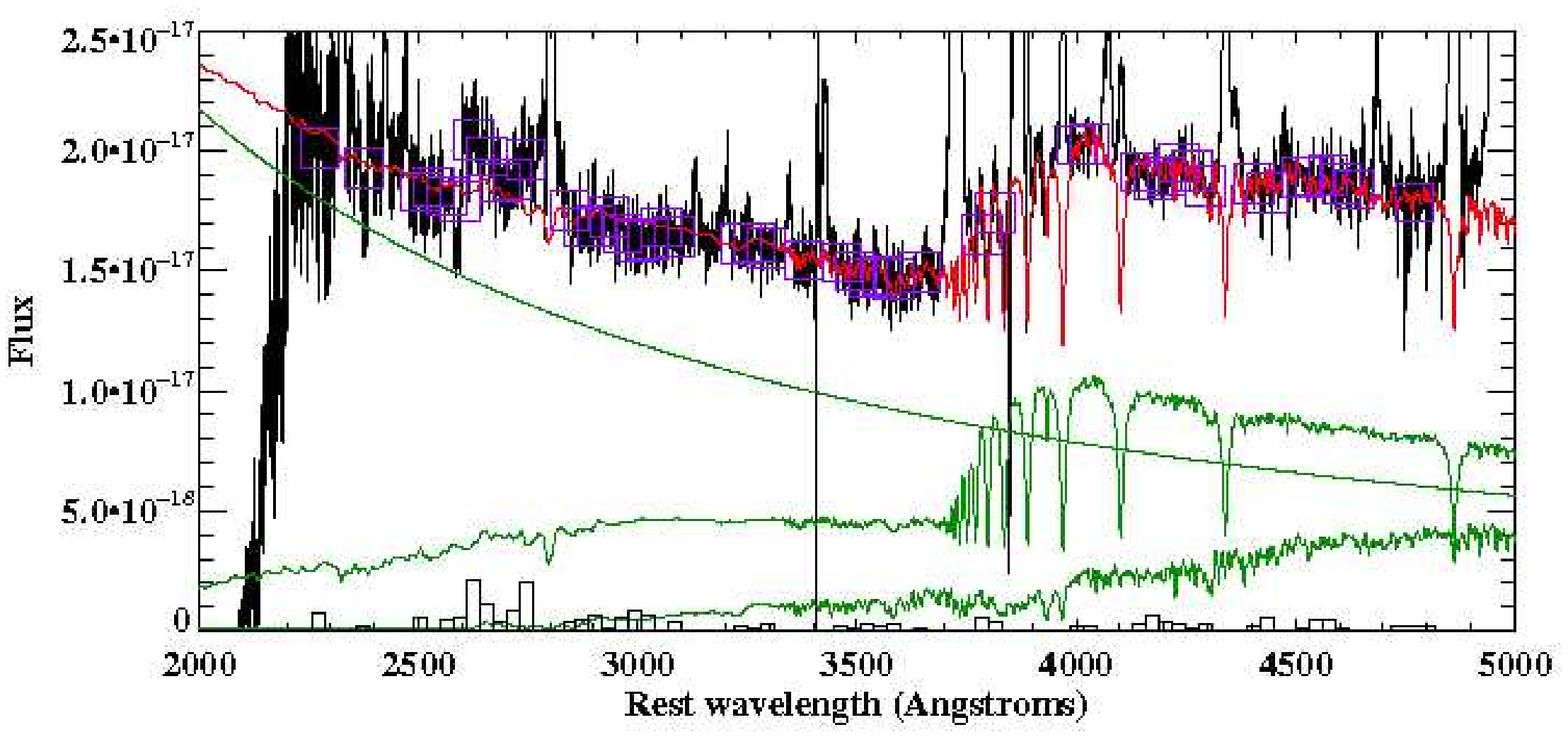,width=9cm,angle=0.} &
\psfig{file=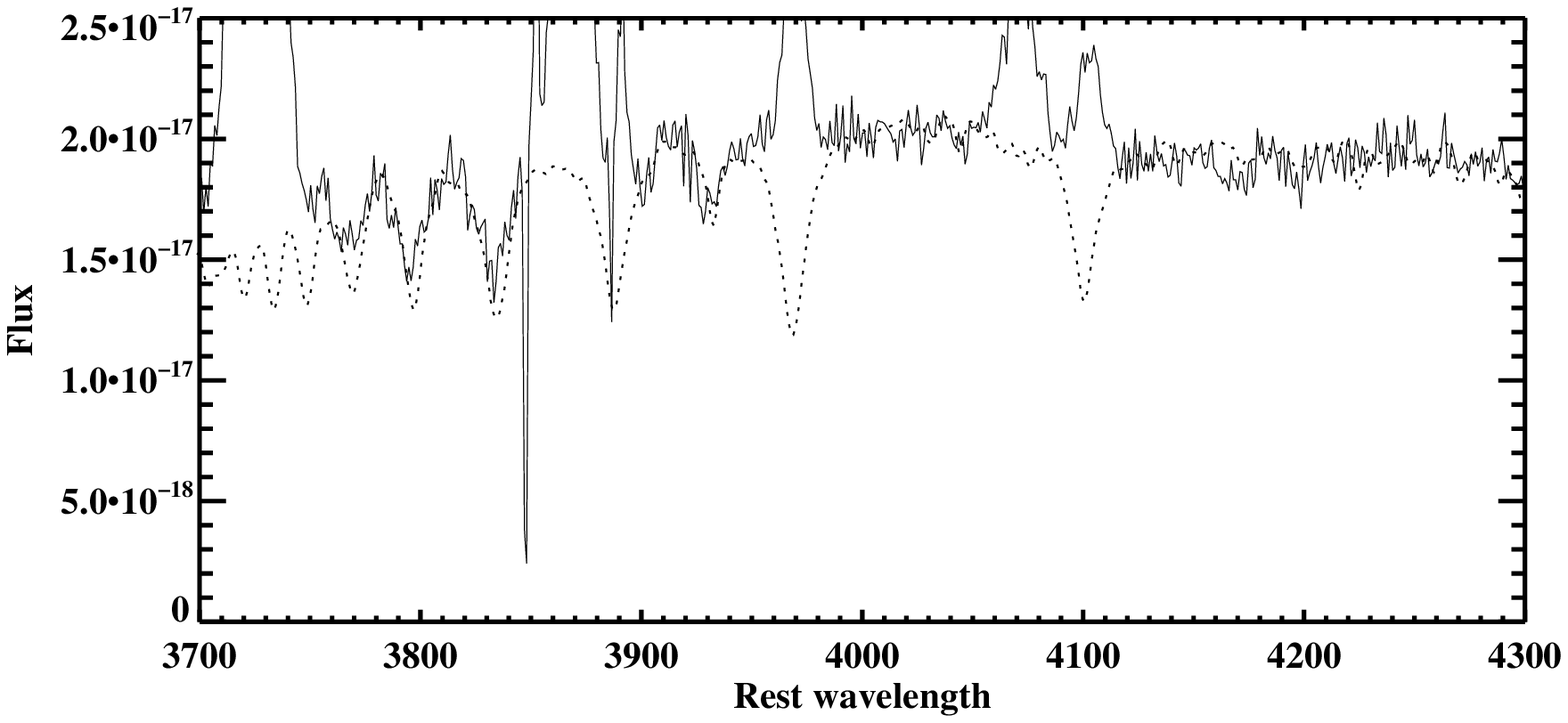,width=9cm,angle=0.}\\
\multicolumn{2}{c}{\bf NGC 612 (nuc): 12.5 Gyr OSP + 0.05 Gyr YSP
  (\ebv~= 1.2)} \\
\hspace*{-1cm}\psfig{file=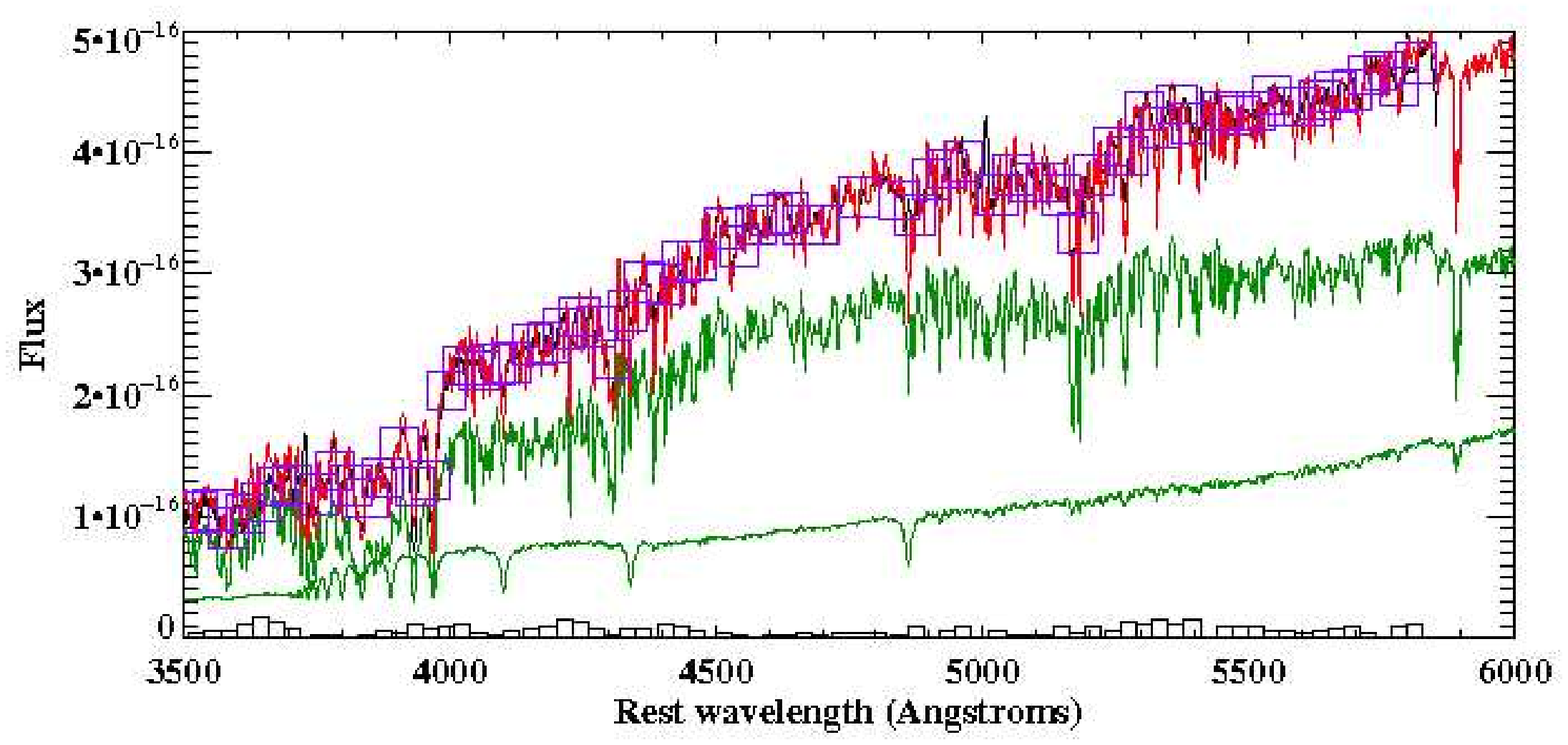,width=9cm,angle=0.} &
\psfig{file=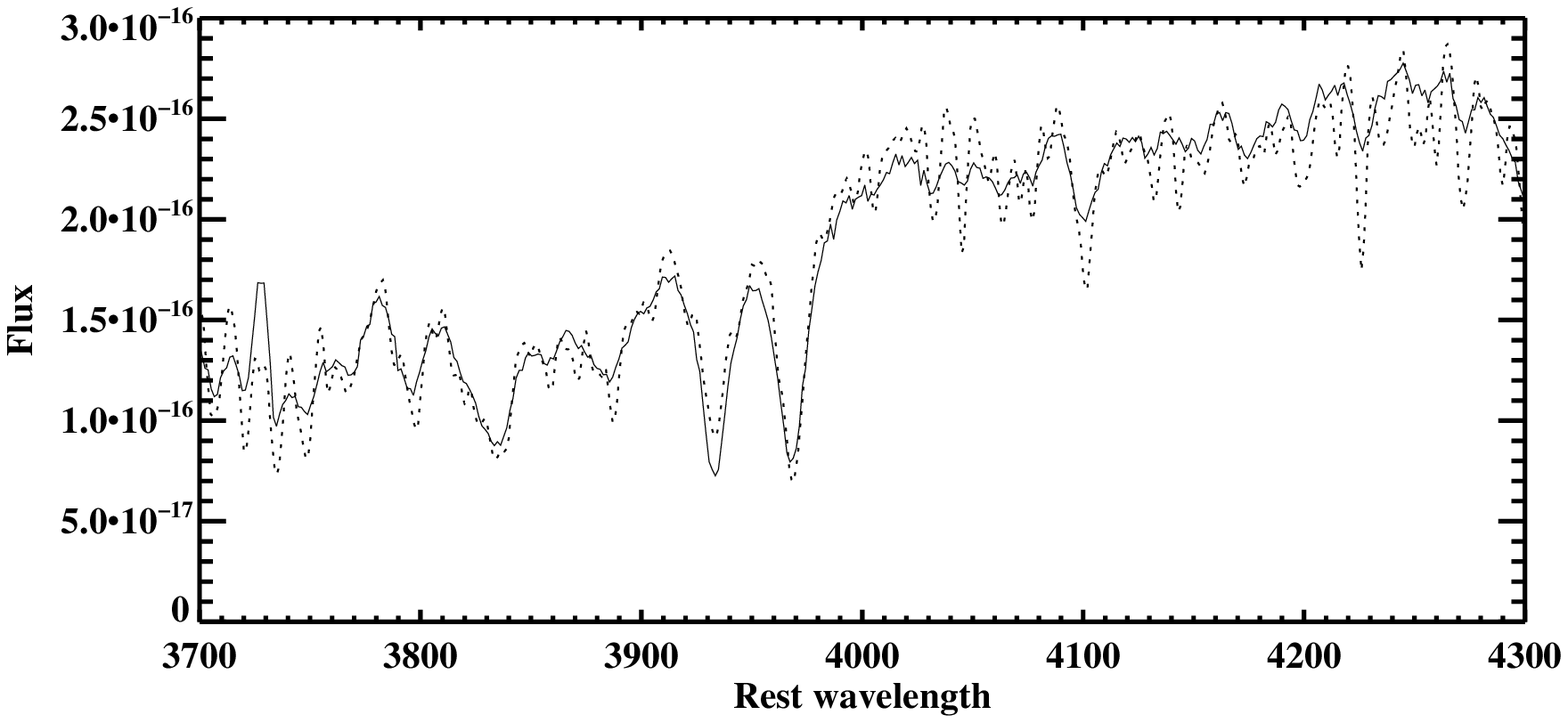,width=9cm,height=4.5cm,angle=0.}\\
\multicolumn{2}{c}{\bf NGC 612: (A): 12.5 Gyr OSP + 0.1 Gyr YSP
  (\ebv~= 0.9)} \\
\hspace*{-1cm}\psfig{file=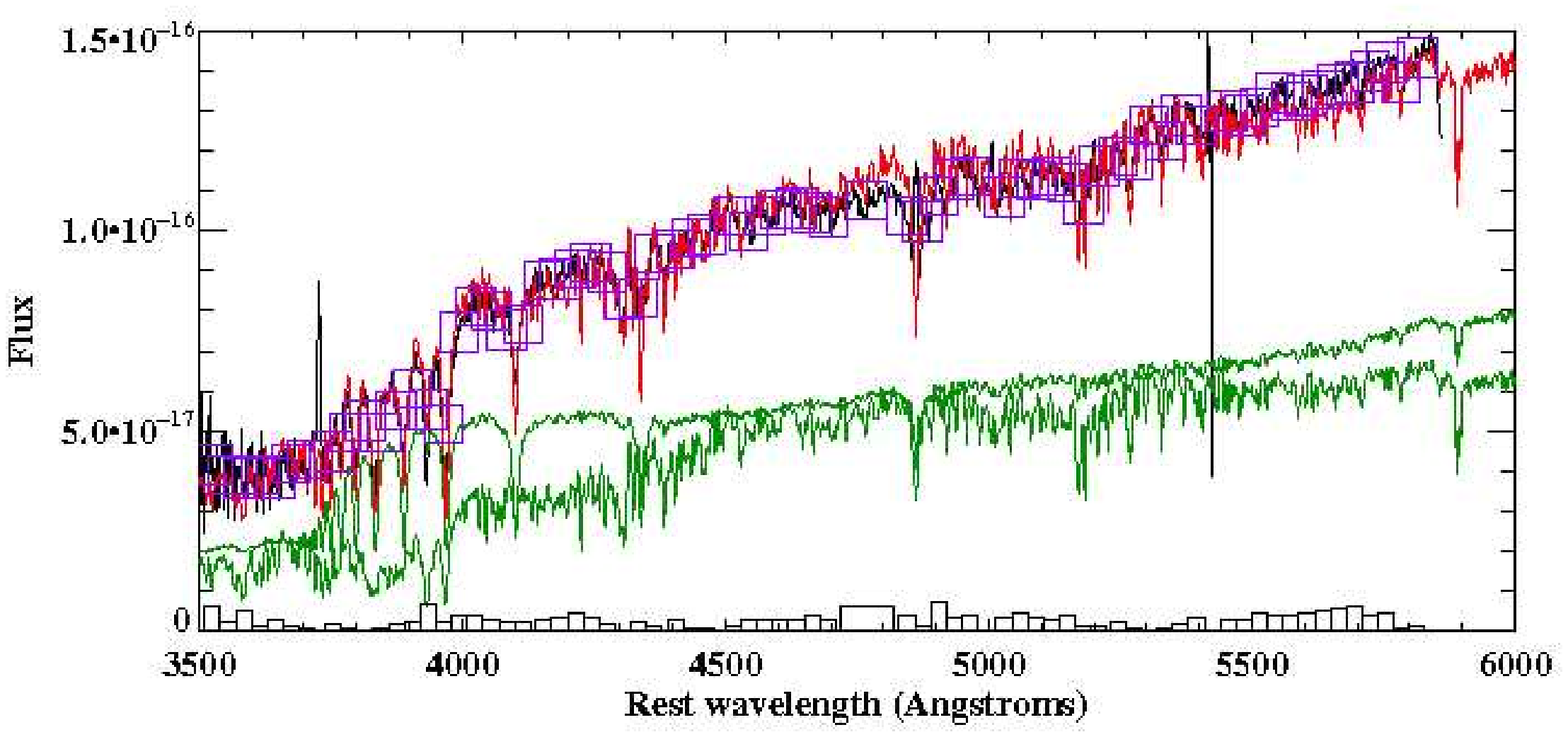,width=9cm,angle=0.} &
\psfig{file=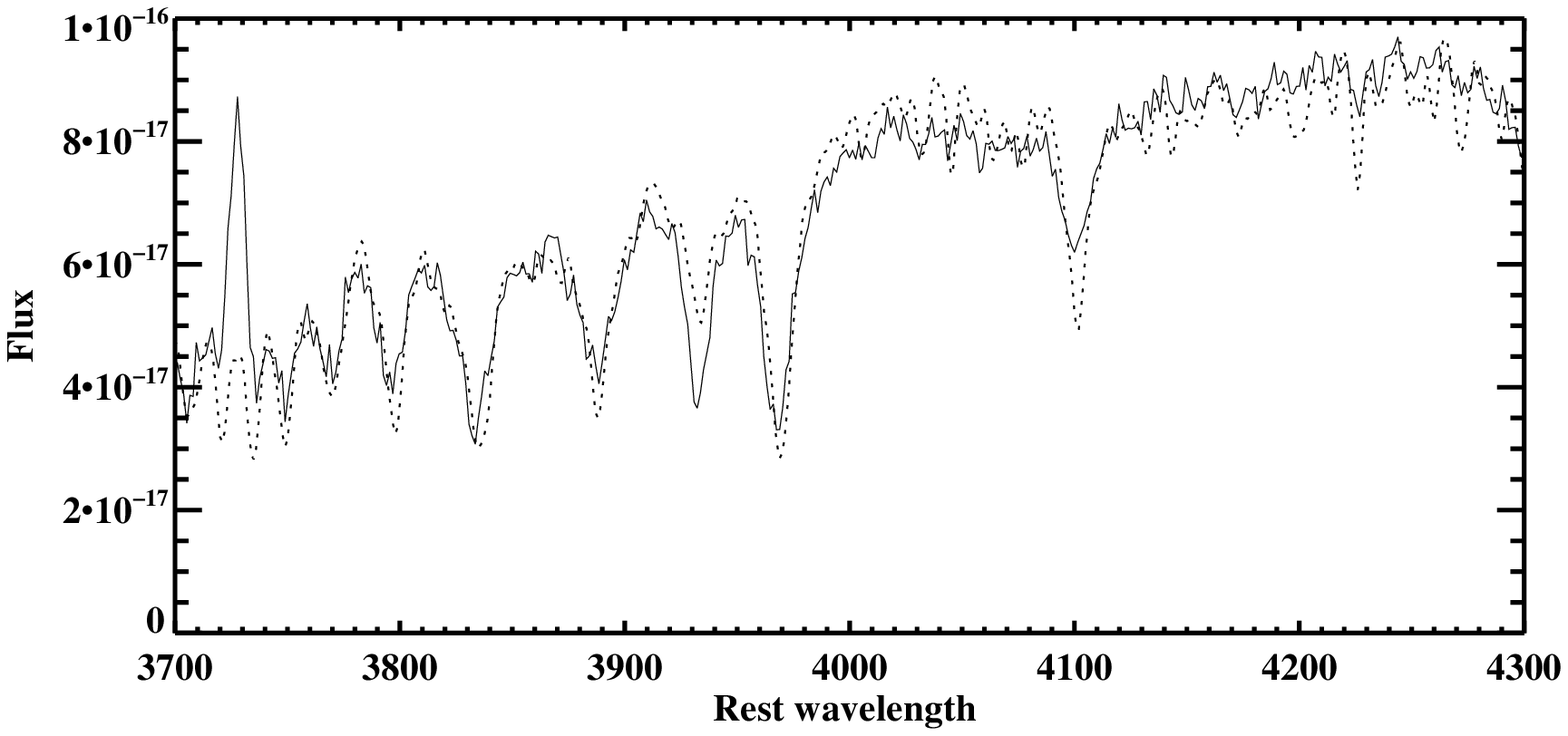,width=9cm,height=4.5cm,angle=0.}\\
\multicolumn{2}{c}{\bf NGC 612: (B): 12.5 Gyr OSP + 0.1 Gyr YSP
  (\ebv~= 0.6)} \\
\hspace*{-1cm}\psfig{file=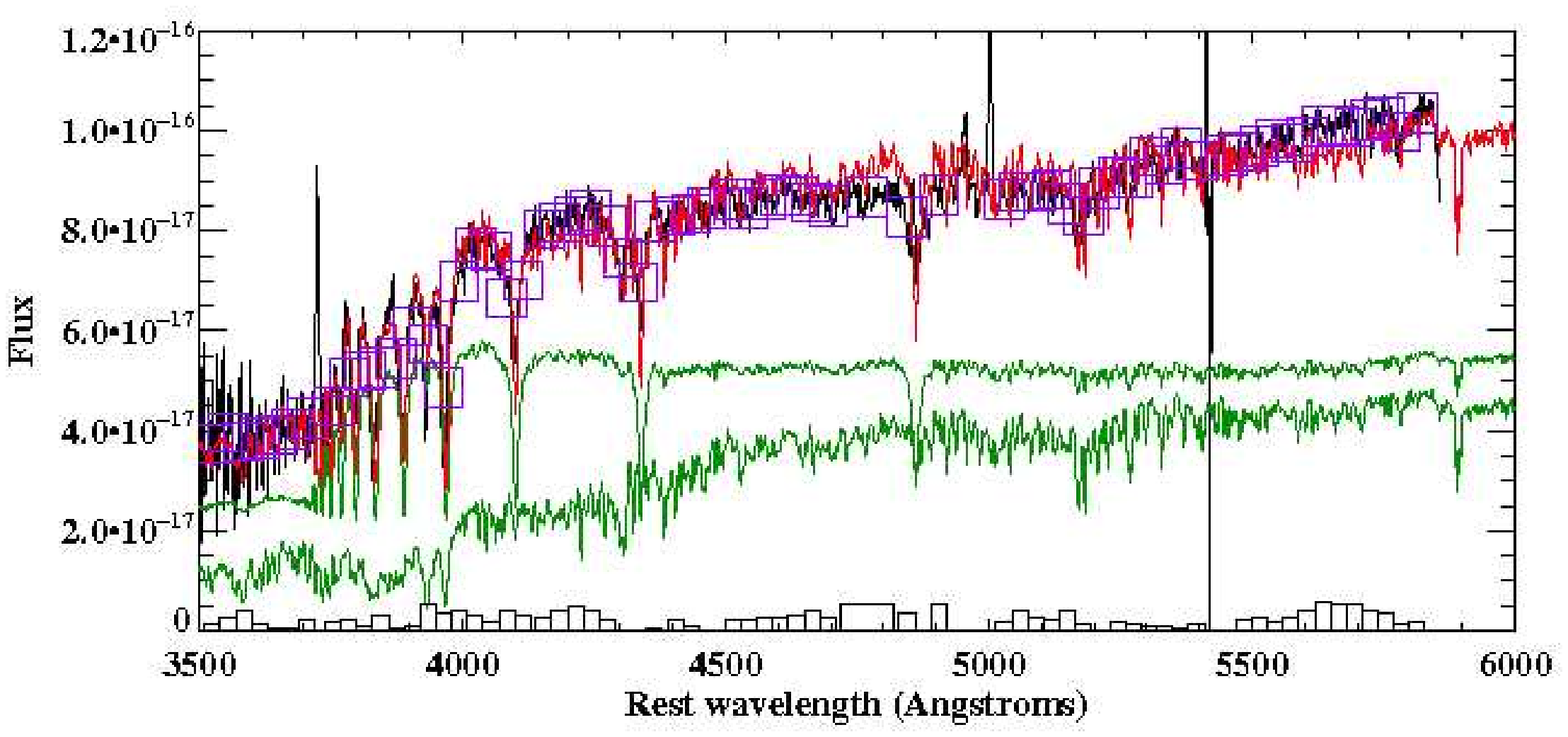,width=9cm,angle=0.} &
\psfig{file=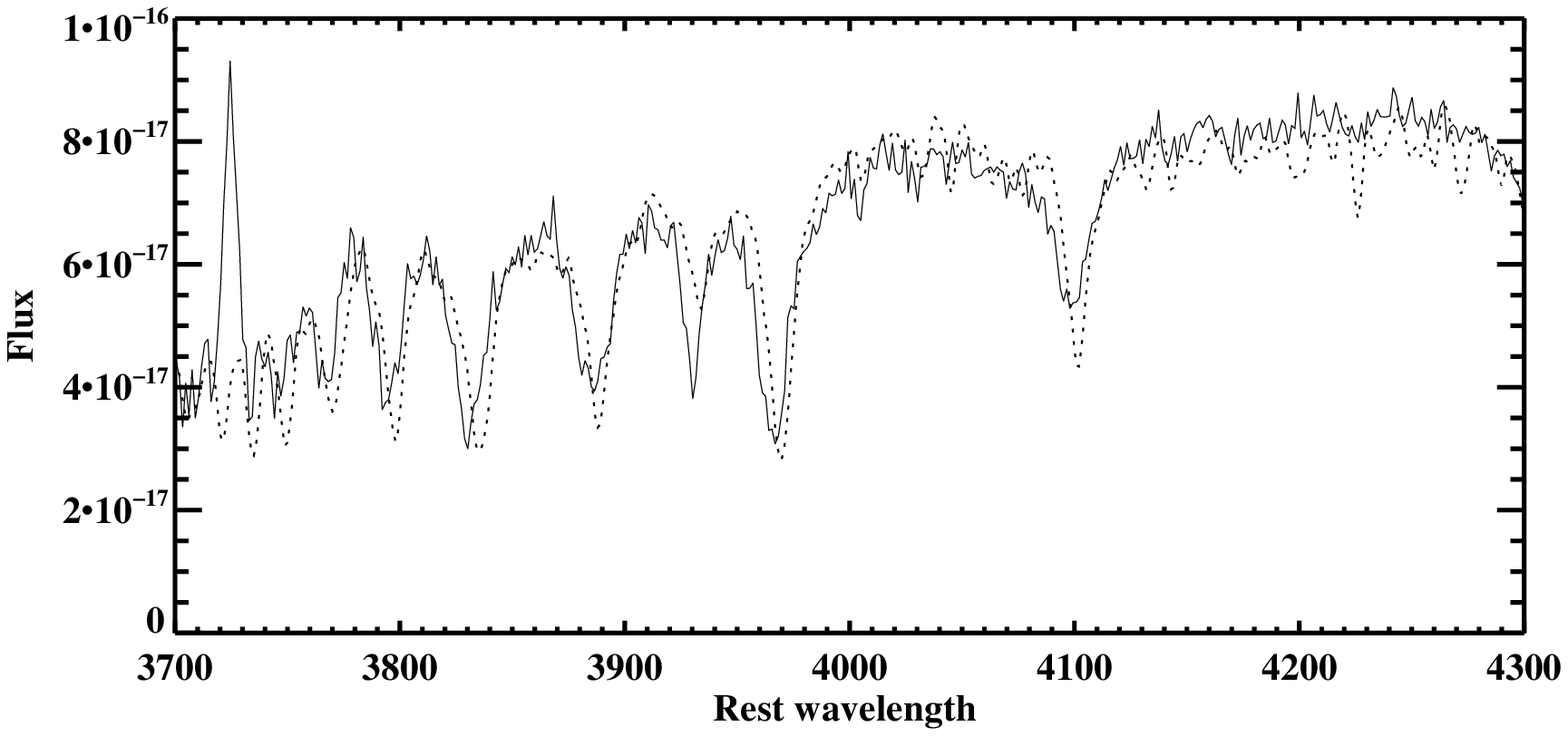,width=9cm,height=4.5cm,angle=0.}\\
\end{tabular}
\caption[]{SEDs and detailed fits {\it continued}
}
\label{fig:SED}
\end{minipage}
\end{figure*} 
\setcounter{figure}{3}
\begin{figure*}
\begin{minipage}{170mm}
\begin{tabular}{cc}
\multicolumn{2}{c}{\bf NGC 612: (C): 12.5 Gyr OSP + 0.1 Gyr YSP
  (\ebv~= 0.4)} \\
\hspace*{-1cm}\psfig{file=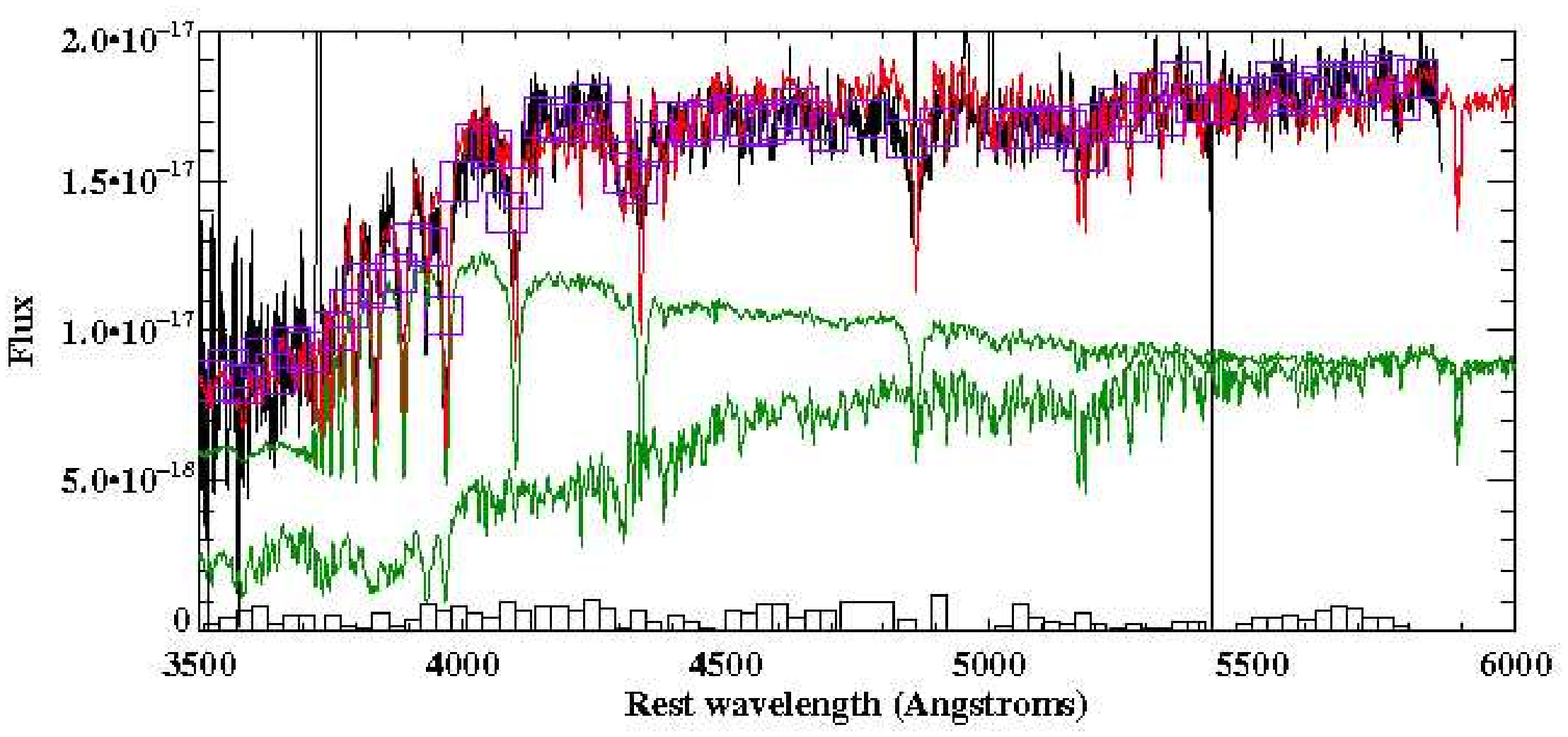,width=9cm,angle=0.} &
\psfig{file=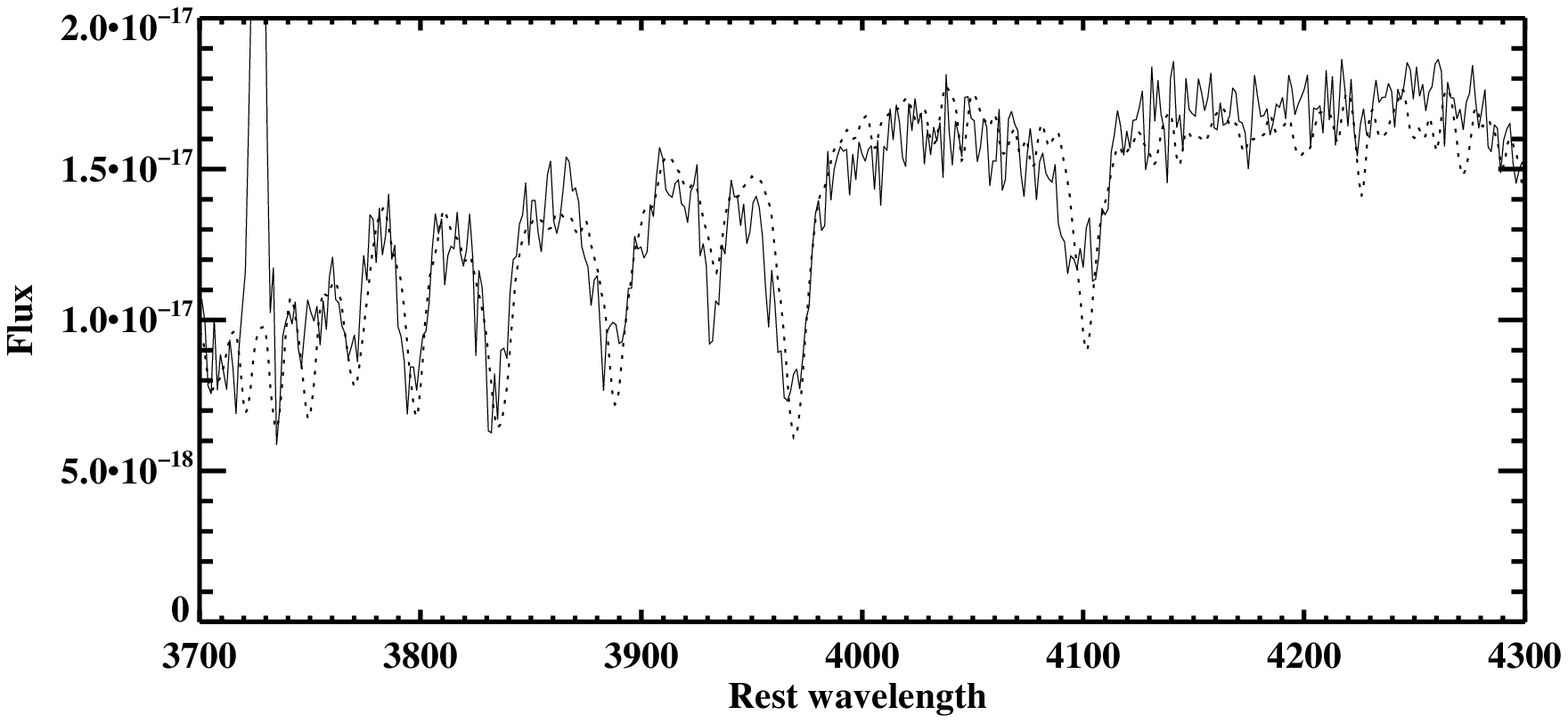,width=9cm,height=4.5cm,angle=0.}\\
\multicolumn{2}{c}{\bf NGC 612: (D): 12.5 Gyr OSP + 0.05 Gyr YSP
  (\ebv~= 0.6)} \\
\hspace*{-1cm}\psfig{file=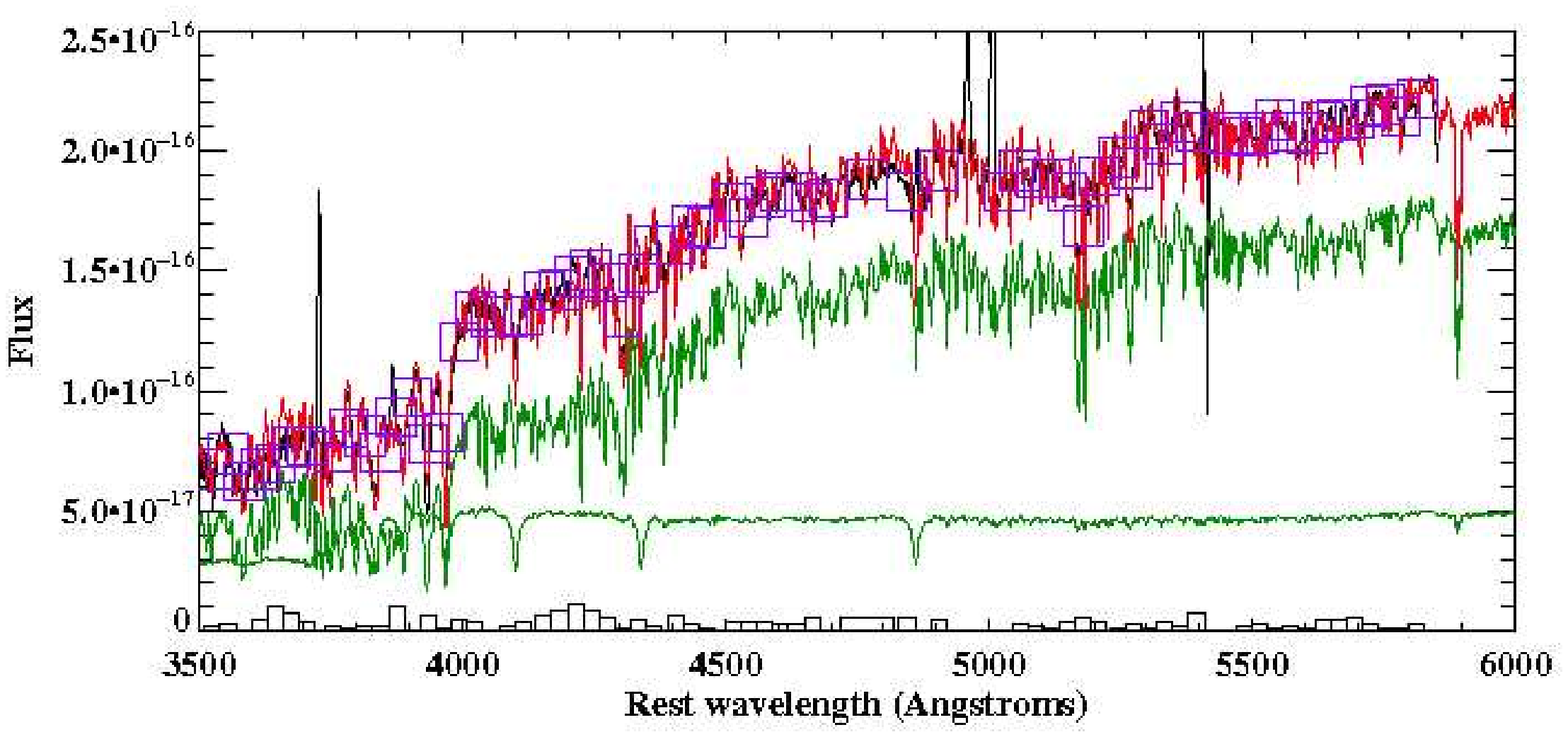,width=9cm,angle=0.} &
\psfig{file=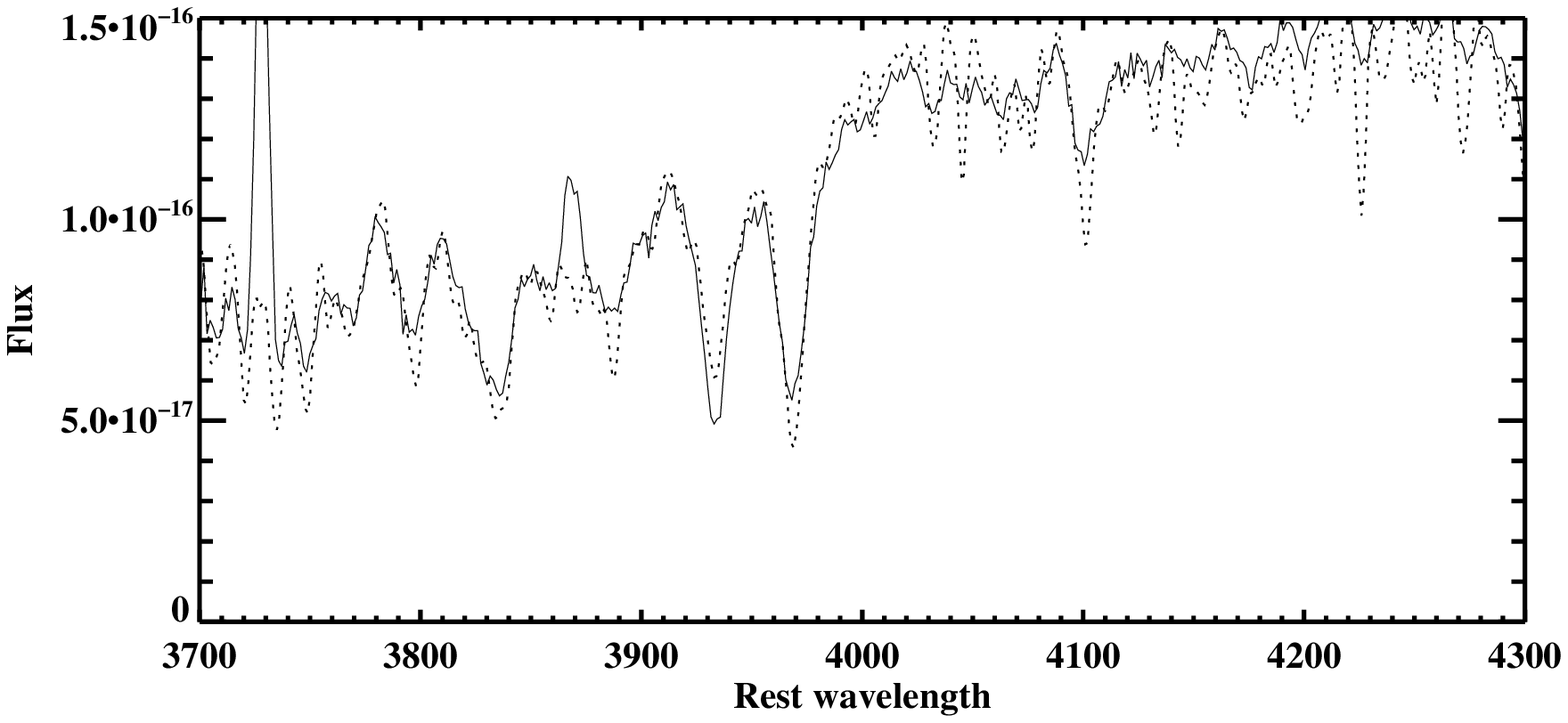,width=9cm,height=4.5cm,angle=0.}\\
\multicolumn{2}{c}{\bf NGC 612: (E): 12.5 Gyr OSP + 0.05 Gyr YSP
  (\ebv~= 0.4)} \\
\hspace*{-1cm}\psfig{file=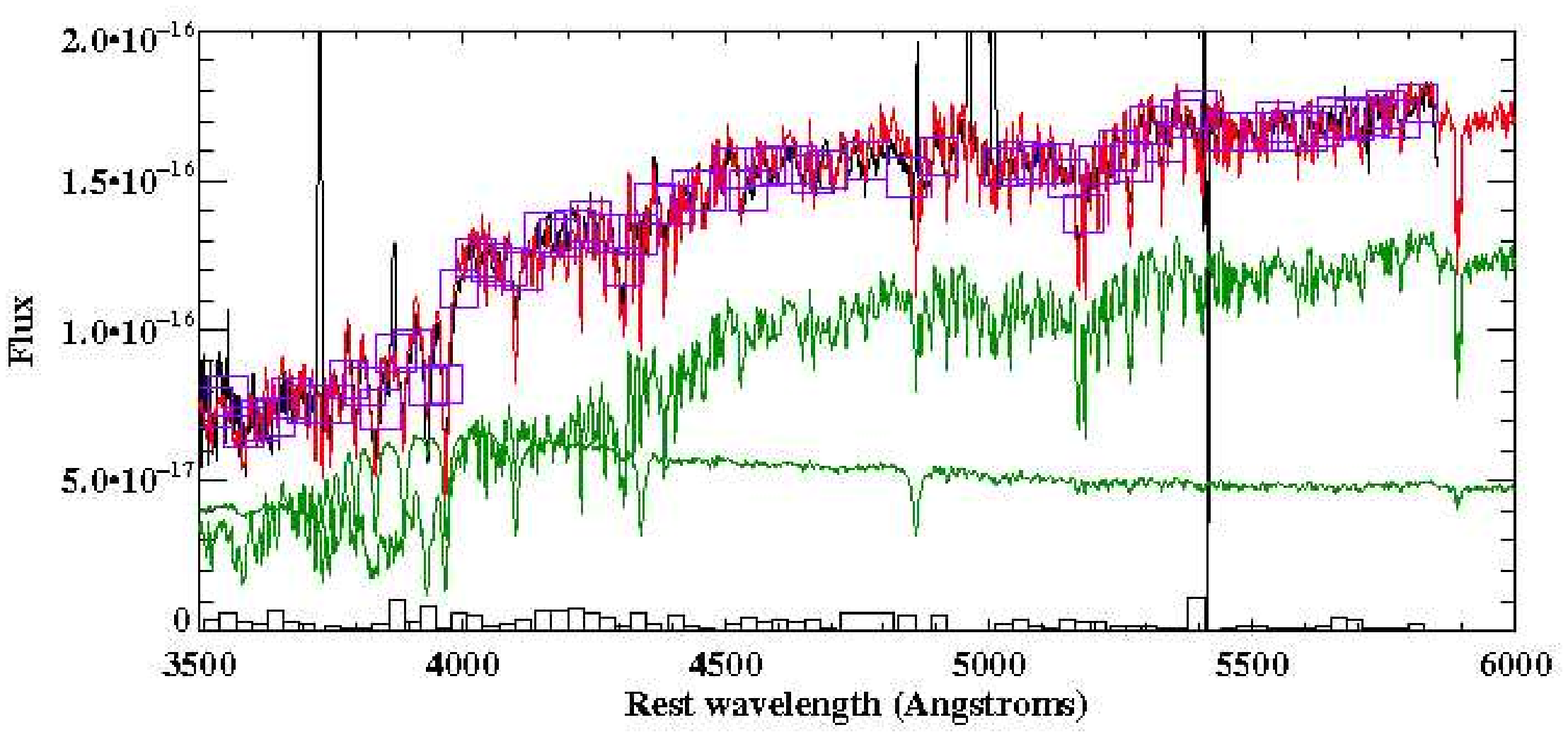,width=9cm,angle=0.} &
\psfig{file=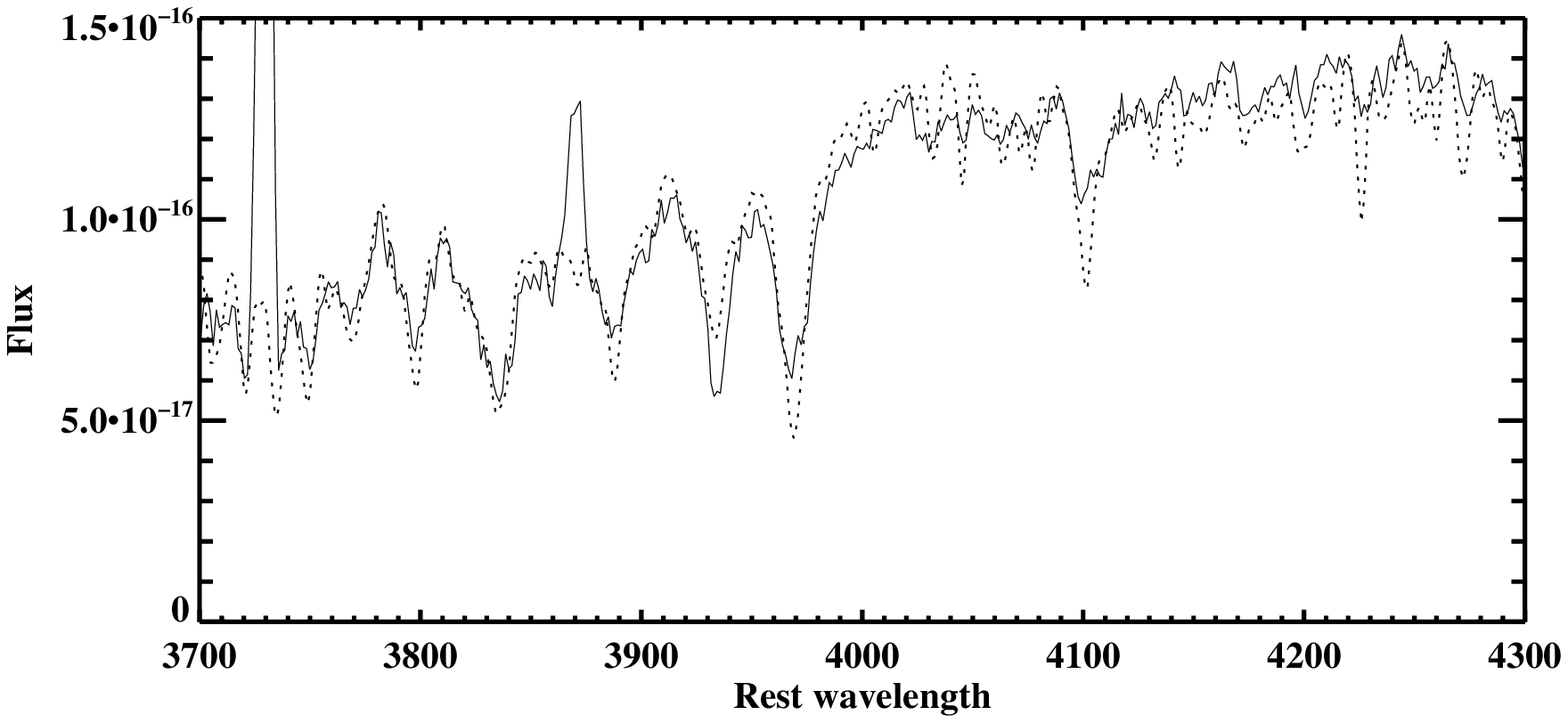,width=9cm,height=4.5cm,angle=0.}\\
\multicolumn{2}{c}{\bf NGC 612: (F): 12.5 Gyr OSP + 0.1 Gyr YSP
  (\ebv~= 0.0)} \\
\hspace*{-1cm}\psfig{file=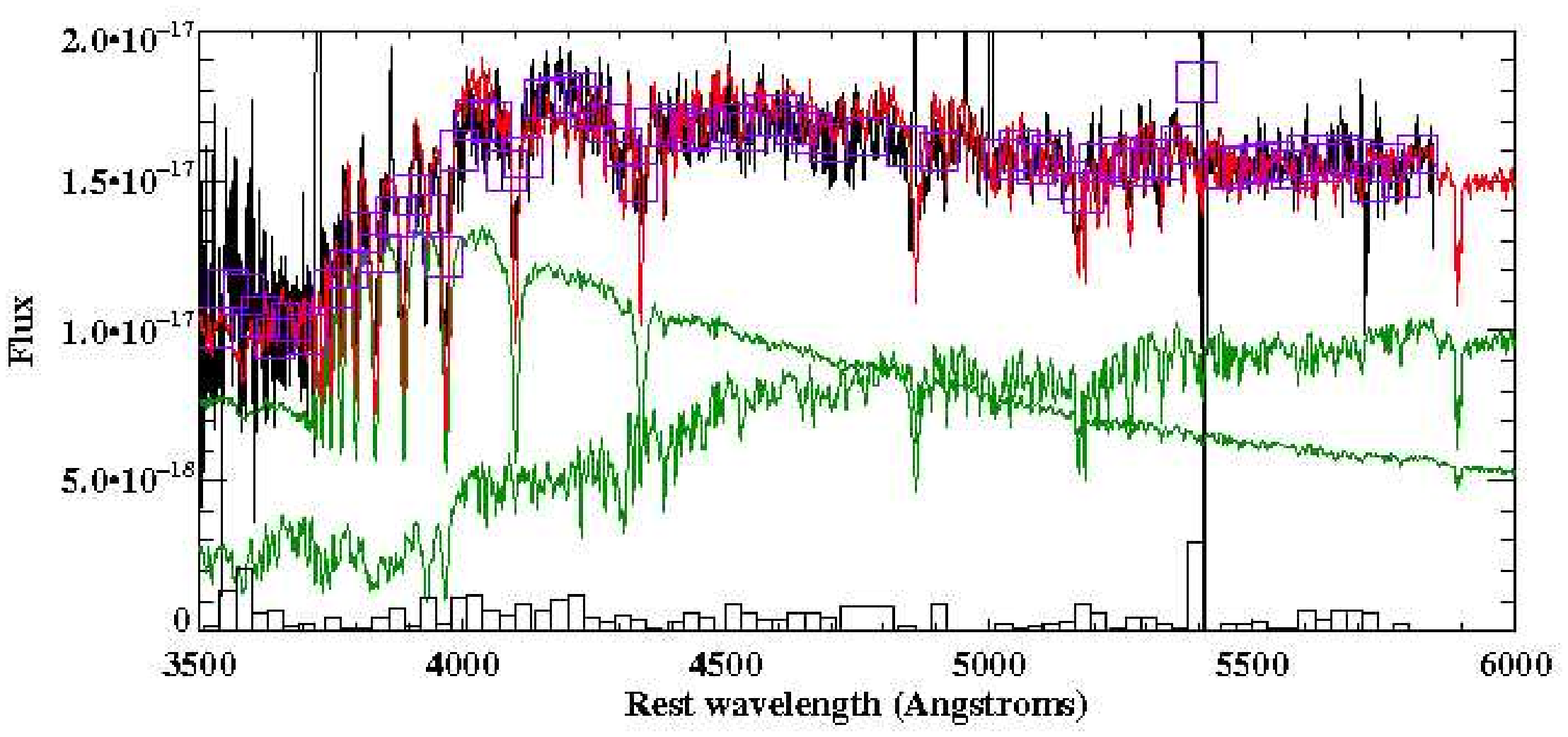,width=9cm,angle=0.} &
\psfig{file=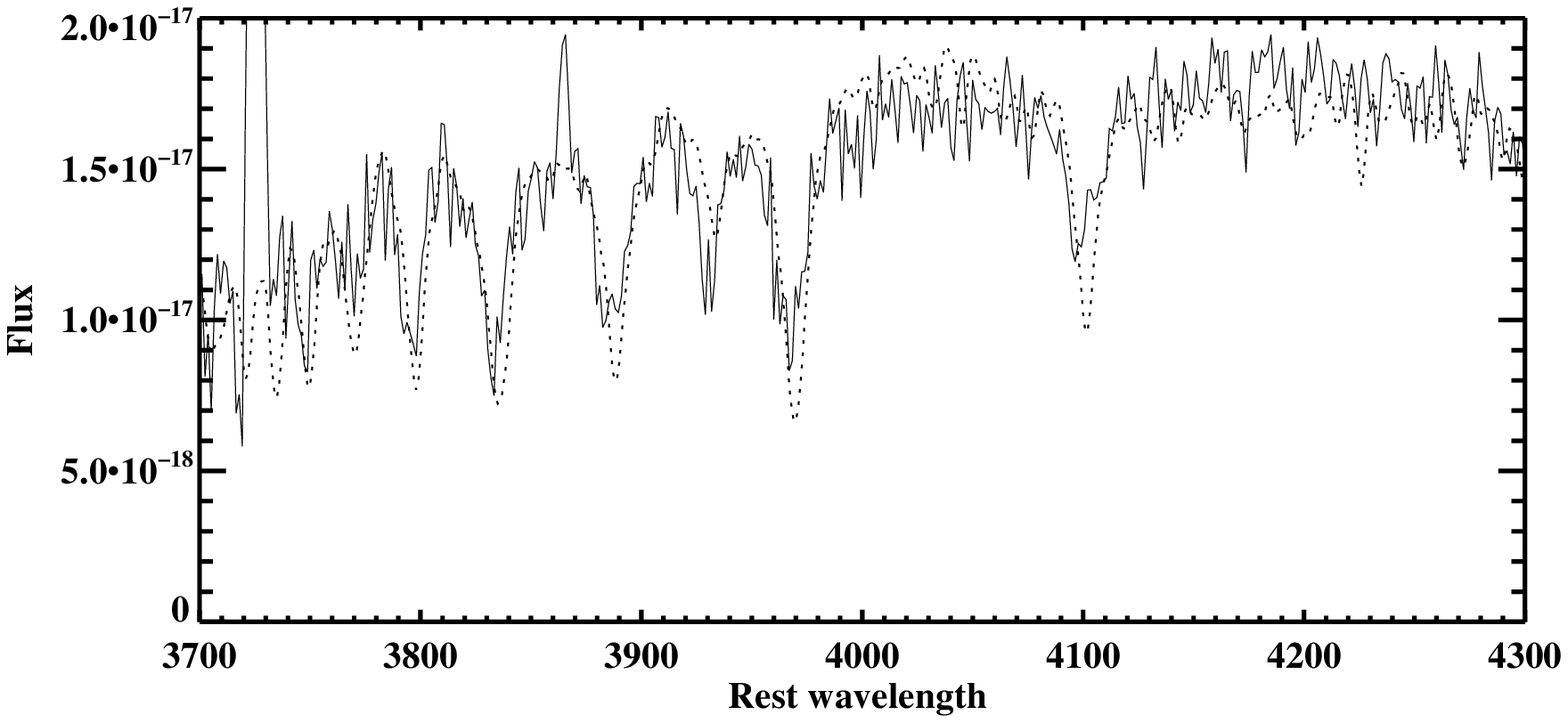,width=9cm,height=4.5cm,angle=0.}\\
\end{tabular}
\caption[]{SEDs and detailed fits {\it continued}
}
\label{fig:SED}
\end{minipage}
\end{figure*}

\begin{table*}
\begin{minipage}{180mm}
\caption[]{Best fitting models to the galaxy continua. Columns are: $(a)$
  aperture (see Figure {\ref{fig:d4000}} for sizes and positions;
  $(b)$ percentage contribution of the subtracted nebular
  continuum {\it by flux} in the bin 3540-3640\AA~(i.e. corresponding
  to the region of the {\it maximum} nebular continuum contribution)
  reddened (where applicable) using the 
  \protect\citet{seaton79} reddening law; $(c)$ 
percentage of OSP (12.5 Gyr); $(d)$ percentage of YSP;
  $(e)$ age of YSP in Gyr; $(f)$ degree of reddening required for the
  YSP; $(g)$ percentage of power law component; $(h)$ spectral index
  of power law; $(i)$  percentage of total stellar mass attributed to
  the  YSP component;
$(j)$ best reduced chi squared obtained. \newline
$^{\star}$ denotes galaxies best modelled with a different age of OSP: PKS
  0039-44 (7 Gyr), PKS 0409-75 (7 Gyr). \newline
$\dagger$ $all$ ages of YSP give an acceptible fit (\chisq~and
  absorption lines) but there exists a clearly defined minimum in the
  \chisq~space -- see text. }
\label{tab:results}
\begin{center}
\begin{tabular}{llrrrrrrrrr}\hline\\
& & \multicolumn{1}{c}{Nebular} &
\multicolumn{1}{c}{Old} & \multicolumn{1}{c}{YSP} &
\multicolumn{1}{c}{YSP age} & \multicolumn{1}{c}{\ebv} &
\multicolumn{1}{c}{Power law} &
\multicolumn{1}{c}{Power law} &\multicolumn{1}{c}{YSP mass}& \multicolumn{1}{c}{\chisq} \\
 &&\multicolumn{1}{c}{\%} & \multicolumn{1}{c}{\%} & \multicolumn{1}{c}{\%} &
\multicolumn{1}{c}{Gyr}& & \multicolumn{1}{c}{\%} &
\multicolumn{1}{c}{$\alpha$} &  \multicolumn{1}{c}{\%} & \\
\multicolumn{2}{c}{$(a)$} & \multicolumn{1}{c}{$(b)$} & 
\multicolumn{1}{c}{$(c)$} & \multicolumn{1}{c}{$(d)$} & 
\multicolumn{1}{c}{$(e)$} & \multicolumn{1}{c}{$(f)$} & 
\multicolumn{1}{c}{$(g)$} & \multicolumn{1}{c}{$(h)$}
&\multicolumn{1}{c}{$(i)$} & \multicolumn{1}{c}{$(j)$}  \\\hline \\ 
3C 218 & nuc & $<$1& \bf 45 &\bf  50 & \bf 0.05 &\bf  0.4 &\bf  -- & \bf -- & \bf 4.1 &\bf 1.2 \\
\\
3C 236 & nuc& $<$1& \bf 75-100 & \bf 10-30 & \bf 0.04-1.0 & \bf 1.6-0.6& \bf -- & \bf -- & \bf 20-32&\bf $<$1.0\\
& &$<$1  & 87 & -- & -- & -- & 12 & 3.6 & -- &1.1\\
\\
3C 285 &nuc & 1.8 & \bf 72-68 &  \bf33-39 &  \bf0.1-0.5 & \bf 0.2-0.0 &  \bf-- &  \bf-- &  \bf2.0-2.5& \bf0.7-1.4\\
\\
3C 321 & SE nuc & 5.8&  \bf 15-55 & \bf30-60 & \bf0.1-1.0 &\bf 0.3-0.1 & \bf10-20 &
\bf0.0-0.6 & \bf1-34&\bf$\sim$0.3\\
 & NW  nuc&  17.7& \bf12-67 & \bf25-75 & \bf0.1-1.4 & \bf0.3-0.1 &
\bf5-15 & \bf-0.5 -- 0.8 & \bf2-54&\bf0.4-0.5\\
 & mid &0.3 & \bf30-50 & \bf40-60 & \bf0.4-1.0 &\bf0.0-0.1 & \bf 5-8 & \bf-1.6 -- -1.7 &\bf 4-15&\bf0.2-0.5\\
 & ext &25.2 &\bf 20-36 &\bf 54-72 & \bf0.7-1.0 & \bf0.0 &\bf 5-6 & \bf-1.2 -- -1.3 &\bf 8-23&\bf0.7-0.9\\
\\
3C 381 & nuc  & 1.7 & \bf  80 & \bf  --&\bf  -- & \bf  -- & \bf  20 & \bf  1.65 & \bf  --&\bf  0.9 \\
 & & 1.7 &  82 & 17 & 0.02 &1.0 & -- & -- & 4 &0.9 \\
\\
3C 433 &SW nuc & 4.5 &\bf 67-80 & \bf30-20 & \bf0.03-0.1 & \bf0.8-0.4 & \bf-- & \bf-- &\bf 5-30 &
\bf0.6-1.0 \\
& & 4.5 &  74 &  -- &  -- &  -- &  24 &  1.0 &  --& 1.4\\
& & 4.5 & 70-20 & 20-70 & 0.05-1.4$\dagger$ & 0.7-0.3 $\dagger$& 5-20 & $\gtrsim$
0.0& 6-80&$\lesssim$0.6\\
 & NE nuc & 4.5 &\bf 65-95 &\bf 5-30& \bf 0.03-1.4&\bf 0.3-0.9&-- &
--&\bf 1-15 & \bf0.5-0.9 \\ 
& & 4.5 &  100  & -- & -- & -- & --& -- & -- &  1.1 \\ 
 & & 4.5 & 95-65 & 4-30 & 0.03-1.6 & 0.9-0.3 & -- & -- & 1-16 & 0.5-0.9\\
 &mid  & $<$1& \bf 80-95 & \bf 10-25 & \bf 0.05-1.0 & \bf 0.7-0.0
& \bf -- & \bf -- & \bf 1-4&\bf 0.9-1.1\\
\\
PKS 0023-26 &nuc & 22.5 & \bf 40-54 & \bf 50-60 & \bf 0.03-0.05 & \bf 0.7-0.9 & \bf -- & \bf -- & \bf 9-18&\bf 0.9-1.1\\ 
 & & 22.5 & $<$50 & 10-60 & 0.05-1.2 & $<$1.0 & $<$50 & $\gtrsim$ 0.0
& & 0.6-0.9 \\ 
\\
PKS 0039-44$^{\star}$  & nuc &  11.7 &  \bf 57 &\bf  -- & \bf -- &\bf  --& \bf 40 & \bf -0.5 &\bf -- & \bf 0.4\\
\\
PKS 0409-75$^{\star}$ &nuc &6.0& \bf 55-80 &\bf 45-25 &\bf 0.01-0.04 &\bf 0.9-0.5 &
\bf-- & \bf-- & \bf9-4&
\bf0.8 \\
 & & 6.0 & 60 & -- & -- & -- & 45 & 0.86 & -- & 1.2 \\
\\
PKS 1932-46 & nuc & 21.9& \bf59 &\bf -- &\bf -- & \bf-- &\bf 41 & \bf0.23 &\bf --& \bf0.8 \\
 & & 21.9 & 0-60 & 10-80 & $all\dagger$ & $all\dagger$& 5-40 & $>$-0.6 &$<$100 &0.4-0.8 \\
\\
PKS 2135-20 & nuc & 13.9& \bf24-0 & \bf77-45 &\bf0.04-0.6 &\bf 0.5-0.2 &
\bf3-42 & \bf-4
-- -1 & \bf16-100& \bf0.4-0.6\\
\\
NGC 612 &nuc & $<$1& \bf 77-73 & \bf 25-28 & \bf 0.05-0.1 &\bf  1.4-1.2 & \bf -- & \bf -- & \bf 19-28&\bf 0.5-0.3\\
 &A &  $<$1& \bf 47-51 & \bf 56-54 & \bf 0.05-0.1 & \bf 0.9-1.0 &\bf  -- & \bf -- & \bf 27-28&
\bf  0.5-0.6\\
 &B &  $<$1& \bf 30-45 & \bf 70-60 & \bf 0.05-0.1 & \bf 0.8-0.6& \bf -- &\bf  -- &\bf 14-29 &\bf 0.6 \\
 &C &  $<$1& \bf 30-45 & \bf 70-60 & \bf 0.05-0.1 & \bf 0.6-0.4 & \bf -- & \bf -- & \bf 15-7&\bf 0.5\\
 &D &  $<$1& \bf 70-80 &\bf 30-23 &\bf  0.04-0.2 & \bf 0.8-0.3 &\bf  -- & \bf -- & \bf 7-2&\bf 0.4-0.6\\
 &E &  $<$1& \bf 66-69& \bf 37-35 &\bf  0.04-0.05 & \bf 0.5-0.4 &\bf  -- & \bf -- &\bf  3-2&\bf 0.5-0.4\\
 &F &  $<$1& \bf 25-50& \bf 75-54 & \bf 0.05-0.1 & \bf 0.4-0.0 & \bf -- & \bf -- & \bf 10-1&\bf 0.5-0.6\\\hline
\end{tabular}
\end{center}
\end{minipage}
\end{table*}
NGC 612 is the nearest radio source in our sample ($z$ = 0.030) and,
unusually for a powerful radio source, is optically identified with a peculiar
lenticular/S0 galaxy with a prominent  dust lane and stellar disk
\citep{westerlund66,ekers78}. NGC 612 also contains a galaxy-scale
disk of neutral hydrogen (H{\,\small I}) gas, which follows
the optical disk and dust-lane of the galaxy {\citealt{emontsphd};
  Emonts et al. 2007 in prep). 
 Several authors have noted similarities
to Centaurus A (e.g. \citealt{westerlund66,ekers78}). The double radio
source spans $\sim$ 14 arcmin ($\sim$ 0.4 Mpc) and 
is oriented approximately E-W,   perpendicular to the dust 
lane. Due to its proximity, NGC 612 is
well resolved spatially. Line and continuum emission is observed to
large radii, although the emission lines (e.g. {[O II]}, {[O III]},
{[N II]} and H$\alpha$) are weak
\citep{westerlund66,goss80,tadhunter93}. In addition, as well as the
usual absorption features associated with an OSP, the
higher Balmer lines are detected in absorption 
(\citealt {tadhunter93}, see also Figure {\ref{fig:SED}}), suggesting a
significant YSP component. \citet{raimann05} have recently
modelled the stellar 
populations in NGC 612 using five distinct components, finding evidence
that $>$ 30\% of the light in the nuclear and extended apertures is
attributable to young stars with age 0.003-1 Gyr with the 1Gyr
component being the dominant YSP.

The slit position chosen for the current
study was aligned parallel to the large-scale
dust lane and stellar disk. Figure {\ref{fig:d4000}} demonstrates 
that the UV
excess extends out to a radius of at least 25 arcsec from the nucleus
in both directions. Interestingly, the UV excess
{\it increases} with distance 
from the nucleus -- the nucleus itself has the largest $D4000$
ratio. This is in contrast to the other spatially resolved sources in
the sample which generally have either a constant $D4000$ ratio across
the galaxy or a peak in the UV excess in the nucleus. In addition to
modelling the nuclear aperture, we have also extracted 6 extended bins
in the disk of the galaxy. All apertures are highlighted in
Figure {\ref{fig:d4000}}. 

{\bf The nucleus}. 
It is not possible to model the nuclear aperture of NGC 612 with an
OSP, with or without an power-law component (\chisq~$>$ 2) and so a 
YSP component is 
essential to model the SED. Figure {\ref{fig:contours}} shows the
\chisq~space for two component 
(OSP plus YSP) models. in which the YSP is reddened using both the
\citet{seaton79} and \citet{calzetti00} reddening laws. We find that
use of \citet{calzetti00} rather than the \citet{seaton79}
reddening law for the YSP makes little difference to our results (see
Section {\ref{sect:reddening}}  therefore
for the analysis presented here, we shall focus on the results obtained
using the \citet{seaton79} law.

Two component fits (OSP plus reddened YSP)
provide viable fits for ages 0.1-2.0 Gyr  with two
distinct minima at $\sim$ 0.1 Gyr (\ebv~$\sim$ 1.2) and $\sim$ 0.9 Gyr
(\ebv~$\sim$ 0.6) with \chisq~= 0.29 and 0.31 respectively (Figure
{\ref{fig:contours}}).  
 Hence, as with
many of the apertures, it is impossible to distinguish between the
models on the basis of the \chisq~to the overall SED alone.

Based on the detailed fits, models including YSPs with ages 0.03-0.1
Gyr (\ebv~$\sim$ 1.4-1.2)  provide the best fit to the absorption
features and continuum in the range 3700-4300\AA, with contributions
of 80-75\% and 20-28\% for the 12.5 Gyr and YSP components
respectively. Models including YSPs with ages  $>$ 0.2 Gyr tend
to either over-estimate the continuum in the region
4000-4100\AA~and/or over-estimate the strength of the Ca K line,
especially as the age of the YSP increases. Figure 
{\ref{fig:SED}} shows the SED and Balmer line fits for the 12.5 OSP
plus 0.05 Gyr YSP with \ebv~= 1.2 and \chisq~= 0.3. 

The fits were not significantly improved when including a power-law
component. This is not surprising given that
NGC 612 has no strong evidence for a power law component -- the emission
lines are weak and previous work has not discovered significant
polarised light (optical polarisation $\lesssim$ 2\%;
\citealt{brindle90}).

{\bf Extended apertures}
As discussed above, NGC 612 is spatially well resolved and the UV
excess covers the entire spatial exent of the galaxy. Further, 
given that an AGN component is not likely to be
significant in the nuclear regions, it is also logical to assume the
extended regions also do not have a power-law component. The analysis
therefore 
focusses on modelling the extended aperture SEDs using a combination
of OSP and reddened YSP components.

As in the nuclear aperture, a pure OSP fails to provide an adequate
fit to the SED in all of the extended apertures (\chisq~$>$ 6) and a
YSP component is required. Indeed, including a YSP component
significantly improves the fits in all extended apertures -- solutions
are found with \chisq~ $<$ 1, see Figure {\ref{fig:contours}}.

The minimum chi-squred plots for apertures A, B, D and E show two minima
at $\sim$0.05 -- 0.1 Gyr (\ebv~$\sim$0.5 -- 1.0) and $\sim$1.0 Gyr
(\ebv~$\sim$0.0 -- 0.3).  However, consistent with fits to
the spectra of other objects, the older ages (second minimum) can be ruled out due to the
over prediction of the strength of CaII~K, and failure to provide an
adequate continuum fit in the region around 3700-4300\AA. The cases
of apertures C and F are more clear-cut in the sense that only models
with relatively young ages (0.04 -- 0.1~Gyr) and moderate
reddening ($0.1 < E(B-V) < 0.2$) provide adequate fits to the SEDs.
It is notable, however, that although the YSP models with ages $\sim$0.04 -- 0.1
provide a good fit to the continuum SEDs and the higher Balmer absorption lines,
they significantly underpredict the
strength of the CaII~K absorption feature in all the extended apertures.
One plausible explanation for this is that CaII~K absorption 
due to intervening
ISM is important for these apertures. Such an explanation
is consistent with the relatively
high reddening deduced for the YSP in these extended apertures, as well
as the alignment of the slit with the large-scale dust lane.

It is also worth noting that, in apertures C and F, the absorption line
profiles are complex and appear to have a split profile. At these
positions, there is an enhancement in the continuum brightness on the
2D spectra and, in the case of aperture C, a `blob' in the galaxy
halo. 
This may be the signature of stellar populations which are offset in
velocity e.g. a young star-forming cluster in the halo which is moving
differently to the older stars in the galaxy. However, the data
presented here have insufficient resolution and S/N to investigate
this issue and we have assumed the absorption profiles can be modelled
by up to two stellar populations at the same velocity.

{\bf Summary for NGC 612}. The high spatial resolution 
of the continuum structure in NGC 612 allows us to note some
interesting trends in the stellar populations in this radio
galaxy: the model fits for all apertures are consistent with
a combination of an OSP and a reddened YSP of age $\sim$0.04 -- 0.1~Gyr, 
with the contribution of the YSP decreasing towards the nucleus --- as
expected for an SO galaxy. However, although we cannot absolutely rule out a 
 contribution by an older ($\sim$1~Gyr) reddened YSP in the nuclear
 regions, our spectra are not consistent with the presence of the
 dominant $\sim$~1Gyr unreddened YSP deduced by \citet{raimann05}.

\section{Discussion -- Determining the properties of the YSP in radio galaxies }

The results presented in this paper provide a good illustration
of the challenges faced when determining
the propoerties of the stellar populations in AGN host galaxies.
The issues that need to be considered are discussed in the following sections.

\subsection{\bf AGN-related components.}
\setcounter{figure}{8}
\begin{figure}
\centerline{\psfig{file=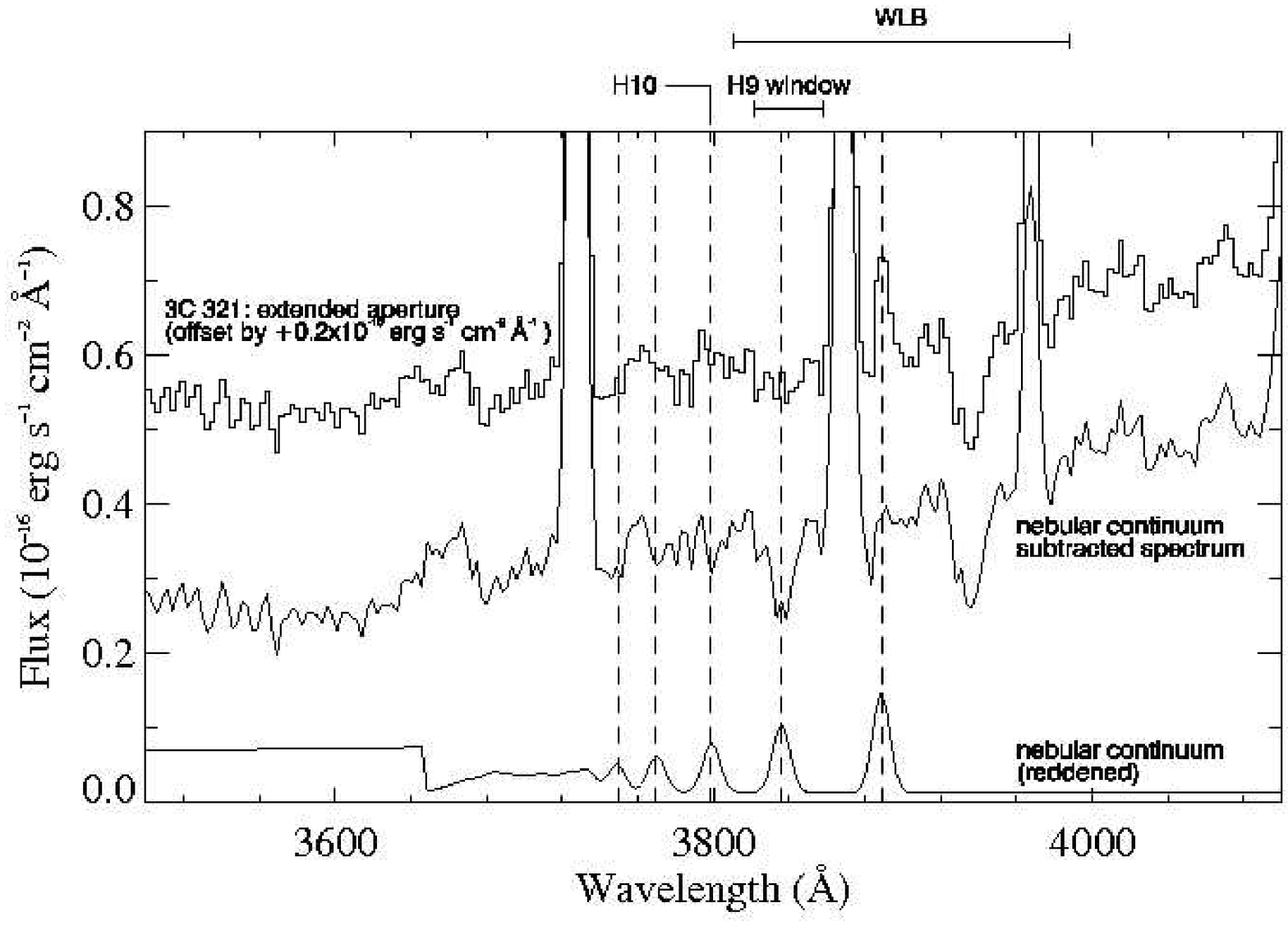,width=9cm,angle=0.}}
\caption[The nebular continuum.]{The effect of the nebular 
correction  in the extended aperture of 3C 321. The different spectra
are labelled in the plot and the dotted lines are to aid the eye for
comparisons between the original spectrum, the nebular model and
the nebular model subtracted spectrum. 
In this aperture, the
nebular component contributes $\sim$25\% of the flux in the region
3540-3640\AA~and $\sim$12\% of the flux in the H9 window
(3822-3858\AA) defined by \protect\citet{raimann05}. The location of
the H9 window and the WLB window (3810-3988\AA), both defined by
\citet{raimann05} and the position of H10 (used by
\citealt{aretxaga01}) are marked on the plot. } 
\label{fig:nebcont}
\end{figure}

As stressed in previous studies
(e.g. \citealt{tadhunter96,tadhunter02}), AGN continuum and emission
line components have a 
major effect on the the continuum spectra of radio galaxies at optical
and UV wavelengths. 
In the current study, scattered AGN light is important for at least two objects
(PKS 0039-44, 3C 321), a further three objects show good evidence for
directly transmitted 
AGN light (PKS 1932-46, PKS 2135-20, 3C 381), and nebular continuum and
emission line infilling 
of absorption lines is important in most objects (see Table {\ref{fig:SED}}).  To
illustrate
this point Figure {\ref{fig:nebcont}} shows an example of the
correction for both nebular 
continuum and the emission line infilling of the higher order Balmer
lines in the case of 
the extended aperture of 3C 321. Not only does the nebular continuum
make a large contribution to the UV continuum below the Balmer break in this
case ($\sim$25\%), but infilling of
the absorption lines is also important: the higher order Balmer lines
(e.g. H9, H10) only 
become visible in absorption following the subtraction of our nebular
model (based on 
H$\beta$). Clearly, failure to take into account
the AGN-related components may lead to the wrong solution in terms of
modelling the YSP.

\subsection{\bf Reddening of the YSP.}
\label{sect:reddening}
\begin{center}
\begin{figure}
\setcounter{figure}{9}
\begin{tabular}{cc}
\centerline{\psfig{file=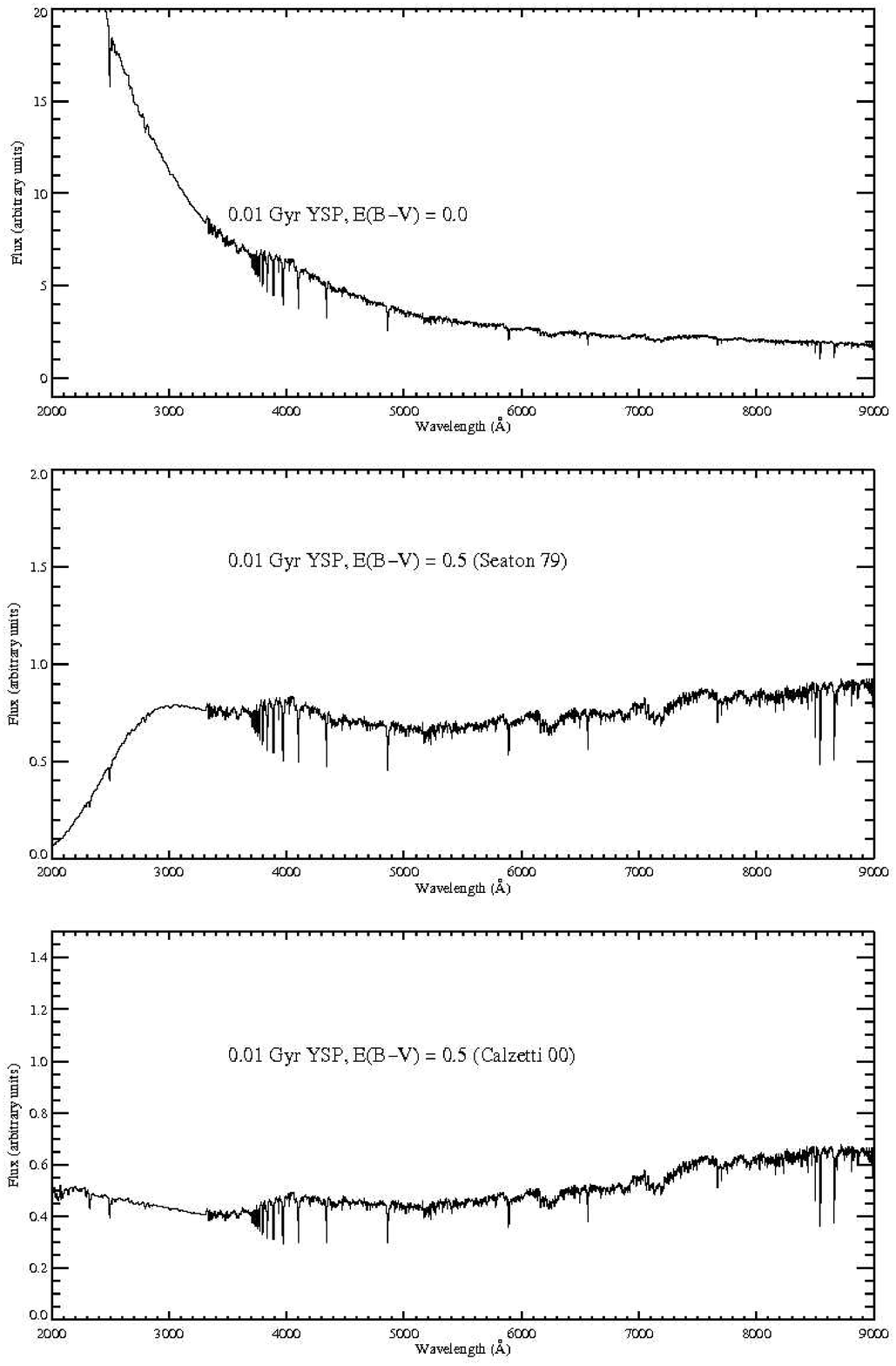,width=9cm,angle=0}}
\end{tabular}
\caption[Extinction law comparison.]{Extinction law comparisons. 
The unreddened 0.01 Gyr YSP from BC03 is plotted (top) and is reddened by
E(B-V) = 0.5 using the {\protect\citet{seaton79}} extinction law
(middle) and the {\protect\citet{calzetti00}} extinction law
(bottom). The extinction laws differ in two respects. First, the
re-normalisation factor, $R$ ($R$ = 3.1 for \protect\citet{seaton79}, $R$
= 4.1 for \protect\citet{calzetti00}) which affects the overall level
of the slope and second, the shape in the region 2000-2500\AA~(the
{\protect\citet{seaton79}} has a prominent `hump' at 2175\AA, a
feature observed in the Galaxy but rarely in extragalactic sources,
see Section \protect\ref{sect:reddening}) which affects the SED slope
in the UV. The \protect\citet{calzetti00} curve also has an overall
steeper slope towards the red. Hence, it is important to also consider the
effect of using a particular reddening law will have on the results
when observing sources into the UV.}
\label{fig:reddening}
\end{figure}
\end{center}

We often find that YSP models with significant reddening are required in order to 
obtain good fits to the spectra of many of the objects/apertures in our
sample. Although the importance of reddening for the YSP has been
discounted in some previous studies (e.g. \citealt{raimann05}),
the detection of significant reddening in the current work is not
surprising given the  
morphological evidence for dusty circumnuclear
environments from HST images of
several nearby radio galaxies (e.g. \citealt{dekoff00,allen02,madrid06}). It is
also notable that, while the reddening is often
most important in the circumnuclear regions, it can also be
significant in the extended 
haloes and disks of the galaxies (e.g. NGC 612). 

Figure {\ref{fig:reddening}} highlights the differences in shape
between the \citet{seaton79} and \citet{calzetti00} reddening laws. At
optical wavelengths, the \citet{calzetti00} curve rises more steeply
towards longer wavelengths than the \citet{seaton79} law. This
difference in shape causes the changes in the contour plots for NGC
612 (Figure {\ref{fig:contours}}) -- a steeper reddening law
(e.g. \citealt{calzetti00})  will
require less reddening in terms of \ebv~than a shallower
curve (e.g. \citealt{seaton79}) to model the SED with the same age
YSP. The difference in the optical 
wavelength range is subtle and so does not pose a major problem for
our results as demonstrated by Figure {\ref{fig:contours}}. However, in
the UV, the curves are markedly different with a prominent hump in the
\citet{seaton79} curve not observed in the \citet{calzetti00} curve. 
Hence, the choice of reddening curve  requires more careful
consideration for spectroscopic studies of higher redshift  
objects that sample the rest-frame UV.

\subsection{\bf Uniqueness of solution.}
\label{sect:uniqueness}
Perhaps the greatest problem faced when modelling the YSP in radio galaxies is the 
degeneracy in the solutions (see also \citealt{tadhunter05}). Despite
the fact that we have adopted the pragmatic approach of modelling the spectra with
the minimum number of components necessary (OSP, YSP, power-law), in some cases
a wide range of YSP age/reddening combinations can give acceptable fits to the SEDs
and absorption features.
Unsurprisingly, the degeneracy problem is at its worst for objects in which the
YSP component makes a small contribution to the optical/UV continuum relative to
the OSP and/or AGN components; the most clear-cut solutions are obtained for the
objects/apertures in which the YSP contributes $>$40\% of the total
flux in the normalising 
bin. While it is possible to remove some of this degeneracy through detailed
examination of model fits to absorption lines that are not affected by emission line
infilling (e.g. CaII~K, G-band, MgIb and, in some case, higher order Balmer lines),
in a number of cases it has proved necessary to use auxilary information (e.g.
UV polarisation, HST point source, power-law slope) to distinguish between
model solutions. Such information is particularly valuable when
distinguishing between 
a very young YSP that is highly reddened, and a scattered or direct AGN component.

It is also important to stress that, because it is not possible to
completely eliminate  
systematic errors in the flux calibration, and the exact choice of bins will also
affect the precise \chisq~values of fits, the best solutions
numerically (in terms of \chisq~or other numerical measures of
goodness of fit), are  
not necessarily the best solutions in terms of the detailed fits to continuum and
absorption lines.

Overall, it is clear that considerable care is required in order
to avoid degeneracies in the YSP
solutions; the greater the number of stellar components included in the fits the 
more serious
the degeneracy problem is likely to become, especially considering
that the different YSP components may be reddened by different
amounts. This is why we prefer to model our data with the minimum number of stellar 
components, rather than assume several components at the outset.

\section{Summary of the main results}

Despite the challenges inherent in studies of this type, we have been
able to extract useful information about the detailed properties of
the YSP in nine out of the twelve radio galaxies in our sample. For the
remaining three galaxies (PKS 0039-44, PKS 1932-46 and 3C 381) we failed to find
strong evidence for a YSP based on 
our spectroscopy, although this does not rule out a contribution from
a YSP at some level. Indeed, because  
of the dominance of the scattered AGN component in PKS 0039-44, and the
direct AGN component in PKS 1932-46 and 3C 381, 
it would be difficult to detect a YSP that contributes as much as
$\sim$20\% of the total light, and a greater proportion of the stellar
light, in such objects. We now summarise the main results for the YSP
in the remaining ten objects for which we have been able to extract
useful information. 

\subsection{Spatial extent of the UV excess}

A striking aspect of this study is that, for all of the objects that
are well-resolved, the UV  
excesses are spatially extended, covering the full measurable extents
of the galaxies. Similar results 
were found for the four galaxies studied by
\citet{tadhunter05} \& \citet{emonts06}. 
 Moreover, in all cases in which we have
sufficiently high S/N to model the spectra of the off-nuclear emission
in detail we find clear evidence for young stellar populations. It is
also notable that for the two double nucleus systems in our  sample,
both of the nuclei in each system show a UV excess. Together, these   
results suggest that where starbursts occur in radio galaxies, they
tend to be galaxy-wide rather 
than confined to the immediate vicinities of the nuclei hosting the
AGN. 

\subsection{Ages of the YSP}

The ages of the YSP have the potential to provide key information for
understanding the order of events which triggered the AGN in  galaxy
mergers (see \citealt{tadhunter05}). With regard to the YSP age
determination we can divide the 
current sample into four categories.
\begin{enumerate} 
\item{\bf Objects with indeterminate YSP ages: 3C 236, PKS 2135-20,
3C 285.} In these cases the contribution of the YSP is
  relatively minor over the wavelength range covered by the
  observations, and, although 
we believe that there is good evidence for YSPs in these cases, it has not
proved possible to determine the YSP ages with any accuracy: all YSP
ages ranging from moderately young ($\sim$0.04 -- 0.1~Gyr) to
intermediate ($\sim$0.5 -- 1.0~Gyr) are possible in these objects. 

\item{\bf Objects with intermediate YSP ages (0.3 -- 3.0~Gyr): 3C 321.}
  In contrast to some recent studies 
(e.g. \citealt{tadhunter05,emonts06,raimann05}) we find unequivocal
evidence for intermediate age YSPs in only one object: 3C 321. In that
case the YSP in the extended apertures (which makes a 
relatively large contribution to the total optical flux: 40-70\%) 
have age $\sim$0.7-1.0~Gyr. Although the model results for the other
apertures in 3C 321 are less clear-cut, fits  to the spectrum of the
main (SE) nucleus are also consistent with an intermediate YSP age. 

\item{\bf Objects with moderately young YSP (0.05 -- 0.2~Gyr): 3C 218,
  NGC 612, 3C 433.} Because of the relatively large proportional contribution
  of the YSP, weak AGN, and clearly detected Balmer absorption lines,
  it 
has proved possible to pin down the YSP ages accurately in 3C 218 and
NGC 612. In both of these cases 
the YSP are moderately young. 

\item{\bf Objects with very young YSP ($<$0.05~Gyr): PKS 0023-26,
PKS 0409-75.} A new finding from this study is the existence
  of a group of powerful radio galaxies with very young, reddened
  YSP. The clearest case from the SED fits is PKS 0023-26, in which we
  detect the higher Balmer lines directly in absorption, but, based on
  consideration of auxiliary data, PKS 0409-75 is also a
  strong candidate for an object containing 
a very young YSP. It is not surprising that the young stellar
populations in these cases are also heavily reddened, since starbursts
are expected to be heavily dust enshrouded in the early stages.  
\end{enumerate}

\subsection{Reddening of the YSP and broad-band colours}
A large fraction ($\sim$70\%) of the objects for which we have good
information on the YSP  show evidence for 
significant reddening in their nuclear regions (\ebv~$\gtrsim$ 0.4),
and in one case -- NGC 612 -- we also find evidence for substantial
reddening in the off-nuclear regions. Since many studies of AGN host
galaxies are based on the broad-band colours, it is interesting to
consider the impact that such reddening has 
on the broad-band continuum colours of the objects in our sample. In
Table \ref{tab:bvcolours} we show the (B-V) colours 
derived from our spectra for the objects in our sample along with the
three objects in Tadhunter et al. (2005). These colours have been
derived by measuring the ratio of the fluxes measured over 500\AA\,
bins at 
the central wavelengths of the B and V filters in the rest frames of
the objects.  For reference, quiescent elliptical galaxies have
$B-V$ $\sim$ 0.9-1.0. 

From Table \ref{tab:bvcolours} it is clear that some galaxies have
E(B-V) colours that 
are significantly bluer than quiescent elliptical galaxies, as
expected in the case of major contributions 
from unreddened YSP and/or scattered AGN components. However, many of
the sources 
have broad-band colours that are similar to, or significant redder
than, quiescent elliptical 
galaxies, despite the clear detection of UV-excesses based on 4000\AA,
break measurements and, in 
some cases, strong evidence for
YSPs from spectral fitting and the detection of Balmer absorption
lines. These results serve to 
demonstrate that considerable caution is required when making
conclusions about the stellar  
populations in AGN host galaxies on the basis of broad-band colours alone.

\begin{table}
\caption[]{$B-V$ colours of the nuclei in the rest frame, after
  correcting for Galactic extinction.  To avoid contamination from the
  emission lines, the colours 
  were measured using the best fitting models to the SED. For
  comparison, we include the $B-V$ colour of a 12.5 Gyr stellar
  population from the \protect\citet{bruzual03} models.}
\begin{center}
\begin{tabular}{l l  l  l l }\hline
Object & $B-V$ & & Object & $B-V$ \\ \hline
3C 218  & 0.67&&PKS 0023-26& 0.93\\
3C 236  & 1.05&&PKS 0039-44& 0.86\\
3C 285  & 0.75&&PKS 0409-75& 0.84\\
3C 321  & 0.71&&PKS 1932-46& 0.88\\
3C 381  & 1.00&&PKS 2135-20& 0.52\\
3C 433  & 0.88&&NGC 612 & 1.03\\\hline
12.5 Gyr & 0.92\\
\hline
\label{tab:bvcolours}
\end{tabular}
\end{center}
\end{table}

\subsection{The mass contribution of the YSP}

Another significant feature of our results is that the YSP typically
make up a significant fraction of 
the total stellar mass in the regions sampled by the spectroscopic
slits ($\sim$1 -- 35\%). Note that this is true even for some of the
objects in which the measured age of the YSP is relatively young. In
such cases the YSP are often heavily reddened, therefore, although the
YSP contribute a relatively small fraction of the total light, they
can represent a significant fraction of the total stellar mass.

\section{Conclusions}

We have modelled deep wide spectral coverage data for a sample of 12 powerful
radio galaxies with previous evidence for young stellar
populations. The results  demonstrate the importance of accounting for
AGN contamination and reddening 
when determining the detailed properties of the stellar
populations. Moreover, in cases where 
the YSP makes a relatively small contribution to the optical/UV
continuum, it has 
proved difficult to determine the precise ages of the YSP. We find
that the YSP with 
well-determined properties cover a wide range of post-starburst ages
(0.03 -- 1.5~Gyr) 
and reddening ($0.0 < E(B-V) < 1.5$), and account for a significant
fraction of the total 
stellar mass in the regions covered by our spectra (1 --- 50\%). The
implications of these results 
for our understanding of the evolutionary status of radio source host 
galaxies, and the heating
mechanism of the warm/cool dust in AGN, will be discussed in
forthcoming papers (Tadhunter et al. 
2007a,b).

\section*{\sc Acknowledgements}
JH, KJI \& JR acknowledge financial support from PPARC;
KAW acknowledges support from the Royal Society. 
The William Herschel Telescope is operated on the
island of La Palma by the Isaac Newton Group in the Spanish
Observatorio del Roque de los Muchachos of the Instituto de
Astrofisica de Canarias. This research has
made use of the NASA/IPAC Extragalactic Database (NED) which is
operated by the Jet Propulsion Laboratory, California Institute of
Technology, under contract with the National Aeronautics and 
Space Administration. Based on observations made with ESO Telescopes
at the  Paranal Observatory under programme ID 71.B-0616(A).

\bibliographystyle{mn2e}
\bibliography{abbrev,refs}

\end{document}